\documentclass[
reprint,
superscriptaddress,
nofootinbib,
amsmath,amssymb,
aps,
]{revtex4-2}

\usepackage{caption}
\usepackage{graphicx}% Include figure files
\usepackage{verbatim}
\usepackage{dcolumn}% Align table columns on decimal point
\usepackage{bm}% bold math
\usepackage{subcaption}
\usepackage{hyperref}% add hypertext capabilities
\usepackage{stackengine}
 \hypersetup{
     colorlinks= true,
     linkcolor = red,
     citecolor = red,      
     }
\usepackage{tikz-feynman,contour}
\usepackage{amsmath}

\newcommand{\bubble}{\rotatebox[origin=c]{-90}{$\between$}}

\newcommand\stackequal[2]{%
  \mathrel{\stackunder[2pt]{\stackon[4pt]{=}{$\scriptscriptstyle#1$}}{%
  $\scriptscriptstyle#2$}}}

\allowdisplaybreaks

\begin{document}

\preprint{APS/123-QED}
\title{The spinning self-force EFT: 1SF waveform recursion relation and Compton scattering}% 

\author{Dogan Akpinar}\email{dogan.akpinar@ed.ac.uk}
\affiliation{Higgs Centre for Theoretical Physics, School of Physics and Astronomy, University of Edinburgh, EH9 3FD, UK}%
\author{Vittorio del Duca}\email{Vittorio.DelDuca@lnf.infn.it} 
\affiliation{INFN, Laboratori Nazionali di Frascati, 00044 Frascati (RM), Italy }%
\author{Riccardo Gonzo}\email{rgonzo@ed.ac.uk} 
\affiliation{Higgs Centre for Theoretical Physics, School of Physics and Astronomy, University of Edinburgh, EH9 3FD, UK}%

\begin{abstract}
Building on recent approaches, we develop an effective field theory for the interaction of spinning particles modeling Kerr black holes within the gravitational self-force expansion. To incorporate dimensional regularization into this framework, we analyze the higher-dimensional metric arising from the minimal coupling solution, comparing it against the Myers-Perry black hole and its particle description. We then derive the 1SF self-force effective action up to quadratic order in the spin expansion, identifying a new type of spinning recoil term that arises from integrating out the heavy dynamics. Next, we study the 1SF metric perturbation both from the traditional self-force perspective and through the diagrammatic background field expansion, making contact with the radiative waveform. This leads us to consider a novel recursion relation for the curved space 1SF Compton amplitude, which we study up to one-loop in the wave regime and compare with the flat space one-loop Compton for Kerr up to quadratic order in spin. Finally, we investigate the 1SF spinning Compton amplitude in the eikonal regime, clarifying how strong-field effect---such as the location of the separatrix---emerge from the resummation of the perturbative weak-field expansion.
\end{abstract}

\maketitle

%----------------------------------------------------------------------
\section{Motivation and introduction}\label{sec:intro}
%----------------------------------------------------------------------
Given the growing sensitivity of the LIGO-Virgo-KAGRA network and the advent of next-generation gravitational wave detectors, there is a pressing call for high-precision theoretical waveform templates. However, numerical simulations of compact binary mergers remain computationally intensive, often obscuring the simplicity of the underlying two-body dynamics.

In an attempt to address this challenge, a host of perturbative approaches have been developed to solve the two-body problem in general relativity (GR). These include the Post-Newtonian (PN) expansion \cite{Blanchet:2002av,Foffa:2013qca}, the Post-Minkowskian (PM) theory \cite{Bern:2019crd,Buonanno:2022pgc,Bjerrum-Bohr:2022blt} and the gravitational self-force (GSF) theory \cite{Barack:2018yvs,Pound:2021qin,Poisson:2011nh}. Recently established PM methods rely on the weak-field expansion and apply only to widely separated bodies, while GSF techniques are applicable in the strong gravity regime but are valid where one body is much smaller than the other. It is therefore essential to combine the results from these methods. On one hand, resumming PM observables in the small-mass-ratio regime may shed light on the analytic structure of GSF \cite{Damour:2022ybd,Barack:2023oqp,Rettegno:2023ghr,Buonanno:2024vkx}, where most results are numerical. On the other hand, incorporating GSF insights into PM computations could extend weak-field methods to the strong-field regime, i.e. beyond their domain of validity \cite{Barack:2022pde,Bini:2024icd,Long:2024ltn}. 

Motivated by this, a new effective field theory approach for massive particles in GR has been proposed recently in the mass ratio expansion both in the amplitude \cite{Kosmopoulos:2023bwc} and worldline \cite{Cheung:2023lnj,Cheung:2024byb} formulations. A crucial ingredient in this story is the identification of the metric generated by a spinless point particle with the Schwarzschild black hole solution \cite{Duff:1973zz,Neill:2013wsa,Bjerrum-Bohr:2018xdl,KoemansCollado:2018hss,Jakobsen:2020ksu,Mougiakakos:2020laz}, a fact that has been recently proved non-perturbatively \cite{Mougiakakos:2024nku,Damgaard:2024fqj}. Interestingly, it was also possible to integrate out the heavy dynamics to obtain a non-local effective action for the light-body dynamics.

Including spin effects is crucial for astrophysical black holes, and treating Kerr black holes as elementary particles \cite{Vaidya:2014kza,Vines:2017hyw,Arkani-Hamed:2017jhn,Chung:2018kqs,Guevara:2018wpp,Guevara:2019fsj,Arkani-Hamed:2019ymq,Maybee:2019jus,Bern:2020buy,Aoude:2021oqj,Alessio:2023kgf} has led to remarkable progress in pushing high-precision PM calculations of spinning binary systems 
\cite{Kosmopoulos:2021zoq,Liu:2021zxr,Chen:2021kxt,Menezes:2022tcs,Bern:2022kto,Bautista:2023szu,Bohnenblust:2024hkw,Chen:2024mmm,Akpinar:2024meg,FebresCordero:2022jts,Jakobsen:2023ndj,Jakobsen:2022fcj,Damgaard:2022jem,Luna:2023uwd,Gonzo:2024zxo,Akpinar:2025bkt,Alaverdian:2025jtw}. An essential ingredient in this endeavor is the construction of the gravitational Compton amplitude for a massive spinning particle interacting with
two gravitons \cite{Bjerrum-Bohr:2016hpa,Chung:2018kqs,Guevara:2018wpp,Johansson:2019dnu,Aoude:2020onz,Falkowski:2020aso,Bautista:2021wfy,Chiodaroli:2021eug,Chen:2022clh,Aoude:2022trd,Bautista:2022wjf,Bjerrum-Bohr:2023jau,Cangemi:2022bew,Cangemi:2023bpe,Scheopner:2023rzp,Bjerrum-Bohr:2023iey,Bautista:2023sdf,Azevedo:2024rrf,Vazquez-Holm:2025ztz}, which describes the propagation of gravitons in a Kerr spacetime. This is analogous to the traditional self-force approach \cite{Galley:2008ih,Detweiler:2002mi}, where the field at 1SF order is determined---besides the geodesic---by the graviton propagator in the background spacetime.

In this paper, we extend the self-force effective field theory to the case of spinning particles, thereby modeling Kerr black holes using the $\mathcal{N}=2$ supersymmetric worldline formulation developed in Refs.~\cite{Gibbons:1993ap,Bastianelli:2005vk,Mogull:2020sak} (see also Refs.~\cite{Bonocore:2024uxk,Haddad:2024ebn}). In section \ref{sec:MP_metric}, we will introduce the higher-dimensional metric arising from the minimal coupling solution of a spinning point particle in $d$ dimensions~\cite{Gambino:2024uge}, comparing it with the Myers-Perry black hole \cite{Myers:2011yc,Myers:1986un,Gambino:2024uge,Bianchi:2024shc}. This will allow us to use dimensional regularization techniques to set up the spinning self-force EFT in section \ref{sec:1SF_action}, which we will obtain at 1SF order. We will then show how to derive new non-local spinning recoil operators by integrating out the heavy dynamics, extending the previous construction in Refs.~\cite{Cheung:2023lnj,Cheung:2024byb}. The consequences of the derived 1SF effective action will then be discussed in section \ref{sec:1SFmetric}, first from the traditional self-force perspective and then from the diagrammatic background field method \cite{DeWitt:1967ub,tHooft:1974toh,Abbott:1980hw,Abbott:1981ke,Boulware:1968zz,Goldberger:2004jt,Porto:2016pyg,Donoghue:2017pgk,Goldberger:2022rqf}. Finally, in Section \ref{sec:1SFCompton}, we compute the 1SF Compton amplitude appearing in the recursion relation, both in the wave and eikonal regimes. We compare it with the flat space Compton amplitude for Kerr at loop level and demonstrate how resumming the weak-field series allows to make a connection with strong-field effects.

%----------------------------------------------------------------------
\textit{Conventions---}Throughout this paper we work in the negative metric signature with $\epsilon_{0123} = 1$, $\kappa^2 = 32 \pi G_{\mathrm{N}}$ and $c = 1$.
We adopt the convention $\hat{\delta}^{(d)}(\cdot) = (2\pi)^d \delta^{(d)}(\cdot)$ and $\hat{\mathrm{d}}^{d}k = \mathrm{d}^d k/(2\pi)^d$. We write index symmetrization and anti-symmetrization as $2x^{(\mu} y^{\nu)} = x^\mu y^\nu + x^\nu y^\mu$ and $2x^{[\mu} y^{\nu]} = x^\mu y^\nu - x^\nu y^\mu$. We denote curved space amplitudes by $\mathcal{M}$ with polarization tensors $\boldsymbol{\varepsilon}_{\mu \nu} \!=\! \boldsymbol{\varepsilon}_\mu \boldsymbol{\varepsilon}_\nu$, while flat space amplitudes are denoted by $\mathcal{A}$ with polarization tensors $\varepsilon_{\mu \nu} \!=\! \varepsilon_\mu \varepsilon_\nu$.
%----------------------------------------------------------------------
\vspace{-9pt}
%----------------------------------------------------------------------
\section{Higher-dimensional rotating metrics from spinning point particles}
\label{sec:MP_metric}
%----------------------------------------------------------------------
The Kerr metric in Kerr-Schild form reads
\begin{align}
\hspace{-6pt}\mathrm{~d}& s^2=  \mathrm{d} t^2-\mathrm{d} x^2-\mathrm{d} y^2-\mathrm{d} z^2 \nonumber \\
&\hspace{-0.7cm} \qquad -\frac{2 G_\mathrm{N} m_\mathrm{H} R(a_\mathrm{H},\vec{x})^3}{R(a_\mathrm{H},\vec{x})^4+a_\mathrm{H}^2 z^2}\Big[\mathrm{~d} t+\frac{R(a_\mathrm{H},\vec{x})(x \mathrm{~d} x+y \mathrm{~d} y)}{a_\mathrm{H}^2+R(a_\mathrm{H},\vec{x})^2} \nonumber \\
&\hspace{-0.7cm} \qquad\qquad\qquad\qquad +\frac{a_\mathrm{H}(y \mathrm{~d} x-x \mathrm{~d} y)}{a_\mathrm{H}^2+R(a_\mathrm{H},\vec{x})^2}+\frac{z \mathrm{~d} z}{R(a_\mathrm{H},\vec{x})} \Big]^2\,,
\label{eq:metricKerr}
\end{align}
where the spin vector $a_\mathrm{H}^{\mu}$ is chosen to be aligned along the $z-$direction, $a_\mathrm{H} =J_{\rm phys}/m_\mathrm{H}$ is the (normalized) spin parameter and $R(a_\mathrm{H},\vec{x})$ is determined by the constraint
\begin{align}
x^2+y^2+z^2=R(a_\mathrm{H},\vec{x})^2+a_\mathrm{H}^2\left[1-\frac{z^2}{R(a_\mathrm{H},\vec{x})^2}\right] \,.
\label{eq:R_ellipsoid}
\end{align}
Unlike the spinless case, there is no proof that this metric is generated by a spinning point particle of mass $m_{\mathrm{H}}$ and spin $a_\mathrm{H}$ beyond the leading order in $G_\mathrm{N}$~\cite{Vines:2017hyw}, but we will explicitly check the consistency of our calculations at quadratic order in the spin expansion. In a covariant way, we can define a basis of 4 vectors in which the Kerr metric is naturally expanded: the four-velocity $v_\mathrm{H}^{\mu}$, the spacetime coordinate $x^{\mu}$, the spin vector $a_\mathrm{H}^{\mu}$ and a convenient Levi-Civita dependent vector $L_{\mu}$ 
\begin{align}
    L_{\mu} = \epsilon_{\mu \nu \alpha \beta} v_\mathrm{H}^{\nu} a_\mathrm{H}^{\alpha} r^{\beta} \,,
    \label{eq:Lvector}
\end{align}
where $r^{\mu}$ is obtained through the action of a spatial projector $P^{\mu \nu}$ on the spacetime coordinate $x^{\mu}$
\begin{align}
\label{eq:projector_spatial}
   \hspace{-8pt} r^{\mu} = P^{\mu}_{\,\,\,\nu} x^{\nu} = x^{\mu} - (x \cdot v_\mathrm{H}) v_\mathrm{H}^{\mu}\,, \,\,\, P^{\mu \nu} = \eta^{\mu \nu} - v_\mathrm{H}^{\mu} v_\mathrm{H}^{\nu}\,.
\end{align}
In this basis, after transforming \eqref{eq:metricKerr} into spherical coordinates $(t,r,\theta,\phi)$, we can write the Kerr metric as
\begin{align}
\label{eq:Kerr_covariant}
& \bar{g}_{\mu \nu} = \eta_{\mu \nu} -\frac{2 G_\mathrm{N} m_\mathrm{H} R^3 l_{\mu} l_{\nu}}{\left(R^2+a_\mathrm{H}^2\right)^2 \left((r \cdot a_\mathrm{H})^2+R^4\right)}\,, \\
l_{\mu} &=v_{\mathrm{H} \mu}\, (R^2+a_\mathrm{H}^2) -r_{\mu}\, R- L_{\mu}+a_{\mathrm{H}\mu}\,(r \cdot a_\mathrm{H})/R\,,
\end{align}
where the $R$ in \eqref{eq:R_ellipsoid} is a function of $r=|\vec{x}|$, $a_\mathrm{H}$ and $x \cdot a_\mathrm{H}$
\begin{align}
\hspace{-10pt}R(a_\mathrm{H},\vec{x}) \!=\! \frac{1}{\sqrt{2}} \sqrt{r^2-a_\mathrm{H}^2 +\sqrt{4 (r \cdot a_\mathrm{H})^2+\left(r^2 - a_\mathrm{H}^2\right)^2}} \,.
\end{align}
The Kerr-Schild gauge \eqref{eq:Kerr_covariant} of the metric is particularly convenient for the PM expansion because it manifestly reduces to the flat one in the limit $G_\mathrm{N} \to 0$, while other sets of coordinates (such as Boyer-Lindquist) usually retain their spurious spin dependence in such a limit.

In order to harness effective field theory tools in the mass ratio expansion, we need to employ dimensional regularization. This necessarily requires one to define a rotating metric which generalizes the Kerr one to higher dimensions. In general, there are various spinning black hole solutions in $d>4$ dimensions \cite{Emparan:2008eg} and they are not all smoothly connected to Kerr \cite{Gambino:2024uge,Bianchi:2024shc}. A natural choice is the $d=2+2n$ Myers-Perry metric of Ref.~\cite{Myers:1986un,Myers:2011yc}, which depends on $n$ different spinning parameters $a_{\rm H,i}$ for each of the $n$ spatial planes. Introducing the physical mass $M_{\rm phys}$ and angular momentum components $J_{\rm phys,i}$
\begin{align}
    M_{\rm phys}&=\frac{(d-2) \Omega_{d-2}}{8 \pi} m_{\rm H}\,, \nonumber \\
    J_{\rm phys,i}&=\frac{2}{d-2} M_{\rm phys} a_{\rm H,i}\,,\quad i=1,\dots n
\end{align}
where $\Omega_{d-2} =2 \pi ^{\frac{d-1}{2}}/ \Gamma \left((d-1)/2\right)$ is the angular measure, it is natural to expect that the simply rotating case with $J_{\rm phys}=J_{\rm phys,1}$ and $J_{\rm phys,i} = 0$ for $i=2,\dots n$ is related to Kerr in the $d=4$ limit. To establish a covariant form of the Myers-Perry metric, it is useful to introduce a $d$ dimensional spin vector $a^{d}_{\mathrm{H}\mu}$ such that the following spin tensor decomposition holds
\begin{align}
\label{eq:spin_tensor}
S_{\mu \nu} &= m_\mathrm{H} \epsilon^{(1)}_{\mu \nu \alpha \beta} v_\mathrm{H}^{\alpha} (a^d_\mathrm{H})^{\beta}\,, \\
\epsilon^{(1)}_{\mu \nu \alpha \beta} &= \ (-1)^{n+1}\epsilon_{\mu \nu \alpha \beta \widehat{i_1 j_1} \dots i_k j_k \dots i_n j_n} \nonumber \\
        & \qquad \qquad \times \widehat{e^{i_1} e^{j_1}} \cdots e^{i_k} e^{j_k} \cdots e^{i_n} e^{j_n}\,, \nonumber
\end{align}
where $(e_{i_k},e_{j_k})$ are the unit vectors in the $k$-plane and the hat notation indicates that the corresponding indices and vectors are missing. Correspondingly, we also define the $n=1$ rotation vector in $d=2+2n$ dimensions
\begin{align}
L^{(1)}_{\mu} &= \epsilon^{(1)}_{\mu \nu \alpha \beta} v_\mathrm{H}^{\nu} (a^d_\mathrm{H})^{\alpha} r^{\beta}\,,
\label{eq:angular}
\end{align}
which agrees with \eqref{eq:Lvector} for a single plane in $d=4$. Moreover, we can define the transverse vector 
\begin{align}
\hspace{-8pt}\rho^{\perp}_{\mu} = r_{\mu} + (r \cdot e_{i_1}) e_{i_1\mu} + (r \cdot e_{j_1}) e_{j_1\mu}  + \frac{(r \cdot a^d_\mathrm{H})}{(a^d_\mathrm{H})^2} a^d_{\mathrm{H}\mu} \,,
\label{eq:transverse_rho}
\end{align}
which vanishes identically as $d \to 4$, provided that the physical 4-dimensional subspace is formed by the vectors $(v_\mathrm{H \mu},e_{i_1 \mu},e_{j_1 \mu},a^d_{\mathrm{H}\mu})$. Following appendix E of Ref.~\cite{Frolov:2017kze}, we define the simply rotating Myers-Perry metric 
\begin{align}
\label{eq:Myers-Perry}
&\hspace{-8pt}\bar{g}^{\rm MP}_{\mu \nu} = \eta_{\mu \nu}- \frac{2 G_\mathrm{N} m_\mathrm{H} \,R_d^{7-d} \, l^{(d)}_{\mu} l^{(d)}_{\nu}}{\left(R_d^2+(a^d_\mathrm{H})^2\right)^{2} \left((r \cdot a^d_\mathrm{H})^2+R_d^4\right) } \,, \\
&\hspace{-8pt}l^{(d)}_{\mu} =v_{\mathrm{H} \mu}\, (R_d^2+(a^d_\mathrm{H})^2) -r_{\mu}\, R_d - L^{(1)}_{\mu} \nonumber \\
&\qquad\qquad\quad +\frac{1}{R_d} \left((r \cdot a^d_\mathrm{H}) a^d_{\mathrm{H}\mu} - (a^d_\mathrm{H})^2 \rho^{\perp}_{\mu}\right)\,, \! \nonumber\\
&r^2 =R_d(a^d_\mathrm{H},\vec{x})^2+(a^d_\mathrm{H})^2-\frac{(r \cdot a^d_\mathrm{H})^2+(a^d_\mathrm{H})^2 (\rho^{\perp})^2}{R_d(a^d_\mathrm{H},\vec{x})^2}\,, \nonumber 
\end{align} 
where we have introduced the coordinate system
\begin{align}
x_k &= r_k \sin (\theta) \cos (\phi_k), \, y_k = r_k \sin (\theta) \sin (\phi_k), \,\, k=1,\dots, n\nonumber \\
&\qquad \qquad z= r \cos (\theta) \,, \quad r^2=|\vec{x}|^2=\sum_{k=1}^n r^2_k \,.
\label{eq:coord_dim}
\end{align}
Notice that the $d$-dimensional metric \eqref{eq:Myers-Perry} identically reduces to Kerr in $d \to 4$, as claimed by Refs.~\cite{Myers:1986un,Emparan:2008eg}. 

Although it remains an open question to show how the metric \eqref{eq:Myers-Perry} is related to the metric produced by a spinning point particle in $d$ dimensions at all PM orders, it has been shown \cite{Gambino:2024uge,Bianchi:2024shc} that the Myers-Perry solution---unlike the Kerr black hole one---corresponds to a particular choice of the dimension dependent non-minimal couplings $(C_1^{\rm MP}(d),C_2^{\rm MP}(d))$ from the effective field theory perspective\footnote{The definition of these couplings in the spinning worldline theory will be clarified in Section \ref{sec:1SF_action}.}. Interestingly, the simplest higher-dimensional rotating metric arising from the minimal coupling solution $(C^{\rm MC}_1=1,C^{\rm MC}_2=0)$ is instead only known perturbatively \cite{Gambino:2024uge}, and little is understood about its full non-linear structure. For the purpose of this work, we only need its structure up to one-loop and quadratic in spin order, which in harmonic gauge is
\begin{align}
\label{eq:MC_metric}
& \hspace{1cm} \bar{g}^{\rm MC}_{\mu \nu} = \eta_{\mu \nu}  +\sum_{i = 1}^{9} h^{(i)}_{\rm MC}(m_{\mathrm{H}},a^d_{\mathrm{H}},r)\, c^{(i)}_{\mu \nu}\,,  \\
&\vec{c}_{\mu \nu} = \big\{v_{\mathrm{H}\mu} v_{\mathrm{H}\nu},\, v_{\mathrm{H}(\mu} L^{(1)}_{\nu)},\, L^{(1)}_\mu L^{(1)}_\nu,\, e_{i_1 \mu} e_{i_1 \nu},\, e_{j_1 \mu} e_{j_1 \nu}, \nonumber \\
&\qquad\qquad\qquad\qquad\ r_{\mu} r_{\nu},\,  e_{i_1 (\mu} r_{\nu)},\, e_{j_1 (\mu} r_{\nu)},\,
\eta_{\mu \nu} \big\}\,, \nonumber
\end{align} 
where the explicit coefficients are listed in appendix \ref{sec:app2}.

%----------------------------------------------------------------------
\section{The spinning self-force EFT}\label{sec:1SF_action}
%----------------------------------------------------------------------
Here we use the formalism developed in Refs.~\cite{Cheung:2023lnj,Cheung:2024byb} (see also Ref.~\cite{Kosmopoulos:2023bwc}) to describe the classical two-body dynamics in the self-force expansion
\begin{align}
    \mathcal{S}=\mathcal{S}_{\mathrm{GR}} + \mathcal{S}_{\mathrm{L}} + \mathcal{S}_{\mathrm{H}} \,,
    \label{eq:self-force_action}
\end{align}
where $\mathcal{S}_{\mathrm{GR}}$ is the gravitational (gauge-fixed) Einstein-Hilbert action and $\mathcal{S}_{\mathrm{L}}$ (resp. $\mathcal{S}_{\mathrm{H}}$) stands for the action describing light (resp. heavy) matter degrees of freedom. The gravitational action is defined as
\begin{align}
\mathcal{S}_{\mathrm{GR}}\! &=\!-\frac{1}{16 \pi G_\mathrm{N}} \int \mathrm{d}^4 x \sqrt{-g} R+\frac{\lambda^2}{32 \pi G_\mathrm{N}}\int \mathrm{d}^4 x \sqrt{-\bar{g}} F^\mu F_\mu\,,  \nonumber  \\
&  F_\mu =\overline{\nabla}^\nu \delta \boldsymbol{g}_{\mu \nu}-\frac{1}{2} \overline{\nabla}_\mu \delta \boldsymbol{g}\,, \quad  \delta \boldsymbol{g} = \bar{g}^{\mu \nu} \delta \boldsymbol{g}_{\mu \nu} \,,
\label{eq:GR-action}
\end{align}
where we have chosen the background field gauge for the perturbation $\lambda \delta \boldsymbol{g}_{\mu \nu} = g_{\mu \nu} - \bar{g}_{\mu \nu}$ with $\bar{g}_{\mu \nu}$ being the background metric of a Schwarzschild or a Kerr black hole. In what follows we will first review the self-force EFT for spinless bodies, and then extend the discussion to that of spinning bodies.
\subsection*{Review of the spinless self-force EFT}
Let us start by reviewing the spinless case discussed in Refs.~\cite{Cheung:2023lnj,Cheung:2024byb}. The matter action is defined as
\begin{align}
    \hspace{-6pt} \mathcal{S}_{\mathrm{m}}  =- \sum_{i = \mathrm{L},\mathrm{H}} \mathcal{S}_i =- \sum_{i = \mathrm{L},\mathrm{H}} m_i \int \mathrm{d}\tau \, \frac{1}{2} g_{\mu \nu} \dot{x}_i^\mu \dot{x}_i^\nu \,,
    \label{eq:matter-action}
\end{align}
where $\tau$ is the proper time, $m_i$ is the mass and $x_i^\mu(\tau)$ is the trajectory of the $i$-th particle obeying the on-shell constraint $\dot{x}_i^2 = 1$. As usual, the equation of motion is simply the geodesic equation 
\begin{equation}
    \ddot{x}_i^\mu + \Gamma^{\mu}_{~\rho \sigma}(x_i)\dot{x}_i^\rho\dot{x}_i^\sigma = 0\,,
    \label{eq:GeodesicEqn}
\end{equation}
where $\Gamma^{\mu}_{~\rho \sigma}$ is the Christoffel connection
\begin{equation}
    \Gamma^{\mu}_{~\rho \sigma} = \frac{1}{2}g^{\mu \alpha}(\partial_\rho g_{\alpha \sigma} + \partial_\sigma g_{\rho \alpha} - \partial_\alpha g_{\rho \sigma})\,.
\end{equation}
The goal is then to perform the GSF expansion, which is simply an expansion in the small mass ratio $\lambda = m_\mathrm{L}/m_\mathrm{H}$. At leading order in $\lambda$, we evaluate the action \eqref{eq:self-force_action} on the Schwarzschild-Tangherlini metric $\bar{g}_{\mu \nu}$ (i.e. taking the limit $a^d_\mathrm{H} \to 0$ in \eqref{eq:Myers-Perry}) and enforce that it is sourced by the heavy particle stress tensor supported on the straight-line trajectory $\bar{x}_\mathrm{H}^{\mu}(\tau) = v_\mathrm{H}^{\mu} \tau$
\begin{equation}
    \begin{split}
        &\qquad \bar{R}_{\mu \nu}-\frac{1}{2} \bar{g}_{\mu \nu} \bar{R}=8 \pi G_\mathrm{N} \bar{T}_{\mathrm{H} \mu \nu}\,, \\
        & \bar{T}^{\mu \nu}_{\mathrm{H}}  = \frac{ m_\mathrm{H}}{\sqrt{-\bar{g}}} \int \mathrm{d}\tau \delta^{(4)}\left(x-v_\mathrm{H} \tau\right) v_\mathrm{H}^\mu v_\mathrm{H}^\nu \,,
    \end{split}
    \label{eq:heavy_source_eom}
\end{equation}
so that the light particle moves in the geodesic trajectory 
\begin{equation}
    \ddot{\bar x}_{\mathrm{L}}^\mu + \bar{\Gamma}^{\mu}_{~\rho \sigma}(\bar x_\mathrm{L})\dot{\bar x}_{\mathrm{L}}^\rho\dot{\bar x}_\mathrm{L}^\sigma = 0\,.
\end{equation}
Already at this order, we notice that dimensional regularization is extremely helpful to handle divergent terms involving the metric evaluated on the heavy trajectory $\bar{g}_{\mu \nu}(\bar{x}_\mathrm{H})$; these self-energy contributions can be absorbed with a mass counterterm, effectively setting \cite{Cheung:2023lnj,Cheung:2024byb}
\begin{equation}
   \bar{g}_{\mu \nu}(\bar{x}_\mathrm{H}) \to \eta_{\mu \nu}\,, \quad \bar{\Gamma} ^{\mu}_{~\rho \sigma}(\bar x_\mathrm{H}), \bar{R}_{\mu \nu \alpha \beta}(\bar{x}_\mathrm{H})  \to 0 \,.
   \label{eq:dim-reg_xH}
\end{equation}

At the first self-force (1SF) order, we perturb the metric and worldline trajectories by corrections \cite{Cheung:2023lnj,Cheung:2024byb}
\begin{align}
    x_i^\mu(\tau) &= \bar{x}_i^\mu(\tau) + \lambda \,\delta x_i^\mu(\tau)\,, \nonumber \\
    g_{\mu \nu}(x) &= \bar{g}_{\mu \nu}(x) + \lambda \,\delta \boldsymbol{g}_{\mu \nu}(x)\,,
    \label{eq:SF_exp1}
\end{align}
where $\lambda$ is made explicit in the corrections. Using \eqref{eq:SF_exp1}, we further expand the metric around the trajectory
\begin{align}
    &g_{\mu \nu}(x_i) = g_{\mu \nu}(\bar{x}_i) + \lambda \delta x_i^\alpha \partial_\alpha g_{\mu \nu}(\bar{x}_i) \nonumber \\
    & \qquad \qquad + \frac{\lambda^2}{2} \delta x_i^\alpha \delta x_i^\beta \partial_\alpha \partial_\beta g_{\mu \nu}(\bar{x}_i) + \mathcal{O}(\lambda^3)\,.
    \label{eq:SF_exp2}
\end{align}
Substituting \eqref{eq:SF_exp1} and \eqref{eq:SF_exp2} into the combined matter \eqref{eq:matter-action} and graviton \eqref{eq:GR-action} action  (see Ref.~\cite{Cheung:2024byb} and appendix \ref{sec:app1} for more details) we recover the 1SF contribution 
\begin{align}
\label{eq:1SF_action}
    \mathcal{S}^{\mathrm{1SF}} &= \mathcal{S}^{\mathrm{1SF}}_{\mathrm{GR}} + \mathcal{S}^{\mathrm{1SF}}_{\mathrm{L}} +\mathcal{S}^{\mathrm{1SF}}_{\mathrm{H}}, \\
    \mathcal{S}^{\mathrm{1SF}}_{\mathrm{GR}} &= \frac{-\lambda^2}{32 \pi G_\mathrm{N}} \int \mathrm{d}^4 x \, \sqrt{-\bar{g}} \nonumber \\
    & \hspace{-10pt}\times \Big\{ \left(\frac{1}{2} \bar{g}^{\mu \nu} \bar{g}^{\alpha \gamma} \bar{g}^{\beta \delta} -\frac{1}{4} \bar{g}^{\mu \nu} \bar{g}^{\alpha \beta} \bar{g}^{\gamma \delta} \right)  \overline{\nabla}_\mu \delta \boldsymbol{g}_{\alpha \beta} \overline{\nabla}_\nu \delta \boldsymbol{g}_{\gamma \delta} \nonumber \\
    &\hspace{-10pt} \quad + \Big[\frac{1}{4} \bar{R} (2 \bar{g}^{\xi \zeta} \bar{g}^{\xi' \zeta'} - \bar{g}^{\xi \xi'} \bar{g}^{\zeta \zeta'} ) \nonumber \\
    &\hspace{-10pt} \quad+ \bar{R}_{\rho \gamma} (\bar{g}^{\rho \zeta} \bar{g}^{\gamma \zeta'} \bar{g}^{\xi \xi'} - \bar{g}^{\rho \xi} \bar{g}^{\gamma \zeta} \bar{g}^{\xi' \zeta'} )\nonumber \\
    &\hspace{-10pt} \qquad\qquad - \bar{R}_{\rho \gamma \sigma \lambda} \bar{g}^{\rho \zeta} \bar{g}^{\sigma \zeta'} \bar{g}^{\gamma \xi} \bar{g}^{\lambda \xi'}\Big] \delta \boldsymbol{g}_{\zeta \zeta'} \delta \boldsymbol{g}_{\xi \xi'} \Big\}\,, \nonumber \\
    \mathcal{S}^{\mathrm{1SF}}_{\mathrm{L}} &= -m_{\mathrm{H}} \lambda^2 \int \mathrm{d}\tau \frac{1}{2} \dot{\bar{x}}_\mathrm{L}^\mu \dot{\bar{x}}_\mathrm{L}^\nu \delta \boldsymbol{g}_{\mu \nu}\left(\bar{x}_\mathrm{L}\right), \nonumber \\
    \mathcal{S}^{\mathrm{1SF}}_{\mathrm{H}} &= -m_\mathrm{H} \lambda^2 \int \mathrm{d}\tau \Big\{\frac{1}{2} \delta \dot{x}_\mathrm{H}^2 -\delta x_\mathrm{H}^\rho \dot{\bar{x}}_\mathrm{H}^\mu \dot{\bar{x}}_\mathrm{H}^\nu \delta \Gamma_{\rho \mu \nu}\left(\bar{x}_\mathrm{H}\right)\Big\},   \nonumber 
\end{align}
where we have used the equations of motion at 0SF order, dropped the non-dynamical contributions and defined the variation in the connection
\begin{align}
\delta \Gamma^{\rho}_{~\mu \nu} & = \Gamma^{\rho}_{~\mu \nu} - \bar{\Gamma}^{\rho}_{~\mu \nu} \nonumber\\
    & = \frac{1}{2}\bar{g}^{\rho \alpha}(\overline{\nabla}_\mu \delta \boldsymbol{g}_{\nu \alpha}+ \overline{\nabla}_\nu \delta \boldsymbol{g}_{\mu \alpha} - \overline{\nabla}_\alpha \delta \boldsymbol{g}_{\mu \nu}) \,.
\end{align}
It is worth noticing that the distributional nature of the source (i.e. the heavy point particle) does not allow one to set $\bar{R}_{\mu \nu} $ and $\bar{R}$ to zero~\cite{Kosmopoulos:2023bwc,Cheung:2023lnj,Cheung:2024byb}. Instead, one should consider the Ricci tensor and scalar in the sense of distributions in order to have a consistent theory of interacting point particles~\cite{Geroch:1986jjl,Balasin:1993fn}, where their value is fixed by \eqref{eq:heavy_source_eom} at the 1SF order\footnote{Taking the trace of Einstein's equation, we obtain the equation $R = -8 \pi G_\mathrm{N} g_{\mu \nu} T^{\mu \nu}$, which immediately fixes both $R$ and $R^{\alpha \beta}$ in terms of the stress tensor.}. At this point it is important to note that we can formally integrate out the $\delta x_\mathrm{H}$ perturbation 
\begin{align}
\label{eq:deltax_EOM}
\delta \ddot{x}^{\mu}_\mathrm{H}  + \delta \Gamma^{\mu}_{\,\alpha \beta}(\bar{x}_\mathrm{H}) \dot{\bar{x}}^{\alpha}_\mathrm{H} \dot{\bar{x}}^{\beta}_\mathrm{H} = 0 \,, \nonumber 
\end{align}
to obtain ($1/\partial_{\tau}$ stands for the integration over $\tau$)
\begin{align}
    \delta x_{\mathrm{H} \mu} = -\frac{1}{\partial_{\tau}^2} \Big[ \delta \Gamma_{\mu \alpha \beta}(\bar{x}_\mathrm{H}) \dot{\bar{x}}^{\alpha}_\mathrm{H} \dot{\bar{x}}^{\beta}_\mathrm{H} \Big] \,. 
\end{align}
Therefore, at the level of the path integral, we obtain an effective 1SF self-force action written only in terms of the graviton and the light particle dynamics 
\begin{align}
\label{eq:path-integral}
&\int \mathcal{D} [g] \,\mathcal{D} [x_\mathrm{L}]\, \mathcal{D} [x_\mathrm{H}] e^{i \mathcal{S}[x_\mathrm{L},x_\mathrm{H},g]/\hbar} \Big|_{\mathrm{1SF}} \\
& =  \int \mathcal{D} [\delta \boldsymbol{g}]\, \mathcal{D} [\delta x_\mathrm{L}]\, \mathcal{D} [\delta x_\mathrm{H}] \, e^{i (\mathcal{S}^{\mathrm{1SF}}_{\mathrm{GR}}+\mathcal{S}^{\mathrm{1SF}}_\mathrm{L}+\mathcal{S}^{\mathrm{1SF}}_\mathrm{H})[\delta x_\mathrm{H},\delta x_\mathrm{L},\delta \boldsymbol{g}] /\hbar}  \nonumber \\
&  =  \int \mathcal{D} [\delta \boldsymbol{g}]\,  e^{i (\mathcal{S}^{\mathrm{1SF}}_{\mathrm{GR}}+\mathcal{S}^{\mathrm{1SF}}_\mathrm{L}+\mathcal{S}^{\mathrm{1SF}}_{\mathrm{recoil}})[\delta \boldsymbol{g}] /\hbar}\,, \nonumber
\end{align}
where we introduced a non-local recoil operator \cite{Cheung:2023lnj,Cheung:2024byb}
\begin{align}
\label{eq:recoil-spinless}
&\mathcal{S}^{\mathrm{1SF}}_{\mathrm{recoil}} = -i \hbar \log\left[\int \mathcal{D} [\delta x_\mathrm{H}]  e^{i \mathcal{S}^{\mathrm{1SF}}_\mathrm{H}[\delta x_\mathrm{L},\delta x_\mathrm{H},\delta \boldsymbol{g}] /\hbar} \right]   \\ 
&=  -\frac{m_\mathrm{H}}{2}\int \mathrm{d} \tau\,\left[\dot{\bar{x}}_\mathrm{H}^\alpha\dot{\bar{x}}_\mathrm{H}^\beta \delta\Gamma^{\mu}_{~\alpha \beta}(\bar{x}_\mathrm{H})\right] \frac{1}{\partial_\tau^2}\left[\dot{\bar{x}}_\mathrm{H}^\gamma \dot{\bar{x}}_\mathrm{H}^\delta\delta\Gamma_{\mu \gamma \delta}(\bar{x}_\mathrm{H}) \right]\,. \nonumber
\end{align}
In this way, the 1SF effective action can be written as
\begin{align}
\label{eq:1SF_action2}
    \mathcal{S}^{\mathrm{1SF}}_{\mathrm{eff}}[\delta \boldsymbol{g};\{\bar{x}_\mathrm{L}\}] &\equiv \mathcal{S}^{\mathrm{1SF}}_{\mathrm{GR}} + \mathcal{S}^{\mathrm{1SF}}_{\mathrm{L}} +\mathcal{S}^{\mathrm{1SF}}_{\mathrm{recoil}} \,,
\end{align}
which is a function of a single dynamical field $\delta \boldsymbol{g}^{\mu \nu}$ but implicitly depends on the background trajectory $\bar{x}_\mathrm{L}^{\mu}$.
\subsection*{The spinning self-force EFT}
Having reviewed and understood the self-force expansion for the spinless case, we are now ready to extend this EFT to the spinning case. In doing so, we will restrict our analysis up to quadratic order in spin, taking advantage of the $\mathcal{N}=2$ SUSY worldline model  \cite{Gibbons:1993ap,Bastianelli:2005vk,Jakobsen:2021zvh}; higher spin orders can be included systematically, for example following Refs.~\cite{Levi:2015msa,Saketh:2022wap,Ben-Shahar:2023djm,Haddad:2024ebn}. The matter action reads
\begin{align}
\label{eq:spinning-matter-action}
    &\hspace{-8pt} \mathcal{S}^{\mathrm{spin}}_{\mathrm{m}} =- \sum_{i = \mathrm{L},\mathrm{H}} m_i \int \mathrm{d}\tau \Big\{\frac{1}{2} g_{\mu \nu} \dot{x}_i^\mu \dot{x}_i^\nu+i m_i \bar{\psi}_{i, a} \frac{D \psi_i^a}{\mathrm{d}\tau}  \\
    & \qquad + C_1 \frac{m_i^2}{2} R_{a b c d} \bar{\psi}_i^a \psi_i^b \bar{\psi}_i^c \psi_i^d \nonumber \\ 
    & \qquad + C_2 \frac{m_i^2}{2} R_{a \mu b \nu} \dot{x}_i^\mu \dot{x}_i^\nu \bar{\psi}_i^a \psi_i^b P_{c d} \bar{\psi}_i^c \psi_i^d + \mathcal{O}(S_i^3)\Big\}\, ,\nonumber
\end{align}
where Latin indices denote a locally flat spacetime, $\bar{\psi}_i$ and $\psi_i$ are complex Grassmann variables, $P_{a b} = \eta_{a b}-e_{a \mu} e_{b \nu} (\dot{x}^\mu \dot{x}^\nu)/\dot{x}^2$ is a convenient projector and 
\begin{align}
    \frac{D \psi_i^a}{\mathrm{d}\tau} = \dot{\psi}_i^a + \dot{x}_i^\mu \omega_{\mu}^{~ab}(x_i) \psi_{i,b}\,,
\end{align}
with the spin connection $\omega_{\mu}^{~ab}$. Note that we embed fields defined in the local spacetime into the global spacetime using the vielbein $e^{\mu}_{~a}(x)$, e.g. $\psi_i^\mu(x) = e^{\mu}_{~a}(x) \psi_i^a$. In this way, the vielbein and the spin tensor are defined as
\begin{align}
\label{eq:spin_massscaling}
    g_{\mu \nu}=e_\mu^a e_\nu^b \eta_{a b} \,, \quad S_i^{\mu \nu}=-2 i m_i e_a^\mu e_b^\nu \bar{\psi}_i^{[a} \psi_i^{b]}\,,
\end{align}
and the spin connection reads
\begin{equation}
    \omega_{\mu}^{~ab}  = - \omega_{\mu}^{~ba} = e^{a}_{~\nu}(\partial_\mu e^{\nu b} + \Gamma^{\nu}_{~\mu \lambda}e^{\lambda b})\,.
\end{equation}
In general, the spinning matter action~\eqref{eq:spinning-matter-action} admits two independent couplings $(C_1,C_2)$ at quadrupolar order. For a Kerr black hole in $d=4$, only the minimal coupling solution with $C_1=1$ and $C_2=0$ is relevant. The coefficient $C_2$ therefore encodes spin-induced finite-size corrections beyond the Kerr case~\cite{Thorne:1980ru,Porto:2008jj,Vines:2016unv,Jakobsen:2021zvh}. 

The Myers-Perry metric~\eqref{eq:Myers-Perry} and the minimally coupled one~\eqref{eq:MC_metric} represent distinct higher-dimensional extensions of the Kerr solution, and thus correspond to different choices of the Wilson coefficients $ (C_1, C_2) $ at quadratic in spin order in the effective action~\cite{Gambino:2024uge}. As the name suggests, the minimal coupling metric is associated with the universal choice
\begin{align}
\bar{g}^{\rm MC}_{\mu \nu} \leftrightarrow (C_1^{\rm MC} = 1,\; C_2^{\rm MC} = 0)\,,
\end{align}
independently of the spacetime dimension $d$. In contrast, the Myers-Perry solution corresponds to non-minimal, dimension-dependent values (see appendix \ref{sec:app5})
\begin{align}
\label{eq:nonminimal_MP}
\hspace{-7pt}\bar{g}^{\rm MP}_{\mu \nu} \leftrightarrow \left(C_1^{\rm MP}(d)\!=\!\frac{d-2}{2 (d-1)},\; C_2^{\rm MP}(d)\!=\!\frac{ d(d-2)}{4 (d-1)}\right)\,.
\end{align}
Although the Myers-Perry metric identically reduces to the Kerr solution in $d = 4$ (because $C_1^{\rm MP} + C_2^{\rm MP} \to 1$ as $d \to 4$), the presence of additional, independent stress-energy multipoles in $d > 4$ ~\cite{Gambino:2024uge,Bianchi:2024shc}---beyond the usual mass and current multipoles---formally requires both couplings when using dimensional regularization.

In our application of worldline EFT tools to the gravitational self-force problem, we therefore choose to work with the simplest minimally coupled rotating metric~\eqref{eq:MC_metric} and set $C_1 = 1$ and  $C_2 = 0$ in \eqref{eq:spinning-matter-action}. It is then straightforward to derive the equations of motion by treating $\bar{\psi}_i$ and $\psi_i$ as independent variables
\begin{align}
     &\dot{\psi}^a_i + \dot{x}_i^\mu \omega_{\mu}^{~ab}(x_i) \psi_{i,b} -i m_i R_{abcd}(x_i)\psi_i^b\bar{\psi}_i^c \psi_i^d = 0\,, \label{eq:psiEOM}\\
     &\ddot{x}_i^\mu + \Gamma^{\mu}_{~\alpha \beta}(x_i)  \dot{x}_i^\alpha \dot{x}_i^\beta - \frac{1}{2}  R_{\alpha \beta~\nu}^{~~~\mu}(x_i)  \dot{x}_i^\nu S_i^{\alpha \beta} \label{eq:xEOM}\\
     & \hspace{8mm} \qquad  \qquad \qquad + \frac{1}{8} \nabla_{x_i}^{\mu}  R_{\alpha \beta \rho \lambda}(x_i) S_i^{\alpha \beta}  S_i^{\rho \lambda} = 0\,,  \nonumber  
\end{align}
where the equation of motion for $\bar{\psi}_i$ is given by the conjugate of \eqref{eq:psiEOM}. An important consequence of the mass dependence in the spin tensor \eqref{eq:spin_massscaling} is that we expect every power of the spin tensor on the light body (resp. heavy body) to be suppressed (resp. enhanced) compared to the spinless term. Thus, we consider an explicit spin expansion of $\bar{x}_\mathrm{L}^\mu$ and $\psi^a_\mathrm{L}$ through quadratic in spin
\begin{align}
\label{eq:light-trajectory_spin}
    x_\mathrm{L}^\mu(\tau) &= \bar{x}_{\mathrm{L}}^{(0)\mu}(\tau) + \lambda \bar{x}_{\mathrm{L}}^{(1)\mu}(\tau) + \lambda^2 \bar{x}_{\mathrm{L}}^{(2)\mu}(\tau) + \lambda \delta x_\mathrm{L}^{\mu}(\tau)\,, \nonumber \\
     \psi^a_\mathrm{L}(\tau) &= \Psi^{(0)a}_{\mathrm{L}}(\tau) + \lambda \Psi^{(1)a}_{\mathrm{L}}(\tau) + \lambda \delta \psi^a_\mathrm{L}(\tau)\,,
\end{align}
where $\bar{x}_\mathrm{L}^{\mu}(\tau) \equiv \bar{x}_{\mathrm{L}}^{(0)\mu}(\tau) + \lambda \bar{x}_{\mathrm{L}}^{(1)\mu}(\tau) + \lambda^2 \bar{x}_{\mathrm{L}}^{(2)\mu}(\tau)$ and $\Psi^a_\mathrm{L}(\tau) \equiv \Psi^{(0)a}_{\mathrm{L}}(\tau) + \lambda \Psi^{(1)a}_{\mathrm{L}}(\tau)$ satisfy
\begin{align}
     &\hspace{-5pt}\dot{\Psi}^a_\mathrm{L} + \dot{\bar{x}}_\mathrm{L}^\mu \bar{\omega}_{\mu}^{~ab}(\bar{x}_\mathrm{L}) \Psi_{\mathrm{L},b} -i m_\mathrm{L} \bar{R}_{abcd}(\bar{x}_\mathrm{L})\Psi_\mathrm{L}^b\bar{\Psi}_\mathrm{L}^c \Psi_\mathrm{L}^d = 0\,, \nonumber \\
     &\hspace{-5pt}\ddot{\bar{x}}_\mathrm{L}^\mu + \bar{\Gamma}^{\mu}_{~\alpha \beta}(\bar{x}_\mathrm{L})  \dot{\bar{x}}_\mathrm{L}^\alpha \dot{\bar{x}}_\mathrm{L}^\beta - \frac{1}{2}  \bar{R}_{\alpha \beta~\nu}^{~~~\mu}(\bar{x}_\mathrm{L})  \dot{\bar{x}}_\mathrm{L}^\nu \bar{S}_\mathrm{L}^{\alpha \beta} \nonumber\\
     & \hspace{4mm} \qquad  \qquad \qquad + \frac{1}{8} \overline{\nabla}_{\bar{x}_\mathrm{L}}^{\mu}  \bar{R}_{\alpha \beta \rho \lambda}(\bar{x}_\mathrm{L}) \bar{S}_\mathrm{L}^{\alpha \beta}  \bar{S}_\mathrm{L}^{\rho \lambda} = 0\,,
     \label{eq:background_eq}
\end{align}
with the background value of the light particle spin tensor defined as $\bar{S}_\mathrm{L}^{\mu \nu}=-2 i m_\mathrm{L} \bar{\Psi}_\mathrm{L}^{[\mu} \Psi_\mathrm{L}^{\nu]}$. On the other hand, for the heavy particle, we keep all spin contributions
\begin{align}
    x_\mathrm{H}^\mu(\tau) &= \bar{x}_{\mathrm{H}}^\mu(\tau)  + \lambda \delta x_\mathrm{H}^{\mu}(\tau)\,, \nonumber \\
     \psi^a_\mathrm{H}(\tau) &= \Psi^a_\mathrm{H}(\tau) + \lambda \delta \psi^a_\mathrm{H}(\tau)\,,
\end{align}
given that the renormalization of the self-energy corrections (using dimensional regularization to handle divergent terms on the worldline) in \eqref{eq:dim-reg_xH} already implies
\begin{align}
     &\ddot{\bar x}_\mathrm{H}^\mu = \dot{\Psi}^a_\mathrm{H} = 0\,.
\end{align}
Notice that by expanding the $d$-dimensional spinning metric \eqref{eq:MC_metric} around $d=4-2 \epsilon$, we effectively recover the Kerr metric, consistently with the fact that our spinning particle lives in a four-dimensional spacetime. Therefore, we obtain the effective action at 0SF order 
\begin{align}
     \mathcal{S}^{\mathrm{spin},\mathrm{0SF}}_{\mathrm{m}}  &=- m_\mathrm{H} \lambda \int \mathrm{d}\tau \, \frac{1}{2} \bar{g}_{\mu \nu}(\bar{x}_{\mathrm{L}}^{(0)}) \dot{x}_{\mathrm{L}}^{(0)\mu} \dot{x}_{\mathrm{L}}^{(0)\nu} \\
      & - m_\mathrm{H} \int \mathrm{d}\tau \, \left[\frac{1}{2} \eta_{\mu \nu} \dot{\bar x}_\mathrm{H}^\mu \dot{\bar x}_\mathrm{H}^\nu+i m_\mathrm{H} \bar{\Psi}_{\mathrm{H} a} \dot{\Psi}_\mathrm{H}^a \right]\,, \nonumber 
    \label{eq:matter_spin_action}
\end{align}
where the 0SF trajectory for the heavy and light body is
\begin{align}
     &\bar{x}_\mathrm{H}^\mu(\tau) = v_\mathrm{H}^{\mu} \tau\,, \qquad \psi^a_\mathrm{H}(\tau) = \Psi^a_\mathrm{H}\,,\\
     &\, \ddot{\bar x}_{\mathrm{L}}^{(0)\mu} + \bar{\Gamma}^{\mu}_{~\alpha \beta}(\bar{x}_{\mathrm{L}}^{(0)})  \dot{\bar x}_{\mathrm{L}}^{(0)\alpha} \dot{\bar x}_{\mathrm{L}}^{(0)\beta} = 0\,, \nonumber
\end{align}
together with $\bar{S}_\mathrm{H}^{\mu \nu}=-2 i m_\mathrm{H} \bar{\Psi}_\mathrm{H}^{[\mu} \Psi_\mathrm{H}^{\nu]}$. As stressed earlier, it is crucial to note that the mass scaling \eqref{eq:spin_massscaling} forces us to treat the light body as spinless at 0SF order, while the heavy particle carries a tower of spinning contributions.

Moving on to the 1SF order, both the spinless and linear in spin contributions of $\bar{x}_\mathrm{L}^\mu$ will be relevant, as well as the leading contribution of $\Psi^{a}_{\mathrm{L}}$ and $\bar{\Psi}^{a}_{\mathrm{L}}$. To perform the GSF expansion, we expand the spin connection and Riemann tensor around their background value
\begin{equation}
    \begin{split}
        & \omega_{\mu}^{~ab}(x) = \bar{\omega}_{\mu}^{~ab}(x) +\sum_{n = 1}^{\infty} \lambda^{n} \delta \omega^{(n)ab}_{\mu} (x)\,, \\
        & R_{abcd}(x) = \bar{R}_{abcd}(x) + \sum_{n = 1}^{\infty} \lambda^{n} \delta R^{(n)}_{abcd}(x)\;,
    \end{split}
    \label{eq:expansion_omegaR}
\end{equation}
where the variations admit an expansion in $\delta \boldsymbol{g}_{\mu \nu}$ with $(n)$ denoting the corresponding order. Combining the linear contribution in $\delta \boldsymbol{g}_{\mu \nu}$ from the matter \eqref{eq:spinning-matter-action} and graviton \eqref{eq:GR-action} actions, together with the following identities (see appendix \ref{sec:app1})
\begin{align}
  \label{eq: linear-order-identities}
  & m_\mathrm{H}\int \mathrm{d} \tau \, \dot{\bar{x}}_\mathrm{H}^\mu \bar{\Psi}_\mathrm{H}^a \delta \omega^{(1)}_{\mu a b}(\bar{x}_\mathrm{H}) \Psi_\mathrm{H}^b  \\
  & = -\frac{i}{2}\int \mathrm{d}^4 x \int \mathrm{d}\tau\, \delta \boldsymbol{g}_{\mu \nu}  v_\mathrm{H}^{(\mu}  (\bar{S}_\mathrm{H} \cdot \partial_{x})^{\nu)} \delta^4\left(x-v_\mathrm{H} \tau\right)\,, \nonumber \\
  & m_\mathrm{H}^2\int \mathrm{d} \tau \, \delta R^{(1)}_{a b c d}(\bar{x}_\mathrm{H}) \bar{\Psi}_\mathrm{H}^a \Psi_\mathrm{H}^b \bar{\Psi}_\mathrm{H}^c \Psi_\mathrm{H}^d  \\
  & = \frac{1}{2}\int \mathrm{d}^4 x \int \mathrm{d}\tau\, \delta \boldsymbol{g}_{\mu \nu} (\bar{S}_\mathrm{H} \cdot \partial_{x})^{(\mu}  (\bar{S}_\mathrm{H} \cdot \partial_{x})^{\nu)}  \delta^4\left(x-v_\mathrm{H} \tau\right)\,, \nonumber 
\end{align}
we obtain Einstein's equation \eqref{eq:heavy_source_eom} with
\begin{align}
\label{eq:stresstensor_spin}
    & \bar{T}^{\mu \nu}_{\mathrm{H},{\mathrm{spin}}} =\frac{m_\mathrm{H}}{\sqrt{-\bar{g}}} \int \mathrm{d}\tau \Big[v_\mathrm{H}^{\mu} v_\mathrm{H}^{\nu} +  v_\mathrm{H}^{(\mu}  (\bar{S}_\mathrm{H} \cdot \partial_{x})^{\nu)}   \\
    &\qquad\qquad\quad + \frac{1}{2} (\bar{S}_\mathrm{H} \cdot \partial_{x})^{(\mu}  (\bar{S}_\mathrm{H} \cdot \partial_{x})^{\nu)}\Big]  \delta^4\left(x-v_\mathrm{H} \tau\right)\,. \nonumber
\end{align}
The spinning extension of the 1SF effective action in \eqref{eq:1SF_action2} takes the form (see appendix \ref{sec:app1} for details)
\begin{align}
\label{eq:1SF_action_spin}
    &\mathcal{S}^{\mathrm{spin},\mathrm{1SF}} = \mathcal{S}^{\mathrm{1SF}}_{\mathrm{GR}} + \mathcal{S}^{\mathrm{spin},\mathrm{1SF}}_{\mathrm{L}}  + \mathcal{S}^{\mathrm{spin},\mathrm{1SF}}_{\mathrm{H}}\,, \\
     &\mathcal{S}^{\mathrm{spin},\mathrm{1SF}}_{\mathrm{L}} = -m_\mathrm{H} \lambda^2 \int \mathrm{d}\tau\Big\{ \frac{1}{2} \dot{\bar{x}}_{\mathrm{L}}^{(0)\mu} \dot{\bar{x}}_{\mathrm{L}}^{(0)\nu} \delta \boldsymbol{g}_{\mu \nu}(\bar{x}_{\mathrm{L}}^{(0)}) \nonumber \\
    &\qquad+ \frac{1}{2} \dot{\bar{x}}_{\mathrm{L}}^{(0)\mu} \dot{\bar{x}}_{\mathrm{L}}^{(0)\nu} \bar{x}_{\mathrm{L}}^{(1)\rho} \partial_{\rho} g_{\mu \nu}(\bar{x}_{\mathrm{L}}^{(0)})+ \dot{\bar{x}}_{\mathrm{L}}^{(1)\mu} \dot{\bar{x}}_{\mathrm{L}}^{(0)\nu} \bar{g}_{\mu \nu}(\bar{x}_{\mathrm{L}}^{(0)}) \nonumber \\
    &\qquad  +i m_\mathrm{H} \bar{\Psi}^{(0)}_{\mathrm{L}a} \dot{\Psi}_{\mathrm{L}}^{(0)a}+i m_\mathrm{H}\dot{\bar{x}}_{\mathrm{L}}^{(0)\mu} \bar{\Psi}^{(0)a}_{\mathrm{L}} \omega_{\mu a b}(\bar{x}_{\mathrm{L}}^{(0)}) \Psi^{(0)b}_{\mathrm{L}} \Big\} ,\nonumber \\
    &\mathcal{S}^{\mathrm{spin},\mathrm{1SF}}_{\mathrm{H}} = -m_\mathrm{H} \lambda^2 \int \mathrm{d}\tau\Big\{ \frac{1}{2} \delta \dot{x}_\mathrm{H}^2 -\delta x_\mathrm{H}^\rho \dot{\bar{x}}_\mathrm{H}^\mu \dot{\bar{x}}_\mathrm{H}^\nu \delta \Gamma_{\rho \mu \nu}\left(\bar{x}_\mathrm{H}\right)\nonumber  \\
    &\qquad +i m_\mathrm{H} \delta \bar{\psi}_{\mathrm{H} a} \delta \dot{\psi}^a_{\mathrm{H}} + i m_\mathrm{H}\delta \dot{x}_\mathrm{H}^{\mu} \bar{\Psi}_{\mathrm{H} a} \delta \omega_{\mu}^{(1)a b}(\bar{x}_\mathrm{H}) \Psi_{\mathrm{H} b} \nonumber \\
    & \qquad + i m_\mathrm{H} \dot{\bar{x}}_\mathrm{H}^{\mu} \delta x_\mathrm{H}^{\alpha} \bar{\Psi}_{\mathrm{H} a} \partial_{\alpha} \delta \omega_{\mu}^{(1)a b}(\bar{x}_\mathrm{H}) \Psi_{\mathrm{H} b} \nonumber \\
    & \qquad + i m_\mathrm{H}\dot{\bar{x}}_\mathrm{H}^{\mu} \left(\delta \bar{\psi}_{\mathrm{H} a}  \Psi_{\mathrm{H} b} + \bar{\Psi}_{\mathrm{H} a} \delta \psi_{\mathrm{H} b} \right) \delta \omega_{\mu}^{(1)a b}(\bar{x}_\mathrm{H}) \nonumber \\
    & \qquad + m_\mathrm{H}^2\delta R^{(1)}_{a b c d}(\bar{x}_\mathrm{H})  \left(\delta \bar{\psi}_\mathrm{H}^a \Psi_\mathrm{H}^b \bar{\Psi}_\mathrm{H}^c \Psi_\mathrm{H}^d + \bar{\Psi}_\mathrm{H}^a \delta \psi_\mathrm{H}^b \bar{\Psi}_\mathrm{H}^c \Psi_\mathrm{H}^d\right)\nonumber \\[0.1cm]
    & \qquad + \frac{m_\mathrm{H}^2}{2}\delta x_\mathrm{H}^{\mu} \partial_{\mu} \delta R^{(1)}_{a b c d}(\bar{x}_\mathrm{H}) \bar{\Psi}_\mathrm{H}^a \Psi_\mathrm{H}^b \bar{\Psi}_\mathrm{H}^c \Psi_\mathrm{H}^d \nonumber \\
    & \qquad  + i m_\mathrm{H} \dot{\bar{x}}_\mathrm{H}^\mu \bar{\Psi}_{\mathrm{H} a} \delta \omega_{\mu}^{(2)ab}(\bar{x}_\mathrm{H}) \Psi_{\mathrm{H} b} \nonumber \\
    & \qquad + \frac{m_\mathrm{H}^2}{2}\delta R^{(2)}_{abcd}(\bar{x}_\mathrm{H})\bar{\Psi}_\mathrm{H}^a \Psi_\mathrm{H}^b \bar{\Psi}_\mathrm{H}^c \Psi_\mathrm{H}^d \Big\}\,, \nonumber
\end{align}
where we have dropped non-dynamical terms contributing to the heavy dynamics by virtue of dimensional regularization, and we have used the 0SF equations of motion for the trajectory, the Grassmann fields and the graviton field. Notice that the last two terms contributing to the heavy dynamics are contact terms that arise by expanding \eqref{eq:expansion_omegaR} to quadratic order. Importantly, the equations of motion for the $(\delta x_\mathrm{H}^{\mu},\delta \psi^a_{\mathrm{H}})$ fluctuations 
\begin{align}
&\delta \dot{\psi}^a_{\mathrm{H}} + \dot{\bar{x}}_\mathrm{H}^{\mu} \delta \omega_{\mu}^{(1) a b}(\bar{x}_\mathrm{H}) \Psi_{\mathrm{H} b}  -i m_\mathrm{H}\delta R^{(1) a}_{\,\,\, b c d}(\bar{x}_\mathrm{H})  \Psi^b_{\mathrm{H}} \bar{\Psi}_\mathrm{H}^c \Psi_\mathrm{H}^d \!=\! 0\,,\! \nonumber  \\
    &\delta \ddot{x}_{\mathrm{H} \mu} + \delta \Gamma_{\mu \alpha \beta}(\bar{x}_\mathrm{H}) \dot{\bar{x}}^{\alpha}_\mathrm{H} \dot{\bar{x}}^{\beta}_\mathrm{H} -i m_\mathrm{H}\dot{\bar{x}}_\mathrm{H}^{\alpha} \bar{\Psi}_{\mathrm{H} a} \partial_{\mu} \delta \omega_{\alpha}^{(1) a b}(\bar{x}_\mathrm{H}) \Psi_{\mathrm{H} b}  \nonumber \\
    & \qquad \quad +i m_\mathrm{H} \frac{\mathrm{d}}{\mathrm{d} \tau} \left(\bar{\Psi}_{\mathrm{H} a} \delta \omega_{\mu}^{(1) a b}(\bar{x}_\mathrm{H}) \Psi_{\mathrm{H} b} \right)\nonumber \\
    & \qquad \quad -\frac{m_\mathrm{H}^2}{2} \partial_{\mu} \delta R^{(1)}_{a b c d}(\bar{x}_\mathrm{H}) \bar{\Psi}_\mathrm{H}^a \Psi_\mathrm{H}^b \bar{\Psi}_\mathrm{H}^c \Psi_\mathrm{H}^d = 0 \,, 
    \label{eq:deltaxpsi_spin_EOM}
\end{align}
can be formally solved as in the spinless case
\begin{align}
    &\hspace{-10pt}\delta \psi^a_{\mathrm{H}} = - \frac{1}{\partial_{\tau}} \Big[ \dot{\bar{x}}_\mathrm{H}^{\mu} \delta \omega_{\mu}^{(1)a b}(\bar{x}_\mathrm{H}) \Psi_{\mathrm{H} b} - i m_\mathrm{H}\delta R^{(1) a}_{\,\,\, b c d}(\bar{x}_\mathrm{H})  \Psi^b_{\mathrm{H}} \bar{\Psi}_\mathrm{H}^c \Psi_\mathrm{H}^d \Big] \,, \nonumber \\
    &\hspace{-13pt} \delta x_{\mathrm{H} \mu} = -\frac{1}{\partial_{\tau}^2} \Big[ \delta \Gamma_{\mu \alpha \beta}(\bar{x}_\mathrm{H}) \dot{\bar{x}}^{\alpha}_\mathrm{H} \dot{\bar{x}}^{\beta}_\mathrm{H}  \nonumber \\
    &\qquad\qquad\qquad-i m_\mathrm{H}\dot{\bar{x}}_\mathrm{H}^{\alpha} \bar{\Psi}_{\mathrm{H} a} \partial_{\mu} \delta \omega_{\alpha}^{(1) a b}(\bar{x}_\mathrm{H}) \Psi_{\mathrm{H} b} \nonumber \\
    &\qquad\qquad\qquad +i m_\mathrm{H}\frac{\mathrm{d}}{\mathrm{d} \tau} \left(\bar{\Psi}_{\mathrm{H} a} \delta \omega_{\mu}^{(1) a b}(\bar{x}_\mathrm{H}) \Psi_{\mathrm{H} b} \right) \nonumber \\
    &\qquad\qquad\qquad -\frac{m_\mathrm{H}^2}{2} \partial_{\mu} \delta R^{(1)}_{a b c d}(\bar{x}_\mathrm{H}) \bar{\Psi}_\mathrm{H}^a \Psi_\mathrm{H}^b \bar{\Psi}_\mathrm{H}^c \Psi_\mathrm{H}^d \Big] \,. 
\end{align}
Following \eqref{eq:path-integral} and integrating over both $\delta x^{\mu}_\mathrm{H}$ and $\delta \psi^a_\mathrm{H}$, $\delta \bar{\psi}^a_\mathrm{H}$ for the heavy particle dynamics
\begin{align}
&\exp(i \mathcal{S}^{\mathrm{spin},\mathrm{1SF}}_{\mathrm{recoil}} /\hbar) =\int \mathcal{D} [\delta x_\mathrm{H}]  \mathcal{D} [\delta \psi^a_\mathrm{H}] \mathcal{D} [\delta \bar{\psi}^a_\mathrm{H}]\\
&\qquad \qquad \quad \times e^{i \mathcal{S}^{\mathrm{1SF}}_\mathrm{H}[\delta x_\mathrm{L},\delta \psi^a_\mathrm{L},\delta \bar{\psi}^a_\mathrm{L},\delta x_\mathrm{H},\delta \psi^a_\mathrm{H},\delta \bar{\psi}^a_\mathrm{H},\delta \boldsymbol{g}] /\hbar}\,,  \nonumber 
\end{align}
we obtain a novel closed-form solution of the spinning recoil operator at 1SF order 
\begin{align}
\label{eq:1SF_recoil_spin}
    &\hspace{-6pt}\mathcal{S}^{\mathrm{spin},\mathrm{1SF}}_{\mathrm{recoil}}  = -\frac{m_\mathrm{H} \lambda^2}{2} \int \mathrm{d}\tau \Big\{\mathcal{X}_\mu \partial_\tau^{-2} \mathcal{X}^\mu + 2i \bar{\mathcal{Y}}_a \partial_\tau^{-1} \mathcal{Y}^a \\
    &\qquad\qquad\qquad+ 2i m_\mathrm{H} \dot{\bar{x}}_\mathrm{H}^\mu \bar{\Psi}_{\mathrm{H} a} \delta \omega_{\mu}^{(2)ab}(\bar{x}_\mathrm{H}) \Psi_{\mathrm{H} b} \nonumber \\
    &\qquad\qquad\qquad+ m_\mathrm{H}^2 \delta R^{(2)}_{abcd}(\bar{x}_{\mathrm{H}})\bar{\Psi}_\mathrm{H}^a \Psi_\mathrm{H}^b \bar{\Psi}_\mathrm{H}^c \Psi_\mathrm{H}^d \Big\}\,, \nonumber
\end{align}
where we have defined the following quantities
\begin{align}
        \mathcal{X}_\mu = \ & \dot{\bar{x}}_\mathrm{H}^\alpha \dot{\bar{x}}_\mathrm{H}^\beta \delta \Gamma_{\mu \alpha \beta}(\bar{x}_\mathrm{H})+ im_\mathrm{H}  \frac{\mathrm{d}}{\mathrm{d}\tau}\left( \bar{\Psi}_{\mathrm{H} a} \delta \omega_{\mu}^{(1)a b}(\bar{x}_\mathrm{H}) \Psi_{\mathrm{H} b}\right)  \nonumber \\
    & - im_\mathrm{H}  \dot{\bar{x}}_\mathrm{H}^{\alpha} \bar{\Psi}_{\mathrm{H} a} \partial_{\mu} \delta \omega_{\alpha}^{(1)a b}(\bar{x}_\mathrm{H}) \Psi_{\mathrm{H} b} \\
    & - \frac{m_\mathrm{H}^2}{2}\partial_{\mu} \delta R_{a b c d}^{(1)}(\bar{x}_\mathrm{H}) \bar{\Psi}_\mathrm{H}^a \Psi_\mathrm{H}^b \bar{\Psi}_\mathrm{H}^c \Psi_\mathrm{H}^d \,, \; \nonumber \\
    \bar{\mathcal{Y}}^a = \ & \dot{\bar{x}}_\mathrm{H}^{\mu} \delta \omega_{\mu}^{(1) a b}(\bar{x}_\mathrm{H})\bar{\Psi}_{\mathrm{H} b}  + im_\mathrm{H}  \delta R^{(1) a}_{\,\,\, b c d}(\bar{x}_\mathrm{H}) \bar{\Psi}_\mathrm{H}^b \Psi_\mathrm{H}^c \bar{\Psi}_\mathrm{H}^d \,, \nonumber \\
    \mathcal{Y}^a = \ & \dot{\bar{x}}_\mathrm{H}^{\mu} \delta \omega_{\mu}^{(1) a b}(\bar{x}_\mathrm{H})\Psi_{\mathrm{H} b}  - im_\mathrm{H}  \delta R^{(1) a}_{\,\,\,~b c d}(\bar{x}_\mathrm{H}) \Psi_\mathrm{H}^b \bar{\Psi}_\mathrm{H}^c \Psi_\mathrm{H}^d \,. \nonumber 
\end{align}
In this way, we have derived the effective 1SF action
\begin{align}
\label{eq:1SF_action_spin}
    \hspace{-8pt}\mathcal{S}^{\mathrm{spin},\mathrm{1SF}}_{\mathrm{eff}}[\delta \boldsymbol{g}; \{\bar{x}_\mathrm{L}, \Psi_\mathrm{L}\}] &\!\equiv \!\mathcal{S}^{\mathrm{1SF}}_{\mathrm{GR}} \!+\! \mathcal{S}^{\mathrm{spin},\mathrm{1SF}}_{\mathrm{L}} \!+\!\mathcal{S}^{\mathrm{spin},\mathrm{1SF}}_{\mathrm{recoil}} \,,
\end{align}
which, as in the spinless case, is a function of $\delta \boldsymbol{g}^{\mu \nu}$ only, but has implicit dependence on the linear in spin background trajectory $(\bar{x}_\mathrm{L}^{\mu},\Psi^a_\mathrm{L})$.
Ideally, one would like to have a closed-form solution in the time-domain in order to evaluate this effective action. Interestingly, this is one of the few example of integrable models: because of the separability of the equations \cite{Witzany:2019nml,Ramond:2022vhj,Ramond:2024ozy,Gonzo:2024zxo,Witzany:2024ttz} the analytic solution for $(\bar{x}_\mathrm{L}^{\mu},\Psi^a_\mathrm{L})$ exists \cite{Skoupy:2024uan}, albeit in a parametric form. For practical applications in the scattering regime, we can also solve \eqref{eq:background_eq} in the PM expansion as in Refs.~\cite{Cheung:2023lnj,Cheung:2024byb}
\begin{align}
\bar{x}_{\mathrm{L}}^{\mu}(\tau) &= b^{\mu} + v_\mathrm{L}^{\mu} \tau +  (\bar{x}^{\mu}_{\mathrm{L}})|_{\mathcal{O}(G_\mathrm{N})}(\tau) + (\bar{x}^{\mu}_{\mathrm{L}})|_{\mathcal{O}(G_\mathrm{N}^2)}(\tau) + \dots\,, \nonumber \\
\Psi_{\mathrm{L}}^{a}(\tau) &=  \Psi^{a}_{\mathrm{L}} + (\Psi^{a}_{\mathrm{L}})|_{\mathcal{O}(G_\mathrm{N})}(\tau) + \dots\,.
\label{eq:LB_trajectory}
\end{align}

\subsection*{All-order self-force expansion: strong vs weak field metric perturbation}
\label{eq:strongvsweak}

In this section, we will discuss the generic dependence of the self-force EFT on the dynamical fields and how the self-force expansion for the background field method relates with the usual $\lambda$ expansion of the weak field metric perturbation. We first strip off the $\lambda$ dependence on the $n$SF effective action, defining 
\begin{align}
\mathcal{S}^{\mathrm{nSF}} = \lambda^{n+1} \mathbf{S}^{\mathrm{nSF}}\,,
\end{align}
for each graviton, recoil and light contribution. In general, we expect that the self-force effective action at higher SF orders will contain also the dynamical variables $(\delta x_\mathrm{L}^{\mu},\delta \Psi^a_\mathrm{L})$ and---after integrating out the heavy particle dynamics---will be of the form \cite{Cheung:2024byb}
\newpage
\begin{align}
\label{eq:scaling_nSF}
&\hspace{-15pt}\mathcal{S}^{\mathrm{spin}}_{\mathrm{eff}}[\delta \boldsymbol{g},\delta x_\mathrm{L},\delta \Psi_\mathrm{L}] \\
&= \lambda^2 \left(\mathbf{S}^{\mathrm{spin},\mathrm{1SF}}_{\mathrm{GR}+\mathrm{recoil}}[(\delta \boldsymbol{g})^2]  +\mathbf{S}^{\mathrm{spin},\mathrm{1SF}}_{\mathrm{L}}[\delta \boldsymbol{g}] \right)  \nonumber \\
&+ \lambda^3 \left(\mathbf{S}^{\mathrm{spin},\mathrm{2SF}}_{\mathrm{GR}+\mathrm{recoil}}[(\delta \boldsymbol{g})^3]  + \mathbf{S}^{\mathrm{spin},\mathrm{2SF}}_{\mathrm{L}}[\delta \boldsymbol{g},\delta x_\mathrm{L},\delta \Psi^a_\mathrm{L}] \right) \nonumber \\
&+  \dots + \lambda^{n+1} \Big(\mathbf{S}^{\mathrm{spin},\mathrm{nSF}}_{\mathrm{GR}+\mathrm{recoil}}[(\delta \boldsymbol{g})^{n+1}] \nonumber \\
& \qquad \qquad \qquad \quad + \mathbf{S}^{\mathrm{spin},\mathrm{nSF}}_{\mathrm{L}}[\delta \boldsymbol{g},\delta x_\mathrm{L},\delta \Psi^a_\mathrm{L}] \Big) + \dots\,, \nonumber
\end{align}
where we have grouped together the graviton and recoil terms, emphasizing their dependence on $n+1$ graviton fields at $n$SF order. 

The action \eqref{eq:scaling_nSF} is invariant under diffeomorphisms of the background metric $\bar{g}_{\mu \nu}$. The equations of motion for the graviton field $\delta \boldsymbol{g}_{\mu \nu}$ are then organized as an expansion on the background spacetime---usually called BH perturbation theory---where the field $\delta \boldsymbol{g}_{\mu \nu}$ itself is treated as a small perturbation (non-linear higher order terms are suppressed in $\lambda$). In the weak field approach, instead, the metric perturbation $\kappa h_{\mu \nu}$ defined in the PM or PN theory is defined as
\begin{align}
\label{eq:shift_metric}
g_{\mu \nu} &= \bar{g}_{\mu \nu} + \lambda \delta \boldsymbol{g}_{\mu \nu} \nonumber \\
&= \eta_{\mu \nu} + \underbrace{(\bar{g}_{\mu \nu} - \eta_{\mu \nu}) + \lambda \delta \boldsymbol{g}_{\mu \nu}}_{\equiv\kappa h_{\mu \nu}} \,,
\end{align}
and it transforms as a tensor in flat spacetime $\eta_{\mu \nu}$. Introducing the external background field
\begin{align}
\label{eq:background_metric}
\bar{H}_{\mu \nu}&=  \left(\bar{g}_{\mu \nu} - \eta_{\mu \nu}\right)=\sum_{k=1}^{\infty} (G_\mathrm{N} m_\mathrm{H})^k \bar{H}^{(k)}_{\mu \nu} \,,
\end{align}
where the coefficients $\bar{H}^{(k)}_{\mu \nu}$ are independent of $G_\mathrm{N} m_\mathrm{H}$, we can restore the implicit dependence on $\bar{g}_{\mu \nu}$ in \eqref{eq:scaling_nSF}, giving
\begin{align}
\label{eq:scaling_weak}
& \mathcal{S}^{\mathrm{spin}}_{\mathrm{eff}}[\delta \boldsymbol{g},\delta x_\mathrm{L},\delta \Psi_\mathrm{L};\{\bar{H}^k\}] \\
&= \lambda^2 \left(\mathbf{S}^{\mathrm{spin},\mathrm{1SF}}_{\mathrm{GR}+\mathrm{recoil}}[(\delta \boldsymbol{g})^2; \{\bar{H}^k\}]  +\mathbf{S}^{\mathrm{spin},\mathrm{1SF}}_{\mathrm{L}}[\delta \boldsymbol{g};\{\bar{H}^k\}] \right)  \nonumber \\
& + \lambda^3 \Big(\mathbf{S}^{\mathrm{spin},\mathrm{2SF}}_{\mathrm{GR}+\mathrm{recoil}}[(\delta \boldsymbol{g})^3;\{\bar{H}^k\}]  \nonumber \\
& \qquad \qquad \qquad \quad + \mathbf{S}^{\mathrm{spin},\mathrm{2SF}}_{\mathrm{L}}[\delta \boldsymbol{g},\delta x_\mathrm{L},\delta \Psi^a_\mathrm{L};\{\bar{H}^k\}] \Big)\nonumber \\
&+  \dots + \lambda^{n+1} \Big(\mathbf{S}^{\mathrm{spin},\mathrm{nSF}}_{\mathrm{GR}+\mathrm{recoil}}[(\delta \boldsymbol{g})^{n+1};\{\bar{H}^k\}] \nonumber \\
& \qquad \qquad \qquad \quad + \mathbf{S}^{\mathrm{spin},\mathrm{nSF}}_{\mathrm{L}}[\delta \boldsymbol{g},\delta x_\mathrm{L},\delta \Psi^a_\mathrm{L};\{\bar{H}^k\}] \Big) + \dots\,. \nonumber
\end{align}
The upshot of this analysis is that we can compute $\delta \boldsymbol{g}_{\mu \nu}$ either with strong-field tools (as an expansion in $\lambda$ from \eqref{eq:scaling_nSF}) or with weak-field perturbative tools (as a double expansion in $(G_\mathrm{N},\lambda)$ from \eqref{eq:scaling_weak} in powers of the background field \eqref{eq:background_metric}). Instead, computing directly $h_{\mu \nu}$ from \eqref{eq:scaling_weak} would require to consider higher SF terms, spoiling the power counting in $\lambda$. The relation between $\delta \boldsymbol{g}_{\mu \nu}$ and $h_{\mu \nu}$ will be the main subject of sections \ref{sec:1SFmetric} and \ref{sec:1SFCompton}. 

%----------------------------------------------------------------------
\section{1SF metric perturbation: traditional vs diagrammatic approach}
\label{sec:1SFmetric}
%----------------------------------------------------------------------
In this section, we study the dynamical consequences of \eqref{eq:1SF_action_spin} both from the traditional self-force approach and in the diagrammatic expansion using the background field method. First, we introduce the equation for the 1SF metric perturbation for particle-generated (non-vacuum) spacetimes, including the contribution from the matter-mediated force. Then we introduce the background field method to compute the 1SF metric from our effective action for scattering orbits, discussing the waveform recursion relation and the relevance of the 1SF Compton amplitude for the resummation.

%----------------------------------------------------------------------
\subsection*{1SF metric for particle-generated spacetimes}
%----------------------------------------------------------------------

It is interesting to explore the classical equation of motion for the metric perturbation $\delta \boldsymbol{g}_{\mu \nu}$, which is obtained by extremizing the 1SF effective action.
In terms of the trace-reversed perturbation
\begin{align}
\delta \boldsymbol{g}^{\mathrm{tr}}_{\mu \nu} = \delta \boldsymbol{g}_{\mu \nu} - \frac{1}{2} \bar{g}_{\mu \nu} \delta \boldsymbol{g}\,,
\label{eq:metric_tr}
\end{align}
we obtain the compact expression
\begin{align}
&\hspace{-7pt}\overline{\nabla}_{\rho} \overline{\nabla}^{\rho} \delta \boldsymbol{g}^{\mathrm{tr}}_{\mu \nu}- \bar{R}^{\alpha \beta} (\bar{g}_{\mu \nu} \delta \boldsymbol{g}^{\mathrm{tr}}_{\alpha \beta} - \bar{g}_{\alpha \beta} \delta \boldsymbol{g}^{\mathrm{tr}}_{\mu \nu} + 2 \bar{g}_{\beta (\mu} \delta \boldsymbol{g}^{\mathrm{tr}}_{\nu) \alpha} ) \nonumber \\
&\,+ 2 \bar{R}_{\mu \; \; \nu}^{\; \; \alpha \; \;\beta}  \delta \boldsymbol{g}^{\mathrm{tr}}_{\alpha \beta}  =- 16 \pi G_\mathrm{N} (T_{\mathrm{L}\, \mu \nu}^{(0)} + \mathcal{F}_{\mu \nu}^{\mathrm{recoil}}[\delta \boldsymbol{g}^{\mathrm{tr}}])\,, 
\label{eq:deltag_1SF}
\end{align}
where $T_{\mathrm{L}\,\mu \nu}^{(0)}$ and $\mathcal{F}_{\mu \nu}^{\mathrm{recoil}}[\delta \boldsymbol{g}^{\mathrm{tr}}]$ are defined from
\begin{align}
&T_{\mathrm{L} \,\mu \nu}^{(0)} = \frac{m_\mathrm{L} }{\sqrt{-\bar{g}}} \int \mathrm{d} \tau \,\delta^{(4)}(x-\bar{x}_\mathrm{L}^{(0)}(\tau)) \, \dot{x}_\mu \dot{x}_\nu\,, \\
&\qquad \qquad \mathcal{F}_{\mu \nu}^{\mathrm{recoil}}[\delta \boldsymbol{g}^{\mathrm{tr}}] = \frac{2}{\sqrt{-\bar{g}}} \frac{\delta \mathcal{S}^{\mathrm{spin},\mathrm{1SF}}_{\mathrm{recoil}}}{\delta (\delta \boldsymbol{g}^{\mu \nu})} \,.
\end{align}
Notice that $F_{\mu \nu}^{\mathrm{recoil}}[\delta \boldsymbol{g}^{\mathrm{tr}}]$ is linear in the trace-reversed perturbation, alongside being supported on the spatial heavy particle trajectory $\propto \delta^{(3)}(\vec{x}-\vec{v}_\mathrm{H} \tau)$. Moreover, the term proportional to $\bar{R}^{\alpha \beta}$ in \eqref{eq:deltag_1SF} is on a similar footing, and can be written as  
\begin{align}
&\hspace{-6pt}\mathcal{F}_{\mu \nu}^{\mathrm{source}}[\delta \boldsymbol{g}^{\mathrm{tr}}] = \bar{R}^{\alpha \beta} (\bar{g}_{\mu \nu} \delta \boldsymbol{g}^{\mathrm{tr}}_{\alpha \beta} - \bar{g}_{\alpha \beta} \delta \boldsymbol{g}^{\mathrm{tr}}_{\mu \nu} + 2 \bar{g}_{\beta (\mu} \delta \boldsymbol{g}^{\mathrm{tr}}_{\nu) \alpha} ) \nonumber \\
&\qquad \qquad = 8 \pi G_\mathrm{N} \left(T_\mathrm{H}^{\alpha \beta} - \frac{1}{2} \bar{g}^{\alpha \beta} \,\bar{g}^{\rho \lambda} T_{\mathrm{H} \rho \lambda} \right) \nonumber \\
&\quad \qquad \qquad \times (\bar{g}_{\mu \nu} \delta \boldsymbol{g}^{\mathrm{tr}}_{\alpha \beta} - \bar{g}_{\alpha \beta} \delta \boldsymbol{g}^{\mathrm{tr}}_{\mu \nu} + 2 \bar{g}_{\beta (\mu} \delta \boldsymbol{g}^{\mathrm{tr}}_{\nu) \alpha} )\,,
\end{align}
because of the heavy particle equation of motion \eqref{eq:heavy_source_eom}. It is then natural to write \eqref{eq:deltag_1SF} in the equivalent form
\begin{align}
&\overline{\nabla}_{\rho} \overline{\nabla}^{\rho} \delta \boldsymbol{g}^{\mathrm{tr}}_{\mu \nu}\!+\!2 \bar{R}_{\mu \; \; \nu}^{\; \; \alpha \; \;\beta}  \delta \boldsymbol{g}^{\mathrm{tr}}_{\alpha \beta}\nonumber \\
&\qquad \qquad=- 16 \pi G_\mathrm{N} (T_{\mathrm{L}\, \mu \nu}^{(0)}\!+\!\Delta T^{\mathrm{contact}}_{\mu \nu}[\delta \boldsymbol{g}^{\mathrm{tr}}])\,, \nonumber \\
&\quad \Delta T^{\mathrm{contact}}_{\mu \nu}[\delta \boldsymbol{g}^{\mathrm{tr}}] = \mathcal{F}_{\mu \nu}^{\mathrm{source}}[\delta \boldsymbol{g}^{\mathrm{tr}}] + \mathcal{F}_{\mu \nu}^{\mathrm{recoil}}[\delta \boldsymbol{g}^{\mathrm{tr}}]\,,
\label{eq:deltag_1SF_new}
\end{align}
where we anticipate that $\Delta T^{\mathrm{contact}}_{\mu \nu}[\delta \boldsymbol{g}^{\mathrm{tr}}]$ corresponds to local contact terms in the amplitude picture. As we will discuss later, these contact terms provide a unique prescription for connecting the self-force expansion with the point-particle description.

Observe that \eqref{eq:deltag_1SF_new} is strikingly similar to the traditional formulation of the 1SF equation for metric perturbations (see eq.~(22) of Ref.~\cite{Barack:2009ux}) in vacuum, albeit being defined here for a particle-generated spacetime where an additional ``matter-mediated'' contribution $\Delta T^{\mathrm{contact}}_{\mu \nu}[\delta \boldsymbol{g}^{\mathrm{tr}}]$ appears~\cite{Pfenning:2000zf,Gralla:2021qaf}. Given that the only spacetime region where the non-perturbative black hole vacuum and non-vacuum solution can differ in their description is the source of such a potential, it is unsurprising that the contact terms are supported on the spatial heavy particle trajectory. In principle, it should be possible to provide an exact relation between the solution of \eqref{eq:deltag_1SF_new} (perhaps by defining the regular and singular field in the spirit of the Detweiler-Whiting prescription \cite{Detweiler:2000gt,Detweiler:2002mi,Poisson:2011nh}) and the corresponding one for vacuum spacetimes with the usual BH boundary conditions at the horizon and at infinity. Consequently, this would provide an exact map from GSF solutions to PN and PM results. We leave such a tantalizing perspective for a future analysis.

%----------------------------------------------------------------------
\subsection*{1SF metric from the background field method}
%----------------------------------------------------------------------

In this section, we develop a diagrammatic approach to compute the 1SF metric perturbation \eqref{eq:deltag_1SF_new} in the scattering regime using the background field method. Following Refs.~\cite{Boulware:1968zz,Goldberger:2004jt,Iteanu:2024dvx}, we define the generating functional 
\begin{align}
    e^{i \mathbf{W}^\mathrm{1SF}[J]} &= \int \mathcal{D} [\delta \boldsymbol{g}] \,\exp\Big[i \mathbf{S}^{\mathrm{spin},\mathrm{1SF}}_{\mathrm{eff}}[\delta \boldsymbol{g};\{\bar{H}^k\}] \nonumber \\
    &\qquad \qquad + i \int \mathrm{d}^4 y\, \sqrt{-\bar{g}} \mathbf{J}^{\mu \nu}(y) \delta \boldsymbol{g}_{\mu \nu}(y) \Big]\,,
\end{align}
where $\mathbf{J}^{\mu \nu}$ is a source term which is eventually set to zero. At 1SF order, we then study the expectation value
\begin{align}
\langle \delta \boldsymbol{g}_{\mu \nu}(x) \rangle \Big|_{\mathrm{1SF}} = \frac{1}{\sqrt{-\bar{g}}} \frac{\delta \mathbf{W}^{\mathrm{1SF}}[J]}{i \delta \mathbf{J}^{\mu \nu}(x)} \Big|_{\mathbf{J}=0}\,,
\label{eq:1SF_metric_weak}
\end{align}
which, as explained earlier, is independent of $(\delta x^{\mu}_\mathrm{L},\delta \Psi^a_\mathrm{L})$. Note that in our calculation we neglect all quantum effects---which involve graviton loops---enforcing the classical saddle point approximation as explained in Ref.~\cite{Goldberger:2004jt}. 

To compute \eqref{eq:1SF_metric_weak}, we then extract the perturbative Feynman rules from our 1SF effective action \eqref{eq:1SF_action_spin} up to quadratic order in spin. As usual, we perform a Fourier transform of the dynamical and background fields
\begin{align}
   \lambda \,\delta  \boldsymbol{g}_{\mu \nu}(x) &= \int \hat{\mathrm{d}}^4 k \, e^{-ik\cdot x} \,\kappa \,\mathbf{\tilde{g}}_{\mu \nu}(k)\,, \nonumber \\
   \bar{H}_{\mu \nu}(x) &= \int \hat{\mathrm{d}}^4 k \, e^{-ik\cdot x} \tilde{H}_{\mu \nu}(k)\,. 
   \label{eq:Fourier_transform}
\end{align}
First we compute the vertex for the light-particle source
\begin{align}
      \begin{tikzpicture}[baseline=(current bounding box.center)]
              \begin{feynman}[every blob={/tikz/fill=gray!30,/tikz/inner sep=3pt}]
                \vertex[blob] (m) at (-1.5,0) {$L$};
                \vertex (a) at (0,0) {$\mu \nu $};
                \diagram* {(m) -- [boson, text=black, momentum={$k$}] (a)};
              \end{feynman} 
    \end{tikzpicture} 
         = i \mathcal{V}^{\mu \nu}_{\mathrm{L}}(k) = i \frac{\delta \mathbf{S}^{\mathrm{1SF}}_{\mathrm{L}}}{\delta \mathbf{\tilde{g}}_{\mu \nu}(k)} \,,  
\end{align}
which gives the stress tensor contribution for the spinless light-body trajectory in Kerr spacetime, as observed also in Ref.~\cite{Cheung:2024byb}. Then we study the graviton curved space propagator, which is defined as
\begin{align}
    G^{\rho \sigma \gamma \lambda}(k_1,k_2) &= \frac{\delta \mathbf{S}^{\mathrm{1SF}}_{\mathrm{GR}}}{\delta \mathbf{\tilde{g}}_{\rho \sigma}(k_1) \delta \mathbf{\tilde{g}}_{\gamma \lambda}(k_2)}\,.
\end{align}
To evaluate this explicitly we use the minimal coupling metric \eqref{eq:MC_metric} and expand all background quantities (Christoffel, Riemann, etc.) in $\mathcal{S}^{\mathrm{1SF}}_{\mathrm{GR}}$ around the flat metric in powers of the background field \eqref{eq:background_metric}. The leading term is the free graviton propagator in de Donder gauge 
\begin{equation}
    \hspace{-0.5cm}\begin{tikzpicture}[baseline=(current bounding box.center)]
              \begin{feynman}[every blob={/tikz/fill=gray!30,/tikz/inner sep=3pt}]
                \vertex (a) at (-1.5,0) {$\rho \sigma $};
                \vertex (b) at (1.5,0) {$\gamma \lambda$};
                \diagram* {(a) -- [boson, text=black, momentum={$k$}] (b)};
              \end{feynman} 
    \end{tikzpicture} 
       \!\!\! = i G^{(0) \rho \sigma \gamma \lambda}(k) = \frac{i\, \mathcal{P}^{(d) \rho \sigma \gamma \lambda}}{2 k^2}   \,,
\end{equation}
where $\mathcal{P}^{(d) \rho \sigma \gamma \lambda}$ is the $d$-dimensional graviton projector
\begin{align}
    \label{eq:projector}
    \mathcal{P}_{\mu \nu}^{(d)\,\alpha \beta} = \delta_{(\mu}{ }^{(\alpha} \delta_{\nu)}{ }^{\beta)}-\frac{1}{d-2} \eta_{\mu \nu} \eta^{\alpha \beta}\,.
\end{align}
The interacting piece gives
\begin{align}
\label{eq:Gint}
   &\hspace{-12pt}G_{\mathrm{int}}^{\rho \sigma \gamma \lambda}(k_1,k_2) = \hat{\delta}(v_\mathrm{H} \cdot k_1 + v_\mathrm{H} \cdot k_2) \nonumber \\
   &\quad \times \int \mathrm{d}^{d-1} \vec{x} \,\mathrm{e}^{-i (\vec{k}_1+\vec{k}_2) \cdot \vec{x}} f_{\mathcal{G}}^{\rho \sigma \gamma \lambda}(\vec{x},a^d_{\mathrm{H}})\,, 
\end{align}
where $f_{\mathcal{G}}^{\rho \sigma \gamma \lambda}(\vec{x},a^d_{\mathrm{H}})$ is a rational polynomial function. We then perform the tensor decomposition and expand this function up to quadratic order in spin, allowing one to compute the Fourier transform of the integral basis via
\begin{align}
    \label{eq:integral_tensor}
    &\!\int \mathrm{d}^{d-1} \vec{x} \,\mathrm{e}^{-i \vec{q} \cdot \vec{x}} r^{\mu_1} \cdots r^{\mu_{n}} g_{\mathcal{G}}(|\vec{x}|) \\
    &\,=(-i)^{n} P^{\mu_1 \nu_1} \cdots P^{\mu_{n} \nu_{n}}  \frac{\partial^{n}}{\partial q^{\nu_1} \ldots \partial q^{\nu_{n}}} \!\int \mathrm{d}^{d-1} \vec{x} \,\mathrm{e}^{-i \vec{q} \cdot \vec{x}}  g_{\mathcal{G}}(|\vec{x}|)\,, \nonumber
\end{align}
where $g_{\mathcal{G}}(|\vec{x}|)$ denotes a generic coefficient in the decomposition with the spatial $r^{\mu}$ and $P^{\mu \nu}$ defined in \eqref{eq:projector_spatial}. Finally, the leftover scalar integrals in \eqref{eq:integral_tensor} can be directly evaluated in spherical coordinates, giving the vertex
\begin{equation}
    \begin{tikzpicture}[baseline=(current bounding box.center)]
              \begin{feynman}[every blob={/tikz/fill=gray!30,/tikz/inner sep=3pt}]
                \vertex[blob] (m) at (0,0) {$G$};
                \vertex (a) at (-1.5,0) {$\rho \sigma $};
                \vertex (b) at (1.5,0) {$\gamma \lambda$};
                \diagram* {(a) -- [boson, text=black, momentum={$k_1$}] (m) -- [boson, text=black, reversed momentum={$k_2$}] (b)};
              \end{feynman} 
    \end{tikzpicture} 
        = i G_{\mathrm{int}}^{\rho \sigma \gamma \lambda}(k_1,k_2) \,.
\end{equation}
The remaining vertex for the recoil insertion 
\begin{equation}
    \begin{tikzpicture}[baseline=(current bounding box.center)]
              \begin{feynman}[every blob={/tikz/fill=gray!30,/tikz/inner sep=3pt}]
                \vertex[blob] (m) at (0,0) {$H$};
                \vertex (a) at (-1.5,0) {$\rho \sigma $};
                \vertex (b) at (1.5,0) {$\gamma \lambda$};
                \diagram* {(a) -- [boson, text=black, momentum={$k_1$}] (m) -- [boson, text=black, reversed momentum={$k_2$}] (b)};
              \end{feynman} 
    \end{tikzpicture} 
        = i H^{\rho \sigma \gamma \lambda}(k_1,k_2)\,.
\end{equation}
is readily evaluated in terms of the 1SF recoil action
\begin{align}
   H^{\rho \sigma \gamma \lambda}(k_1,k_2) &= \frac{\delta \mathbf{S}^{\mathrm{1SF}}_{\mathrm{recoil}}}{\delta \mathbf{\tilde{g}}_{\rho \sigma}(k_1) \delta \mathbf{\tilde{g}}_{\gamma \lambda}(k_2)}\,.
\end{align}
The result can be conveniently organized in terms of spin
\begin{align}
    H^{\rho \sigma \gamma \lambda}(k_1,k_2) &=  32 \pi G_\mathrm{N} \hat{\delta}(v_\mathrm{H} \cdot k_1 + v_\mathrm{H} \cdot k_2) \nonumber \\
    & \times m_{\mathrm{H}} \frac{\sum_{n = 0}^{2}H_{n}^{\rho \sigma \gamma \lambda}(k_1,k_2)}{(v_\mathrm{H} \cdot k_1)(v_\mathrm{H} \cdot k_2)}\,,
\end{align}
where for the spinless dynamics we obtain
\begin{align}
    H^{\rho \sigma \gamma \lambda}_0(k_1,k_2) = N^{\mu \rho \sigma}(k_1)N_{\mu}^{~\gamma \lambda}(k_2) \;,
\end{align}
for the linear-in-spin dynamics we find
\begin{align}
    & H^{\rho \sigma \gamma \lambda}_1(k_1,k_2) = \frac{i}{4} \{ 2(v_\mathrm{H} \cdot k_2)N^{\gamma\rho \sigma}(k_1)(\bar{S}_\mathrm{H} \cdot k_2)^\lambda  \\[-0.1cm]
    & - 2(v_\mathrm{H} \cdot k_2)N^{\rho \gamma \lambda}(k_2)(\bar{S}_\mathrm{H} \cdot k_1)^\sigma + (v_\mathrm{H} \cdot k_2) v_\mathrm{H}^\sigma v_\mathrm{H}^\lambda M^{\rho \gamma}(k_1, k_2) 
 \nonumber \\
    & - 2k_{2\mu}v_\mathrm{H}^\gamma N^{\mu\rho \sigma}( k_1)(\bar{S}_\mathrm{H} \cdot k_2)^\lambda + 2k_{1\mu}v_\mathrm{H}^\rho N^{\mu \gamma \lambda}( k_2)(\bar{S}_\mathrm{H} \cdot k_1)^\sigma  \nonumber \\
    &  + (v_\mathrm{H} \cdot k_1)(v_\mathrm{H} \cdot k_2) [K^{\rho \sigma \gamma \lambda}(k_2) +  K^{ \gamma \lambda \rho \sigma}(k_1)]\}\;, \nonumber
\end{align}
and, finally, for the quadratic-in-spin dynamics we have
\begin{align}
    & H^{\rho \sigma \gamma \lambda}_2(k_1,k_2) = \frac{1}{8} \{2k_{2\mu} N^{\mu \rho \sigma}(k_1) (\bar{S}_\mathrm{H} \cdot k_2)^\lambda(\bar{S}_\mathrm{H} \cdot k_1)^\gamma \nonumber \\
    & \hspace{0.25cm}+2k_{1\mu} N^{\mu \gamma \lambda}(k_2) (\bar{S}_\mathrm{H} \cdot k_1)^\rho(\bar{S}_\mathrm{H} \cdot k_2)^\sigma \nonumber \\
    & \hspace{0.25cm}- 2(v_\mathrm{H} \cdot k_1)(v_\mathrm{H} \cdot k_2)(\bar{S}_\mathrm{H} \cdot k_1)^\rho (\bar{S}_{\mathrm{H}} \cdot k_2)^\gamma \eta^{\sigma \lambda} \nonumber \\
    & \hspace{0.25cm}- 2(v_\mathrm{H} \cdot k_2) [k_2^\sigma v_\mathrm{H}^\lambda - k_1^\lambda v_\mathrm{H}^\sigma](\bar{S}_\mathrm{H} \cdot k_1)^\rho (\bar{S}_\mathrm{H} \cdot k_2)^\gamma \\
    & \hspace{0.25cm}- 2(k_1 \cdot k_2) v_\mathrm{H}^\sigma v_\mathrm{H}^\lambda (\bar{S}_\mathrm{H} \cdot k_1)^\rho (\bar{S}_\mathrm{H} \cdot k_2)^\gamma \nonumber \\
    & \hspace{0.25cm}- 2(v_\mathrm{H} \cdot k_2)[v_\mathrm{H}^\lambda M^{\sigma \gamma}(k_1,k_2)(\bar{S}_\mathrm{H} \cdot k_1)^\rho \nonumber\\
    & \hspace{0.25cm}- v_\mathrm{H}^\sigma M^{\rho \lambda}(k_2,k_1)(\bar{S}_\mathrm{H} \cdot k_2)^\gamma] \nonumber \\ 
    & \hspace{0.25cm}-(v_\mathrm{H} \cdot k_1)(v_\mathrm{H} \cdot k_2)[Q^{\rho \sigma \gamma \lambda}(k_1,k_2)+Q^{\gamma \lambda \rho \sigma}(k_2,k_1) \nonumber\\
    & \hspace{0.25cm}+ J^{\rho \sigma \gamma \lambda}(k_2) + J^{\gamma \lambda \rho \sigma}(k_1) - M^{\rho \kappa}(k_1, k_2) \bar{S}_\mathrm{H}^{\sigma \lambda}]\}\nonumber\,.
\end{align}
Notice that, to make the expressions more compact, we have introduced the following tensors
\begin{align}
     \hspace{-0.25cm} 2N^{\alpha \mu \nu}(k) &= (k \cdot v_\mathrm{H})[v_\mathrm{H}^\mu \eta^{\nu \alpha}+v_\mathrm{H}^\nu \eta^{\mu \alpha}] - v_\mathrm{H}^\mu v_\mathrm{H}^\nu k^\alpha\,, \nonumber \\
     \hspace{-0.25cm}M^{\rho \gamma}(k_1, k_2) &= (k_1 \cdot k_2) \bar{S}_\mathrm{H}^{\rho \gamma} - k_1^\gamma (\bar{S}_\mathrm{H} \cdot k_2)^\rho \nonumber \\
    & + k_2^\rho (\bar{S}_\mathrm{H} \cdot k_1)^\gamma + (k_1 \cdot \bar{S}_\mathrm{H} \cdot k_2) \eta^{\rho \gamma}\,, \nonumber\\
      \hspace{-0.25cm}2K^{\rho \sigma \gamma \lambda}(k) &=  2v_\mathrm{H}^\gamma \eta^{\rho \lambda} (\bar{S}_\mathrm{H} \cdot k)^\sigma - 2v_\mathrm{H}^\lambda k ^\rho \bar{S}_\mathrm{H}^{\sigma \gamma} \nonumber \\
    & + (v_\mathrm{H} \cdot k) \eta^{\rho \lambda}\bar{S}_\mathrm{H}^{\sigma \gamma}\,, \\
     \hspace{-0.25cm}2Q^{\rho \sigma \gamma \lambda}(k_1,k_2)& = 2(\bar{S}_\mathrm{H}\cdot k_{12})^\gamma (\bar{S}_\mathrm{H} \cdot k_2)^\sigma \eta^{\rho \lambda} \nonumber \\
    & - 2(\bar{S}_\mathrm{H}\cdot k_{12})^\lambda k_2^\rho \eta^{\sigma \gamma} - (k_1 \cdot \bar{S}_\mathrm{H} \cdot k_2) \bar{S}_\mathrm{H}^{\sigma \gamma} \eta^{\rho \lambda}\,,\nonumber  \\
     \hspace{-0.25cm} J^{\rho \sigma \gamma \lambda}(k)& = (\bar{S}_\mathrm{H} \cdot k)^\sigma (\bar{S}_\mathrm{H} \cdot k)^\gamma \eta^{\rho \lambda} - k^\rho (\bar{S}_\mathrm{H} \cdot k)^\kappa \bar{S}_\mathrm{H}^{\sigma \lambda}\,, \nonumber
\end{align}
where $k^{\mu}_{12} = k^{\mu}_1 + k^{\mu}_2$. From this we define the graviton-graviton effective vertex
\begin{align}
  \mathcal{G}_2^{\rho \sigma \gamma \lambda}(k_1,k_2) = G_{\mathrm{int}}^{\rho \sigma \gamma \lambda}(k_1,k_2) + H^{\rho \sigma \gamma \lambda}(k_1,k_2)\,,
\end{align}
\begin{align}
        \begin{tikzpicture}[baseline=(current bounding box.center)]
              \begin{feynman}[every blob={/tikz/fill=gray!30,/tikz/inner sep=3pt}]
                \vertex[blob] (m) at (0,0) {$\mathcal{G}_2$};
                \vertex (a) at (-1.5,0) {$\rho \sigma $};
                \vertex (b) at (1.5,0) {$\gamma \lambda$};
                \diagram* {(a) -- [boson, text=black, momentum={$k_1$}] (m) -- [boson, text=black, reversed momentum={$k_2$}] (b)};
              \end{feynman} 
    \end{tikzpicture} 
       \equiv  
       \begin{tikzpicture}[baseline=(current bounding box.center)]
              \begin{feynman}[every blob={/tikz/fill=gray!30,/tikz/inner sep=3pt}]
                \vertex[blob] (m) at (0,0) {\small{$G\!+\!H$}};
                \vertex (a) at (-1.5,0) {$\rho \sigma $};
                \vertex (b) at (1.5,0) {$\gamma \lambda$};
                \diagram* {(a) -- [boson, text=black, momentum={$k_1$}] (m) -- [boson, text=black, reversed momentum={$k_2$}] (b)};
              \end{feynman} 
    \end{tikzpicture}  \nonumber
\end{align}
which conveniently repackages the background and recoil contributions together. Importantly, the 1SF recoil contribution exactly matches the spinning deflection and contact contributions in the WQFT formalism~\cite{Jakobsen:2021zvh}. The perturbative Feynman rules discussed so far agree in the spinless limit with Ref.~\cite{Cheung:2024byb}.

%----------------------------------------------------------------------
\subsection*{1SF waveform recursion relation}
\label{sec:1SF_waveform}
%----------------------------------------------------------------------

For the 1SF metric perturbation in \eqref{eq:1SF_metric_weak}, we then only need to consider all diagrams with a single external graviton field, giving the interesting recursion relation 
\begin{align}
\label{eq:1SF_metric}
 &\langle \delta \boldsymbol{g}_{\mu \nu}(x) \rangle \Big|_{\mathrm{1SF}}  =  \int \mathrm{\hat{d}}^d k\, e^{i k \cdot x}\, G^{(0)\mathrm{ret}}_{\mu \nu \alpha \beta}(k) \mathcal{V}^{\alpha \beta}_{\mathrm{L}}(k)  \\
 &\hspace{-5pt}+ \! \int \mathrm{\hat{d}}^d k\, \mathrm{\hat{d}}^d k_1\, e^{i k \cdot x}\,G^{(0)\mathrm{ret}}_{\mu \nu \alpha \beta}(k) \, \mathcal{G}_{2, \mathrm{ret}}^{\alpha \beta \lambda \rho }(k,k_1)\, G^{(0)\mathrm{ret}}_{\lambda \rho \xi \zeta}(k_1) \mathcal{V}^{\xi \zeta}_{\mathrm{L}}(k_1)  \nonumber \\
 &\hspace{-5pt}+ \! \int \mathrm{\hat{d}}^d k\, \mathrm{\hat{d}}^d k_1\,\mathrm{\hat{d}}^d k_2\, e^{i k \cdot x}\, G^{(0)\mathrm{ret}}_{\mu \nu \alpha \beta}(k) \, \mathcal{G}_{2, \mathrm{ret}}^{\alpha \beta \lambda \rho }(k,k_1)\, \nonumber \\
 &\quad \times G^{(0)\mathrm{ret}}_{\lambda \rho \xi \zeta}(k_1) \mathcal{G}_{2, \mathrm{ret}}^{\xi \zeta \phi \theta }(k_1,k_2)  G^{(0)\mathrm{ret}}_{\phi \theta \omega \varphi}(k_2) \mathcal{V}^{\omega \varphi}_{\mathrm{L}}(k_2) + \dots\,, \nonumber
\end{align}
\begin{figure}[t!]
\centering
\includegraphics[scale=0.7]{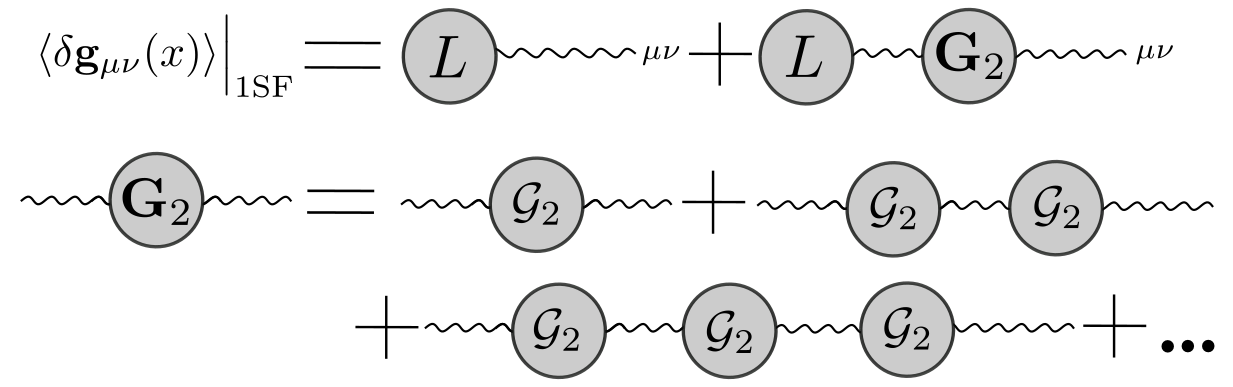}
\caption{The 1SF metric diagrammatic expansion.}
\label{fig:1SF-metric}
\end{figure}
\!\!\!\!\! which is represented diagrammatically in Fig.~\ref{fig:1SF-metric}. Note that here we use retarded boundary conditions for the propagators and the effective vertices, consistently with the fact that we are solving a classical equation in terms of initial boundary conditions. Moreover, given that $\delta \boldsymbol{g}_{\alpha \beta}$ is a perturbation on the background spacetime, it satisfies
\begin{align}
 \left(\delta^{\alpha}_{\mu} \bar{g}^{\beta \gamma} -\frac{1}{2} \bar{g}^{\alpha \beta} \delta^{\gamma}_{\mu}\right) \overline{\nabla}_\gamma \langle \delta \boldsymbol{g}_{\alpha \beta}(x) \rangle = 0 \,, 
 \label{eq:gauge-metric}
\end{align}
which is a non-trivial constraint on the final result. At leading order in the PM expansion, we can use the free light-body trajectory \eqref{eq:LB_trajectory} to compute
\begin{align}
 &\hspace{-6pt}\langle \delta \boldsymbol{g}_{\mu \nu}(x) \rangle \Big|_{\mathrm{1SF},\mathcal{O}(G_\mathrm{N})} \!\!\!\!= -16 \pi G_\mathrm{N}  \int \mathrm{\hat{d}}^d k\, e^{i k \cdot x}\, G^{(0)\mathrm{ret}}_{\mu \nu \alpha \beta}(k) \tilde{T}^{\alpha \beta}_{\mathrm{L}}(k^{\sigma}) \nonumber \\
 &\qquad = 16 \pi G_\mathrm{N} m_{\mathrm{L}} \left(v_{\mathrm{L} \mu} v_{\mathrm{L}\nu} + \frac{\eta_{\mu \nu}}{d-2}\right) \int \mathrm{\hat{d}}^{d-1} \vec{k}\, \frac{e^{-i \vec{k} \cdot \vec{r}_{\mathrm{L}}}}{|\vec{k}|^2} \nonumber \\
 &\qquad \stackrel{d \to 4}{=}\frac{2 G_\mathrm{N} m_{\mathrm{L}}}{|\vec{r}_{\mathrm{L}}|} \left(2 v_{\mathrm{L} \mu} v_{\mathrm{L}\nu} + \eta_{\mu \nu}\right) \,,
 \label{eq:1SF_metric_G}
\end{align}
which is the linearized Schwarzschild metric generated by the light particle source with $r_{\mathrm{L}}^{\mu} = x^{\mu} - (x \cdot v_{\mathrm{L}}) v_{\mathrm{L}}^{\mu}$. 

While \eqref{eq:1SF_metric} provides an off-shell recursion for the metric, we can also consider the on-shell waveform. Performing the LSZ reduction in curved spacetime is subtle, as when considering the external wavefunctions of the asymptotic graviton state we can either take the plane-wave solution $\sim \boldsymbol{\varepsilon}_{\mu \nu}(k) \exp(i k \cdot x)$ \cite{Goldberger:2004jt} or the one that satisfies the linearized equations of motion \cite{Cheung:2022pdk,Adamo:2017nia,Adamo:2023cfp}. Here we consider the first approach, constructing a convenient Newman-Penrose tetrad basis $(n^{\mu},\bar{n}^{\mu},\boldsymbol{\varepsilon}^{\mu},\boldsymbol{\bar{\varepsilon}}^{\mu})$ at null infinity where we project our asymptotic states.

Considering a detector with velocity $t^\mu$ placed at a spatial distance $r$ from the scattering event in the angular direction determined by $\hat{n}^\mu$ (with $\hat{n} \cdot \hat{n}=0$ and $\hat{n} \cdot t=-1$), we define $x^\mu=u t^\mu+r \hat{n}^\nu$ and take the limit $r \to +\infty$ for fixed $(u,\hat{n})$, obtaining the waveform recursion relation
\begin{align}
\label{eq:1SF_waveform}
 \hspace{-13pt}\langle \delta \boldsymbol{g}^{1/|\vec{x}|}_{\mu \nu}(u,\hat{n}) \rangle \Big|_{\mathrm{1SF}} &\!\!\!\!\!\sim \frac{1}{4 \pi |\vec{x}|}  \int \!\mathrm{\hat{d}} \omega  \, e^{-i \omega u} \, \mathcal{P}_{\mu \nu \alpha \beta}^{(d)} \tilde{T}^{\alpha \beta}(\omega \hat{n}) \,,
\end{align}
in terms of the pseudo stress tensor \cite{Maggiore:2007ulw,Jakobsen:2021smu,Mougiakakos:2021ckm}
\begin{align}
\tilde{T}^{\alpha \beta}(\omega \hat{n}) &= \mathcal{V}_{\mathrm{L}}^{\alpha \beta}(\omega \hat{n})  \\
 &\hspace{-20pt}+ \int \mathrm{\hat{d}}^d k_1\,  \, \mathcal{G}_{2, \mathrm{ret}}^{\alpha \beta \lambda \rho }(\omega \hat{n},k_1)\, G^{(0)\mathrm{ret}}_{\lambda \rho \xi \zeta}(k_1) \mathcal{V}^{\xi \zeta}_{\mathrm{L}}(k_1) + \dots\,,  \nonumber
\end{align}
At leading order in $G_\mathrm{N}$ we obtain 
\begin{align}
 \hspace{-10pt}\langle \boldsymbol{\bar{\varepsilon}}^{\mu} \boldsymbol{\bar{\varepsilon}}^{\nu}\delta \boldsymbol{g}^{1/|\vec{x}|}_{\mu \nu}(u,\hat{n}) \rangle \Big|_{\mathrm{1SF},\mathcal{O}(G_\mathrm{N})} = \frac{4 G_\mathrm{N}}{|\vec{x}|} \frac{m_\mathrm{L} \left(\boldsymbol{\bar{\varepsilon}} \cdot v_\mathrm{L}\right)^2}{(\hat{n} \cdot v_\mathrm{L})}\,, 
 \label{eq:linear_memory}
\end{align}
which is the linear memory contribution due to the light particle. At 1SF order, given that our perturbation is on the background spacetime, the BMS frame---which we refer to as the ``intrinsic self-force BMS frame''---is adapted to the background metric. Technically speaking, by including the background metric after the shift  \eqref{eq:shift_metric}, at this order the BMS frame corresponds to the usual amplitude or MPM intrinsic frame \cite{Strominger:2014pwa,Jakobsen:2021smu,Jakobsen:2021lvp,DiVecchia:2022owy,Veneziano:2022zwh,Mougiakakos:2021ckm,Herderschee:2023fxh,Georgoudis:2023eke,Bohnenblust:2023qmy,Adamo:2024oxy,Elkhidir:2024izo,Georgoudis:2024pdz,Bini:2024rsy}. However, in general it might be different beyond 1SF order \cite{Warburton:2024xnr}.

At order $G_\mathrm{N}^2$ and beyond, the 1SF metric \eqref{eq:1SF_metric} becomes sensitive to the graviton 2-point (time-ordered) correlator 
\begin{align}
\hspace{-10pt}\langle \mathcal{T}\delta \boldsymbol{g}_{\mu \nu}(x) \delta \boldsymbol{g}_{\alpha \beta}(y) \rangle \Big|_{\mathrm{1SF}} \!\!\!\!\!=\! \frac{1}{(-\bar{g})} \frac{\delta^2 \mathbf{W}^{\mathrm{1SF}}[J]}{i^2 \delta \mathbf{J}^{\mu \nu}(x) \delta \mathbf{J}^{\alpha \beta}(y)} \Big|_{\mathbf{J}=0}\!\!\,,
\label{eq:graviton_graviton_1SF}
\end{align}
which generalizes the usual definition of the graviton propagator on the background spacetime by including the recoil contributions of the heavy source.  Note that this correlator obeys the gauge condition \eqref{eq:gauge-metric} on either coordinate $(x,y)$ and its associated index group. 

An important amount of work has been done in relating the result of BH perturbation theory in vacuum with the PM expansion at 1SF order using both the MST formalism and the Nekrasov-Shatashvili partition function \cite{Mano:1996mf,Mano:1996gn,Mano:1996vt,Mino:1997bx,Bautista:2021wfy,Bautista:2022wjf,Bautista:2023sdf,Ivanov:2022qqt,Aminov:2023jve,Fucito:2023afe,Ivanov:2024sds,Fucito:2024wlg,Bini:2024icd,Cipriani:2025ikx}. The advantage of sticking with the 1SF waveform---neglecting therefore 2SF contributions and beyond---is that its analytic structure is expected to be simpler \cite{Mino:1997bx,Fucito:2023afe,Georgoudis:2023eke,Fucito:2024wlg,Bini:2024rsy,Cipriani:2025ikx}, as it obeys a single differential equation \eqref{eq:deltag_1SF}. However, there are still three outstanding puzzles. The first is how the boundary conditions for the wave scattering problem for BH vacuum solutions---involving the horizon---are related to the ones for particle-generated backgrounds with the recoil contribution. The second is how to perform the matching with a point-particle description beyond 1SF order. The final puzzle is how to better connect the analytic resummation of the Post-Minkowskian series with strong-field effects both for the 1-point and 2-point function, exploiting the simplicity of the 1SF theory. As an effort in these directions, we will now study the analytic properties of the Compton amplitude at loop level.

%----------------------------------------------------------------------
\section{Kerr Compton amplitude: Post-Minkowskian vs self-force}
\label{sec:1SFCompton}
%----------------------------------------------------------------------

In this section, we study the 1SF Compton amplitude appearing in the waveform recursion relation \eqref{eq:1SF_waveform} and compare it with the one-loop Compton amplitude for Kerr at quadratic order in spin. Physically, for the waveform only the wave regime of the Compton is relevant, which we will explore in detail emphasizing the analytic structure and the relation between the PM and GSF expansion. Finally, to understand the role of the resummation in the recursion relation, we will also consider the geometric optics regime of the 1SF Compton at all loop orders, uncovering the relation between the radius of convergence of the PM expansion and the critical angular momentum corresponding to the photon ring radius.

%----------------------------------------------------------------------
\subsection*{The wave regime: 1SF Compton recursion relation}
\label{sec:1SF_recursion}
%----------------------------------------------------------------------

Let us now compute the 1SF Compton $\mathcal{M}_{1 \to 1}$ defined from $\langle  \mathcal{T} \delta \boldsymbol{g}_{\mu \nu}(x) \delta \boldsymbol{g}_{\alpha \beta}(y) \rangle$ in \eqref{eq:graviton_graviton_1SF}, discussing its relation to the flat space classical Compton amplitude $\mathcal{A}_{2 \to 2}$ derived from the LSZ reduction of $\langle \mathcal{T} h_{\mu \nu}(x) h_{\alpha \beta}(y) \rangle$. The 1SF Compton (resp. flat space Compton) at $\mathcal{O}(G_\mathrm{N}^{n+1})$ is denoted by $\mathcal{M}^{(n)}_{1 \to 1}$ (resp. $\mathcal{A}^{(n)}_{2 \to 2}$) and we strip off the delta function $\hat{\delta}(v_\mathrm{H} \cdot k_1 + v_\mathrm{H} \cdot k_2)$ (resp. $\hat{\delta}^4(k_1 + k_2 + p_{\mathrm{H}} + p'_{\mathrm{H}})$).

As explained in section \ref{eq:strongvsweak}, the corresponding 1-point functions are related by
\begin{align}
\label{eq:shift2}
    \langle \kappa h_{\mu \nu} \rangle = \bar{H}_{\mu \nu} + \langle \lambda \delta \boldsymbol{g}_{\mu \nu} \rangle\,, 
\end{align}
where $\bar{H}_{\mu \nu}$ is the background field. Crucially, this shift has direct implications also for the off-shell 2-point function, although as we will see the on-shell 1SF Compton and the flat space Compton will still agree at loop order. 

Now we turn our attention to the 1SF Compton amplitude $\mathcal{M}_{1 \to 1}(k_1,k_2)$. We start by focusing on the so-called wave or Born regime\footnote{See section II of Ref.~\cite{Correia:2024jgr} for a very nice review.}, which corresponds to the kinematic region $|\vec{q}|,\omega \ll m_\mathrm{H}$ where 
\begin{equation}
    |\vec{q}|=|\vec{k}_1+\vec{k}_2| = \sqrt{-q^2}\,, \quad \omega = v_\mathrm{H} \cdot k_1 = -v_\mathrm{H} \cdot k_2\;,
\end{equation}
denotes the exchanged momentum and frequency of external graviton momenta, respectively. 
\begin{widetext}
   \begin{figure}[t]
    \centering
    \begin{minipage}{.5\textwidth}
         \centering
        \includegraphics[scale=1]{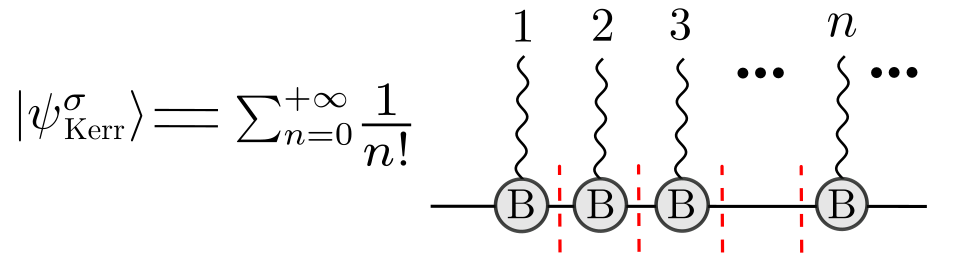}
    \caption{The Kerr background $\left|\psi_{\mathrm{Kerr}}^\sigma\right\rangle$ is described by an off-shell coherent state of virtual gravitons.}
    \label{fig:Kerr_coherent}
    \end{minipage}%
    \begin{minipage}{0.5\textwidth}
        \centering
        \includegraphics[scale=1]{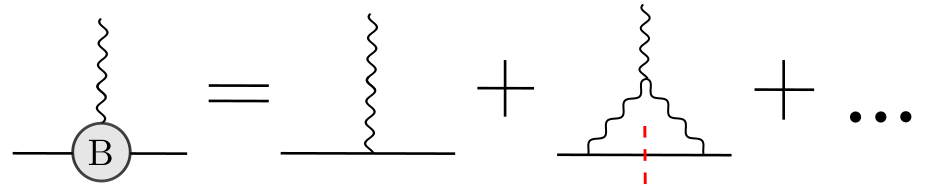}
    \caption{The stress tensor for Kerr spacetime is given by a sum of perturbative diagrams.}
    \label{fig:stres-tensorKerr}
    \end{minipage}
\end{figure}
\end{widetext}
At tree-level, we can directly combine the single insertion of the background and the recoil vertex to obtain
\begin{align}
\label{eq:1SFCompton_tree_res}
    \hspace{-0.25cm}i\mathcal{M}^{(0)}_{1 \to 1}(k_1,k_2)& = \boldsymbol{\varepsilon}_{\mu \nu}(k_1) \mathcal{G}_2^{\mu \nu \alpha \beta}(k_1,k_2)  \boldsymbol{\varepsilon}_{\alpha \beta}(k_2)  \\
    &\hspace{-5pt} \stackrel{\mathcal{O}(G_\mathrm{N})}{=}\frac{4\pi i G_\mathrm{N} m_\mathrm{H} \mathcal{J}_0^2}{|\vec{q}|^2 \left(k_1 \cdot v_\mathrm{H}\right)^2} \sum_{n = 0}^{2}\frac{1}{n!}\left(\frac{\mathcal{J}_1}{\mathcal{J}_0}\right)^n\,, \nonumber 
\end{align}
where the spin multipoles are defined as 
\begin{align}
& \mathcal{J}_0=-2 v_\mathrm{H} \cdot F_1 \cdot F_2 \cdot v_\mathrm{H}\,, \\
& 2i \mathcal{J}_1=2m_\mathrm{H}(k_1 \cdot F_2 \cdot v_\mathrm{H}) (\bar{S}_\mathrm{H} \cdot F_1)\\
&\qquad -m_\mathrm{H}(\left(k_1-k_2\right) \cdot v_\mathrm{H}) (\bar{S}_\mathrm{H} \cdot F_1 \cdot F_2)+(1 \leftrightarrow 2)\,, \nonumber 
\end{align}
in terms of the field strength $F^{\mu \nu}_i(k_i) = 2 k^{[\mu}_i \boldsymbol{\varepsilon}^{\nu]}_i(k_i)$. Therefore, we obtain a correspondence with the usual spinning flat space Compton at tree-level order \cite{DeAngelis:2023lvf,Bautista:2021wfy}
\begin{align}
    \frac{1}{2 m_\mathrm{H}} \mathcal{A}^{(0)}_{2 \to 2}(k_1,k_2)  &\stackequal{\hbar \to 0}{\varepsilon_{\mu \nu} \to \boldsymbol{\varepsilon}_{\mu \nu}} \mathcal{M}^{(0)}_{1 \to 1}(k_1,k_2) \,,
    \label{eq:matching-flat}
\end{align}
where the heavy particle momentum is parametrized as  
\begin{align}
   p^{\mu}_\mathrm{H} = m_\mathrm{H} v^{\mu}_\mathrm{H} - \frac{q^{\mu}}{2} \,, \qquad (p')^{\mu}_\mathrm{H} = -m_\mathrm{H} v^{\mu}_\mathrm{H} -  \frac{q^{\mu}}{2} \,.
\end{align}
Diagrammatically, the recoil contribution can be identified with the effective contact vertex coming from the heavy mass expansion of the following flat space diagrams
\begin{align}
\label{eq:recoilG}
    & \begin{tikzpicture}[baseline=(current bounding box.center)]
              \begin{feynman}[every blob={/tikz/fill=gray!30,/tikz/inner sep=3pt}]
                \vertex[blob] (m) at (0,0) {$H$};
                \vertex (a) at (-1.5,0) {};
                \vertex (b) at (1.5,0) {};
                \diagram* {(a) -- [boson] (m) -- [boson] (b)};
              \end{feynman} 
    \end{tikzpicture} 
    \Bigg|_{\mathcal{O}(G_\mathrm{N})}  \\
    & \hspace{-8pt}\sim \Bigg[\begin{tikzpicture}[baseline=(current bounding box.center)]
      \begin{feynman}
        % Define vertices, their positions, and their labels
        \vertex (v1) at (0,-0.25);
        \vertex (l1) at (1,-0.25) {};
        \vertex (l2) at (0.45,0.75) {};
        \vertex (r1) at (-1,-0.25) {};
        \vertex (r2) at (-0.45,0.75) {};
        % Draw Feynman diagram
        \diagram*{
        (v1) -- [plain] (l1),
        (v1) -- [boson] (r2),
        (r1) -- [plain] (v1),
        (l2) -- [boson] (v1),
        };
      \end{feynman}
    \end{tikzpicture}\!\! + \!\!\begin{tikzpicture}[baseline=(current bounding box.center)]
      \begin{feynman}
        % Define vertices, their positions, and their labels
        \vertex (v1) at (0.45,-0.25);
        \vertex (v2) at (-0.45,-0.25);
        \vertex (l1) at (1,-0.25) {};
        \vertex (l2) at (0.45,0.75) {};
        \vertex (r1) at (-1,-0.25) {};
        \vertex (r2) at (-0.45,0.75) {};
        % Draw Feynman diagram
        \diagram*{
        (v1) -- [plain] (l1),
        (l2) -- [boson] (v2),
        (r1) -- [plain] (v2),
        (r2) -- [boson] (v1),
        (v1) -- [plain] (v2),
        };
      \end{feynman}
    \end{tikzpicture}
    \!\!+\!\! \begin{tikzpicture}[baseline=(current bounding box.center)]
        \begin{feynman}
        % Define vertices, their positions, and their labels
        \vertex (v1) at (0.45,-0.25);
        \vertex (v2) at (-0.45,-0.25);
        \vertex (l1) at (1,-0.25) {};
        \vertex (l2) at (0.45,0.75) {};
        \vertex (r1) at (-1,-0.25) {};
        \vertex (r2) at (-0.45,0.75) {};
        % Draw Feynman diagram
        \diagram*{
        (v1) -- [plain] (l1),
        (l2) -- [boson] (v1),
        (r1) -- [plain] (v2),
        (r2) -- [boson] (v2),
        (v2) -- [plain] (v1),
        };
      \end{feynman}
    \end{tikzpicture} \Bigg]\Bigg|_{\mathcal{O}(m_{\mathrm{H}}^2)}, \nonumber
\end{align}
with the equality holding on-shell. The background insertion, on the other hand, contains an explicit propagator and is given on-shell by the usual t-channel contribution
\begin{align}
    \label{eq:backgroundG}
    & \hspace{-7pt}\begin{tikzpicture}[baseline=(current bounding box.center)]
              \begin{feynman}[every blob={/tikz/fill=gray!30,/tikz/inner sep=3pt}]
                \vertex[blob] (m) at (0,0) {$G$};
                \vertex (a) at (-1.5,0) {};
                \vertex (b) at (1.5,0) {};
                \diagram* {(a) -- [boson] (m) -- [boson] (b)};
              \end{feynman} 
    \end{tikzpicture} 
    \Bigg|_{\mathcal{O}(G_\mathrm{N})}\!\!\!\sim \!\!\! \begin{tikzpicture}[baseline=(current bounding box.center)]
      \begin{feynman}
        % Define vertices, their positions, and their labels
        \vertex (v1) at (-0,-0.25);
        \vertex (v2) at (0,0.3);
        \vertex (l1) at (1,-0.25) {};
        \vertex (l2) at (0.45,0.9) {};
        \vertex (r1) at (-1,-0.25) {};
        \vertex (r2) at (-0.45,0.9) {};
        % Draw Feynman diagram
        \diagram*{
        (v1) -- [plain] (l1),
        (r2) -- [boson] (v2),
        (r1) -- [plain] (v1),
        (l2) -- [boson] (v2),
        (v1) -- [boson] (v2),
        };
      \end{feynman}
    \end{tikzpicture} \Bigg|_{\mathcal{O}(m_{\mathrm{H}}^2)}.
\end{align}
The next object to study is the 1SF Compton amplitude at one-loop, which is given by the combination of the single background vertex at $\mathcal{O}(G_\mathrm{N}^2)$ and the iteration of the combined background and recoil insertion at $\mathcal{O}(G_\mathrm{N})$
\begin{align}
\label{eq:1SF_one-loop}
&i\mathcal{M}^{(1)}_{1 \to 1}(k_1,k_2) = \boldsymbol{\varepsilon}_{\mu \nu}(k_1)  \mathcal{G}_{2}^{\mu \nu \alpha \beta}(k_1,k_2) \boldsymbol{\varepsilon}_{\alpha \beta}(k_2) + I^{\mathrm{s-iter}}_{1 \to 1}\nonumber \\
& I^{\mathrm{s-iter}}_{1 \to 1}=  \int \hat{\mathrm{d}}^d \ell_1\, \boldsymbol{\varepsilon}_{\mu \nu}(k_1) \mathcal{G}_2^{\mu \nu \rho \lambda}(k_1,\ell_1) \nonumber \\[-0.2cm]
& \qquad \qquad \qquad \times \frac{i \mathcal{P}^{(d)}_{\rho \lambda \xi \zeta}}{\ell_1^2 + i \epsilon} \mathcal{G}_2^{\xi \zeta \alpha \beta}(\ell_1,k_2) \boldsymbol{\varepsilon}_{\alpha \beta}(k_2)\,, 
\end{align}
where we parametrize the external graviton legs as
\begin{equation}
    k_1^{\mu} = k^{\mu} + \frac{q^{\mu}}{2}\,, \quad k_2^{\mu} = -k^{\mu} + \frac{q^{\mu}}{2}\,, \quad v_\mathrm{H} \cdot q = 0\,.
\end{equation}
together with the gauge choice $v_\mathrm{H}\cdot \boldsymbol{\varepsilon}_i = 0$.
The 1SF Compton at one-loop is of the form
\begin{align}
&\hspace{-6pt}i\mathcal{M}^{(1)}_{1 \to 1}(k_1,k_2)\! =\!\frac{1}{2 m_{\mathrm{H}}}\!\Big[ \boldsymbol{c}_{\scriptscriptstyle\Box}(k,q,\bar{S}_\mathrm{H}) (\mathcal{I}_{\scriptscriptstyle\Box}(\omega,q) + \mathcal{I}_{\scriptscriptstyle\boxtimes}(\omega,q)) \nonumber \\
  &\qquad\qquad\qquad\qquad \quad +\boldsymbol{c}_{\scriptscriptstyle\bigtriangleup}(k,q,\bar{S}_\mathrm{H}) \, \mathcal{I}_{\scriptscriptstyle\bigtriangleup}(q)   \nonumber \\
  &\qquad\qquad\qquad\qquad \quad +\boldsymbol{c}_{\scriptscriptstyle\bubble}(k,q,\bar{S}_\mathrm{H}) \, \mathcal{I}_{\scriptscriptstyle\bubble}(\omega) \Big] \,,
\label{eq:1SF_one-loop_res}
\end{align}
where we separate the analytic dependence on $q^2$ in the coefficients\footnote{Our notation anticipates a relation to the cuts of the flat space Compton $\mathcal{A}^{(1)}_{2 \to 2}$, which will be discussed later.} from the  master integrals $(\mathcal{I}_{\scriptscriptstyle\Box},\mathcal{I}_{\scriptscriptstyle\boxtimes},\mathcal{I}_{\scriptscriptstyle\bigtriangleup},\mathcal{I}_{\scriptscriptstyle\bubble})$, which are readily evaluated around $d=4-2\epsilon$ 
\begin{align}
    &\hspace{-0.2cm}\mathcal{I}_{\scriptscriptstyle\Box}(q,\omega) + \mathcal{I}_{\scriptscriptstyle\boxtimes}(q,\omega) = i \frac{\bar{\mu}^{2\epsilon}}{2\pi} \int \frac{\hat{\mathrm{d}}^{d-1} \vec{\ell} }{|\vec{k}_1 - \vec{\ell}|^2 |\vec{k}_2 + \vec{\ell}|^2 (\omega^2 - |\vec{\ell}|^2)} \nonumber \\
    &\hspace{2.7cm} = -\frac{i}{16 \pi^2 \omega |\vec{q}|^2} \Bigg[\frac{1}{\epsilon} - \operatorname{log}\frac{|\vec{q}|^2}{\mu^{2}}\Bigg] + \mathcal{O}(\epsilon)\,, \nonumber \\
    &\mathcal{I}_{\scriptscriptstyle\bigtriangleup}(q) =  \bar{\mu}
    ^{2\epsilon} \int \hat{\mathrm{d}}^{d} \ell \,\frac{\delta(v_H \cdot \ell)}{ \ell^2 (\ell-q)^2} = \frac{1}{16 \pi |\vec{q}|} + \mathcal{O}(\epsilon)\,,  \label{eq:master_heavy_mass} \\
    & \mathcal{I}_{\scriptscriptstyle\bubble}(\omega) =  i \bar{\mu}^{2\epsilon} \int \hat{\mathrm{d}}^{d} \ell\, \,\frac{\delta\left(v_{\mathrm{H}} \cdot \ell \right)}{(\ell + k_1)^2} \nonumber \\
    &\hspace{1.1cm}  = -\frac{i \omega}{8\pi^2}\Big[1 + \epsilon\Big(i\pi + 2 - \operatorname{log}\frac{4 \omega^2}{\mu^2}\Big)\Big] + \mathcal{O}(\epsilon^2)\,, \nonumber
\end{align}
where $\bar{\mu}^2 = \mu^2 e^{\gamma_{\text{E}}} /4\pi$ defines the usual modified subtraction scheme when performing dimensional regularization. More precisely, these integrals correspond to cut box, cut triangle and cut bubble master integrals respectively. The value of all coefficients in the wave regime expanded to the leading order in $\epsilon$ can be found in appendix~\ref{sec:app4}. 

We now discuss the interpretation of \eqref{eq:1SF_one-loop_res}. The background contribution arises entirely from the quadratic graviton action \eqref{eq:Gint}, and it has the interpretation of a graviton scattering off a fixed Kerr background at large distances. This is manifest from the equations of motion \eqref{eq:deltag_1SF_new} in the absence of the recoil and the source term, and it can also be verified by constructing the off-shell coherent state describing the Kerr black hole up to quadratic order in spin. Following Refs.~\cite{Adamo:2022ooq,Monteiro:2020plf}, we define (see Fig.~\ref{fig:Kerr_coherent})
\begin{align}
\label{eq:coherent-state-gravitons}
&\hspace{-8pt}\left|\psi_{\mathrm{Kerr}}^\sigma\right\rangle \sim \exp \Big[\int \frac{\hat{\mathrm{d}}^d \ell_1}{\ell_1^2+i \epsilon} \hat{\delta}\left(2 p_H \cdot \ell_1\right) \nonumber \\
& \qquad \qquad \quad \times i \varepsilon^\sigma_{\mu \nu}(\ell_1) \tilde{T}_{\mathrm{H},\mathrm{spin}}^{\mu \nu}(\ell_1) A_\sigma^{\dagger}(\ell_1)\Big]|p_H\rangle \,,  
\end{align}
where $A_\sigma^{\dagger}(\ell_1)$ is a placeholder for the operator creating a virtual graviton and $\tilde{T}_{\mathrm{H},\mathrm{spin}}^{\mu \nu}(\ell_1)$ is the momentum space effective stress tensor for a massive spinning particle. While the linearized term has been discussed earlier \eqref{eq:stresstensor_spin}, building on Refs.~\cite{Jakobsen:2020ksu,Mougiakakos:2020laz,Mougiakakos:2024nku} we can identify higher-loop contributions to the effective stress tensor with the diagrams in Fig.~\ref{fig:stres-tensorKerr} where an off-shell graviton is emitted from the massive spinning particle. Using this intuition, we can verify that the tree-level background contribution arise from gluing a single insertion of the coherent state vertex at $G_\mathrm{N}$ with the wiggly line representing the graviton; see \eqref{eq:backgroundG}. At order $G_\mathrm{N}^2$, we need to combine\footnote{Naively, applying the same logic to the one-loop diagram in Fig.~\ref{fig:stres-tensorKerr}, it would be tempting to identify the single background insertion at $G_\mathrm{N}^2$ with the triangle numerator $c_{\scriptscriptstyle\bigtriangleup}^{{\scriptscriptstyle\bigtriangleup}\text{-cut}} \mathcal{I}_{\scriptscriptstyle\bigtriangleup}(q)$ in \eqref{eq:1SF_one-loop_res}, as was recently observed for the scattering of massless scalar fields in a black hole background \cite{Correia:2024jgr}. However, the story is more intricate for gravitons because of the mixing of various contributions of the metric insertion in the action \eqref{eq:1SF_action}, preventing the identification of the single background insertion with flat space diagrams unless we combine it with its iteration as in \eqref{eq:graviton_iter}. } both the single background insertion and its iteration to get a correspondence with flat space diagrams 
\begin{align}
\label{eq:graviton_iter}
&\begin{tikzpicture}[baseline=(current bounding box.center)]
    \begin{feynman}[every blob={/tikz/fill=gray!30,/tikz/inner sep=3pt}]
                \vertex[blob] (m) at (0,0) {$G$};
                \vertex (a) at (-1.2,0) ;
                \vertex (b) at (1.2,0) ;
                \diagram* {(a) -- [boson, text=black] (m) -- [boson, text=black] (b)};
              \end{feynman} 
\end{tikzpicture}
      +  
      \begin{tikzpicture}[baseline=(current bounding box.center)]
              \begin{feynman}[every blob={/tikz/fill=gray!30,/tikz/inner sep=3pt}]
                \vertex[blob] (m) at (0,0) {$G$};
                \vertex[blob] (n) at (1.6,0) {$G$};
                \vertex (a) at (-1.2,0);
                \vertex (b) at (2.8,0);
                \diagram* {(a) -- [boson, text=black] (m) -- [boson, text=black] (n)-- [boson, text=black] (b)};
              \end{feynman} 
    \end{tikzpicture}  \\
      &\stackrel{\mathcal{O}(G_\mathrm{N}^2)}{\sim}
       \begin{tikzpicture}[baseline=(current bounding box.center)]
      \begin{feynman}[every blob={/tikz/fill=gray!30,/tikz/inner sep=2pt}]
        %Define blobs and their positions
        \vertex[blob] (b1) at (0,0.7) {$\mathcal{\tilde{A}}^{\scalebox{.6}{(0)}}_3$};
        \vertex[blob] (b2) at (0,-0.7) {$\mathrm{B}$};
        % Define vertices, their positions, and their labels
        \vertex (l1) at (-1.5,0.7) {};
        \vertex (l2) at (-1.5,-0.7) {};
        \vertex (r1) at (1.5,0.7) {};
        \vertex (r2) at (1.5,-0.7) {};
        \vertex (b3) at (0,0.6) {};
        \vertex (b4) at (0,-0.6) {};
        % Draw Feynman diagram
        \diagram*{
        (b1) -- [boson] (l1),
        (b1) -- [boson] (r1),
        (l2) -- [plain] (b2),
        (r2) -- [plain] (b2),
        (b1) -- [boson] (b2),
        };
      \end{feynman}
    \end{tikzpicture}  + \begin{tikzpicture}[baseline=(current bounding box.center)]
       \begin{feynman}[every blob={/tikz/fill=gray!30,/tikz/inner sep=2pt}]
        %Define blobs and their positions
        \vertex[blob] (b1) at (-0.75,-0.7) {$\mathrm{B}$};
        \vertex[blob] (b2) at (0.75,-0.7) {$\mathrm{B}$};
        \vertex[blob] (b3) at (0,0.7) {$\mathcal{\tilde{A}}^{\scalebox{.6}{(0)}}_4$};
        % Define vertices, their positions, and their labels
        \vertex (l1) at (-1.75,0.7) {};
        \vertex (l2) at (-1.75,-0.7) {};
        \vertex (r1) at (1.75,0.7) {};
        \vertex (r2) at (1.75,-0.7) {};
        %% Draw dashed line
        \draw [dashed, color = red] ($(0,-0.4)$) -- ($(0,-1)$);
        % Draw Feynman diagram
        \diagram*{
        (b3) -- [boson] (l1),
        (b3) -- [boson] (r1),
        (b1) -- [plain] (l2),
        (b2) -- [plain] (r2),
        (b1) -- [plain] (b2),
        (b1) -- [boson] (b3),
        (b2) -- [boson] (b3),
        (b3) -- [draw=none, text=black] (b1),
        (b3) -- [draw=none, text=black] (b2),
        };
      \end{feynman}
    \end{tikzpicture}   \,, \nonumber
\end{align}
where we introduced the 3-point $\mathcal{\tilde{A}}^{(0)}_3$ and 4-point $\mathcal{\tilde{A}}^{(0)}_4$ graviton tree-level amplitudes (see e.g. Ref.~\cite{Holstein:2006bh}).

The iteration term of the 1SF Compton \eqref{eq:1SF_one-loop}, including both the background and the recoil, then corresponds to the s-channel iteration of the classical piece of the tree-level flat space Compton amplitude $\mathcal{A}^{(0)}_{2 \to 2}$ 
\begin{align}
    \label{eq:s-channel_iden}
      %  I^{\mathrm{s-iter}}_{1 \to 1} = 
      & \begin{tikzpicture}[baseline=(current bounding box.center)]
              \begin{feynman}[every blob={/tikz/fill=gray!30,/tikz/inner sep=3pt}]
                \vertex[blob] (m) at (0,0) {$\mathcal{G}_2$};
                \vertex[blob] (n) at (2,0) {$\mathcal{G}_2$};
                \vertex (a) at (-1.5,0);
                \vertex (b) at (3.5,0);
                \diagram* {(a) -- [boson, text=black, momentum={$k_1$}] (m) -- [boson, text=black, momentum={$\ell_1$}] (n)-- [boson, text=black, reversed momentum={$k_2$}] (b)};
              \end{feynman} 
    \end{tikzpicture}\Bigg|_{\mathcal{O}(G_\mathrm{N}^2)}  \\
      &\hspace{1cm} \sim 
       \begin{tikzpicture}[baseline=(current bounding box.center)]
      \begin{feynman}[every blob={/tikz/fill=gray!30,/tikz/inner sep=2pt}]
        %Define blobs and their positions
        \vertex[blob] (b1) at (-0.75,-1) {$\mathcal{A}^{\scalebox{.6}{(0)}}_{4}$};
        \vertex[blob] (b2) at (1.75,-1) {$\mathcal{A}^{\scalebox{.6}{(0)}}_{4}$};
        % Define vertices, their positions, and their labels
        \vertex (l1) at (-1.75,0) {};
        \vertex (l2) at (-1.75,-1) {};
        \vertex (r1) at (2.75,0) {};
        \vertex (r2) at (2.75,-1) {};
        %% Draw dashed line
        \draw [dashed, color = red] ($(0.46,-0.4)$) -- ($(0.46,-1.2)$);
        % Draw Feynman diagram
        \diagram*{
        (b1) -- [boson, rmomentum'={$k_1$}] (l1),
        (b2) -- [boson, rmomentum={$k_2$}] (r1),
        (b1) -- [plain] (l2),
        (b2) -- [plain] (r2),
        (b1) -- [plain] (b2),
        (b1) -- [boson, quarter left,looseness=0.5] (b2),
        (b2) -- [text=black, rmomentum'={[arrow shorten=0.25,arrow distance=0.7cm] $\ell_1$}] (b1),
        };
      \end{feynman}
    \end{tikzpicture} \Bigg|_{\mathcal{O}(m_\mathrm{H}^3)} \,, \nonumber
\end{align}
in line with the expectations from \eqref{eq:recoilG} and \eqref{eq:backgroundG}. Thus we find a perfect correspondence between the two one-loop Compton amplitudes\footnote{In a previous version of this paper, we found a mismatch at one-loop order related to different contact terms. This has now been fully resolved using the minimal coupling spinning metric in \eqref{eq:MC_metric}.}
\begin{align}
    \frac{1}{2 m_\mathrm{H}} \mathcal{A}^{(1)}_{2 \to 2}(k_1,k_2)  &\stackequal{\hbar \to 0}{\varepsilon_{\mu \nu} \to \boldsymbol{\varepsilon}_{\mu \nu}} \mathcal{M}^{(1)}_{1 \to 1}(k_1,k_2) \,,
    \label{eq:matching-flat2}
\end{align}
generalizing our previous tree-level matching \eqref{eq:matching-flat}. Now we will show this explicitly by performing a direct calculation with on-shell amplitudes.

\begin{figure}[t]
    \centering
    \hspace*{-1.5cm}
    \begin{subfigure}[t]{0.2\textwidth}
    \centering
    \begin{tikzpicture}[baseline=(current bounding box.center)]
      \begin{feynman}[every blob={/tikz/fill=gray!30,/tikz/inner sep=2pt}]
         %Define blobs and their positions
         \vertex[blob] (b1) at (-0.75,-1) {$\mathcal{A}^{\scalebox{.6}{(0)}}_3$};
         \vertex[blob] (b2) at (0.75,-1) {$\mathcal{A}^{\scalebox{.6}{(0)}}_3$};
         \vertex[blob] (b3) at (0,1) {$\tilde{\mathcal{A}}^{\scalebox{.6}{(0)}}_4$};
         % Define vertices, their positions, and their labels
         \vertex (l1) at (-1.75,1) {};
         \vertex (l2) at (-1.75,-1) {};
         \vertex (r1) at (1.75,1) {};
         \vertex (r2) at (1.75,-1) {};
         %% Draw dashed line
         \draw [dashed, color = red] ($(0,-0.7)$) -- ($(0,-1.3)$);
         \draw [dashed, color = red] ($(0.13,-0.2)$) -- ($(0.6,0)$);
         \draw [dashed, color = red] ($(-0.13,-0.2)$) -- ($(-0.6,0)$);
         % Draw Feynman diagram
         \diagram*{
         (b3) -- [boson, rmomentum'={$k_1$}] (l1),
         (b3) -- [boson, rmomentum={$k_2$}] (r1),
         (b1) -- [plain] (l2),
         (b2) -- [plain] (r2),
         (b1) -- [plain] (b2),
         (b1) -- [boson] (b3),
         (b2) -- [boson] (b3),
         (b3) -- [draw=none, text=black, momentum'={[arrow shorten=0.25] $\ell_1$}] (b1),
         (b3) -- [draw=none, text=black, momentum={[arrow shorten=0.25] $\ell_2$}] (b2),
         };
       \end{feynman}
    \end{tikzpicture}
    \end{subfigure}
    \begin{subfigure}[t]{0.2\textwidth}
    \centering
    \begin{tikzpicture}[baseline=(current bounding box.center)]
      \begin{feynman}[every blob={/tikz/fill=gray!30,/tikz/inner sep=2pt}]
         %Define blobs and their positions
         \vertex[blob] (b1) at (-0.75,-1) {$\mathcal{A}^{\scalebox{.6}{(0)}}_{4}$};
         \vertex[blob] (b2) at (1.75,-1) {$\mathcal{A}^{\scalebox{.6}{(0)}}_{4}$};
         % Define vertices, their positions, and their labels
         \vertex (l1) at (-1.75,0) {};
         \vertex (l2) at (-1.75,-1) {};
         \vertex (r1) at (2.75,0) {};
         \vertex (r2) at (2.75,-1) {};
         %% Draw dashed line
         \draw [dashed, color = red] ($(0.46,-0.4)$) -- ($(0.46,-1.1)$);
         % Draw Feynman diagram
         \diagram*{
         (b1) -- [boson, rmomentum'={$k_1$}] (l1),
         (b2) -- [boson, rmomentum={$k_2$}] (r1),
         (b1) -- [plain] (l2),
         (b2) -- [plain] (r2),
         (b1) -- [plain] (b2),
         (b1) -- [boson, quarter left,looseness=0.5] (b2),
         (b2) -- [text=black, rmomentum'={[arrow shorten=0.25,arrow distance=0.7cm] $\ell_1$}] (b1),
         };
       \end{feynman}
    \end{tikzpicture}
    \end{subfigure}
    \caption{The t-channel (left) and s-channel (right) cut configurations contributing to the classical one-loop flat space Compton.}
    \label{fig: one-loop-diag}
\end{figure}

%----------------------------------------------------------------------
\subsection*{1SF Compton vs flat space Compton at one-loop}
\label{sec:one_loop-Compton}
%----------------------------------------------------------------------

The classical one-loop Compton amplitude in flat space can be constructed using generalized unitarity~\cite{Bern:1994zx, Bern:1994cg, Bern:1997sc, Britto:2004nc}. All relevant contributions are entirely fixed by t-channel and s-channel cut configurations, as shown in Fig.~\ref{fig: one-loop-diag}.

Accordingly, the amplitude takes the form
\begin{align}
        \label{eq: flat space-one-loop}
        & \hspace{0.5cm} i \mathcal{A}^{(1)}_{2 \to 2}  = c_{\scriptscriptstyle\Box}(\mathcal{I}_{\scriptscriptstyle\Box} + \mathcal{I}_{\scriptscriptstyle\boxtimes}) + c_{\scriptscriptstyle\bigtriangleup}\mathcal{I}_{\scriptscriptstyle\bigtriangleup}  + c_{\scriptscriptstyle\bubble} \mathcal{I}_{\scriptscriptstyle\bubble}\,, 
\end{align}
where all other contributions are $\mathcal{O}(m_\mathrm{H}^2)$ suppressed. The coefficients expanded around $d = 4 - 2\epsilon$ can be found in the ancillary file, while the integrals values are \eqref{eq:master_heavy_mass}. Here the coefficients $c_{\scriptscriptstyle\Box}$ and $c_{\scriptscriptstyle\bigtriangleup}$ are fixed by the t-channel cut. This requires the 3-point and 4-point graviton tree-level amplitudes~\cite{Holstein:2006bh}, as well as the 3-point amplitude for two massive spinning particles expanded to quadratic order in spin~\cite{Bern:2020buy,Bjerrum-Bohr:2023jau,Johansson:2019dnu}
\begin{equation}
    \begin{split}
        & \quad \qquad \mathcal{A}_3^{(0)}(k) = - i \kappa m_\mathrm{H}^2 (v_\mathrm{H} \cdot \varepsilon) (\text{w} \cdot \varepsilon)\,, \\
        & \hspace{0.3cm} \text{w}^\mu = v_\mathrm{H}^\mu + i \epsilon^{\mu \nu \rho \sigma}k_\nu v_{\mathrm{H}\rho} a_{\mathrm{H} \sigma} + \frac{1}{2}(k \cdot a_\mathrm{H})^2 v_\mathrm{H}^\mu\;,
    \end{split}
\end{equation}
where $k^{\mu}$ is the momentum of the incoming graviton leg and $a^\mu_\mathrm{H}$ is the normalized spin vector associated to the massive spinning legs. Meanwhile, the coefficient $c_{\scriptscriptstyle\bubble}$ is fixed by the s-channel cut, which requires the classical spinning tree-level Compton in \eqref{eq:matching-flat}\footnote{Notice that we have used the heavy-mass expanded Compton \cite{Brandhuber:2021eyq,Brandhuber:2023hhy}, but we have explicitly confirmed in the spinless case that we get the same coefficients with a full quantum calculation.}. The result is then projected into a suitable basis using the {\tt FIRE6}~\cite{Smirnov:2019qkx} program, thereby giving \eqref{eq: flat space-one-loop}. Importantly, both cut configurations fix $c_{\scriptscriptstyle\Box}$, thereby serving as a consistency check for the computation. 

To make a connection with the earlier analysis of the 1SF amplitude, we note that all the master integral coefficients are in agreement with the single insertion of the background vertex and its iteration by formally replacing $\varepsilon_{\mu \nu} \to \boldsymbol{\varepsilon}_{\mu \nu}$. Once again, these coefficients to leading order in $\epsilon$ can be found in appendix \ref{sec:app4}.

Given the simplicity of the background insertion and its dominance in the eikonal regime, we now study the behavior of the 1SF loop-level Compton by expanding the wave regime result in powers of $|\vec{q}| \ll \omega$. At tree-level, we recover the $t$-channel exchange contribution in \eqref{eq:backgroundG}, since all the contact terms coming from the recoil operator are suppressed. At one-loop order, we find a compact expression for the master coefficients at leading order in the expansion $|\vec{q}|/ \omega \ll 1$\footnote{Note that the result exhibits spin shift symmetry, which is an expected property of the amplitude in this regime~\cite{Bern:2022kto,Aoude:2022trd}.}
\begin{align}
    & c_{\scriptscriptstyle\Box}^{\text{eik}} = 512\pi^3 G_\mathrm{N}^2 m_\mathrm{H}^3 \omega^2 (\boldsymbol{\varepsilon}_1 \cdot \boldsymbol{\varepsilon}_2)^2 \Big(2\omega^2  - 2i \omega (l_k \cdot a_\mathrm{H})\nonumber \\
    & \hspace{1cm} + \omega^2[(q\cdot a_\mathrm{H})^2 - q^2 a_\mathrm{H}^2] -q^2 (k \cdot a_\mathrm{H})^2\Big)\,, \\
    & c_{\scriptscriptstyle\bigtriangleup}^{\text{eik}} = -\pi^3 G_\mathrm{N}^2 m_\mathrm{H}^3 (\boldsymbol{\varepsilon}_1 \cdot \boldsymbol{\varepsilon}_2)^2\Big( 480\omega^2 - 640i\omega (l_k \cdot a_\mathrm{H}) \nonumber \\
    & \hspace{1cm} + 380 \omega^2[(q\cdot a_\mathrm{H})^2 - q^2 a_\mathrm{H}^2] - 520 q^2(k \cdot a_\mathrm{H})^2\Big)\,. \nonumber
\end{align}
Unlike in the wave regime, only the triangles contribute classically, since the box represents superclassical iterations of the tree-level eikonal phase and the bubble is quantum suppressed. This is expected also from the optical theorem, as the imaginary part of the one-loop amplitude in the forward limit---which is divergent in this case---is related to the tree-level gravitational Compton cross section~\cite{Bjerrum-Bohr:2016hpa,Bjerrum-Bohr:2017dxw,Gonzo:2023cnv}. The first few subleading orders in the eikonal expansion have also been explored in Ref.~\cite{Chen:2022clh} and can be interpreted as polarization-dependent corrections to the point particle dynamics, although the exponentiation breaks down in the full wave regime as discussed before. Exploiting the simplification of the geometric optics approximation is the goal of our next section.

%----------------------------------------------------------------------
\subsection*{The eikonal regime: an all order resummation}
%----------------------------------------------------------------------

We now turn our attention to the 1SF Compton in the classical geometric optics regime $|\vec{q}| \ll \omega \ll m_\mathrm{H}$. We will show that this corresponds, as expected, to the null geodesic limit for the scattering of gravitons in Kerr. Following Ref.~\cite{Andersson:2020gsj}, we define the WKB approximation 
\begin{align}
&\delta \boldsymbol{g}^{\mathrm{tr}}_{\alpha \beta}(x)  = A_{\alpha \beta}(x, k(x), \hbar) e^{i I(x) / \hbar}\,, \quad k_{\mu}(x) = \overline{\nabla}_{\mu} I(x)\,, \nonumber \\[0.1cm]
&A_{\alpha \beta}(x, k(x), \hbar) =A_{0 \alpha \beta}(x, k(x)) \nonumber \\
& \qquad \qquad \qquad \qquad +\hbar A_{1 \alpha \beta}(x, k(x))+\mathcal{O}\left(\hbar^2\right)\,, 
\label{eq:eikonal_ansatz}
\end{align}
where $I$ is a real scalar function, $A_{\alpha \beta}$ is a complex amplitude\footnote{Notice that this approximation is different to the flat space WKB approximation because we have a position-dependent polarization tensor $A_{\alpha \beta}(x, k(x), \hbar)$ (see also Ref.~\cite{Adamo:2023cfp}), and $\delta \boldsymbol{g}_{\alpha \beta}(x)$ crucially obeys the curved space gauge condition \eqref{eq:gauge-metric}. } and $\hbar$ is a small expansion parameter. Physically, taking $\hbar \to 0$ forces the saddle-point approximation and it is equivalent to having a large classical phase shift. Inserting \eqref{eq:eikonal_ansatz} into the 1SF effective action \eqref{eq:1SF_action_spin}, we obtain
\begin{align}
\label{eq:1SF_action_spin_eik}
    \mathcal{S}^{\mathrm{spin},\mathrm{1SF}} &\sim -\frac{1}{2 \hbar^2} \int \mathrm{d}^4 x\, \sqrt{-\bar{g}} \left[A^{ \alpha \beta} \hat{\mathcal{D}}_{\alpha \beta}{ }^{\gamma \delta}  A_{\gamma \delta} +\mathcal{O}\left(\hbar\right) \right]\,, \nonumber \\
    \hat{\mathcal{D}}_{\alpha \beta}{ }^{\gamma \delta} &= \bar{g}^{\mu \nu} k_\mu(x) k_\nu(x) \left[\delta_\alpha^\gamma \delta_\beta^\delta-\frac{1}{2} \bar{g}_{\alpha \beta} \bar{g}^{\gamma \delta} \right]\,.
\end{align}
The effective action now depends on $I(x)$ (through $k(x)$) and $A_{\mu \nu}(x)$, and we have suppressed contributions which are subleading in the $\hbar \to 0$ limit. Considering the equations of motion for $A_{\mu \nu}$, we then obtain
\begin{align}
    \left(\bar{g}^{\mu \nu} k_\mu(x) k_\nu(x)\right) \left[A_{\alpha \beta}-\frac{1}{2} \bar{g}_{\alpha \beta} \bar{g}^{\gamma \delta} A_{\gamma \delta} \right] = 0\,,
\end{align}
which, discarding the trivial solution $A_{\alpha \beta} = 0$, implies 
\begin{align}
    \bar{g}^{\mu \nu}(x) k_\mu(x) k_\nu(x) = 0\,.
    \label{eq:H-J_particle}
\end{align}
The latter is exactly the Hamilton-Jacobi equation for a point particle moving in the Kerr background $\bar{g}_{\mu \nu}$: indeed, defining the Hamiltonian
\begin{align}
    H(x,p) = \frac{1}{2} \bar{g}^{\mu \nu}(x) p_\mu p_\nu\,,
\end{align}
the condition \eqref{eq:H-J_particle} becomes equivalent to $H(x,\nabla I) = 0$. This is a manifestation of the classical equivalence principle of GR, and it illustrates why the 1SF Compton amplitude in the geometric optics regime agrees with the probe amplitude for a massless scalar in the Kerr background generated by the heavy particle. Building on the amplitude-action relation \cite{Bern:2021yeh,Kol:2021jjc,Adamo:2022ooq,Damgaard:2023ttc}, we obtain 
\begin{align}
 &i \mathcal{M}^{\mathrm{eik}}_{1 \to 1}(k_1,k_2) = 2 m_\mathrm{H} \omega^2 \nonumber \\
    &\quad \times\int \mathrm{d}^{2} b\, e^{-i q \cdot b} \left(e^{\frac{i}{\hbar} (I_r^{>,\epsilon}(\omega,b(J),a_{\mathrm{H}})+ \pi J)}-1\right)\,, 
\end{align}
where we find the resummed in $G_\mathrm{N}$ radial action\footnote{See appendix \ref{sec:app3} for more details about the derivation and the generalization of these ideas to the massive case.}
\begin{align}
    \label{eq:radial_action}
    &\hspace{-7pt}I_{r}^{>,\epsilon}(\omega,J,a_\mathrm{H}) = \frac{4 G_\mathrm{N}  m_{\mathrm{H}} \omega}{\epsilon} \nonumber \\
    &\quad \quad -\pi J {}_3F_2\left(-\frac{1}{2},\frac{1}{6},\frac{5}{6};\frac{1}{2},1;\Lambda^2\right)\\
    &\quad \quad -\frac{3}{4} \sqrt{3} G_\mathrm{N} m_\mathrm{H} \omega\, G_{4,4}^{2,3}\left(
\begin{array}{c}
 -\frac{1}{3},0,\frac{1}{3},1 \\
 0,0,-\frac{1}{2},-\frac{1}{2} \\
\end{array}
|-\Lambda^2\right) \nonumber \\
&\quad \quad-\frac{4 a_\mathrm{H} \omega^2 G_\mathrm{N} m_\mathrm{H}}{J} \, _4F_3\left(\frac{2}{3},1,1,\frac{4}{3};\frac{1}{2},\frac{3}{2},2;\Lambda^2\right)\nonumber \\
&\quad \quad -\frac{5 \pi a_\mathrm{H} \omega^3 G_\mathrm{N}^2 m_\mathrm{H}^2}{J^2} \, _3F_2\left(\frac{7}{6},\frac{3}{2},\frac{11}{6};2,\frac{5}{2};\Lambda^2\right) \!+\! \mathcal{O}(a_\mathrm{H}^2)\,, \nonumber 
\end{align}
written in terms of the variables
\begin{align}
    \Lambda = 3 \sqrt{3} \frac{G_\mathrm{N} m_\mathrm{H} \omega}{J}\,, \qquad  J = b \omega\,.
\end{align}
This result generalizes eq.~(3.8) of Ref.~\cite{Parnachev:2020zbr} to linear order in spin. As expected, the Fourier transform of the $n$-th loop contribution to $I_r^{>}$ is proportional to the 0SF master integral of the ``fan'' diagram \cite{Cheung:2020gbf,Brandhuber:2021eyq}
\begin{align}
\label{fan master def}
    \hspace{-12pt} J^{(L)}(q) &= \int \left( \prod_{i=1} ^L \frac{\mathrm{d}^d k_i}{ \pi ^{d/2}} \frac{\delta(k_i \cdot v_\mathrm{H})}{k_{i} ^2  } \right) \frac{(2\pi)^L }{[\sum_{j=1}^L k_j -q ]^2 } \nonumber \\
    &\hspace{-20pt}  \stackrel{d=4-2 \epsilon}{\simeq} (4 \pi)^{L / 2} \pi^{\frac{L+1}{2}} \frac{\Gamma\left(1-\frac{L}{2} + L \epsilon\right)}{\Gamma\left(\frac{L+1}{2}\right)} q^{L(1-2 \epsilon) -2} \,.
\end{align}
The analytic expression \eqref{eq:radial_action} shows that there is a critical value of the angular momentum 
\begin{align}
    J_{\mathrm{crit}}(\omega) = 3 \sqrt{3} G_\mathrm{N} m_\mathrm{H} \omega\,,
\end{align}
below which (i.e. for $\Lambda>1$) the radial action \eqref{eq:radial_action} develops an imaginary part, as explained also in Refs.~\cite{Fabbrichesi:1993kz,Parnachev:2020zbr}. This is because a null geodesic falls into the black hole for $J <  J_{\mathrm{crit}}(\omega)$, and therefore the initial wave is almost completely absorbed by the black hole in such a regime. The function $J_{\mathrm{crit}}(\omega)$ describes the separatrix between scattering and captured geodesics, which corresponds to the critical orbits which start at infinity and end with an infinite circular whirl at periastron distance (and their time-reversed counterpart) \cite{Barack:2023oqp}. This can be considered a ``smoking gun'' signature of the strong-field nature of a black hole spacetime, and it is pleasing to see such a parameter arising from the perturbative series.

We now extend the consequences of the representation \eqref{eq:radial_action} at the level of observables. The scattering angle is 
\begin{align}
\label{eq:scattering_angle}
 \hspace{-10pt}&\chi(\omega,J,a_\mathrm{H})  = - \frac{\partial I_{r}^{>,\epsilon}(\omega,J,a_\mathrm{H})}{\partial J} \nonumber \\
 &\quad = \pi  \, _2F_1\left(\frac{1}{6},\frac{5}{6};1;\Lambda^2\right) \nonumber \\
 &\quad +\frac{4 G_\mathrm{N} m_{\mathrm{H}} \omega}{J} \, _3F_2\left(\frac{2}{3},1,\frac{4}{3};\frac{3}{2},\frac{3}{2};\Lambda^2\right) \nonumber \\
&\quad -\frac{4 a_\mathrm{H} \omega^2 G_\mathrm{N} m_\mathrm{H}}{J^2} \, _4F_3\left(\frac{2}{3},1,1,\frac{4}{3};\frac{1}{2},\frac{1}{2},2;\Lambda^2\right) \nonumber \\
&\quad -\frac{10 \pi  a_\mathrm{H} \omega^3 G_\mathrm{N}^2 m_\mathrm{H}^2}{J^3} \, _3F_2\left(\frac{7}{6},\frac{3}{2},\frac{11}{6};1,\frac{5}{2};\Lambda^2\right)\,,
\end{align}
while the periastron advance reads, using the scattering-to-bound dictionary \cite{Kalin:2019rwq,Kalin:2019inp,Gonzo:2023goe},
\begin{align}
\label{eq:periastron_adv}
 &\Delta \Phi(\omega,J,a_\mathrm{H}) =  \chi(\omega,J,a_\mathrm{H}) + \chi(\omega,-J,-a_\mathrm{H}) \nonumber \\
 &\quad= 2\pi  \, _2F_1\left(\frac{1}{6},\frac{5}{6};1;\Lambda^2\right) \nonumber \\
&\quad -\frac{20 \pi  a_\mathrm{H} \omega^3 G_\mathrm{N}^2 m_\mathrm{H}^2}{J^3} \, _3F_2\left(\frac{7}{6},\frac{3}{2},\frac{11}{6};1,\frac{5}{2};\Lambda^2\right)\,.
\end{align}
Interestingly, the separatrix $J_{\mathrm{crit}}(\omega)$ provides a critical curve in phase space $(\omega,J,a_\mathrm{H})$ which not only corresponds to the phase transition between scattering and captured geodesics\footnote{Note that we have expanded in the spin parameter $a_{\mathrm{H}}$, therefore linearizing the complicated structure of the Kerr separatrix \cite{Stein:2019buj}.}, but it also provides crucial insights into the resummation through the singular behavior of observables \cite{Long:2024ltn}. Expanding around the separatrix
\begin{align}
    J = J_{\mathrm{crit}}(\omega) + \delta J\,, \quad
\end{align}
we find from \eqref{eq:scattering_angle} and \eqref{eq:periastron_adv}
\begin{align}
\chi(\omega,J,a_\mathrm{H}) &\stackrel{J \to J_{\mathrm{crit}}}{\sim}  -\log\left(\frac{\delta J}{J_{\mathrm{crit}}(\omega)}\right) - \frac{2 a_\mathrm{H} \omega}{\delta J}\,, \nonumber \\
\Delta \Phi(\omega,J,a_\mathrm{H}) &\stackrel{J \to J_{\mathrm{crit}}}{\sim}  -\log\left(\frac{\delta J}{J_{\mathrm{crit}}(\omega)}\right) - \frac{2 a_\mathrm{H} \omega}{\delta J}\,.
\end{align}
The same logarithmic divergence was identified for the scattering case in Refs.~\cite{Damour:2022ybd,Long:2024ltn}; here we find a similar structure for the bound case and we identify the related spin correction for Kerr black holes.  It would be interesting to improve resummation methods for the PM expansion of spinning bound observables, perhaps building on what has been achieved for the scattering case \cite{Damour:2022ybd,Long:2024ltn,Rettegno:2023ghr}.

%----------------------------------------------------------------------
\section{Conclusions and future directions}\label{sec:conclusion}
%----------------------------------------------------------------------

The perturbative amplitude approach has successfully begun integrating into the gravitational wave program to enhance our understanding of compact binary systems. An important milestone in this direction is to explore the resummation of the Post-Minkowskian series, for example adapting tools from the gravitational self-force program. In this work, we have extended the spinless self-force approach in Refs.~\cite{Kosmopoulos:2023bwc,Cheung:2023lnj,Cheung:2024byb} to the case of Kerr black holes up to quadratic in spin order using the worldline formulation introduced in Ref.~\cite{Mogull:2020sak}.

Our first step was to study how to generalize the Kerr metric in $d$ dimensions, analyzing in particular the simply rotating Myers-Perry solution and the minimal coupling metric introduced in \cite{Gambino:2024uge}.  This allows one to set up a dimensional regularization scheme for the spinning self-force EFT, whose action we have derived by expanding the trajectory and the Grassmann variables around their background value for the minimal coupling case $(C_1=1,C_2=0)$. Given that the spin tensor is proportional to the mass, we find that spin effects are suppressed (resp. enhanced) for the light-body (resp. heavy body). 

At 0SF order, we recover the well-known fact that the self-force EFT describes only spinless geodesics in Kerr. At 1SF order, we find that only linear in spin corrections to the light-body degrees of freedom are relevant and---as in the spinless case---we can integrate out the heavy dynamics in terms of a non-local effective action \eqref{eq:1SF_recoil_spin}, which includes novel spinning recoil contributions. 

Remarkably, the 1SF action is only a function of the dynamical graviton field, which allows to study the 1SF metric perturbation from a single equation of motion \eqref{eq:deltag_1SF_new} describing the graviton perturbation in a particle-generated (non-vacuum) spacetime. We compare this with the traditional self-force approach in vacuum, emphasizing the role of the new terms localized on the heavy particle worldline. We then adopt the background field approach to study the metric perturbation and the radiative waveform, clarifying the role of the 2-point function of the graviton field---including both background and recoil contributions---in the resummation.

This led us to consider the curved space 1SF Compton amplitude, defined from the LSZ reduction of $\langle  \mathcal{T} \delta \boldsymbol{g}_{\mu \nu}(x) \delta \boldsymbol{g}_{\alpha \beta}(y) \rangle$, and its analytic properties at loop level in the wave regime. We find that, remarkably, both at tree- and one-loop order, the on-shell curved-space amplitude matches the usual flat space Compton obtained from $\langle \mathcal{T} h_{\mu \nu}(x) h_{\alpha \beta}(y) \rangle$. In particular, for the first time, we compute the classical one-loop Compton at quadratic in spin order with generalized unitarity. Our approach shows that the self-force method provides a direct prescription for the spinning contact terms in the wave scattering scenario, and paves the way for a consistent matching between flat and curved space Compton amplitudes. 

Finally, the importance of the 1SF recursion relation for the Compton amplitude prompt us to consider also a simpler limit, the geometric optics one, where the resummation can be done analytically at all orders in the weak coupling but up to linear in spin order. Surprisingly, a new strong-field scale emerges from the hypergeometric structure of the related scattering and bound observables---the photon ring radius---connected to the phase transition between the scattering and the plunge regime.

A number of questions are left open for future investigations. First, it would be interesting to understand the non-perturbative structure of the minimally coupled spinning metric in generic dimensions. Second, it would be good to develop a spinor-helicity formalism for massive spinning point particles in $d$ dimensions, with the idea of computing the 3-pt amplitude from the on-shell perspective. Third, it would be natural to use our effective framework to compute the spinning radial action and related observables for spinning binaries, along the lines of Refs.~\cite{Cheung:2023lnj,Cheung:2024byb}. A more speculative direction is to further develop the generalization of the Detweiler-Whiting decomposition of the 1SF metric perturbation for non-vacuum spacetime, with the hope of better connecting with PM and PN calculations. Furthermore, it would be interesting to compute analytically the 1SF waveform beyond tree-level, where some simplifications are expected as a consequence of the recursive structure of the 1SF Compton amplitude. The simplicity of the such recursion relation for scalars has been recently exploited in the partial wave basis \cite{Ivanov:2024sds,Caron-Huot:2025tlq}, and we look forward to its extension into the gravitational case. This iterative framework is likely to yield valuable insights into strong-field dynamics, as for the explicit results obtained here, with the hope of generalizing the waveform resummation established in Refs.~\cite{Alessio:2024onn,Fucito:2024wlg,Cipriani:2025ikx}. Finally, we aim to understand the matching of the one-loop Kerr Compton amplitude at higher spin order with black hole perturbation theory, extending the tree-level analysis of Refs.~\cite{Bautista:2021wfy,Bautista:2022wjf,Bautista:2023sdf}. We look forward to these directions and many others, with the hope that the future will bring more analytical insights into the compact binary problem and further connections between amplitudes and gravity.

%----------------------------------------------------------------------
\textit{Note added---} After this work originally appeared in this arXiv, we manage to perform a full flat space calculation of the one-loop spinning Compton amplitude at quadratic in spin order, resolving the previous discrepancy in the comparison with the curved space Compton by using the minimally coupled spinning metric \eqref{eq:MC_metric}. In the spinless limit, our results also agree with a recent one-loop Compton amplitude derivation in Ref.~\cite{Bjerrum-Bohr:2025bqg}.

%----------------------------------------------------------------------
\textit{Acknowledgments---} We wish to thank F. Alessio, R. Aoude, M. Bianchi, G. Brunello, G. Chen, A. de Simone, C. Gambino, H. Johansson, A. Ilderton, R. Monteiro, G. Mogull, A. Pound, K. Rajeev, F. Riccioni, M. Zeng and especially G. Brown for many discussions and comments on the draft. D.A.\ is supported by a STFC studentship. This research was supported in part by grant NSF PHY-2309135 to the Kavli Institute for Theoretical Physics (KITP).

\appendix
%----------------------------------------------------------------------
\section{Simplest rotating metric for a minimally coupled spinning particle}
\label{sec:app2}
%----------------------------------------------------------------------
Here we list the explicit perturbative coefficients of the 2PM rotating metric generated by a minimally coupled particle in $d$ dimensions in harmonic gauge \cite{Gambino:2024uge}
\begin{align}
h^{(1)\mathrm{MC}} &= -4 \Phi(r) + \frac{4 \left(2 d^3 - 22 d^2 + 75 d - 77\right) \Phi^2(r)}{(d - 5)(d - 2)^2} \nonumber \\
&\quad + \frac{2 (d - 3) (d^2 - 15 d + 21) \Phi^2(r) (L^{(1)} \cdot L^{(1)})}{(d - 2) r^4} \nonumber \\
&\quad - \frac{28 (a^d_{\mathrm{H}})^2 (d - 3) \Phi^2(r)}{(d - 2) r^2}, \nonumber \\
h^{(2)\mathrm{MC}} &= -\frac{2 (d - 3) \Phi(r)}{r^2} + \frac{4 (d - 3)^2 \Phi^2(r)}{(d - 2) r^2}, \nonumber \\
h^{(3)\mathrm{MC}} &= -\frac{2 (d - 3)(d - 1) \Phi(r)}{r^4} \nonumber \\
&\quad + \frac{2 (d - 6)(d - 3)(d - 1) \Phi^2(r)}{(d - 2) r^4}, \nonumber \\
h^{(4)\mathrm{MC}} &= \frac{4 (d - 3)^3 \Phi^2(r)}{(d - 5)(d - 2)^2 r^2} - \frac{4 (a^d_{\mathrm{H}})^2 (d - 3) \Phi^2(r)}{(d - 2) r^4}\nonumber \\
&\quad + \frac{2 (d - 3)^2 (d - 1) \Phi^2(r) (L^{(1)} \cdot L^{(1)})}{(d - 2) r^6} , \nonumber \\
h^{(5)\mathrm{MC}} &= \frac{2 (a^d_{\mathrm{H}})^2 (d - 3) \Phi(r)}{r^2} + \frac{2 (a^d_{\mathrm{H}})^2 (d - 3) \Phi^2(r)}{(d - 2) r^2}, \nonumber \\
h^{(6)\mathrm{MC}} &= \frac{4 (d - 3)^3 \Phi^2(r)}{(d - 5)(d - 2)^2 r^2} - \frac{4 (a^d_{\mathrm{H}})^2 (d - 3) \Phi^2(r)}{(d - 2) r^4} \nonumber \\
&\quad + \frac{2 (d - 3)^2 (d - 1) \Phi^2(r) (L^{(1)} \cdot L^{(1)})}{(d - 2) r^6}, \nonumber \\
h^{(7)\mathrm{MC}} &= -\frac{4 (a^d_{\mathrm{H}})^2 (d - 3) \Phi^2(r)\, (r \cdot e_{j_1})}{r^4}, \nonumber \\
h^{(8)\mathrm{MC}} &= -\frac{4 (a^d_{\mathrm{H}})^2 (d - 3) \Phi^2(r)\, (r \cdot e_{j_1})}{r^4}, \nonumber \\
h^{(9)\mathrm{MC}} &= \frac{4 \Phi(r)}{d - 2} + \frac{4 (3 d - 13) \Phi^2(r)}{(d - 5)(d - 2)^2} \nonumber \\
&\quad + \frac{4 (a^d_{\mathrm{H}})^2 (d - 3) \Phi(r)}{(d - 2) r^2} + \frac{24 (a^d_{\mathrm{H}})^2 (d - 3) \Phi^2(r)}{(d - 2)^2 r^2} \nonumber \\
&\quad + \frac{2 (d - 3)(d - 1) \Phi(r)\, (L^{(1)} \cdot L^{(1)})}{(d - 2) r^4} \nonumber \\
&\quad + \frac{2 (d - 3)(9 d - 14) \Phi^2(r) (L^{(1)} \cdot L^{(1)})}{(d - 2)^2 r^4},
\end{align}
where we have defined the radial potential
\begin{align}
    \Phi(r) = \Gamma\left(\frac{d-3}{2}\right) \pi^{(3-d)/ 2} \frac{G_\mathrm{N} m_{\mathrm{H}} }{r^{d-3}}\,.
\end{align}
It is also useful to understand how to compute the Fourier transform $\tilde{h}_{\mu \nu}^{\mathrm{MC}}(q)$ of the metric $h_{\mu \nu}^{\mathrm{MC}}(x)$ in \eqref{eq:MC_metric}, for example in order to evaluate the interacting piece of the graviton propagator \eqref{eq:Gint}. To do so, it is convenient to project the Fourier transform of \eqref{eq:MC_metric} in the basis
\begin{align}
&\vec{\tilde{c}}_{\mu \nu} = \big\{v_{\mathrm{H}\mu} v_{\mathrm{H}\nu},\, v_{\mathrm{H}(\mu} \tilde{L}^{(1)}_{\nu)},\, \tilde{L}^{(1)}_\mu \tilde{L}^{(1)}_\nu,\, e_{i_1 \mu} e_{i_1 \nu},\, e_{j_1 \mu} e_{j_1 \nu}, \nonumber \\
&\qquad\qquad\qquad\qquad\ q_{\mu} q_{\nu},\,  e_{i_1 (\mu} q_{\nu)},\, e_{j_1 (\mu} q_{\nu)},\,
\eta_{\mu \nu} \big\}\,, \nonumber    
\end{align}
where we have defined the $q$-space rotation generator \eqref{eq:angular} 
\begin{align}
\tilde{L}^{(1)}_{\mu} &= \epsilon^{(1)}_{\mu \nu \alpha \beta} v_\mathrm{H}^{\nu} (a^d_\mathrm{H})^{\alpha} q^{\beta}\,.
\label{eq:angularq}
\end{align}
%

%----------------------------------------------------------------------
\section{The non-minimal dimension-dependent couplings from the Myers-Perry solution}
\label{sec:app5}
%----------------------------------------------------------------------

In this appendix, we discuss the extraction of non-minimal couplings at quadratic in spin for the Myers-Perry solution by comparing the linearized $d$-dimensional stress-tensor of Ref.~\cite{Bianchi:2024shc} with the point-particle one. 

On one hand, the linearized stress-tensor for Myers-Perry provided in their eq.~(68) becomes in our notation
\begin{align}
\label{eq:stress-tensor_MP}
&T_{d,\rm MP}^{\mu \nu}(v_\mathrm{H},a^d_\mathrm{H},q)  =m_\mathrm{H} v_\mathrm{H}^\mu v_\mathrm{H}^\nu F_1^{(d)}(\zeta)  \\
&\qquad\qquad\qquad \quad\,-i m_\mathrm{H} v_\mathrm{H}^{(\mu}(\bar{S}_\mathrm{H}  \cdot q)^{\nu)}  F_3^{(d)}(\zeta) \nonumber\\
&\qquad\qquad\qquad \quad\,+m_\mathrm{H} \frac{F_2^{(d)}(\zeta)}{\zeta^2}(\bar{S}_\mathrm{H}  \cdot q)^\mu(\bar{S}_\mathrm{H}  \cdot q)^\nu \,, \nonumber
\end{align}
where the form factors $F_1^{(d)}(\zeta)$, $F_2^{(d)}(\zeta)$ and $F_3^{(d)}(\zeta)$ are given in terms of Bessel functions 
\begin{align}
F_1^{(d)}(\zeta) &= F_2^{(d)}(\zeta) + F_3^{(d)}(\zeta) \,, \\
F_2^{(d)}(\zeta) &= -2^{d-4} (d-2)^{\frac{3-d}{2}} \zeta ^{\frac{5-d}{2}} \Gamma \left(\frac{d-1}{2}\right) J_{\frac{d-1}{2}}\left(\frac{d-2}{2}  \zeta \right) \,, \nonumber\\
F_3^{(d)}(\zeta) &= 2^{d-3} (d-2)^{\frac{3-d}{2}} \zeta ^{\frac{3-d}{2}} \Gamma \left(\frac{d-1}{2}\right) J_{\frac{d-3}{2}}\left(\frac{d-2}{2} \zeta \right) \,, \nonumber
\end{align}
and we have defined the spinning invariant
\begin{align}
\hspace{-7pt}\zeta= \sqrt{-q \cdot \bar{S}_\mathrm{H} \cdot \bar{S}_\mathrm{H} \cdot q}=a^d_{\mathrm{H}} \sqrt{(q \cdot e_{i_1})^2+(q \cdot e_{j_1})^2 }\,.
\end{align}
On the other hand, the worldline matter action \eqref{eq:matter_spin_action} with the non-minimal couplings give rise to the following linearized stress-tensor at spin squared \cite{Jakobsen:2021zvh}
\begin{align}
&T_{d,\rm pp}^{\mu \nu}(v_\mathrm{H},a^d_\mathrm{H},q) = m_\mathrm{H} v_\mathrm{H}^\mu v_\mathrm{H}^\nu- i m_\mathrm{H}  (\bar{S}_\mathrm{H}  \cdot q)^{(\mu} v_\mathrm{H}^{\nu)}\nonumber \\
&\qquad\qquad\qquad\quad- m_\mathrm{H}\frac{C_1}{2}(\bar{S}_\mathrm{H}  \cdot q)^\mu(\bar{S}_\mathrm{H}  \cdot q)^\nu \nonumber \\
&\qquad\qquad\qquad\quad+ m_\mathrm{H} \frac{C_2}{2} v_\mathrm{H}^\mu v_\mathrm{H}^\nu(q \cdot \bar{S}_{\mathrm{H}} \cdot \bar{S}_{\mathrm{H}} \cdot q)\,.
\label{eq:stress-tensor_pp}
\end{align}
Expanding the stress-tensor \eqref{eq:stress-tensor_MP} at spin squared and comparing it with \eqref{eq:stress-tensor_pp}, we find the $d$-dimensional non-minimal couplings
\begin{align}
C_1^{\rm MP}(d)=\frac{d-2}{2 (d-1)}\,, \qquad C_2^{\rm MP}(d)=\frac{ d(d-2)}{4 (d-1)}\,,
\end{align}
as reported in \eqref{eq:nonminimal_MP}. Note that in $d=4$, given that $C_1^{\rm MP}+C_2^{\rm MP} = 1$, the metric is equivalent to the Kerr one (as stressed in Ref.~\cite{Gambino:2024uge}), while in higher-dimensions this is not the case. For example, we recover $C_1^{\rm MP}=3/8$ and $C_2^{\rm MP} = 15/16$ in $d=5$; see eq.~(91) of Ref.~\cite{Gambino:2024uge}. 

%----------------------------------------------------------------------
\section{Details of the GSF expansion of the effective action}
\label{sec:app1}
%----------------------------------------------------------------------

In this appendix, we will discuss the expansion of the effective action in more detail. For simplicity, we will start with the spinless case followed by the spinning one. As discussed in the main text, the GSF expansion involves the use of \eqref{eq:SF_exp1} and \eqref{eq:SF_exp2}, where, to obtain the 1SF contribution, we must expand to $\mathcal{O}(\lambda^2)$. In full generality, we drop the $i=\mathrm{L},\mathrm{H}$ subscripts and consider the expansion of a generic worldline, whose action we denote as $\mathcal{S}_{\text{w}}$. The expansion of the spinless dynamics reads
\begin{align}
    &\mathcal{S}_{\text{w}} = \mathcal{S}^{\lambda^0}_{\text{w}} + \lambda \mathcal{S}^{\lambda}_{\text{w}} + \lambda^2 \mathcal{S}^{\lambda^2}_{\text{w}} + \mathcal{O}(\lambda^3)\,,
\end{align}
where
\begin{align}
    & \mathcal{S}^{\lambda^0}_{\text{w}} = - m \int \mathrm{d}\tau \Big\{ \frac{1}{2} + \frac{1}{2}\bar{g}_{\mu \nu}(\bar{x}) \dot{\bar{x}}^\mu \dot{\bar{x}}^\nu\Big\}\,, \\
    & \mathcal{S}^{\lambda}_{\text{w}} =  - m \int \mathrm{d}\tau \Big\{\frac{1}{2}\delta \boldsymbol{g}_{\mu \nu}(\bar{x}) \dot{\bar{x}}^\mu \dot{\bar{x}}^\nu + \bar{g}_{\mu \nu}(\bar{x})\dot{\bar{x}}^\mu  \delta \dot{x}^\nu\\
    & \hspace{3cm} + \frac{1}{2}\dot{\bar{x}}^\mu \dot{\bar{x}}^\nu\delta x^\alpha \partial_\alpha \bar{g}_{\mu \nu}(\bar{x})\Big\} \,, \nonumber \\
    & \mathcal{S}^{\lambda^2}_{\text{w}} = - m \int \mathrm{d}\tau \Big\{\delta \boldsymbol{g}_{\mu \nu}(\bar{x}) \dot{\bar{x}}^\mu \delta \dot{x}^\nu + \frac{1}{2}\dot{\bar{x}}^\mu \dot{\bar{x}}^\nu \delta x^\alpha \partial_\alpha \bar{g}_{\mu \nu}(\bar{x})\nonumber \\
    & \hspace{1cm}+ \frac{1}{2}\bar{g}_{\mu \nu}(\bar{x}) \delta \dot{x}^\mu \delta \dot{x}^\nu + \dot{\bar{x}}^\mu \delta \dot{x}^\nu \delta x^\alpha \partial_\alpha \bar{g}_{\mu \nu}(\bar{x}) \\
    & \hspace{1cm}+ \frac{1}{4} \dot{\bar{x}}^\mu \dot{\bar{x}}^\nu \delta x^\alpha \delta x^\beta \partial_\alpha \partial_\beta \bar{g}_{\mu \nu}(\bar{x})\Big\} \,. \nonumber
\end{align}
Now we want to simplify these expressions. In particular, performing an integration-by-parts in the second term of $S^{\lambda}_{\text{w}}$ allows us to write 
\begin{align}
    \mathcal{S}^{\lambda}_{\text{w}} =  - m \int \mathrm{d}\tau \Big\{ & \frac{1}{2}\delta \boldsymbol{g}_{\mu \nu}(\bar{x}) \dot{\bar{x}}^\mu \dot{\bar{x}}^\nu  - g_{\mu \nu}(\bar{x}) \ddot{\bar{x}}^\mu \delta x^\nu \\
    & \hspace{1cm} - \delta x^\rho \dot{\bar{x}}^\mu \dot{\bar{x}}^\nu \bar{\Gamma}_{\rho\mu \nu}(\bar{x})\Big\}\,. \nonumber
\end{align}
To simplify $S^{\lambda^2}_{\text{w}}$ we perform an integration-by-parts on the first term, substitute the 0SF geodesic equation, and combine it with the second term to obtain
\begin{align}
    \delta \boldsymbol{g}_{\mu \nu}(\bar{x}) \dot{\bar{x}}^\mu \delta \dot{x}^\nu + \frac{1}{2}\dot{\bar{x}}^\mu  & \dot{\bar{x}}^\nu \delta x^\alpha \partial_\alpha \bar{g}_{\mu \nu}(\bar{x}) \nonumber \\
  &=- \dot{\bar{x}}^\mu \dot{\bar{x}}^\nu \delta x_\rho \delta \Gamma^{\rho}_{~\mu \nu}(\bar{x})\,. 
\end{align}
The remaining terms are more subtle. Using integration-by-parts and adding a total derivative term to the action $S^{\lambda^2}_{\text{w}}$, we can group the expressions together to obtain
\begin{align}
    & \mathcal{S}^{\lambda^2}_{\text{w}} =  -m \int \mathrm{d}\tau \Big\{\frac{1}{2}\bar{g}_{\rho \gamma} (\bar{x})\dot{\bar{x}}^\mu \dot{\bar{x}}^\nu  \overline{\nabla}_\mu \delta x^\rho \overline{\nabla}_\nu \delta x^\lambda \\
    & + \frac{1}{2}\dot{\bar{x}}^\mu \dot{\bar{x}}^\nu \delta x^\rho \delta x^\sigma \bar{R}_{\nu \rho \sigma \mu}(\bar{x}) - \dot{\bar{x}}^\mu \dot{\bar{x}}^\nu \delta x_\rho \delta \Gamma^{\rho}_{~\mu \nu}(\bar{x})\Big\}\,, \nonumber
\end{align}
where we convert $\tau$ derivatives to spatial derivatives using $\mathrm{d}/\mathrm{d}\tau = \dot{\bar{x}}^\mu \partial_\mu$. Doing so, we find agreement with Ref.~\cite{Cheung:2023lnj}.

Now we turn our attention to the spinning case. Following the same steps as before, we write 
\begin{equation}
    \mathcal{S}_{\text{w}}^{\text{spin}} = S^{\text{spin}, \lambda^0}_{\text{w}} + \lambda S^{\text{spin}, \lambda}_{\text{w}} + \lambda^2 \mathcal{S}^{\text{spin}, \lambda^2}_{\text{w}} + \mathcal{O}(\lambda^3)\,,
\end{equation}
where, since the spinless terms are given by $S_{\text{w}}$, we only consider the spinning terms here. Explicitly, we have 
\begin{align}
    & \mathcal{S}^{\text{spin}, \lambda^0}_{\text{w}} = - m^2 \int \mathrm{d}\tau \Big\{ i\bar{\Psi}_a \dot{\Psi}^a + i\dot{\bar{x}}^\mu \bar{\Psi}_a \bar{\omega}_{\mu}^{~ab}(\bar{x}) \Psi_b  \\
    & \hspace{3cm} + \frac{m}{2}\bar{R}_{abcd}(\bar{x})\bar{\Psi}^a \Psi^b\bar{\Psi}^c \Psi^d \Big\}\,, \nonumber\\
    & \mathcal{S}^{\text{spin}, \lambda}_{\text{w}} =  - m^2 \int \mathrm{d}\tau \Big\{  i\delta \bar{\psi}_a \dot{\Psi}^a + i\bar{\Psi}_a \delta \dot{\psi}^a \\
    & \hspace{0.5cm} + i\delta \dot{x}^\mu \bar{\Psi}_a \bar{\omega}_{\mu}^{~ab}(\bar{x}) \Psi_{b} + i \dot{\bar{x}}^\mu[\delta \bar{\psi}_a \Psi_{b} +  \bar{\Psi}_a \delta\psi_{b}]\bar{\omega}_{\mu}^{~ab}(\bar{x})  \nonumber \\
    & \hspace{0.5cm}  + i \dot{\bar{x}}^\mu \bar{\Psi}_a[\delta x^\alpha  \partial_\alpha \bar{\omega}_{\mu}^{~ab}(\bar{x}) + \delta\omega_{\mu}^{~ab}(\bar{x}) \Psi_{b}] \nonumber \\
    & \hspace{0.5cm} + m \bar{R}_{abcd}(\bar{x})[\delta \bar{\psi}^a \Psi^b\bar{\Psi}^c \Psi^d + \bar{\Psi}^a \delta \psi^b\bar{\Psi}^c \Psi^d \nonumber \\
    & \hspace{0.5cm} + \frac{m}{2}[\delta x^\mu \partial_\mu\bar{R}_{abcd}(\bar{x}) + \delta R_{abcd}(\bar{x})]\bar{\Psi}^a \Psi^b\bar{\Psi}^c \Psi^d\Big\} \,, \nonumber\\
    & \hspace{-5pt}\mathcal{S}^{\text{spin},\lambda^2}_{\text{w}}\hspace{-5pt} = - m^2 \int \mathrm{d}\tau \Big\{i \delta \bar{\psi}_a  \delta \dot{\psi}^a +  i \dot{\bar{x}}^\mu \delta \bar{\psi}_a \bar{\omega}_{\mu}^{ab}(\bar{x}) \delta \psi_b   \\
    & \hspace{0.5cm}+ i \dot{\bar{x}}^\mu [\delta \bar{\psi}_a \Psi_{b} + \bar{\Psi}_a \delta\psi_{b}]\delta\omega_{\mu}^{~ab}(\bar{x})  \nonumber\\
    & \hspace{0.5cm}+ i \dot{\bar{x}}^\mu \delta x^\alpha [\bar{\Psi}_a \delta \psi_b + \delta \bar{\psi}_a  \Psi_b]\partial_\alpha \bar{\omega}_{\mu}^{ab}(\bar{x})   \nonumber\\
    & \hspace{0.5cm}+ i \delta \dot{x}^\mu [\bar{\Psi}_a \delta \psi_b +\delta \bar{\psi}_a \Psi_b]\bar{\omega}_{\mu}^{ab}(\bar{x}) \nonumber \\
    & \hspace{0.5cm} + i\delta \dot{x}^\mu \bar{\Psi}_a [\delta\omega_{\mu}^{~ab}(\bar{x}) + \delta x^\alpha \partial_\alpha \bar{\omega}_{\mu}^{ab}(\bar{x})] \Psi_b \nonumber \\
    & \hspace{0.5cm}+ i \dot{\bar{x}}^\mu \delta x^\alpha \bar{\Psi}_a[ \partial_\alpha \delta\omega_{\mu}^{~ab}(\bar{x}) + \frac{1}{2}\delta x^\beta \partial_\alpha \partial_\beta \bar{\omega}_{\mu}^{ab}(\bar{x})] \Psi_b \nonumber \\
    & \hspace{0.5cm}+ m\bar{R}_{abcd}(\bar{x})[\delta \bar{\psi}^a \delta\psi^b \bar{\Psi}^c \Psi^d +  \delta\bar{\psi}^a \Psi^b\bar{\Psi}^c \delta\psi^d] \nonumber\\
    & \hspace{0.5cm}+  \frac{m}{2}\bar{R}_{abcd}(\bar{x})[\delta\bar{\psi}^a \Psi^b \delta\bar{\psi}^c \Psi^d + \bar{\Psi}^a  \delta\psi^b\bar{\Psi}^c \delta\psi^d] \nonumber\\
    & \hspace{0.5cm}+ m\delta x^\mu \partial_\mu \bar{R}_{abcd}(\bar{x})[\delta \bar{\psi}^a \Psi^b \bar{\Psi}^c \Psi^d + \bar{\Psi}^a \delta \psi^b \bar{\Psi}^c \Psi^d] \nonumber \\
    & \hspace{0.5cm} + m\delta R_{abcd}(\bar{x})[\delta \bar{\psi}^a \Psi^b \bar{\Psi}^c \Psi^d + \bar{\Psi}^a \delta\psi^b \bar{\Psi}^c \Psi^d] \nonumber \\
    & \hspace{0.5cm}+ \frac{m}{2}\delta x^\mu \partial_\mu \delta R_{abcd}(\bar{x})\bar{\Psi}^a \Psi^b \bar{\Psi}^c \Psi^d\Big\}\;.\nonumber
\end{align}
Note that we do not manipulate this further for two reasons related to light and heavy dynamics. For the heavy dynamics, many terms vanish because of dimensional regularization. For the light dynamics, higher spin contributions are more suppressed relative to lower spin ones. 

As discussed in the main text, the variation of the spin connection and Riemann tensor exhibit an expansion in $\delta \boldsymbol{g}_{\mu \nu}$. For our purposes, only the first two orders of this expansion evaluated on the position of the heavy particle are relevant. The linear contributions are
\begin{align}
    \label{eq: omega-R-linear}
    \delta \omega_{\mu}^{(1)ab} (\bar{x}_{\mathrm{H}}) & = - \partial^{[a}\delta \boldsymbol{g}^{b]}_{~\mu} (\bar{x}_{\mathrm{H}})\;, \\
        \delta R^{(1)}_{abcd}(\bar{x}_{\mathrm{H}}) 
        & = 2\partial_{[a} \delta \omega^{(1)}_{b] cd}(\bar{x}_{\mathrm{H}}) \,, 
\end{align}
and quadratic contributions read
\begin{align}
    \label{eq: omega-R-quadratic}
    \delta \omega_{\mu}^{(2)ab}(\bar{x}_{\mathrm{H}}) & = - \frac{1}{2}\delta \boldsymbol{g}^{\nu [a} (\bar{x}_{\mathrm{H}})\Big( \partial^{b]} \delta \boldsymbol{g} _{\mu \nu} (\bar{x}_{\mathrm{H}}) \nonumber\\
    & - \partial_\nu \delta \boldsymbol{g}^{b]}_{~~\mu} (\bar{x}_{\mathrm{H}}) + \frac{1}{2}\partial_\mu \delta \boldsymbol{g}^{b]}_{~~\nu} (\bar{x}_{\mathrm{H}})\Big)\;,  \\
    \delta R^{(2)}_{abcd}(\bar{x}_{\mathrm{H}})  
    & = 2 \partial_{[a} \delta \omega^{(2)}_{b] cd}(\bar{x}_{\mathrm{H}}) + 2\delta \omega^{(1)}_{[ace}(\bar{x}_{\mathrm{H}})\delta  \omega^{(1)e}_{b]~~d}(\bar{x}_{\mathrm{H}}) \nonumber\\
     & +\delta \boldsymbol{g}^{\mu}_{~[a}(\bar{x}_{\mathrm{H}})\delta R^{(1)}_{b]\mu cd}(\bar{x}_{\mathrm{H}}) \,,
\end{align}
where terms $\sim \bar{\omega} \delta \omega$ vanish in dimensional regularization. 

In particular, for the heavy particle dynamics, we showed that terms linear in $\delta \boldsymbol{g}_{\mu\nu}$ contribute to the energy-momentum tensor via the identities in~\eqref{eq: linear-order-identities}. Here we  prove these identities, which follow straightforwardly from \eqref{eq: omega-R-linear}. For the first identity in \eqref{eq: linear-order-identities} we have 
\begin{align}
    & m_{\mathrm{H}}\int \mathrm{d} \tau \, \dot{\bar{x}}_{\mathrm{H}}^\mu \bar{\Psi}_{\mathrm{H}}^a \delta \omega^{(1)}_{\mu a b}(\bar{x}_{\mathrm{H}}) \Psi_{\mathrm{H}}^b  \\ 
    & = \frac{i}{2}\int \mathrm{d}^4x \int \mathrm{d}\tau \ \delta^{(4)}(x - v_{\mathrm{H}} \tau) v_{\mathrm{H}}^\mu \bar{S}_{\mathrm{H} ab} (- \partial^{[a}\delta \boldsymbol{g}^{b]}_{~\mu})\nonumber \\
    & = -\frac{i}{2}\int \mathrm{d}^4x \int \mathrm{d}\tau \ \delta \boldsymbol{g}_{\mu \nu}v_{\mathrm{H}}^{(\mu} (\bar{S}_{\mathrm{H}}\cdot \partial_x)^{\nu)} \delta^{(4)}(x - v_{\mathrm{H}} \tau)\nonumber \;,
\end{align}
where we have performed an integration-by-parts in the last equality and defined  $(\bar{S}_{\mathrm{H}}\cdot \partial_x)^\mu = \bar{S}_{\mathrm{H}}^{\mu \nu }\partial_\nu$. Similarly, for the second identity in \eqref{eq: linear-order-identities} we find
\begin{align}
    & m_{\mathrm{H}}^2 \int \mathrm{d} \tau \, \delta R^{(1)}_{a b c d}(\bar{x}_{\mathrm{H}}) \bar{\Psi}_{\mathrm{H}}^a \Psi_{\mathrm{H}}^b \bar{\Psi}_{\mathrm{H}}^c \Psi_{\mathrm{H}}^d  \\
    & = \frac{1}{2}\int d^4x \int \mathrm{d}\tau \ \delta^{4}(x - v_{\mathrm{H}} \tau) \bar{S}^{ab}_{\mathrm{H}} \bar{S}^{cd}_{\mathrm{H}}\partial_b \partial_{d}\delta \boldsymbol{g}_{ac}\nonumber  \\
%        & = \frac{1}{2}\int \mathrm{d}^4x \int \mathrm{d}\tau  \delta \boldsymbol{g}_{\mu \nu} \bar{S}_\mathrm{H}^{(\mu \rho} \bar{S}_\mathrm{H}^{\nu) \sigma} \, \partial_\rho \partial_\sigma \delta^{(4)}(x - v_{\mathrm{H}} \tau)\nonumber \;.
    & = \frac{1}{2}\int \mathrm{d}^4x \int \mathrm{d}\tau \ \delta \boldsymbol{g}_{\mu \nu} (\bar{S}_{\mathrm{H}} \cdot \partial_x)^{(\mu} (\bar{S}_{\mathrm{H}} \cdot \partial_x)^{\nu)}\delta^{(4)}(x - v_{\mathrm{H}} \tau)\nonumber \;.
\end{align}
%

%----------------------------------------------------------------------
\section{Hypergeometric-type functions for the dynamics of a massive particle in Kerr}
\label{sec:app3}
%----------------------------------------------------------------------

In this appendix we explain the derivation of \eqref{eq:radial_action} from the solution of the Hamilton-Jacobi equation 
\begin{align}
    \bar{g}^{\mu \nu}(x) k_\mu(x) k_\nu(x) = m_{\mathrm{L}}^2
\end{align}
for a massive probe particle in Kerr discussed in Ref.~\cite{Gonzo:2024zxo}. Using the perturbative expansion of the spinning radial action up to 6PM order, we have studied the recursive pattern of the PM coefficients in the aligned-spin case. A direct inspection shows that such coefficients follow from an hypergeometric structure, which at all orders in $G_\mathrm{N}$ but restricted to $\mathcal{O}(\bar{S}^0_{\mathrm{L}} \bar{S}^1_{\mathrm{H}})$ is
\begin{align}
    &\hspace{-6pt}I_{r,\mathrm{0SF}}^{>,\epsilon}(y,b(J)) = - \pi J + 2 G_\mathrm{N} m_{\mathrm{L}} m_{\mathrm{H}} \frac{2 y^2 - 1}{\sqrt{y^2-1}} \frac{1}{\epsilon} \\
    &\qquad +\frac{2 G_\mathrm{N} m_{\mathrm{L}} m_{\mathrm{H}}}{\sqrt{y^2 - 1}} \sum_{n=0}^{+\infty} d^{(n)}(y) \left(\frac{G_\mathrm{N} m_{\mathrm{L}} m_{\mathrm{H}}}{J}\right)^{n}\,, \nonumber \\
    &\hspace{-6pt}d^{(n)}(y) = \frac{\cos \left(\frac{\pi n}{2}\right)}{n (n+1) \left(y^2-1\right)^{\frac{n}{2}}} \nonumber \\
    & \qquad\qquad \times \, _2F_1\left(-n-1,\frac{n+2}{2};\frac{1-n}{2};1-y^2\right) \nonumber \\
    & \qquad- \frac{2 a_{\mathrm{H}} y \cos \left(\frac{\pi n}{2}\right) }{(n-1) \left(y^2-1\right)^{\frac{n}{2}-1} J}  \nonumber \\
     & \qquad\qquad \times \, _2F_1\left(-n,\frac{n+4}{2};\frac{3-n}{2};1-y^2\right) \,, \nonumber 
\end{align}
where $y = v_{\mathrm{L}} \cdot v_{\mathrm{H}}$ is the rapidity and $_2F_1\left(a,b,c;x\right)$ is the Gauss hypergeometric function. Expanding at large $y \gg 1$ in the massless limit, and using Euler's identity
\begin{align}
\Gamma(z) \Gamma(1-z)=\frac{\pi}{\sin (\pi z)}\,, 
\end{align}
the expansion of the Gauss hypergeometric function yields a product of Gamma functions, which resum into the structure \eqref{eq:radial_action} after defining the generalized hypergeometric function $_qF_p$ and the Meijer-G function $G_{p,q}^{m,n}$
\begin{align}
    &G_{p, q}^{m, n}\binom{a_1, \ldots, a_p}{b_1, \ldots, b_q}= \frac{r}{2 \pi i} \int \mathrm{d} s\, z^{-s} \, \\
    &\qquad \quad \times \frac{\prod_{i=1}^n \Gamma\left(1-a_i-r s\right) \prod_{j=1}^m  \Gamma\left(b_j+r s\right)}{\prod_{l=n+1}^{p} \Gamma\left(a_{l}+r s\right) \prod_{h=m+1}^{q}\Gamma\left(1-b_{h}-r s\right)} \,, \nonumber \\
&_qF_p(\{a\}_p;\{b\}_q;z) \nonumber \\
&\qquad \qquad = \sum_{k=0}^{\infty}\left(a_1\right)_k \ldots\left(a_p\right)_k /\left(b_1\right)_k \ldots\left(b_q\right)_k z^k / k!\,. \nonumber 
\end{align}
%

%----------------------------------------------------------------------
\section{Box, triangle and bubble coefficients of the one-loop Compton in the wave regime}
\label{sec:app4}
%----------------------------------------------------------------------

In this appendix we provide the master integral coefficients in the wave regime to leading order in the $\epsilon$ expansion, complementing the discussion in the main text. We first observe that the curved \eqref{eq:1SF_one-loop_res} and flat \eqref{eq: flat space-one-loop} space master integral coefficients agree under the substitution 
%$\boldsymbol{\varepsilon}_{\mu \nu} \to \varepsilon_{\mu \nu}$
%
\begin{align}
    \boldsymbol{c}_i \stackrel{\boldsymbol{\varepsilon}_{\mu \nu} \to \varepsilon_{\mu \nu}}{=} c_i\,, \quad i \in \{c_{\scriptscriptstyle\Box}, c_{\scriptscriptstyle\bigtriangleup}, c_{\scriptscriptstyle\bubble}\}\,, 
\end{align}
which are functions of $\omega$, $q^{\mu}$, $\varepsilon_{\mu \nu}$. Furthermore, we define
\begin{align}
     &l_{c}^\mu = \epsilon^{\mu \nu \rho \sigma} q_\nu v_{\mathrm{H}\rho} c_\sigma\,, \qquad l_{cd}^\mu = \epsilon^{\mu \nu \rho \sigma} v_{\mathrm{H}\nu} c_{\rho} d_{\sigma}\,,  \\
     &\qquad \qquad a^{\mu}_{\mathrm{H}} = \frac{1}{2 m_\mathrm{H}} \epsilon^{\mu \nu \rho \sigma} \bar{S}_{\mathrm{H} \nu \rho} v_{\mathrm{H}\sigma} \,, \nonumber
\end{align}
defined for arbitrary vectors $c^{\mu}$ and $d^{\mu}$. Given these conventions and the gauge choice $v_\mathrm{H} \cdot \varepsilon  = 0$, we organize all the coefficients appearing in \eqref{eq: flat space-one-loop} in powers of the spin
\begin{align}        
& c_{\scriptscriptstyle\Box}^{ \text{wave}} = \sum_{n = 0}^{2}c_{\scriptscriptstyle\Box,n} \,, \qquad c_{{\scriptscriptstyle\bigtriangleup}}^{ \text{wave}} = \sum_{n = 0}^{2}c_{{\scriptscriptstyle\bigtriangleup},n} \,,
    \label{eq: one-loop-Compton-wave-coeffs2} \\
& \hspace{1.5cm} c_{{\scriptscriptstyle\bubble}}^{ \text{wave}} = \sum_{n = 0}^{2}c_{{\scriptscriptstyle\bubble},n} \nonumber \,.
\end{align}
Defining $Z = q^2 + 4\omega^2$, we list below the full set of coefficients in \eqref{eq: one-loop-Compton-wave-coeffs2} starting from the spinless case 
\begin{widetext}
\begin{align}
    & \hspace{-0.25cm}\frac{c_{{\scriptscriptstyle\Box},0}}{\pi^3 G_\mathrm{N}^2 m_{\mathrm{H}}^3} 
    = -\frac{2048 \omega^4}{3 Z^4}\Big[ Z^2(q^4 - 24\omega^4)(\varepsilon_1 \cdot \varepsilon_2)^2 - 4Zq^2 (q^2 + 8 \omega^2) (q \cdot \varepsilon_1) (q \cdot \varepsilon_2) (\varepsilon_1 \cdot \varepsilon_2)]   \\
   & \hspace{4.5cm} - 2q^2(q^2 - 8\omega^2)(q \cdot \varepsilon_1)^2 (q \cdot \varepsilon_2)^2 \Big] + \mathcal{O}(\epsilon)\;, \nonumber \\[0.1cm]
      & \hspace{-0.25cm} \frac{c_{{\scriptscriptstyle\bigtriangleup},0}}{\pi^3 G_\mathrm{N}^2 m_{\mathrm{H}}^3} =  \frac{8}{3Z^4}\Big[ Z^2 (q^6 + 180 q^4 \omega^2 + 240 q^2 \omega^4 - 2880 \omega^6)(\varepsilon_1 \cdot \varepsilon_2)^2 \\
   & \hspace{2.5cm} - 4Z (q^6 + 92 q^4 \omega^2 + 432 q^2 \omega^4 - 1728 \omega^8) (q \cdot \varepsilon_1) (q \cdot \varepsilon_2) (\varepsilon_1 \cdot \varepsilon_2)\nonumber \\[0.1cm]
   & \hspace{2.5cm} + 4(q^6 + 4q^4 \omega^2 + 624 q^2 \omega^4 - 576 \omega^6) (q \cdot \varepsilon_1)^2 (q \cdot \varepsilon_2)^2\Big] + \mathcal{O}(\epsilon)\;,  \nonumber \\
   & \hspace{-0.25cm}\frac{c_{{\scriptscriptstyle\bubble},0}}{\pi^3 G_\mathrm{N}^2 m_{\mathrm{H}}^3} 
    = \frac{512 \omega^2}{3 Z^4 \epsilon}\Big[ Z^2(5q^2 + 24\omega^2)(\varepsilon_1 \cdot \varepsilon_2)^2 - 8Z (q^2 + 8 \omega^2) (q \cdot \varepsilon_1) (q \cdot \varepsilon_2) (\varepsilon_1 \cdot \varepsilon_2)]   \\
   & \hspace{4.5cm} - 4 (q^2 - 8\omega^2)(q \cdot \varepsilon_1)^2 (q \cdot \varepsilon_2)^2 \Big] + \mathcal{O}(\epsilon^0)\;, \nonumber 
\end{align}
then proceeding with the linear in spin contributions
\begin{align}
   & \hspace{-0.25cm}\frac{c_{{\scriptscriptstyle\Box},1}}{\pi^3 G_\mathrm{N}^2 m_{\mathrm{H}}^3} 
   = - \frac{1024 i \omega^3}{3Z^4}\Big[ 48 Z^2 \omega^4 (l_k \cdot a_H)(\varepsilon_1 \cdot \varepsilon_2)^2 + 4 Z^2 q^2 \omega^2 (q^2 + 12 \omega^2)(l_{\varepsilon_1 \varepsilon_2} \cdot a_\mathrm{H})( \varepsilon_1 \cdot \varepsilon_2)  \\
   & \hspace{2.5cm} - 2 Z\omega^2(7 q^4+52 q^2 \omega ^2+96 \omega ^4)[(l_{\varepsilon_1} \cdot a_\mathrm{H}) (q \cdot \varepsilon_2) - (l_{\varepsilon_2} \cdot a_\mathrm{H}) (q \cdot \varepsilon_1)]( \varepsilon_1 \cdot \varepsilon_2) \nonumber \\[0.1cm]
   & \hspace{2.5cm} - 4 Z q^2 \omega^2 (5 q^2-12 \omega ^2) [(l_{\varepsilon_1 k} \cdot a_\mathrm{H})(q \cdot \varepsilon_2) + (l_{\varepsilon_2 k} \cdot a_\mathrm{H})(q \cdot \varepsilon_1) ]( \varepsilon_1 \cdot \varepsilon_2) \nonumber \\[0.1cm]
   & \hspace{2.5cm} + 48 Z q^2 \omega^2 (l_{ k} \cdot a_\mathrm{H})(q \cdot \varepsilon_1)(q \cdot \varepsilon_2) + 4 Z q^2 \omega^2 (7 q^2-4 \omega ^2) ( l_{\varepsilon_1 \varepsilon_2} \cdot a_\mathrm{H}) (q \cdot \varepsilon_1)(q \cdot \varepsilon_2) \nonumber \\[0.1cm] 
   & \hspace{2.5cm} - 2 q^2 (3 q^4-32 q^2 \omega ^2+16 \omega ^4)[(l_{\varepsilon_1 k} \cdot a_\mathrm{H})(q \cdot \varepsilon_2) + (l_{\varepsilon_2 k} \cdot a_\mathrm{H})(q \cdot \varepsilon_1)] \nonumber \\[0.1cm] 
   & \hspace{2.5cm} - q^2 (3 q^4+32 q^2 \omega ^2+80 \omega ^4)[(l_{\varepsilon_1} \cdot a_\mathrm{H})(q \cdot \varepsilon_2)-(l_{\varepsilon_2} \cdot a_\mathrm{H})(q \cdot \varepsilon_1)](q \cdot \varepsilon_1)(q \cdot \varepsilon_2)\Big] + \mathcal{O}(\epsilon)\;, \nonumber \\[0.1cm]
    & \hspace{-0.25cm}\frac{c_{{\scriptscriptstyle\bigtriangleup},1}}{\pi^3 G_\mathrm{N}^2 m_{\mathrm{H}}^3} = -\frac{64i\omega}{3Z^4}\Big[6Z^2(q^4 - 80 \omega^2) (l_k \cdot a_\mathrm{H})(\varepsilon_1 \cdot \varepsilon_2)^2 + 2 Z^2 q^2 (q^4 - 24 q^2 \omega^2 - 240 \omega^4) (l_{\varepsilon_1 \varepsilon_2} \cdot a_\mathrm{H})( \varepsilon_1 \cdot \varepsilon_2) \\ 
   & \hspace{2cm} - Z (7 q^6 - 20 q^4 \omega^2 - 432 q^2 \omega^4 - 960 \omega^6) [(l_{\varepsilon_1} \cdot a_\mathrm{H}) (q \cdot \varepsilon_2) - (l_{\varepsilon_2} \cdot a_\mathrm{H}) (q \cdot \varepsilon_1)]( \varepsilon_1 \cdot \varepsilon_2) \nonumber \\[0.1cm]
   & \hspace{2cm} - 4 Z q^2 (q^4 + 240 \omega^4) [(l_{\varepsilon_1 k} \cdot a_\mathrm{H})(q \cdot \varepsilon_2) + (l_{\varepsilon_2 k} \cdot a_\mathrm{H})(q \cdot \varepsilon_1) ]( \varepsilon_1 \cdot \varepsilon_2) \nonumber \\[0.1cm]
   & \hspace{2cm} - 12 Z (q^4 + 24 q^2 \omega^2 - 48 \omega^4)(l_{ k} \cdot a_\mathrm{H})(q \cdot \varepsilon_1)(q \cdot \varepsilon_2)\nonumber \\[0.1cm]
   & \hspace{2cm} - 4 Z q^2(q^4 + 32 q^2 \omega^2 - 144 \omega^4)( l_{\varepsilon_1 \varepsilon_2} \cdot a_\mathrm{H}) (q \cdot \varepsilon_1)(q \cdot \varepsilon_2) \nonumber \\[0.1cm]
   & \hspace{2cm} + 8 q^2(q^4 - 56q^2 \omega^2 + 144 \omega^4)[(l_{\varepsilon_1 k} \cdot a_\mathrm{H})(q \cdot \varepsilon_2) + (l_{\varepsilon_2 k} \cdot a_\mathrm{H})(q \cdot \varepsilon_1)](q \cdot \varepsilon_1)(q \cdot \varepsilon_2) \nonumber \\[0.1cm]
   & \hspace{2cm} + 2 (7q^6 +52 q^4 \omega^2 - 48 q^2 \omega^4 - 576\omega^6)[(l_{\varepsilon_1} \cdot a_\mathrm{H})(q \cdot \varepsilon_2)-(l_{\varepsilon_2} \cdot a_\mathrm{H})(q \cdot \varepsilon_1)](q \cdot \varepsilon_1)(q \cdot \varepsilon_2)\Big] + \mathcal{O}(\epsilon)\,,\nonumber \\[0.1cm]
   & \hspace{-0.25cm}\frac{c_{{\scriptscriptstyle\bubble},1}}{\pi^3 G_\mathrm{N}^2 m_{\mathrm{H}}^3} = -\frac{512i\omega}{3Z^4\epsilon}\Big[6Z^2(q^2 + 8 \omega^2) (l_k \cdot a_\mathrm{H})(\varepsilon_1 \cdot \varepsilon_2)^2 + Z^2 (3q^4 + 16 q^2 \omega^2 - 48 \omega^4) (l_{\varepsilon_1 \varepsilon_2} \cdot a_\mathrm{H})( \varepsilon_1 \cdot \varepsilon_2) \\
   & \hspace{2cm} - 2 Z (3 q^4 + 22 q^2 \omega^2 + 40 \omega^4) [(l_{\varepsilon_1} \cdot a_\mathrm{H}) (q \cdot \varepsilon_2) - (l_{\varepsilon_2} \cdot a_\mathrm{H}) (q \cdot \varepsilon_1)]( \varepsilon_1 \cdot \varepsilon_2) \nonumber \\[0.1cm]
   & \hspace{2cm} + 8 Z \omega^2 (5q^2 - 12 \omega^2) [(l_{\varepsilon_1 k} \cdot a_\mathrm{H})(q \cdot \varepsilon_2) + (l_{\varepsilon_2 k} \cdot a_\mathrm{H})(q \cdot \varepsilon_1) ]( \varepsilon_1 \cdot \varepsilon_2) \nonumber \\[0.1cm]
   & \hspace{2cm} - 96 Z \omega^2 (l_{ k} \cdot a_\mathrm{H})(q \cdot \varepsilon_1)(q \cdot \varepsilon_2) - 8 Z \omega^2(7 q^2 - 4 \omega^2)( l_{\varepsilon_1 \varepsilon_2} \cdot a_\mathrm{H}) (q \cdot \varepsilon_1)(q \cdot \varepsilon_2) \nonumber \\[0.1cm]
   & \hspace{2cm} + 4 (3 q^4 - 32 q^2 \omega^2 + 16 \omega^4)[(l_{\varepsilon_1 k} \cdot a_\mathrm{H})(q \cdot \varepsilon_2) + (l_{\varepsilon_2 k} \cdot a_\mathrm{H})(q \cdot \varepsilon_1)](q \cdot \varepsilon_1)(q \cdot \varepsilon_2) \nonumber \\[0.1cm]
   & \hspace{2cm} + 2 (3q^4 + 32 q^2 \omega^2 + 80\omega^4)[(l_{\varepsilon_1} \cdot a_\mathrm{H})(q \cdot \varepsilon_2)-(l_{\varepsilon_2} \cdot a_\mathrm{H})(q \cdot \varepsilon_1)](q \cdot \varepsilon_1)(q \cdot \varepsilon_2)\Big] + \mathcal{O}(\epsilon^0) \,,\nonumber
\end{align}
\newpage
and finally the quadratic in spin ones
\begin{align}
   & \hspace{-0.25cm}\frac{c_{{\scriptscriptstyle\Box},2}}{\pi^3 G_\mathrm{N}^2 m_{\mathrm{H}}^3} 
   = -\frac{256 \omega^2}{15 Z^6} \Big[Z^3 \omega^2 (15 q^6+172 q^4 \omega ^2-1920 \omega ^6)(q \cdot a_\mathrm{H})^2(\varepsilon_1 \cdot \varepsilon_2)^2  \\
   & \hspace{1.5cm} + 2 Z^3 q^2 \omega^2 (15 q^6+94 q^4 \omega ^2+360 q^2 \omega ^4+960 \omega ^6)a_\mathrm{H}^2(\varepsilon_1 \cdot \varepsilon_2)^2 \nonumber \\[0.1cm]
   & \hspace{1.5cm} + 12 Z^2 q^2 \omega^2 (5 q^6-24 q^4 \omega ^2-80 q^2 \omega ^4+640 \omega ^6) (k \cdot a_\mathrm{H})^2(\varepsilon_1 \cdot \varepsilon_2)^2 \nonumber \\[0.1cm] \nonumber 
   & \hspace{1.5cm} + 4 Z^2 q^4 \omega^2 (5 q^4-104 q^2 \omega ^2-880 \omega ^4)[(\varepsilon_1 \cdot a_\mathrm{H})(q\cdot \varepsilon_2)-(\varepsilon_2 \cdot a_\mathrm{H})(q\cdot \varepsilon_1)](k \cdot a_\mathrm{H})(\varepsilon_1 \cdot \varepsilon_2)\nonumber \\[0.1cm]
   & \hspace{1.5cm} + 4 Z^3 q^2 \omega^2 (5 q^4+42 q^2 \omega ^2+120 \omega ^4)[(\varepsilon_1 \cdot a_\mathrm{H})(q\cdot \varepsilon_2)+(\varepsilon_2 \cdot a_\mathrm{H})(q\cdot \varepsilon_1)](q \cdot a_\mathrm{H})(\varepsilon_1 \cdot \varepsilon_2) \nonumber \\[0.1cm]
   & \hspace{1.5cm} + 4 Z q^2 (5 q^8+20 q^6 \omega ^2+736 q^4 \omega ^4+5696 q^2 \omega ^6-3840 \omega ^8) (k \cdot a_\mathrm{H})^2 (q \cdot \varepsilon_1)(q \cdot \varepsilon_1)(\varepsilon_1 \cdot \varepsilon_2)\nonumber \\[0.1cm]
   & \hspace{1.5cm} - 4 Z^3 q^4 \omega^2(5 q^4+42 q^2 \omega ^2+120 \omega ^4)(\varepsilon_1 \cdot a_\mathrm{H})(\varepsilon_2 \cdot a_\mathrm{H})(\varepsilon_1 \cdot \varepsilon_2) \nonumber \\[0.1cm]
   & \hspace{1.5cm} - Z^2 q^2 (15 q^6+140 q^4 \omega ^2+1376 q^2 \omega ^4+4608 \omega ^6)(q \cdot a_\mathrm{H})^2 (q \cdot \varepsilon_1)(q \cdot \varepsilon_2)(\varepsilon_1 \cdot \varepsilon_2) \nonumber \\[0.1cm]
   & \hspace{1.5cm} + 6 Z^2 q^2 (5 q^8+20 q^6 \omega ^2+8 q^4 \omega ^4-224 q^2 \omega ^6-1280 \omega ^8) a_\mathrm{H}^2 (q \cdot \varepsilon_1)(q \cdot \varepsilon_2)(\varepsilon_1 \cdot \varepsilon_2)\nonumber \\[0.1cm]
   & \hspace{1.5cm} + 8 Z^2 q^2 \omega^4 (5 q^4-64 q^2 \omega ^2-240 \omega ^4)[(q \cdot \varepsilon_1)^2(\varepsilon_2 \cdot a_\mathrm{H})^2+(q \cdot \varepsilon_2)^2(\varepsilon_1 \cdot a_\mathrm{H})^2]\nonumber \\[0.1cm]
   & \hspace{1.5cm} - 4 Z q^2 (15 q^8+100 q^6 \omega ^2+672 q^4 \omega ^4+192 q^2 \omega ^6-1280 \omega ^8)[(\varepsilon_1 \cdot a_\mathrm{H})( q\cdot \varepsilon_2) - (\varepsilon_2 \cdot a_\mathrm{H})( q\cdot \varepsilon_1)](k \cdot a_\mathrm{H}) (q \cdot \varepsilon_1)(q \cdot \varepsilon_2)\nonumber \\[0.1cm]
   & \hspace{1.5cm} + 2 Z^2 q^2 (15 q^6+70 q^4 \omega ^2+32 q^2 \omega ^4-224 \omega ^6)[(\varepsilon_1 \cdot a_\mathrm{H})(q \cdot \varepsilon_2)+(\varepsilon_2 \cdot a_\mathrm{H})(q \cdot \varepsilon_1)](q \cdot a_\mathrm{H})(q \cdot \varepsilon_1) (q \cdot \varepsilon_2)\nonumber \\[0.1cm]
   & \hspace{1.5cm} - 16 Z q^2 \omega^2 (5 q^4-14 q^2 \omega ^2-184 \omega ^4) (q \cdot a_\mathrm{H})^2(q \cdot \varepsilon_1)^2 (q \cdot \varepsilon_2)^2\nonumber \\[0.1cm]
   & \hspace{1.5cm} - 4 Z q^2 (15 q^8+120 q^6 \omega ^2+392 q^4 \omega ^4+448 q^2 \omega ^6-1408 \omega ^8)a_\mathrm{H}^2 (q \cdot \varepsilon_1)^2 (q \cdot \varepsilon_2)^2 \nonumber \\[0.1cm]
   & \hspace{1.5cm} - 32 q^2 \omega^2 (5 q^6-132 q^4 \omega ^2+336 q^2 \omega ^4-64 \omega ^6) (k \cdot a_\mathrm{H})^2 (q \cdot \varepsilon_1)^2 (q \cdot \varepsilon_2)^2 \nonumber \\[0.1cm]
    & \hspace{1.5cm} - 2 Z^2 q^2  (15 q^8+70 q^6 \omega ^2+72 q^4 \omega ^4-736 q^2 \omega ^6-1920 \omega ^8)(\varepsilon_1 \cdot a_\mathrm{H}) (\varepsilon_2 \cdot a_\mathrm{H}) (q \cdot \varepsilon_1) (q \cdot \varepsilon_2)\Big] + \mathcal{O}(\epsilon)\,, \nonumber \\
    &{} \nonumber \\
   & \frac{c_{{\scriptscriptstyle\bigtriangleup},2}}{\pi^3 G_\mathrm{N}^2 m_{\mathrm{H}}^3} = \frac{1}{15Z^6}\Big[Z^3 (351 q^8+7120 q^6 \omega ^2+39840 q^4 \omega ^4-19200 q^2 \omega ^6-364800 \omega ^8)(q \cdot a_\mathrm{H})^2(\varepsilon_1 \cdot \varepsilon_2)^2 \\
   & \hspace{1cm} - Z^3 q^2(591 q^8+4240 q^6 \omega ^2+5280 q^4 \omega ^4-96000 q^2 \omega ^6-364800 \omega ^8)a_\mathrm{H}^2(\varepsilon_1 \cdot \varepsilon_2)^2 \nonumber \\[0.1cm]
   & \hspace{1cm} - 24 Z^2 q^2 (51 q^8+560 q^6 \omega ^2+4000 q^4 \omega ^4-6400 q^2 \omega ^6-83200 \omega ^8) (k \cdot a_\mathrm{H})^2(\varepsilon_1 \cdot \varepsilon_2)^2 \nonumber \\[0.1cm]
   & \hspace{1cm} + 16 Z^2 q^2 (77 q^8+1120 q^6 \omega ^2+2720 q^4 \omega ^4+57600 \omega ^8)[(\varepsilon_1 \cdot a_\mathrm{H})(q\cdot \varepsilon_2)-(\varepsilon_2 \cdot a_\mathrm{H})(q\cdot \varepsilon_1)](k \cdot a_\mathrm{H})(\varepsilon_1 \cdot \varepsilon_2)\nonumber \\[0.1cm]
   & \hspace{1cm} - 8 Z^3 q^2(77 q^6+540 q^4 \omega ^2+240 q^2 \omega ^4-4800 \omega ^6)[(\varepsilon_1 \cdot a_\mathrm{H})(q\cdot \varepsilon_2)+(\varepsilon_2 \cdot a_\mathrm{H})(q\cdot \varepsilon_1)](q \cdot a_\mathrm{H})(\varepsilon_1 \cdot \varepsilon_2) \nonumber \\[0.1cm]
   & \hspace{1cm} + 128 Z q^2 (19 q^8-136 q^6 \omega ^2+1600 q^4 \omega ^4+5760 q^2 \omega ^6-65280 \omega ^8) (k \cdot a_\mathrm{H})^2 (q \cdot \varepsilon_1)(q \cdot \varepsilon_1)(\varepsilon_1 \cdot \varepsilon_2)\nonumber \\[0.1cm]
   & \hspace{1cm} + 8Z^3 q^4 (77 q^6+540 q^4 \omega ^2+240 q^2 \omega ^4-4800 \omega ^6)(\varepsilon_1 \cdot a_\mathrm{H})(\varepsilon_2 \cdot a_\mathrm{H})(\varepsilon_1 \cdot \varepsilon_2)  \nonumber \\[0.1cm]
   & \hspace{1cm} - 4 Z^2 (197 q^8+1616 q^6 \omega ^2+21600 q^4 \omega ^4+26880 q^2 \omega ^6-234240 \omega ^8)(q \cdot a_\mathrm{H})^2 (q \cdot \varepsilon_1)(q \cdot \varepsilon_2)(\varepsilon_1 \cdot \varepsilon_2) \nonumber \\[0.1cm]
   & \hspace{1cm} + 12 Z^2 q^2 (197 q^8+304 q^6 \omega ^2-4640 q^4 \omega ^4-42240 q^2 \omega ^6-142080 \omega ^8) a_\mathrm{H}^2 (q \cdot \varepsilon_1)(q \cdot \varepsilon_2)(\varepsilon_1 \cdot \varepsilon_2)\nonumber \\[0.1cm]
   & \hspace{1cm} - 32Z^2 q^2 (43 q^6 \omega ^2+700 q^4 \omega ^4+6480 q^2 \omega ^6+14400 \omega ^8)[(q \cdot \varepsilon_1)^2(\varepsilon_2 \cdot a_\mathrm{H})^2+(q \cdot \varepsilon_2)^2(\varepsilon_1 \cdot a_\mathrm{H})^2]\nonumber \\[0.1cm]
   & \hspace{1cm} - 32 Z q^2 (77 q^8-288 q^6 \omega ^2-2400 q^4 \omega ^4-40960 q^2 \omega ^6-65280 \omega ^8)[(\varepsilon_1 \cdot a_\mathrm{H})( q\cdot \varepsilon_2) - (\varepsilon_2 \cdot a_\mathrm{H})( q\cdot \varepsilon_1)](k \cdot a_\mathrm{H}) (q \cdot \varepsilon_1)(q \cdot \varepsilon_2)\nonumber \\[0.1cm]
   & \hspace{1cm} + 16 Z^2 q^2 (77 q^6-524 q^4 \omega ^2-5200 q^2 \omega ^4-10560 \omega ^6)[(\varepsilon_1 \cdot a_\mathrm{H})(q \cdot \varepsilon_2)+(\varepsilon_2 \cdot a_\mathrm{H})(q \cdot \varepsilon_1)](q \cdot a_\mathrm{H})(q \cdot \varepsilon_1) (q \cdot \varepsilon_2)\nonumber \\[0.1cm]
   & \hspace{1cm} + 4Z (43 q^8+368 q^6 \omega ^2+25120 q^4 \omega ^4+96000 q^2 \omega ^6-103680 \omega ^8) (q \cdot a_\mathrm{H})^2(q \cdot \varepsilon_1)^2 (q \cdot \varepsilon_2)^2\nonumber \\[0.1cm]
   & \hspace{1cm} - 4 Z q^2 (591 q^8-2416 q^6 \omega ^2-50144 q^4 \omega ^4-244480 q^2 \omega ^6-579840 \omega ^8)a_\mathrm{H}^2 (q \cdot \varepsilon_1)^2 (q \cdot \varepsilon_2)^2 \nonumber \\[0.1cm]
   & \hspace{1cm} + 32 q^2 (q^8-48 q^6 \omega ^2+9312 q^4 \omega ^4-70400 q^2 \omega ^6+57600 \omega ^8) (k \cdot a_\mathrm{H})^2 (q \cdot \varepsilon_1)^2 (q \cdot \varepsilon_2)^2 \nonumber \\[0.1cm]
    & \hspace{1cm} - 16 Z^2 q^2 (77 q^8-696 q^6 \omega ^2-8000 q^4 \omega ^4-36480 q^2 \omega ^6-57600 \omega ^8)(\varepsilon_1 \cdot a_\mathrm{H}) (\varepsilon_2 \cdot a_\mathrm{H}) (q \cdot \varepsilon_1) (q \cdot \varepsilon_2)\Big] + \mathcal{O}(\epsilon)\;, \nonumber \\[0.1cm]
    & \hspace{-0.25cm}\frac{c_{{\scriptscriptstyle\bubble},2}}{\pi^3 G_\mathrm{N}^2 m_{\mathrm{H}}^3} 
   = \frac{64}{15 Z^6\epsilon} \Big[Z^3 (15 q^6+270 q^4 \omega ^2 + 1784 q^2 \omega^4 + 3840 \omega ^6)(q \cdot a_\mathrm{H})^2(\varepsilon_1 \cdot \varepsilon_2)^2  \\
   & \hspace{1.5cm} - 2 Z^3 q^2 (15 q^6+180 q^4 \omega ^2+892 q^2 \omega ^4+1680 \omega ^6)a_\mathrm{H}^2(\varepsilon_1 \cdot \varepsilon_2)^2 \nonumber \\[0.1cm]
   & \hspace{1.5cm} - 12 Z^2 q^2 (5 q^6+70 q^4 \omega ^2+528 q^2 \omega ^4+1440 \omega ^6) (k \cdot a_\mathrm{H})^2(\varepsilon_1 \cdot \varepsilon_2)^2 \nonumber \\[0.1cm] \nonumber 
   & \hspace{1.5cm} + 4 Z^2 (15 q^8+250 q^6 \omega ^2+1232 q^4 \omega ^4+2080 q^2 \omega ^6+3840 \omega ^8)[(\varepsilon_1 \cdot a_\mathrm{H})(q\cdot \varepsilon_2)-(\varepsilon_2 \cdot a_\mathrm{H})(q\cdot \varepsilon_1)](k \cdot a_\mathrm{H})(\varepsilon_1 \cdot \varepsilon_2)\nonumber \\[0.1cm]
   & \hspace{1.5cm} - 2 Z^3(15 q^6+160 q^4 \omega ^2+552 q^2 \omega ^4+480 \omega ^6)[(\varepsilon_1 \cdot a_\mathrm{H})(q\cdot \varepsilon_2)+(\varepsilon_2 \cdot a_\mathrm{H})(q\cdot \varepsilon_1)](q \cdot a_\mathrm{H})(\varepsilon_1 \cdot \varepsilon_2) \nonumber \\[0.1cm]
   & \hspace{1.5cm} + 8 Z (15 q^8+20 q^6 \omega ^2+736 q^4 \omega ^4+5696 q^2 \omega ^6-3840 \omega ^8) (k \cdot a_\mathrm{H})^2 (q \cdot \varepsilon_1)(q \cdot \varepsilon_1)(\varepsilon_1 \cdot \varepsilon_2)\nonumber \\[0.1cm]
   & \hspace{1.5cm} + 2 Z^3 q^2 (15 q^6+160 q^4 \omega ^2+552 q^2 \omega ^4+480 \omega ^6)(\varepsilon_1 \cdot a_\mathrm{H})(\varepsilon_2 \cdot a_\mathrm{H})(\varepsilon_1 \cdot \varepsilon_2) \nonumber \\[0.1cm]
   & \hspace{1.5cm} - 2 Z^2 (15 q^6+140 q^4 \omega ^2+1376 q^2 \omega ^4+4608 \omega ^6)(q \cdot a_\mathrm{H})^2 (q \cdot \varepsilon_1)(q \cdot \varepsilon_2)(\varepsilon_1 \cdot \varepsilon_2) \nonumber \\[0.1cm]
   & \hspace{1.5cm} + 24 Z^2 q^2 (5 q^6+50 q^4 \omega ^2+244 q^2 \omega ^4+528 \omega ^6) a_\mathrm{H}^2 (q \cdot \varepsilon_1)(q \cdot \varepsilon_2)(\varepsilon_1 \cdot \varepsilon_2)\nonumber \\[0.1cm]
   & \hspace{1.5cm} + 16 Z^2 (5 q^4 \omega ^4-64 q^2 \omega ^6-240 \omega ^8)^2[(q \cdot \varepsilon_1)^2(\varepsilon_2 \cdot a_\mathrm{H})^2+(q \cdot \varepsilon_2)^2(\varepsilon_1 \cdot a_\mathrm{H})^2] \nonumber \\[0.1cm]
   & \hspace{1.5cm} - 8 Z (15 q^8+100 q^6 \omega ^2+672 q^4 \omega ^4+192 q^2 \omega ^6-1280 \omega ^8)[(\varepsilon_1 \cdot a_\mathrm{H})( q\cdot \varepsilon_2) - (\varepsilon_2 \cdot a_\mathrm{H})( q\cdot \varepsilon_1)](k \cdot a_\mathrm{H}) (q \cdot \varepsilon_1)(q \cdot \varepsilon_2)\nonumber \\[0.1cm]
   & \hspace{1.5cm} + 4 Z^2 (15 q^6+70 q^4 \omega ^2+32 q^2 \omega ^4-224 \omega ^6)[(\varepsilon_1 \cdot a_\mathrm{H})(q \cdot \varepsilon_2)+(\varepsilon_2 \cdot a_\mathrm{H})(q \cdot \varepsilon_1)](q \cdot a_\mathrm{H})(q \cdot \varepsilon_1) (q \cdot \varepsilon_2)\nonumber \\[0.1cm]
   & \hspace{1.5cm} - 32 Z \omega^2 (5 q^4-14 q^2 \omega ^2-184 \omega ^4) (q \cdot a_\mathrm{H})^2(q \cdot \varepsilon_1)^2 (q \cdot \varepsilon_2)^2\nonumber \\[0.1cm]
   & \hspace{1.5cm} - 8 Z (15 q^8+120 q^6 \omega ^2+392 q^4 \omega ^4+448 q^2 \omega ^6-1408 \omega ^8)a_\mathrm{H}^2 (q \cdot \varepsilon_1)^2 (q \cdot \varepsilon_2)^2 \nonumber \\[0.1cm]
   & \hspace{1.5cm} + 64 \omega^2 (5 q^6-132 q^4 \omega ^2+336 q^2 \omega ^4-64 \omega ^6) (k \cdot a_\mathrm{H})^2 (q \cdot \varepsilon_1)^2 (q \cdot \varepsilon_2)^2 \nonumber \\[0.1cm]
    & \hspace{1.5cm} - 4 Z^2  (15 q^8+70 q^6 \omega ^2+72 q^4 \omega ^4-736 q^2 \omega ^6-1920 \omega ^8)(\varepsilon_1 \cdot a_\mathrm{H}) (\varepsilon_2 \cdot a_\mathrm{H}) (q \cdot \varepsilon_1) (q \cdot \varepsilon_2)\Big] + \mathcal{O}(\epsilon^0)\,. \nonumber
\end{align}
\end{widetext}

\newpage

\bibliography{referencesSF}% Produces the bibliography via BibTeX.

\begin{thebibliography}{168}%
\makeatletter
\providecommand \@ifxundefined [1]{%
 \@ifx{#1\undefined}
}%
\providecommand \@ifnum [1]{%
 \ifnum #1\expandafter \@firstoftwo
 \else \expandafter \@secondoftwo
 \fi
}%
\providecommand \@ifx [1]{%
 \ifx #1\expandafter \@firstoftwo
 \else \expandafter \@secondoftwo
 \fi
}%
\providecommand \natexlab [1]{#1}%
\providecommand \enquote  [1]{``#1''}%
\providecommand \bibnamefont  [1]{#1}%
\providecommand \bibfnamefont [1]{#1}%
\providecommand \citenamefont [1]{#1}%
\providecommand \href@noop [0]{\@secondoftwo}%
\providecommand \href [0]{\begingroup \@sanitize@url \@href}%
\providecommand \@href[1]{\@@startlink{#1}\@@href}%
\providecommand \@@href[1]{\endgroup#1\@@endlink}%
\providecommand \@sanitize@url [0]{\catcode `\\12\catcode `\$12\catcode
  `\&12\catcode `\#12\catcode `\^12\catcode `\_12\catcode `\%12\relax}%
\providecommand \@@startlink[1]{}%
\providecommand \@@endlink[0]{}%
\providecommand \url  [0]{\begingroup\@sanitize@url \@url }%
\providecommand \@url [1]{\endgroup\@href {#1}{\urlprefix }}%
\providecommand \urlprefix  [0]{URL }%
\providecommand \Eprint [0]{\href }%
\providecommand \doibase [0]{https://doi.org/}%
\providecommand \selectlanguage [0]{\@gobble}%
\providecommand \bibinfo  [0]{\@secondoftwo}%
\providecommand \bibfield  [0]{\@secondoftwo}%
\providecommand \translation [1]{[#1]}%
\providecommand \BibitemOpen [0]{}%
\providecommand \bibitemStop [0]{}%
\providecommand \bibitemNoStop [0]{.\EOS\space}%
\providecommand \EOS [0]{\spacefactor3000\relax}%
\providecommand \BibitemShut  [1]{\csname bibitem#1\endcsname}%
\let\auto@bib@innerbib\@empty
%</preamble>
\bibitem [{\citenamefont {Blanchet}(2002)}]{Blanchet:2002av}%
  \BibitemOpen
  \bibfield  {author} {\bibinfo {author} {\bibfnamefont {L.}~\bibnamefont
  {Blanchet}},\ }\bibfield  {title} {\bibinfo {title} {{Gravitational radiation
  from postNewtonian sources and inspiraling compact binaries}},\ }\href
  {https://doi.org/10.12942/lrr-2002-3} {\bibfield  {journal} {\bibinfo
  {journal} {Living Rev. Rel.}\ }\textbf {\bibinfo {volume} {5}},\ \bibinfo
  {pages} {3} (\bibinfo {year} {2002})},\ \Eprint
  {https://arxiv.org/abs/gr-qc/0202016} {arXiv:gr-qc/0202016} \BibitemShut
  {NoStop}%
\bibitem [{\citenamefont {Foffa}\ and\ \citenamefont
  {Sturani}(2014)}]{Foffa:2013qca}%
  \BibitemOpen
  \bibfield  {author} {\bibinfo {author} {\bibfnamefont {S.}~\bibnamefont
  {Foffa}}\ and\ \bibinfo {author} {\bibfnamefont {R.}~\bibnamefont
  {Sturani}},\ }\bibfield  {title} {\bibinfo {title} {{Effective field theory
  methods to model compact binaries}},\ }\href
  {https://doi.org/10.1088/0264-9381/31/4/043001} {\bibfield  {journal}
  {\bibinfo  {journal} {Class. Quant. Grav.}\ }\textbf {\bibinfo {volume}
  {31}},\ \bibinfo {pages} {043001} (\bibinfo {year} {2014})},\ \Eprint
  {https://arxiv.org/abs/1309.3474} {arXiv:1309.3474 [gr-qc]} \BibitemShut
  {NoStop}%
\bibitem [{\citenamefont {Bern}\ \emph {et~al.}(2019)\citenamefont {Bern},
  \citenamefont {Cheung}, \citenamefont {Roiban}, \citenamefont {Shen},
  \citenamefont {Solon},\ and\ \citenamefont {Zeng}}]{Bern:2019crd}%
  \BibitemOpen
  \bibfield  {author} {\bibinfo {author} {\bibfnamefont {Z.}~\bibnamefont
  {Bern}}, \bibinfo {author} {\bibfnamefont {C.}~\bibnamefont {Cheung}},
  \bibinfo {author} {\bibfnamefont {R.}~\bibnamefont {Roiban}}, \bibinfo
  {author} {\bibfnamefont {C.-H.}\ \bibnamefont {Shen}}, \bibinfo {author}
  {\bibfnamefont {M.~P.}\ \bibnamefont {Solon}},\ and\ \bibinfo {author}
  {\bibfnamefont {M.}~\bibnamefont {Zeng}},\ }\bibfield  {title} {\bibinfo
  {title} {{Black Hole Binary Dynamics from the Double Copy and Effective
  Theory}},\ }\href {https://doi.org/10.1007/JHEP10(2019)206} {\bibfield
  {journal} {\bibinfo  {journal} {JHEP}\ }\textbf {\bibinfo {volume} {10}},\
  \bibinfo {pages} {206}},\ \Eprint {https://arxiv.org/abs/1908.01493}
  {arXiv:1908.01493 [hep-th]} \BibitemShut {NoStop}%
\bibitem [{\citenamefont {Buonanno}\ \emph {et~al.}(2022)\citenamefont
  {Buonanno}, \citenamefont {Khalil}, \citenamefont {O'Connell}, \citenamefont
  {Roiban}, \citenamefont {Solon},\ and\ \citenamefont
  {Zeng}}]{Buonanno:2022pgc}%
  \BibitemOpen
  \bibfield  {author} {\bibinfo {author} {\bibfnamefont {A.}~\bibnamefont
  {Buonanno}}, \bibinfo {author} {\bibfnamefont {M.}~\bibnamefont {Khalil}},
  \bibinfo {author} {\bibfnamefont {D.}~\bibnamefont {O'Connell}}, \bibinfo
  {author} {\bibfnamefont {R.}~\bibnamefont {Roiban}}, \bibinfo {author}
  {\bibfnamefont {M.~P.}\ \bibnamefont {Solon}},\ and\ \bibinfo {author}
  {\bibfnamefont {M.}~\bibnamefont {Zeng}},\ }\bibfield  {title} {\bibinfo
  {title} {{Snowmass White Paper: Gravitational Waves and Scattering
  Amplitudes}},\ }in\ \href@noop {} {\emph {\bibinfo {booktitle} {{Snowmass
  2021}}}}\ (\bibinfo {year} {2022})\ \Eprint
  {https://arxiv.org/abs/2204.05194} {arXiv:2204.05194 [hep-th]} \BibitemShut
  {NoStop}%
\bibitem [{\citenamefont {Bjerrum-Bohr}\ \emph {et~al.}(2022)\citenamefont
  {Bjerrum-Bohr}, \citenamefont {Damgaard}, \citenamefont {Plante},\ and\
  \citenamefont {Vanhove}}]{Bjerrum-Bohr:2022blt}%
  \BibitemOpen
  \bibfield  {author} {\bibinfo {author} {\bibfnamefont {N.~E.~J.}\
  \bibnamefont {Bjerrum-Bohr}}, \bibinfo {author} {\bibfnamefont {P.~H.}\
  \bibnamefont {Damgaard}}, \bibinfo {author} {\bibfnamefont {L.}~\bibnamefont
  {Plante}},\ and\ \bibinfo {author} {\bibfnamefont {P.}~\bibnamefont
  {Vanhove}},\ }\bibfield  {title} {\bibinfo {title} {{The SAGEX review on
  scattering amplitudes Chapter 13: Post-Minkowskian expansion from scattering
  amplitudes}},\ }\href {https://doi.org/10.1088/1751-8121/ac7a78} {\bibfield
  {journal} {\bibinfo  {journal} {J. Phys. A}\ }\textbf {\bibinfo {volume}
  {55}},\ \bibinfo {pages} {443014} (\bibinfo {year} {2022})},\ \Eprint
  {https://arxiv.org/abs/2203.13024} {arXiv:2203.13024 [hep-th]} \BibitemShut
  {NoStop}%
\bibitem [{\citenamefont {Barack}\ and\ \citenamefont
  {Pound}(2019)}]{Barack:2018yvs}%
  \BibitemOpen
  \bibfield  {author} {\bibinfo {author} {\bibfnamefont {L.}~\bibnamefont
  {Barack}}\ and\ \bibinfo {author} {\bibfnamefont {A.}~\bibnamefont {Pound}},\
  }\bibfield  {title} {\bibinfo {title} {{Self-force and radiation reaction in
  general relativity}},\ }\href {https://doi.org/10.1088/1361-6633/aae552}
  {\bibfield  {journal} {\bibinfo  {journal} {Rept. Prog. Phys.}\ }\textbf
  {\bibinfo {volume} {82}},\ \bibinfo {pages} {016904} (\bibinfo {year}
  {2019})},\ \Eprint {https://arxiv.org/abs/1805.10385} {arXiv:1805.10385
  [gr-qc]} \BibitemShut {NoStop}%
\bibitem [{\citenamefont {Pound}\ and\ \citenamefont
  {Wardell}(2021)}]{Pound:2021qin}%
  \BibitemOpen
  \bibfield  {author} {\bibinfo {author} {\bibfnamefont {A.}~\bibnamefont
  {Pound}}\ and\ \bibinfo {author} {\bibfnamefont {B.}~\bibnamefont
  {Wardell}},\ }\bibfield  {title} {\bibinfo {title} {{Black hole perturbation
  theory and gravitational self-force}}\ }\href
  {https://doi.org/10.1007/978-981-15-4702-7\_38-1}
  {10.1007/978-981-15-4702-7\_38-1} (\bibinfo {year} {2021}),\ \Eprint
  {https://arxiv.org/abs/2101.04592} {arXiv:2101.04592 [gr-qc]} \BibitemShut
  {NoStop}%
\bibitem [{\citenamefont {Poisson}\ \emph {et~al.}(2011)\citenamefont
  {Poisson}, \citenamefont {Pound},\ and\ \citenamefont
  {Vega}}]{Poisson:2011nh}%
  \BibitemOpen
  \bibfield  {author} {\bibinfo {author} {\bibfnamefont {E.}~\bibnamefont
  {Poisson}}, \bibinfo {author} {\bibfnamefont {A.}~\bibnamefont {Pound}},\
  and\ \bibinfo {author} {\bibfnamefont {I.}~\bibnamefont {Vega}},\ }\bibfield
  {title} {\bibinfo {title} {{The Motion of point particles in curved
  spacetime}},\ }\href {https://doi.org/10.12942/lrr-2011-7} {\bibfield
  {journal} {\bibinfo  {journal} {Living Rev. Rel.}\ }\textbf {\bibinfo
  {volume} {14}},\ \bibinfo {pages} {7} (\bibinfo {year} {2011})},\ \Eprint
  {https://arxiv.org/abs/1102.0529} {arXiv:1102.0529 [gr-qc]} \BibitemShut
  {NoStop}%
\bibitem [{\citenamefont {Damour}\ and\ \citenamefont
  {Rettegno}(2023)}]{Damour:2022ybd}%
  \BibitemOpen
  \bibfield  {author} {\bibinfo {author} {\bibfnamefont {T.}~\bibnamefont
  {Damour}}\ and\ \bibinfo {author} {\bibfnamefont {P.}~\bibnamefont
  {Rettegno}},\ }\bibfield  {title} {\bibinfo {title} {{Strong-field scattering
  of two black holes: Numerical relativity meets post-Minkowskian gravity}},\
  }\href {https://doi.org/10.1103/PhysRevD.107.064051} {\bibfield  {journal}
  {\bibinfo  {journal} {Phys. Rev. D}\ }\textbf {\bibinfo {volume} {107}},\
  \bibinfo {pages} {064051} (\bibinfo {year} {2023})},\ \Eprint
  {https://arxiv.org/abs/2211.01399} {arXiv:2211.01399 [gr-qc]} \BibitemShut
  {NoStop}%
\bibitem [{\citenamefont {Barack}\ \emph {et~al.}(2023)\citenamefont {Barack}
  \emph {et~al.}}]{Barack:2023oqp}%
  \BibitemOpen
  \bibfield  {author} {\bibinfo {author} {\bibfnamefont {L.}~\bibnamefont
  {Barack}} \emph {et~al.},\ }\bibfield  {title} {\bibinfo {title} {{Comparison
  of post-Minkowskian and self-force expansions: Scattering in a scalar charge
  toy model}},\ }\href {https://doi.org/10.1103/PhysRevD.108.024025} {\bibfield
   {journal} {\bibinfo  {journal} {Phys. Rev. D}\ }\textbf {\bibinfo {volume}
  {108}},\ \bibinfo {pages} {024025} (\bibinfo {year} {2023})},\ \Eprint
  {https://arxiv.org/abs/2304.09200} {arXiv:2304.09200 [hep-th]} \BibitemShut
  {NoStop}%
\bibitem [{\citenamefont {Rettegno}\ \emph {et~al.}(2023)\citenamefont
  {Rettegno}, \citenamefont {Pratten}, \citenamefont {Thomas}, \citenamefont
  {Schmidt},\ and\ \citenamefont {Damour}}]{Rettegno:2023ghr}%
  \BibitemOpen
  \bibfield  {author} {\bibinfo {author} {\bibfnamefont {P.}~\bibnamefont
  {Rettegno}}, \bibinfo {author} {\bibfnamefont {G.}~\bibnamefont {Pratten}},
  \bibinfo {author} {\bibfnamefont {L.~M.}\ \bibnamefont {Thomas}}, \bibinfo
  {author} {\bibfnamefont {P.}~\bibnamefont {Schmidt}},\ and\ \bibinfo {author}
  {\bibfnamefont {T.}~\bibnamefont {Damour}},\ }\bibfield  {title} {\bibinfo
  {title} {{Strong-field scattering of two spinning black holes: Numerical
  relativity versus post-Minkowskian gravity}},\ }\href
  {https://doi.org/10.1103/PhysRevD.108.124016} {\bibfield  {journal} {\bibinfo
   {journal} {Phys. Rev. D}\ }\textbf {\bibinfo {volume} {108}},\ \bibinfo
  {pages} {124016} (\bibinfo {year} {2023})},\ \Eprint
  {https://arxiv.org/abs/2307.06999} {arXiv:2307.06999 [gr-qc]} \BibitemShut
  {NoStop}%
\bibitem [{\citenamefont {Buonanno}\ \emph {et~al.}(2024)\citenamefont
  {Buonanno}, \citenamefont {Jakobsen},\ and\ \citenamefont
  {Mogull}}]{Buonanno:2024vkx}%
  \BibitemOpen
  \bibfield  {author} {\bibinfo {author} {\bibfnamefont {A.}~\bibnamefont
  {Buonanno}}, \bibinfo {author} {\bibfnamefont {G.~U.}\ \bibnamefont
  {Jakobsen}},\ and\ \bibinfo {author} {\bibfnamefont {G.}~\bibnamefont
  {Mogull}},\ }\bibfield  {title} {\bibinfo {title} {{Post-Minkowskian theory
  meets the spinning effective-one-body approach for two-body scattering}},\
  }\href {https://doi.org/10.1103/PhysRevD.110.044038} {\bibfield  {journal}
  {\bibinfo  {journal} {Phys. Rev. D}\ }\textbf {\bibinfo {volume} {110}},\
  \bibinfo {pages} {044038} (\bibinfo {year} {2024})},\ \Eprint
  {https://arxiv.org/abs/2402.12342} {arXiv:2402.12342 [gr-qc]} \BibitemShut
  {NoStop}%
\bibitem [{\citenamefont {Barack}\ and\ \citenamefont
  {Long}(2022)}]{Barack:2022pde}%
  \BibitemOpen
  \bibfield  {author} {\bibinfo {author} {\bibfnamefont {L.}~\bibnamefont
  {Barack}}\ and\ \bibinfo {author} {\bibfnamefont {O.}~\bibnamefont {Long}},\
  }\bibfield  {title} {\bibinfo {title} {{Self-force correction to the
  deflection angle in black-hole scattering: A scalar charge toy model}},\
  }\href {https://doi.org/10.1103/PhysRevD.106.104031} {\bibfield  {journal}
  {\bibinfo  {journal} {Phys. Rev. D}\ }\textbf {\bibinfo {volume} {106}},\
  \bibinfo {pages} {104031} (\bibinfo {year} {2022})},\ \Eprint
  {https://arxiv.org/abs/2209.03740} {arXiv:2209.03740 [gr-qc]} \BibitemShut
  {NoStop}%
\bibitem [{\citenamefont {Bini}\ \emph
  {et~al.}(2024{\natexlab{a}})\citenamefont {Bini}, \citenamefont {Geralico},
  \citenamefont {Kavanagh}, \citenamefont {Pound},\ and\ \citenamefont
  {Usseglio}}]{Bini:2024icd}%
  \BibitemOpen
  \bibfield  {author} {\bibinfo {author} {\bibfnamefont {D.}~\bibnamefont
  {Bini}}, \bibinfo {author} {\bibfnamefont {A.}~\bibnamefont {Geralico}},
  \bibinfo {author} {\bibfnamefont {C.}~\bibnamefont {Kavanagh}}, \bibinfo
  {author} {\bibfnamefont {A.}~\bibnamefont {Pound}},\ and\ \bibinfo {author}
  {\bibfnamefont {D.}~\bibnamefont {Usseglio}},\ }\bibfield  {title} {\bibinfo
  {title} {{Post-Minkowskian self-force in the low-velocity limit: Scalar field
  scattering}},\ }\href {https://doi.org/10.1103/PhysRevD.110.064050}
  {\bibfield  {journal} {\bibinfo  {journal} {Phys. Rev. D}\ }\textbf {\bibinfo
  {volume} {110}},\ \bibinfo {pages} {064050} (\bibinfo {year}
  {2024}{\natexlab{a}})},\ \Eprint {https://arxiv.org/abs/2406.15878}
  {arXiv:2406.15878 [gr-qc]} \BibitemShut {NoStop}%
\bibitem [{\citenamefont {Long}\ \emph {et~al.}(2024)\citenamefont {Long},
  \citenamefont {Whittall},\ and\ \citenamefont {Barack}}]{Long:2024ltn}%
  \BibitemOpen
  \bibfield  {author} {\bibinfo {author} {\bibfnamefont {O.}~\bibnamefont
  {Long}}, \bibinfo {author} {\bibfnamefont {C.}~\bibnamefont {Whittall}},\
  and\ \bibinfo {author} {\bibfnamefont {L.}~\bibnamefont {Barack}},\
  }\bibfield  {title} {\bibinfo {title} {{Black hole scattering near the
  transition to plunge: Self-force and resummation of post-Minkowskian
  theory}},\ }\href {https://doi.org/10.1103/PhysRevD.110.044039} {\bibfield
  {journal} {\bibinfo  {journal} {Phys. Rev. D}\ }\textbf {\bibinfo {volume}
  {110}},\ \bibinfo {pages} {044039} (\bibinfo {year} {2024})},\ \Eprint
  {https://arxiv.org/abs/2406.08363} {arXiv:2406.08363 [gr-qc]} \BibitemShut
  {NoStop}%
\bibitem [{\citenamefont {Kosmopoulos}\ and\ \citenamefont
  {Solon}(2023)}]{Kosmopoulos:2023bwc}%
  \BibitemOpen
  \bibfield  {author} {\bibinfo {author} {\bibfnamefont {D.}~\bibnamefont
  {Kosmopoulos}}\ and\ \bibinfo {author} {\bibfnamefont {M.~P.}\ \bibnamefont
  {Solon}},\ }\bibfield  {title} {\bibinfo {title} {{Gravitational Self Force
  from Scattering Amplitudes in Curved Space}},\ }\href@noop {} {\  (\bibinfo
  {year} {2023})},\ \Eprint {https://arxiv.org/abs/2308.15304}
  {arXiv:2308.15304 [hep-th]} \BibitemShut {NoStop}%
\bibitem [{\citenamefont {Cheung}\ \emph {et~al.}(2023)\citenamefont {Cheung},
  \citenamefont {Parra-Martinez}, \citenamefont {Rothstein}, \citenamefont
  {Shah},\ and\ \citenamefont {Wilson-Gerow}}]{Cheung:2023lnj}%
  \BibitemOpen
  \bibfield  {author} {\bibinfo {author} {\bibfnamefont {C.}~\bibnamefont
  {Cheung}}, \bibinfo {author} {\bibfnamefont {J.}~\bibnamefont
  {Parra-Martinez}}, \bibinfo {author} {\bibfnamefont {I.~Z.}\ \bibnamefont
  {Rothstein}}, \bibinfo {author} {\bibfnamefont {N.}~\bibnamefont {Shah}},\
  and\ \bibinfo {author} {\bibfnamefont {J.}~\bibnamefont {Wilson-Gerow}},\
  }\bibfield  {title} {\bibinfo {title} {{Effective Field Theory for Extreme
  Mass Ratios}},\ }\href@noop {} {\  (\bibinfo {year} {2023})},\ \Eprint
  {https://arxiv.org/abs/2308.14832} {arXiv:2308.14832 [hep-th]} \BibitemShut
  {NoStop}%
\bibitem [{\citenamefont {Cheung}\ \emph {et~al.}(2024)\citenamefont {Cheung},
  \citenamefont {Parra-Martinez}, \citenamefont {Rothstein}, \citenamefont
  {Shah},\ and\ \citenamefont {Wilson-Gerow}}]{Cheung:2024byb}%
  \BibitemOpen
  \bibfield  {author} {\bibinfo {author} {\bibfnamefont {C.}~\bibnamefont
  {Cheung}}, \bibinfo {author} {\bibfnamefont {J.}~\bibnamefont
  {Parra-Martinez}}, \bibinfo {author} {\bibfnamefont {I.~Z.}\ \bibnamefont
  {Rothstein}}, \bibinfo {author} {\bibfnamefont {N.}~\bibnamefont {Shah}},\
  and\ \bibinfo {author} {\bibfnamefont {J.}~\bibnamefont {Wilson-Gerow}},\
  }\bibfield  {title} {\bibinfo {title} {{Gravitational scattering and beyond
  from extreme mass ratio effective field theory}},\ }\href
  {https://doi.org/10.1007/JHEP10(2024)005} {\bibfield  {journal} {\bibinfo
  {journal} {JHEP}\ }\textbf {\bibinfo {volume} {10}},\ \bibinfo {pages}
  {005}},\ \Eprint {https://arxiv.org/abs/2406.14770} {arXiv:2406.14770
  [hep-th]} \BibitemShut {NoStop}%
\bibitem [{\citenamefont {Duff}(1973)}]{Duff:1973zz}%
  \BibitemOpen
  \bibfield  {author} {\bibinfo {author} {\bibfnamefont {M.~J.}\ \bibnamefont
  {Duff}},\ }\bibfield  {title} {\bibinfo {title} {{Quantum Tree Graphs and the
  Schwarzschild Solution}},\ }\href {https://doi.org/10.1103/PhysRevD.7.2317}
  {\bibfield  {journal} {\bibinfo  {journal} {Phys. Rev. D}\ }\textbf {\bibinfo
  {volume} {7}},\ \bibinfo {pages} {2317} (\bibinfo {year} {1973})}\BibitemShut
  {NoStop}%
\bibitem [{\citenamefont {Neill}\ and\ \citenamefont
  {Rothstein}(2013)}]{Neill:2013wsa}%
  \BibitemOpen
  \bibfield  {author} {\bibinfo {author} {\bibfnamefont {D.}~\bibnamefont
  {Neill}}\ and\ \bibinfo {author} {\bibfnamefont {I.~Z.}\ \bibnamefont
  {Rothstein}},\ }\bibfield  {title} {\bibinfo {title} {{Classical Space-Times
  from the S Matrix}},\ }\href
  {https://doi.org/10.1016/j.nuclphysb.2013.09.007} {\bibfield  {journal}
  {\bibinfo  {journal} {Nucl. Phys. B}\ }\textbf {\bibinfo {volume} {877}},\
  \bibinfo {pages} {177} (\bibinfo {year} {2013})},\ \Eprint
  {https://arxiv.org/abs/1304.7263} {arXiv:1304.7263 [hep-th]} \BibitemShut
  {NoStop}%
\bibitem [{\citenamefont {Bjerrum-Bohr}\ \emph {et~al.}(2018)\citenamefont
  {Bjerrum-Bohr}, \citenamefont {Damgaard}, \citenamefont {Festuccia},
  \citenamefont {Plant\'e},\ and\ \citenamefont
  {Vanhove}}]{Bjerrum-Bohr:2018xdl}%
  \BibitemOpen
  \bibfield  {author} {\bibinfo {author} {\bibfnamefont {N.~E.~J.}\
  \bibnamefont {Bjerrum-Bohr}}, \bibinfo {author} {\bibfnamefont {P.~H.}\
  \bibnamefont {Damgaard}}, \bibinfo {author} {\bibfnamefont {G.}~\bibnamefont
  {Festuccia}}, \bibinfo {author} {\bibfnamefont {L.}~\bibnamefont
  {Plant\'e}},\ and\ \bibinfo {author} {\bibfnamefont {P.}~\bibnamefont
  {Vanhove}},\ }\bibfield  {title} {\bibinfo {title} {{General Relativity from
  Scattering Amplitudes}},\ }\href
  {https://doi.org/10.1103/PhysRevLett.121.171601} {\bibfield  {journal}
  {\bibinfo  {journal} {Phys. Rev. Lett.}\ }\textbf {\bibinfo {volume} {121}},\
  \bibinfo {pages} {171601} (\bibinfo {year} {2018})},\ \Eprint
  {https://arxiv.org/abs/1806.04920} {arXiv:1806.04920 [hep-th]} \BibitemShut
  {NoStop}%
\bibitem [{\citenamefont {Koemans~Collado}\ \emph {et~al.}(2018)\citenamefont
  {Koemans~Collado}, \citenamefont {Di~Vecchia}, \citenamefont {Russo},\ and\
  \citenamefont {Thomas}}]{KoemansCollado:2018hss}%
  \BibitemOpen
  \bibfield  {author} {\bibinfo {author} {\bibfnamefont {A.}~\bibnamefont
  {Koemans~Collado}}, \bibinfo {author} {\bibfnamefont {P.}~\bibnamefont
  {Di~Vecchia}}, \bibinfo {author} {\bibfnamefont {R.}~\bibnamefont {Russo}},\
  and\ \bibinfo {author} {\bibfnamefont {S.}~\bibnamefont {Thomas}},\
  }\bibfield  {title} {\bibinfo {title} {{The subleading eikonal in
  supergravity theories}},\ }\href {https://doi.org/10.1007/JHEP10(2018)038}
  {\bibfield  {journal} {\bibinfo  {journal} {JHEP}\ }\textbf {\bibinfo
  {volume} {10}},\ \bibinfo {pages} {038}},\ \Eprint
  {https://arxiv.org/abs/1807.04588} {arXiv:1807.04588 [hep-th]} \BibitemShut
  {NoStop}%
\bibitem [{\citenamefont {Jakobsen}(2020)}]{Jakobsen:2020ksu}%
  \BibitemOpen
  \bibfield  {author} {\bibinfo {author} {\bibfnamefont {G.~U.}\ \bibnamefont
  {Jakobsen}},\ }\bibfield  {title} {\bibinfo {title}
  {{Schwarzschild-Tangherlini Metric from Scattering Amplitudes}},\ }\href
  {https://doi.org/10.1103/PhysRevD.102.104065} {\bibfield  {journal} {\bibinfo
   {journal} {Phys. Rev. D}\ }\textbf {\bibinfo {volume} {102}},\ \bibinfo
  {pages} {104065} (\bibinfo {year} {2020})},\ \Eprint
  {https://arxiv.org/abs/2006.01734} {arXiv:2006.01734 [hep-th]} \BibitemShut
  {NoStop}%
\bibitem [{\citenamefont {Mougiakakos}\ and\ \citenamefont
  {Vanhove}(2021)}]{Mougiakakos:2020laz}%
  \BibitemOpen
  \bibfield  {author} {\bibinfo {author} {\bibfnamefont {S.}~\bibnamefont
  {Mougiakakos}}\ and\ \bibinfo {author} {\bibfnamefont {P.}~\bibnamefont
  {Vanhove}},\ }\bibfield  {title} {\bibinfo {title}
  {{Schwarzschild-Tangherlini metric from scattering amplitudes in various
  dimensions}},\ }\href {https://doi.org/10.1103/PhysRevD.103.026001}
  {\bibfield  {journal} {\bibinfo  {journal} {Phys. Rev. D}\ }\textbf {\bibinfo
  {volume} {103}},\ \bibinfo {pages} {026001} (\bibinfo {year} {2021})},\
  \Eprint {https://arxiv.org/abs/2010.08882} {arXiv:2010.08882 [hep-th]}
  \BibitemShut {NoStop}%
\bibitem [{\citenamefont {Mougiakakos}\ and\ \citenamefont
  {Vanhove}(2024)}]{Mougiakakos:2024nku}%
  \BibitemOpen
  \bibfield  {author} {\bibinfo {author} {\bibfnamefont {S.}~\bibnamefont
  {Mougiakakos}}\ and\ \bibinfo {author} {\bibfnamefont {P.}~\bibnamefont
  {Vanhove}},\ }\bibfield  {title} {\bibinfo {title} {{Schwarzschild metric
  from Scattering Amplitudes to all orders in $G_N$}},\ }\href@noop {} {\
  (\bibinfo {year} {2024})},\ \Eprint {https://arxiv.org/abs/2405.14421}
  {arXiv:2405.14421 [hep-th]} \BibitemShut {NoStop}%
\bibitem [{\citenamefont {Damgaard}\ and\ \citenamefont
  {Lee}(2024)}]{Damgaard:2024fqj}%
  \BibitemOpen
  \bibfield  {author} {\bibinfo {author} {\bibfnamefont {P.~H.}\ \bibnamefont
  {Damgaard}}\ and\ \bibinfo {author} {\bibfnamefont {K.}~\bibnamefont {Lee}},\
  }\bibfield  {title} {\bibinfo {title} {{Schwarzschild Black Hole from
  Perturbation Theory to All Orders}},\ }\href
  {https://doi.org/10.1103/PhysRevLett.132.251603} {\bibfield  {journal}
  {\bibinfo  {journal} {Phys. Rev. Lett.}\ }\textbf {\bibinfo {volume} {132}},\
  \bibinfo {pages} {251603} (\bibinfo {year} {2024})},\ \Eprint
  {https://arxiv.org/abs/2403.13216} {arXiv:2403.13216 [hep-th]} \BibitemShut
  {NoStop}%
\bibitem [{\citenamefont {Vaidya}(2015)}]{Vaidya:2014kza}%
  \BibitemOpen
  \bibfield  {author} {\bibinfo {author} {\bibfnamefont {V.}~\bibnamefont
  {Vaidya}},\ }\bibfield  {title} {\bibinfo {title} {{Gravitational spin
  Hamiltonians from the S matrix}},\ }\href
  {https://doi.org/10.1103/PhysRevD.91.024017} {\bibfield  {journal} {\bibinfo
  {journal} {Phys. Rev. D}\ }\textbf {\bibinfo {volume} {91}},\ \bibinfo
  {pages} {024017} (\bibinfo {year} {2015})},\ \Eprint
  {https://arxiv.org/abs/1410.5348} {arXiv:1410.5348 [hep-th]} \BibitemShut
  {NoStop}%
\bibitem [{\citenamefont {Vines}(2018)}]{Vines:2017hyw}%
  \BibitemOpen
  \bibfield  {author} {\bibinfo {author} {\bibfnamefont {J.}~\bibnamefont
  {Vines}},\ }\bibfield  {title} {\bibinfo {title} {{Scattering of two spinning
  black holes in post-Minkowskian gravity, to all orders in spin, and
  effective-one-body mappings}},\ }\href
  {https://doi.org/10.1088/1361-6382/aaa3a8} {\bibfield  {journal} {\bibinfo
  {journal} {Class. Quant. Grav.}\ }\textbf {\bibinfo {volume} {35}},\ \bibinfo
  {pages} {084002} (\bibinfo {year} {2018})},\ \Eprint
  {https://arxiv.org/abs/1709.06016} {arXiv:1709.06016 [gr-qc]} \BibitemShut
  {NoStop}%
\bibitem [{\citenamefont {Arkani-Hamed}\ \emph {et~al.}(2021)\citenamefont
  {Arkani-Hamed}, \citenamefont {Huang},\ and\ \citenamefont
  {Huang}}]{Arkani-Hamed:2017jhn}%
  \BibitemOpen
  \bibfield  {author} {\bibinfo {author} {\bibfnamefont {N.}~\bibnamefont
  {Arkani-Hamed}}, \bibinfo {author} {\bibfnamefont {T.-C.}\ \bibnamefont
  {Huang}},\ and\ \bibinfo {author} {\bibfnamefont {Y.-t.}\ \bibnamefont
  {Huang}},\ }\bibfield  {title} {\bibinfo {title} {{Scattering amplitudes for
  all masses and spins}},\ }\href {https://doi.org/10.1007/JHEP11(2021)070}
  {\bibfield  {journal} {\bibinfo  {journal} {JHEP}\ }\textbf {\bibinfo
  {volume} {11}},\ \bibinfo {pages} {070}},\ \Eprint
  {https://arxiv.org/abs/1709.04891} {arXiv:1709.04891 [hep-th]} \BibitemShut
  {NoStop}%
\bibitem [{\citenamefont {Chung}\ \emph {et~al.}(2019)\citenamefont {Chung},
  \citenamefont {Huang}, \citenamefont {Kim},\ and\ \citenamefont
  {Lee}}]{Chung:2018kqs}%
  \BibitemOpen
  \bibfield  {author} {\bibinfo {author} {\bibfnamefont {M.-Z.}\ \bibnamefont
  {Chung}}, \bibinfo {author} {\bibfnamefont {Y.-T.}\ \bibnamefont {Huang}},
  \bibinfo {author} {\bibfnamefont {J.-W.}\ \bibnamefont {Kim}},\ and\ \bibinfo
  {author} {\bibfnamefont {S.}~\bibnamefont {Lee}},\ }\bibfield  {title}
  {\bibinfo {title} {{The simplest massive S-matrix: from minimal coupling to
  Black Holes}},\ }\href {https://doi.org/10.1007/JHEP04(2019)156} {\bibfield
  {journal} {\bibinfo  {journal} {JHEP}\ }\textbf {\bibinfo {volume} {04}},\
  \bibinfo {pages} {156}},\ \Eprint {https://arxiv.org/abs/1812.08752}
  {arXiv:1812.08752 [hep-th]} \BibitemShut {NoStop}%
\bibitem [{\citenamefont {Guevara}\ \emph
  {et~al.}(2019{\natexlab{a}})\citenamefont {Guevara}, \citenamefont
  {Ochirov},\ and\ \citenamefont {Vines}}]{Guevara:2018wpp}%
  \BibitemOpen
  \bibfield  {author} {\bibinfo {author} {\bibfnamefont {A.}~\bibnamefont
  {Guevara}}, \bibinfo {author} {\bibfnamefont {A.}~\bibnamefont {Ochirov}},\
  and\ \bibinfo {author} {\bibfnamefont {J.}~\bibnamefont {Vines}},\ }\bibfield
   {title} {\bibinfo {title} {{Scattering of Spinning Black Holes from
  Exponentiated Soft Factors}},\ }\href
  {https://doi.org/10.1007/JHEP09(2019)056} {\bibfield  {journal} {\bibinfo
  {journal} {JHEP}\ }\textbf {\bibinfo {volume} {09}},\ \bibinfo {pages}
  {056}},\ \Eprint {https://arxiv.org/abs/1812.06895} {arXiv:1812.06895
  [hep-th]} \BibitemShut {NoStop}%
\bibitem [{\citenamefont {Guevara}\ \emph
  {et~al.}(2019{\natexlab{b}})\citenamefont {Guevara}, \citenamefont
  {Ochirov},\ and\ \citenamefont {Vines}}]{Guevara:2019fsj}%
  \BibitemOpen
  \bibfield  {author} {\bibinfo {author} {\bibfnamefont {A.}~\bibnamefont
  {Guevara}}, \bibinfo {author} {\bibfnamefont {A.}~\bibnamefont {Ochirov}},\
  and\ \bibinfo {author} {\bibfnamefont {J.}~\bibnamefont {Vines}},\ }\bibfield
   {title} {\bibinfo {title} {{Black-hole scattering with general spin
  directions from minimal-coupling amplitudes}},\ }\href
  {https://doi.org/10.1103/PhysRevD.100.104024} {\bibfield  {journal} {\bibinfo
   {journal} {Phys. Rev. D}\ }\textbf {\bibinfo {volume} {100}},\ \bibinfo
  {pages} {104024} (\bibinfo {year} {2019}{\natexlab{b}})},\ \Eprint
  {https://arxiv.org/abs/1906.10071} {arXiv:1906.10071 [hep-th]} \BibitemShut
  {NoStop}%
\bibitem [{\citenamefont {Arkani-Hamed}\ \emph {et~al.}(2020)\citenamefont
  {Arkani-Hamed}, \citenamefont {Huang},\ and\ \citenamefont
  {O'Connell}}]{Arkani-Hamed:2019ymq}%
  \BibitemOpen
  \bibfield  {author} {\bibinfo {author} {\bibfnamefont {N.}~\bibnamefont
  {Arkani-Hamed}}, \bibinfo {author} {\bibfnamefont {Y.-t.}\ \bibnamefont
  {Huang}},\ and\ \bibinfo {author} {\bibfnamefont {D.}~\bibnamefont
  {O'Connell}},\ }\bibfield  {title} {\bibinfo {title} {{Kerr black holes as
  elementary particles}},\ }\href {https://doi.org/10.1007/JHEP01(2020)046}
  {\bibfield  {journal} {\bibinfo  {journal} {JHEP}\ }\textbf {\bibinfo
  {volume} {01}},\ \bibinfo {pages} {046}},\ \Eprint
  {https://arxiv.org/abs/1906.10100} {arXiv:1906.10100 [hep-th]} \BibitemShut
  {NoStop}%
\bibitem [{\citenamefont {Maybee}\ \emph {et~al.}(2019)\citenamefont {Maybee},
  \citenamefont {O'Connell},\ and\ \citenamefont {Vines}}]{Maybee:2019jus}%
  \BibitemOpen
  \bibfield  {author} {\bibinfo {author} {\bibfnamefont {B.}~\bibnamefont
  {Maybee}}, \bibinfo {author} {\bibfnamefont {D.}~\bibnamefont {O'Connell}},\
  and\ \bibinfo {author} {\bibfnamefont {J.}~\bibnamefont {Vines}},\ }\bibfield
   {title} {\bibinfo {title} {{Observables and amplitudes for spinning
  particles and black holes}},\ }\href
  {https://doi.org/10.1007/JHEP12(2019)156} {\bibfield  {journal} {\bibinfo
  {journal} {JHEP}\ }\textbf {\bibinfo {volume} {12}},\ \bibinfo {pages}
  {156}},\ \Eprint {https://arxiv.org/abs/1906.09260} {arXiv:1906.09260
  [hep-th]} \BibitemShut {NoStop}%
\bibitem [{\citenamefont {Bern}\ \emph {et~al.}(2021)\citenamefont {Bern},
  \citenamefont {Luna}, \citenamefont {Roiban}, \citenamefont {Shen},\ and\
  \citenamefont {Zeng}}]{Bern:2020buy}%
  \BibitemOpen
  \bibfield  {author} {\bibinfo {author} {\bibfnamefont {Z.}~\bibnamefont
  {Bern}}, \bibinfo {author} {\bibfnamefont {A.}~\bibnamefont {Luna}}, \bibinfo
  {author} {\bibfnamefont {R.}~\bibnamefont {Roiban}}, \bibinfo {author}
  {\bibfnamefont {C.-H.}\ \bibnamefont {Shen}},\ and\ \bibinfo {author}
  {\bibfnamefont {M.}~\bibnamefont {Zeng}},\ }\bibfield  {title} {\bibinfo
  {title} {{Spinning black hole binary dynamics, scattering amplitudes, and
  effective field theory}},\ }\href
  {https://doi.org/10.1103/PhysRevD.104.065014} {\bibfield  {journal} {\bibinfo
   {journal} {Phys. Rev. D}\ }\textbf {\bibinfo {volume} {104}},\ \bibinfo
  {pages} {065014} (\bibinfo {year} {2021})},\ \Eprint
  {https://arxiv.org/abs/2005.03071} {arXiv:2005.03071 [hep-th]} \BibitemShut
  {NoStop}%
\bibitem [{\citenamefont {Aoude}\ and\ \citenamefont
  {Ochirov}(2021)}]{Aoude:2021oqj}%
  \BibitemOpen
  \bibfield  {author} {\bibinfo {author} {\bibfnamefont {R.}~\bibnamefont
  {Aoude}}\ and\ \bibinfo {author} {\bibfnamefont {A.}~\bibnamefont
  {Ochirov}},\ }\bibfield  {title} {\bibinfo {title} {{Classical observables
  from coherent-spin amplitudes}},\ }\href
  {https://doi.org/10.1007/JHEP10(2021)008} {\bibfield  {journal} {\bibinfo
  {journal} {JHEP}\ }\textbf {\bibinfo {volume} {10}},\ \bibinfo {pages}
  {008}},\ \Eprint {https://arxiv.org/abs/2108.01649} {arXiv:2108.01649
  [hep-th]} \BibitemShut {NoStop}%
\bibitem [{\citenamefont {Alessio}(2024)}]{Alessio:2023kgf}%
  \BibitemOpen
  \bibfield  {author} {\bibinfo {author} {\bibfnamefont {F.}~\bibnamefont
  {Alessio}},\ }\bibfield  {title} {\bibinfo {title} {{Kerr binary dynamics
  from minimal coupling and double copy}},\ }\href
  {https://doi.org/10.1007/JHEP04(2024)058} {\bibfield  {journal} {\bibinfo
  {journal} {JHEP}\ }\textbf {\bibinfo {volume} {04}},\ \bibinfo {pages}
  {058}},\ \Eprint {https://arxiv.org/abs/2303.12784} {arXiv:2303.12784
  [hep-th]} \BibitemShut {NoStop}%
\bibitem [{\citenamefont {Kosmopoulos}\ and\ \citenamefont
  {Luna}(2021)}]{Kosmopoulos:2021zoq}%
  \BibitemOpen
  \bibfield  {author} {\bibinfo {author} {\bibfnamefont {D.}~\bibnamefont
  {Kosmopoulos}}\ and\ \bibinfo {author} {\bibfnamefont {A.}~\bibnamefont
  {Luna}},\ }\bibfield  {title} {\bibinfo {title} {{Quadratic-in-spin
  Hamiltonian at $ \mathcal{O} $(G$^{2}$) from scattering amplitudes}},\ }\href
  {https://doi.org/10.1007/JHEP07(2021)037} {\bibfield  {journal} {\bibinfo
  {journal} {JHEP}\ }\textbf {\bibinfo {volume} {07}},\ \bibinfo {pages}
  {037}},\ \Eprint {https://arxiv.org/abs/2102.10137} {arXiv:2102.10137
  [hep-th]} \BibitemShut {NoStop}%
\bibitem [{\citenamefont {Liu}\ \emph {et~al.}(2021)\citenamefont {Liu},
  \citenamefont {Porto},\ and\ \citenamefont {Yang}}]{Liu:2021zxr}%
  \BibitemOpen
  \bibfield  {author} {\bibinfo {author} {\bibfnamefont {Z.}~\bibnamefont
  {Liu}}, \bibinfo {author} {\bibfnamefont {R.~A.}\ \bibnamefont {Porto}},\
  and\ \bibinfo {author} {\bibfnamefont {Z.}~\bibnamefont {Yang}},\ }\bibfield
  {title} {\bibinfo {title} {{Spin Effects in the Effective Field Theory
  Approach to Post-Minkowskian Conservative Dynamics}},\ }\href
  {https://doi.org/10.1007/JHEP06(2021)012} {\bibfield  {journal} {\bibinfo
  {journal} {JHEP}\ }\textbf {\bibinfo {volume} {06}},\ \bibinfo {pages}
  {012}},\ \Eprint {https://arxiv.org/abs/2102.10059} {arXiv:2102.10059
  [hep-th]} \BibitemShut {NoStop}%
\bibitem [{\citenamefont {Chen}\ \emph
  {et~al.}(2022{\natexlab{a}})\citenamefont {Chen}, \citenamefont {Chung},
  \citenamefont {Huang},\ and\ \citenamefont {Kim}}]{Chen:2021kxt}%
  \BibitemOpen
  \bibfield  {author} {\bibinfo {author} {\bibfnamefont {W.-M.}\ \bibnamefont
  {Chen}}, \bibinfo {author} {\bibfnamefont {M.-Z.}\ \bibnamefont {Chung}},
  \bibinfo {author} {\bibfnamefont {Y.-t.}\ \bibnamefont {Huang}},\ and\
  \bibinfo {author} {\bibfnamefont {J.-W.}\ \bibnamefont {Kim}},\ }\bibfield
  {title} {\bibinfo {title} {{The 2PM Hamiltonian for binary Kerr to quartic in
  spin}},\ }\href {https://doi.org/10.1007/JHEP08(2022)148} {\bibfield
  {journal} {\bibinfo  {journal} {JHEP}\ }\textbf {\bibinfo {volume} {08}},\
  \bibinfo {pages} {148}},\ \Eprint {https://arxiv.org/abs/2111.13639}
  {arXiv:2111.13639 [hep-th]} \BibitemShut {NoStop}%
\bibitem [{\citenamefont {Menezes}\ and\ \citenamefont
  {Sergola}(2022)}]{Menezes:2022tcs}%
  \BibitemOpen
  \bibfield  {author} {\bibinfo {author} {\bibfnamefont {G.}~\bibnamefont
  {Menezes}}\ and\ \bibinfo {author} {\bibfnamefont {M.}~\bibnamefont
  {Sergola}},\ }\bibfield  {title} {\bibinfo {title} {{NLO deflections for
  spinning particles and Kerr black holes}},\ }\href
  {https://doi.org/10.1007/JHEP10(2022)105} {\bibfield  {journal} {\bibinfo
  {journal} {JHEP}\ }\textbf {\bibinfo {volume} {10}},\ \bibinfo {pages}
  {105}},\ \Eprint {https://arxiv.org/abs/2205.11701} {arXiv:2205.11701
  [hep-th]} \BibitemShut {NoStop}%
\bibitem [{\citenamefont {Bern}\ \emph {et~al.}(2023)\citenamefont {Bern},
  \citenamefont {Kosmopoulos}, \citenamefont {Luna}, \citenamefont {Roiban},\
  and\ \citenamefont {Teng}}]{Bern:2022kto}%
  \BibitemOpen
  \bibfield  {author} {\bibinfo {author} {\bibfnamefont {Z.}~\bibnamefont
  {Bern}}, \bibinfo {author} {\bibfnamefont {D.}~\bibnamefont {Kosmopoulos}},
  \bibinfo {author} {\bibfnamefont {A.}~\bibnamefont {Luna}}, \bibinfo {author}
  {\bibfnamefont {R.}~\bibnamefont {Roiban}},\ and\ \bibinfo {author}
  {\bibfnamefont {F.}~\bibnamefont {Teng}},\ }\bibfield  {title} {\bibinfo
  {title} {{Binary Dynamics through the Fifth Power of Spin at O(G2)}},\ }\href
  {https://doi.org/10.1103/PhysRevLett.130.201402} {\bibfield  {journal}
  {\bibinfo  {journal} {Phys. Rev. Lett.}\ }\textbf {\bibinfo {volume} {130}},\
  \bibinfo {pages} {201402} (\bibinfo {year} {2023})},\ \Eprint
  {https://arxiv.org/abs/2203.06202} {arXiv:2203.06202 [hep-th]} \BibitemShut
  {NoStop}%
\bibitem [{\citenamefont {Bautista}(2023)}]{Bautista:2023szu}%
  \BibitemOpen
  \bibfield  {author} {\bibinfo {author} {\bibfnamefont {Y.~F.}\ \bibnamefont
  {Bautista}},\ }\bibfield  {title} {\bibinfo {title} {{Dynamics for
  super-extremal Kerr binary systems at O(G2)}},\ }\href
  {https://doi.org/10.1103/PhysRevD.108.084036} {\bibfield  {journal} {\bibinfo
   {journal} {Phys. Rev. D}\ }\textbf {\bibinfo {volume} {108}},\ \bibinfo
  {pages} {084036} (\bibinfo {year} {2023})},\ \Eprint
  {https://arxiv.org/abs/2304.04287} {arXiv:2304.04287 [hep-th]} \BibitemShut
  {NoStop}%
\bibitem [{\citenamefont {Bohnenblust}\ \emph
  {et~al.}(2024{\natexlab{a}})\citenamefont {Bohnenblust}, \citenamefont
  {Cangemi}, \citenamefont {Johansson},\ and\ \citenamefont
  {Pichini}}]{Bohnenblust:2024hkw}%
  \BibitemOpen
  \bibfield  {author} {\bibinfo {author} {\bibfnamefont {L.}~\bibnamefont
  {Bohnenblust}}, \bibinfo {author} {\bibfnamefont {L.}~\bibnamefont
  {Cangemi}}, \bibinfo {author} {\bibfnamefont {H.}~\bibnamefont {Johansson}},\
  and\ \bibinfo {author} {\bibfnamefont {P.}~\bibnamefont {Pichini}},\
  }\bibfield  {title} {\bibinfo {title} {{Binary Kerr black-hole scattering at
  2PM from quantum higher-spin Compton}},\ }\href@noop {} {\  (\bibinfo {year}
  {2024}{\natexlab{a}})},\ \Eprint {https://arxiv.org/abs/2410.23271}
  {arXiv:2410.23271 [hep-th]} \BibitemShut {NoStop}%
\bibitem [{\citenamefont {Chen}\ and\ \citenamefont
  {Wang}(2025)}]{Chen:2024mmm}%
  \BibitemOpen
  \bibfield  {author} {\bibinfo {author} {\bibfnamefont {G.}~\bibnamefont
  {Chen}}\ and\ \bibinfo {author} {\bibfnamefont {T.}~\bibnamefont {Wang}},\
  }\bibfield  {title} {\bibinfo {title} {{Dynamics of spinning binary at
  2PM}},\ }\href {https://doi.org/10.1007/JHEP12(2024)213} {\bibfield
  {journal} {\bibinfo  {journal} {JHEP}\ }\textbf {\bibinfo {volume} {12}},\
  \bibinfo {pages} {213}},\ \Eprint {https://arxiv.org/abs/2406.09086}
  {arXiv:2406.09086 [hep-th]} \BibitemShut {NoStop}%
\bibitem [{\citenamefont {Akpinar}\ \emph
  {et~al.}(2025{\natexlab{a}})\citenamefont {Akpinar}, \citenamefont
  {Febres~Cordero}, \citenamefont {Kraus}, \citenamefont {Ruf},\ and\
  \citenamefont {Zeng}}]{Akpinar:2024meg}%
  \BibitemOpen
  \bibfield  {author} {\bibinfo {author} {\bibfnamefont {D.}~\bibnamefont
  {Akpinar}}, \bibinfo {author} {\bibfnamefont {F.}~\bibnamefont
  {Febres~Cordero}}, \bibinfo {author} {\bibfnamefont {M.}~\bibnamefont
  {Kraus}}, \bibinfo {author} {\bibfnamefont {M.~S.}\ \bibnamefont {Ruf}},\
  and\ \bibinfo {author} {\bibfnamefont {M.}~\bibnamefont {Zeng}},\ }\bibfield
  {title} {\bibinfo {title} {{Spinning black hole scattering at $ \mathcal{O}
  $(G$^{3}$S$^{2}$): Casimir terms, radial action and hidden symmetry}},\
  }\href {https://doi.org/10.1007/JHEP03(2025)126} {\bibfield  {journal}
  {\bibinfo  {journal} {JHEP}\ }\textbf {\bibinfo {volume} {03}},\ \bibinfo
  {pages} {126}},\ \Eprint {https://arxiv.org/abs/2407.19005} {arXiv:2407.19005
  [hep-th]} \BibitemShut {NoStop}%
\bibitem [{\citenamefont {Febres~Cordero}\ \emph {et~al.}(2023)\citenamefont
  {Febres~Cordero}, \citenamefont {Kraus}, \citenamefont {Lin}, \citenamefont
  {Ruf},\ and\ \citenamefont {Zeng}}]{FebresCordero:2022jts}%
  \BibitemOpen
  \bibfield  {author} {\bibinfo {author} {\bibfnamefont {F.}~\bibnamefont
  {Febres~Cordero}}, \bibinfo {author} {\bibfnamefont {M.}~\bibnamefont
  {Kraus}}, \bibinfo {author} {\bibfnamefont {G.}~\bibnamefont {Lin}}, \bibinfo
  {author} {\bibfnamefont {M.~S.}\ \bibnamefont {Ruf}},\ and\ \bibinfo {author}
  {\bibfnamefont {M.}~\bibnamefont {Zeng}},\ }\bibfield  {title} {\bibinfo
  {title} {{Conservative Binary Dynamics with a Spinning Black Hole at O(G3)
  from Scattering Amplitudes}},\ }\href
  {https://doi.org/10.1103/PhysRevLett.130.021601} {\bibfield  {journal}
  {\bibinfo  {journal} {Phys. Rev. Lett.}\ }\textbf {\bibinfo {volume} {130}},\
  \bibinfo {pages} {021601} (\bibinfo {year} {2023})},\ \Eprint
  {https://arxiv.org/abs/2205.07357} {arXiv:2205.07357 [hep-th]} \BibitemShut
  {NoStop}%
\bibitem [{\citenamefont {Jakobsen}\ \emph {et~al.}(2023)\citenamefont
  {Jakobsen}, \citenamefont {Mogull}, \citenamefont {Plefka}, \citenamefont
  {Sauer},\ and\ \citenamefont {Xu}}]{Jakobsen:2023ndj}%
  \BibitemOpen
  \bibfield  {author} {\bibinfo {author} {\bibfnamefont {G.~U.}\ \bibnamefont
  {Jakobsen}}, \bibinfo {author} {\bibfnamefont {G.}~\bibnamefont {Mogull}},
  \bibinfo {author} {\bibfnamefont {J.}~\bibnamefont {Plefka}}, \bibinfo
  {author} {\bibfnamefont {B.}~\bibnamefont {Sauer}},\ and\ \bibinfo {author}
  {\bibfnamefont {Y.}~\bibnamefont {Xu}},\ }\bibfield  {title} {\bibinfo
  {title} {{Conservative Scattering of Spinning Black Holes at Fourth
  Post-Minkowskian Order}},\ }\href
  {https://doi.org/10.1103/PhysRevLett.131.151401} {\bibfield  {journal}
  {\bibinfo  {journal} {Phys. Rev. Lett.}\ }\textbf {\bibinfo {volume} {131}},\
  \bibinfo {pages} {151401} (\bibinfo {year} {2023})},\ \Eprint
  {https://arxiv.org/abs/2306.01714} {arXiv:2306.01714 [hep-th]} \BibitemShut
  {NoStop}%
\bibitem [{\citenamefont {Jakobsen}\ and\ \citenamefont
  {Mogull}(2022)}]{Jakobsen:2022fcj}%
  \BibitemOpen
  \bibfield  {author} {\bibinfo {author} {\bibfnamefont {G.~U.}\ \bibnamefont
  {Jakobsen}}\ and\ \bibinfo {author} {\bibfnamefont {G.}~\bibnamefont
  {Mogull}},\ }\bibfield  {title} {\bibinfo {title} {{Conservative and
  Radiative Dynamics of Spinning Bodies at Third Post-Minkowskian Order Using
  Worldline Quantum Field Theory}},\ }\href
  {https://doi.org/10.1103/PhysRevLett.128.141102} {\bibfield  {journal}
  {\bibinfo  {journal} {Phys. Rev. Lett.}\ }\textbf {\bibinfo {volume} {128}},\
  \bibinfo {pages} {141102} (\bibinfo {year} {2022})},\ \Eprint
  {https://arxiv.org/abs/2201.07778} {arXiv:2201.07778 [hep-th]} \BibitemShut
  {NoStop}%
\bibitem [{\citenamefont {Damgaard}\ \emph {et~al.}(2022)\citenamefont
  {Damgaard}, \citenamefont {Hoogeveen}, \citenamefont {Luna},\ and\
  \citenamefont {Vines}}]{Damgaard:2022jem}%
  \BibitemOpen
  \bibfield  {author} {\bibinfo {author} {\bibfnamefont {P.~H.}\ \bibnamefont
  {Damgaard}}, \bibinfo {author} {\bibfnamefont {J.}~\bibnamefont {Hoogeveen}},
  \bibinfo {author} {\bibfnamefont {A.}~\bibnamefont {Luna}},\ and\ \bibinfo
  {author} {\bibfnamefont {J.}~\bibnamefont {Vines}},\ }\bibfield  {title}
  {\bibinfo {title} {{Scattering angles in Kerr metrics}},\ }\href
  {https://doi.org/10.1103/PhysRevD.106.124030} {\bibfield  {journal} {\bibinfo
   {journal} {Phys. Rev. D}\ }\textbf {\bibinfo {volume} {106}},\ \bibinfo
  {pages} {124030} (\bibinfo {year} {2022})},\ \Eprint
  {https://arxiv.org/abs/2208.11028} {arXiv:2208.11028 [hep-th]} \BibitemShut
  {NoStop}%
\bibitem [{\citenamefont {Luna}\ \emph {et~al.}(2024)\citenamefont {Luna},
  \citenamefont {Moynihan}, \citenamefont {O'Connell},\ and\ \citenamefont
  {Ross}}]{Luna:2023uwd}%
  \BibitemOpen
  \bibfield  {author} {\bibinfo {author} {\bibfnamefont {A.}~\bibnamefont
  {Luna}}, \bibinfo {author} {\bibfnamefont {N.}~\bibnamefont {Moynihan}},
  \bibinfo {author} {\bibfnamefont {D.}~\bibnamefont {O'Connell}},\ and\
  \bibinfo {author} {\bibfnamefont {A.}~\bibnamefont {Ross}},\ }\bibfield
  {title} {\bibinfo {title} {{Observables from the spinning eikonal}},\ }\href
  {https://doi.org/10.1007/JHEP08(2024)045} {\bibfield  {journal} {\bibinfo
  {journal} {JHEP}\ }\textbf {\bibinfo {volume} {08}},\ \bibinfo {pages}
  {045}},\ \Eprint {https://arxiv.org/abs/2312.09960} {arXiv:2312.09960
  [hep-th]} \BibitemShut {NoStop}%
\bibitem [{\citenamefont {Gonzo}\ and\ \citenamefont
  {Shi}(2024)}]{Gonzo:2024zxo}%
  \BibitemOpen
  \bibfield  {author} {\bibinfo {author} {\bibfnamefont {R.}~\bibnamefont
  {Gonzo}}\ and\ \bibinfo {author} {\bibfnamefont {C.}~\bibnamefont {Shi}},\
  }\bibfield  {title} {\bibinfo {title} {{Scattering and Bound Observables for
  Spinning Particles in Kerr Spacetime with Generic Spin Orientations}},\
  }\href {https://doi.org/10.1103/PhysRevLett.133.221401} {\bibfield  {journal}
  {\bibinfo  {journal} {Phys. Rev. Lett.}\ }\textbf {\bibinfo {volume} {133}},\
  \bibinfo {pages} {221401} (\bibinfo {year} {2024})},\ \Eprint
  {https://arxiv.org/abs/2405.09687} {arXiv:2405.09687 [hep-th]} \BibitemShut
  {NoStop}%
\bibitem [{\citenamefont {Akpinar}\ \emph
  {et~al.}(2025{\natexlab{b}})\citenamefont {Akpinar}, \citenamefont
  {Febres~Cordero}, \citenamefont {Kraus}, \citenamefont {Smirnov},\ and\
  \citenamefont {Zeng}}]{Akpinar:2025bkt}%
  \BibitemOpen
  \bibfield  {author} {\bibinfo {author} {\bibfnamefont {D.}~\bibnamefont
  {Akpinar}}, \bibinfo {author} {\bibfnamefont {F.}~\bibnamefont
  {Febres~Cordero}}, \bibinfo {author} {\bibfnamefont {M.}~\bibnamefont
  {Kraus}}, \bibinfo {author} {\bibfnamefont {A.}~\bibnamefont {Smirnov}},\
  and\ \bibinfo {author} {\bibfnamefont {M.}~\bibnamefont {Zeng}},\ }\bibfield
  {title} {\bibinfo {title} {{A First Look at Quartic-in-Spin Binary Dynamics
  at Third Post-Minkowskian Order}},\ }\href@noop {} {\  (\bibinfo {year}
  {2025}{\natexlab{b}})},\ \Eprint {https://arxiv.org/abs/2502.08961}
  {arXiv:2502.08961 [hep-th]} \BibitemShut {NoStop}%
\bibitem [{\citenamefont {Alaverdian}\ \emph {et~al.}(2025)\citenamefont
  {Alaverdian}, \citenamefont {Bern}, \citenamefont {Kosmopoulos},
  \citenamefont {Luna}, \citenamefont {Roiban}, \citenamefont {Scheopner},\
  and\ \citenamefont {Teng}}]{Alaverdian:2025jtw}%
  \BibitemOpen
  \bibfield  {author} {\bibinfo {author} {\bibfnamefont {M.}~\bibnamefont
  {Alaverdian}}, \bibinfo {author} {\bibfnamefont {Z.}~\bibnamefont {Bern}},
  \bibinfo {author} {\bibfnamefont {D.}~\bibnamefont {Kosmopoulos}}, \bibinfo
  {author} {\bibfnamefont {A.}~\bibnamefont {Luna}}, \bibinfo {author}
  {\bibfnamefont {R.}~\bibnamefont {Roiban}}, \bibinfo {author} {\bibfnamefont
  {T.}~\bibnamefont {Scheopner}},\ and\ \bibinfo {author} {\bibfnamefont
  {F.}~\bibnamefont {Teng}},\ }\bibfield  {title} {\bibinfo {title}
  {{Observables and Unconstrained Spin Tensor Dynamics in General Relativity
  from Scattering Amplitudes}},\ }\href@noop {} {\  (\bibinfo {year} {2025})},\
  \Eprint {https://arxiv.org/abs/2503.03739} {arXiv:2503.03739 [hep-th]}
  \BibitemShut {NoStop}%
\bibitem [{\citenamefont {Bjerrum-Bohr}\ \emph {et~al.}(2016)\citenamefont
  {Bjerrum-Bohr}, \citenamefont {Donoghue}, \citenamefont {Holstein},
  \citenamefont {Plante},\ and\ \citenamefont
  {Vanhove}}]{Bjerrum-Bohr:2016hpa}%
  \BibitemOpen
  \bibfield  {author} {\bibinfo {author} {\bibfnamefont {N.~E.~J.}\
  \bibnamefont {Bjerrum-Bohr}}, \bibinfo {author} {\bibfnamefont {J.~F.}\
  \bibnamefont {Donoghue}}, \bibinfo {author} {\bibfnamefont {B.~R.}\
  \bibnamefont {Holstein}}, \bibinfo {author} {\bibfnamefont {L.}~\bibnamefont
  {Plante}},\ and\ \bibinfo {author} {\bibfnamefont {P.}~\bibnamefont
  {Vanhove}},\ }\bibfield  {title} {\bibinfo {title} {{Light-like Scattering in
  Quantum Gravity}},\ }\href {https://doi.org/10.1007/JHEP11(2016)117}
  {\bibfield  {journal} {\bibinfo  {journal} {JHEP}\ }\textbf {\bibinfo
  {volume} {11}},\ \bibinfo {pages} {117}},\ \Eprint
  {https://arxiv.org/abs/1609.07477} {arXiv:1609.07477 [hep-th]} \BibitemShut
  {NoStop}%
\bibitem [{\citenamefont {Johansson}\ and\ \citenamefont
  {Ochirov}(2019)}]{Johansson:2019dnu}%
  \BibitemOpen
  \bibfield  {author} {\bibinfo {author} {\bibfnamefont {H.}~\bibnamefont
  {Johansson}}\ and\ \bibinfo {author} {\bibfnamefont {A.}~\bibnamefont
  {Ochirov}},\ }\bibfield  {title} {\bibinfo {title} {{Double copy for massive
  quantum particles with spin}},\ }\href
  {https://doi.org/10.1007/JHEP09(2019)040} {\bibfield  {journal} {\bibinfo
  {journal} {JHEP}\ }\textbf {\bibinfo {volume} {09}},\ \bibinfo {pages}
  {040}},\ \Eprint {https://arxiv.org/abs/1906.12292} {arXiv:1906.12292
  [hep-th]} \BibitemShut {NoStop}%
\bibitem [{\citenamefont {Aoude}\ \emph {et~al.}(2020)\citenamefont {Aoude},
  \citenamefont {Haddad},\ and\ \citenamefont {Helset}}]{Aoude:2020onz}%
  \BibitemOpen
  \bibfield  {author} {\bibinfo {author} {\bibfnamefont {R.}~\bibnamefont
  {Aoude}}, \bibinfo {author} {\bibfnamefont {K.}~\bibnamefont {Haddad}},\ and\
  \bibinfo {author} {\bibfnamefont {A.}~\bibnamefont {Helset}},\ }\bibfield
  {title} {\bibinfo {title} {{On-shell heavy particle effective theories}},\
  }\href {https://doi.org/10.1007/JHEP05(2020)051} {\bibfield  {journal}
  {\bibinfo  {journal} {JHEP}\ }\textbf {\bibinfo {volume} {05}},\ \bibinfo
  {pages} {051}},\ \Eprint {https://arxiv.org/abs/2001.09164} {arXiv:2001.09164
  [hep-th]} \BibitemShut {NoStop}%
\bibitem [{\citenamefont {Falkowski}\ and\ \citenamefont
  {Machado}(2021)}]{Falkowski:2020aso}%
  \BibitemOpen
  \bibfield  {author} {\bibinfo {author} {\bibfnamefont {A.}~\bibnamefont
  {Falkowski}}\ and\ \bibinfo {author} {\bibfnamefont {C.~S.}\ \bibnamefont
  {Machado}},\ }\bibfield  {title} {\bibinfo {title} {{Soft Matters, or the
  Recursions with Massive Spinors}},\ }\href
  {https://doi.org/10.1007/JHEP05(2021)238} {\bibfield  {journal} {\bibinfo
  {journal} {JHEP}\ }\textbf {\bibinfo {volume} {05}},\ \bibinfo {pages}
  {238}},\ \Eprint {https://arxiv.org/abs/2005.08981} {arXiv:2005.08981
  [hep-th]} \BibitemShut {NoStop}%
\bibitem [{\citenamefont {Bautista}\ \emph
  {et~al.}(2023{\natexlab{a}})\citenamefont {Bautista}, \citenamefont
  {Guevara}, \citenamefont {Kavanagh},\ and\ \citenamefont
  {Vines}}]{Bautista:2021wfy}%
  \BibitemOpen
  \bibfield  {author} {\bibinfo {author} {\bibfnamefont {Y.~F.}\ \bibnamefont
  {Bautista}}, \bibinfo {author} {\bibfnamefont {A.}~\bibnamefont {Guevara}},
  \bibinfo {author} {\bibfnamefont {C.}~\bibnamefont {Kavanagh}},\ and\
  \bibinfo {author} {\bibfnamefont {J.}~\bibnamefont {Vines}},\ }\bibfield
  {title} {\bibinfo {title} {{Scattering in black hole backgrounds and
  higher-spin amplitudes. Part I}},\ }\href
  {https://doi.org/10.1007/JHEP03(2023)136} {\bibfield  {journal} {\bibinfo
  {journal} {JHEP}\ }\textbf {\bibinfo {volume} {03}},\ \bibinfo {pages}
  {136}},\ \Eprint {https://arxiv.org/abs/2107.10179} {arXiv:2107.10179
  [hep-th]} \BibitemShut {NoStop}%
\bibitem [{\citenamefont {Chiodaroli}\ \emph {et~al.}(2022)\citenamefont
  {Chiodaroli}, \citenamefont {Johansson},\ and\ \citenamefont
  {Pichini}}]{Chiodaroli:2021eug}%
  \BibitemOpen
  \bibfield  {author} {\bibinfo {author} {\bibfnamefont {M.}~\bibnamefont
  {Chiodaroli}}, \bibinfo {author} {\bibfnamefont {H.}~\bibnamefont
  {Johansson}},\ and\ \bibinfo {author} {\bibfnamefont {P.}~\bibnamefont
  {Pichini}},\ }\bibfield  {title} {\bibinfo {title} {{Compton black-hole
  scattering for s \ensuremath{\leq} 5/2}},\ }\href
  {https://doi.org/10.1007/JHEP02(2022)156} {\bibfield  {journal} {\bibinfo
  {journal} {JHEP}\ }\textbf {\bibinfo {volume} {02}},\ \bibinfo {pages}
  {156}},\ \Eprint {https://arxiv.org/abs/2107.14779} {arXiv:2107.14779
  [hep-th]} \BibitemShut {NoStop}%
\bibitem [{\citenamefont {Chen}\ \emph
  {et~al.}(2022{\natexlab{b}})\citenamefont {Chen}, \citenamefont {Chung},
  \citenamefont {Huang},\ and\ \citenamefont {Kim}}]{Chen:2022clh}%
  \BibitemOpen
  \bibfield  {author} {\bibinfo {author} {\bibfnamefont {W.-M.}\ \bibnamefont
  {Chen}}, \bibinfo {author} {\bibfnamefont {M.-Z.}\ \bibnamefont {Chung}},
  \bibinfo {author} {\bibfnamefont {Y.-t.}\ \bibnamefont {Huang}},\ and\
  \bibinfo {author} {\bibfnamefont {J.-W.}\ \bibnamefont {Kim}},\ }\bibfield
  {title} {\bibinfo {title} {{Gravitational Faraday effect from on-shell
  amplitudes}},\ }\href {https://doi.org/10.1007/JHEP12(2022)058} {\bibfield
  {journal} {\bibinfo  {journal} {JHEP}\ }\textbf {\bibinfo {volume} {12}},\
  \bibinfo {pages} {058}},\ \Eprint {https://arxiv.org/abs/2205.07305}
  {arXiv:2205.07305 [hep-th]} \BibitemShut {NoStop}%
\bibitem [{\citenamefont {Aoude}\ \emph {et~al.}(2022)\citenamefont {Aoude},
  \citenamefont {Haddad},\ and\ \citenamefont {Helset}}]{Aoude:2022trd}%
  \BibitemOpen
  \bibfield  {author} {\bibinfo {author} {\bibfnamefont {R.}~\bibnamefont
  {Aoude}}, \bibinfo {author} {\bibfnamefont {K.}~\bibnamefont {Haddad}},\ and\
  \bibinfo {author} {\bibfnamefont {A.}~\bibnamefont {Helset}},\ }\bibfield
  {title} {\bibinfo {title} {{Searching for Kerr in the 2PM amplitude}},\
  }\href {https://doi.org/10.1007/JHEP07(2022)072} {\bibfield  {journal}
  {\bibinfo  {journal} {JHEP}\ }\textbf {\bibinfo {volume} {07}},\ \bibinfo
  {pages} {072}},\ \Eprint {https://arxiv.org/abs/2203.06197} {arXiv:2203.06197
  [hep-th]} \BibitemShut {NoStop}%
\bibitem [{\citenamefont {Bautista}\ \emph
  {et~al.}(2023{\natexlab{b}})\citenamefont {Bautista}, \citenamefont
  {Guevara}, \citenamefont {Kavanagh},\ and\ \citenamefont
  {Vines}}]{Bautista:2022wjf}%
  \BibitemOpen
  \bibfield  {author} {\bibinfo {author} {\bibfnamefont {Y.~F.}\ \bibnamefont
  {Bautista}}, \bibinfo {author} {\bibfnamefont {A.}~\bibnamefont {Guevara}},
  \bibinfo {author} {\bibfnamefont {C.}~\bibnamefont {Kavanagh}},\ and\
  \bibinfo {author} {\bibfnamefont {J.}~\bibnamefont {Vines}},\ }\bibfield
  {title} {\bibinfo {title} {{Scattering in black hole backgrounds and
  higher-spin amplitudes. Part II}},\ }\href
  {https://doi.org/10.1007/JHEP05(2023)211} {\bibfield  {journal} {\bibinfo
  {journal} {JHEP}\ }\textbf {\bibinfo {volume} {05}},\ \bibinfo {pages}
  {211}},\ \Eprint {https://arxiv.org/abs/2212.07965} {arXiv:2212.07965
  [hep-th]} \BibitemShut {NoStop}%
\bibitem [{\citenamefont {Bjerrum-Bohr}\ \emph {et~al.}(2023)\citenamefont
  {Bjerrum-Bohr}, \citenamefont {Chen},\ and\ \citenamefont
  {Skowronek}}]{Bjerrum-Bohr:2023jau}%
  \BibitemOpen
  \bibfield  {author} {\bibinfo {author} {\bibfnamefont {N.~E.~J.}\
  \bibnamefont {Bjerrum-Bohr}}, \bibinfo {author} {\bibfnamefont
  {G.}~\bibnamefont {Chen}},\ and\ \bibinfo {author} {\bibfnamefont
  {M.}~\bibnamefont {Skowronek}},\ }\bibfield  {title} {\bibinfo {title}
  {{Classical spin gravitational Compton scattering}},\ }\href
  {https://doi.org/10.1007/JHEP06(2023)170} {\bibfield  {journal} {\bibinfo
  {journal} {JHEP}\ }\textbf {\bibinfo {volume} {06}},\ \bibinfo {pages}
  {170}},\ \Eprint {https://arxiv.org/abs/2302.00498} {arXiv:2302.00498
  [hep-th]} \BibitemShut {NoStop}%
\bibitem [{\citenamefont {Cangemi}\ \emph {et~al.}(2023)\citenamefont
  {Cangemi}, \citenamefont {Chiodaroli}, \citenamefont {Johansson},
  \citenamefont {Ochirov}, \citenamefont {Pichini},\ and\ \citenamefont
  {Skvortsov}}]{Cangemi:2022bew}%
  \BibitemOpen
  \bibfield  {author} {\bibinfo {author} {\bibfnamefont {L.}~\bibnamefont
  {Cangemi}}, \bibinfo {author} {\bibfnamefont {M.}~\bibnamefont {Chiodaroli}},
  \bibinfo {author} {\bibfnamefont {H.}~\bibnamefont {Johansson}}, \bibinfo
  {author} {\bibfnamefont {A.}~\bibnamefont {Ochirov}}, \bibinfo {author}
  {\bibfnamefont {P.}~\bibnamefont {Pichini}},\ and\ \bibinfo {author}
  {\bibfnamefont {E.}~\bibnamefont {Skvortsov}},\ }\bibfield  {title} {\bibinfo
  {title} {{Kerr Black Holes From Massive Higher-Spin Gauge Symmetry}},\ }\href
  {https://doi.org/10.1103/PhysRevLett.131.221401} {\bibfield  {journal}
  {\bibinfo  {journal} {Phys. Rev. Lett.}\ }\textbf {\bibinfo {volume} {131}},\
  \bibinfo {pages} {221401} (\bibinfo {year} {2023})},\ \Eprint
  {https://arxiv.org/abs/2212.06120} {arXiv:2212.06120 [hep-th]} \BibitemShut
  {NoStop}%
\bibitem [{\citenamefont {Cangemi}\ \emph {et~al.}(2024)\citenamefont
  {Cangemi}, \citenamefont {Chiodaroli}, \citenamefont {Johansson},
  \citenamefont {Ochirov}, \citenamefont {Pichini},\ and\ \citenamefont
  {Skvortsov}}]{Cangemi:2023bpe}%
  \BibitemOpen
  \bibfield  {author} {\bibinfo {author} {\bibfnamefont {L.}~\bibnamefont
  {Cangemi}}, \bibinfo {author} {\bibfnamefont {M.}~\bibnamefont {Chiodaroli}},
  \bibinfo {author} {\bibfnamefont {H.}~\bibnamefont {Johansson}}, \bibinfo
  {author} {\bibfnamefont {A.}~\bibnamefont {Ochirov}}, \bibinfo {author}
  {\bibfnamefont {P.}~\bibnamefont {Pichini}},\ and\ \bibinfo {author}
  {\bibfnamefont {E.}~\bibnamefont {Skvortsov}},\ }\bibfield  {title} {\bibinfo
  {title} {{Compton Amplitude for Rotating Black Hole from QFT}},\ }\href
  {https://doi.org/10.1103/PhysRevLett.133.071601} {\bibfield  {journal}
  {\bibinfo  {journal} {Phys. Rev. Lett.}\ }\textbf {\bibinfo {volume} {133}},\
  \bibinfo {pages} {071601} (\bibinfo {year} {2024})},\ \Eprint
  {https://arxiv.org/abs/2312.14913} {arXiv:2312.14913 [hep-th]} \BibitemShut
  {NoStop}%
\bibitem [{\citenamefont {Scheopner}\ and\ \citenamefont
  {Vines}(2024)}]{Scheopner:2023rzp}%
  \BibitemOpen
  \bibfield  {author} {\bibinfo {author} {\bibfnamefont {T.}~\bibnamefont
  {Scheopner}}\ and\ \bibinfo {author} {\bibfnamefont {J.}~\bibnamefont
  {Vines}},\ }\bibfield  {title} {\bibinfo {title} {{Dynamical implications of
  the Kerr multipole moments for spinning black holes}},\ }\href
  {https://doi.org/10.1007/JHEP12(2024)060} {\bibfield  {journal} {\bibinfo
  {journal} {JHEP}\ }\textbf {\bibinfo {volume} {12}},\ \bibinfo {pages}
  {060}},\ \Eprint {https://arxiv.org/abs/2311.18421} {arXiv:2311.18421
  [gr-qc]} \BibitemShut {NoStop}%
\bibitem [{\citenamefont {Bjerrum-Bohr}\ \emph {et~al.}(2024)\citenamefont
  {Bjerrum-Bohr}, \citenamefont {Chen},\ and\ \citenamefont
  {Skowronek}}]{Bjerrum-Bohr:2023iey}%
  \BibitemOpen
  \bibfield  {author} {\bibinfo {author} {\bibfnamefont {N.~E.~J.}\
  \bibnamefont {Bjerrum-Bohr}}, \bibinfo {author} {\bibfnamefont
  {G.}~\bibnamefont {Chen}},\ and\ \bibinfo {author} {\bibfnamefont
  {M.}~\bibnamefont {Skowronek}},\ }\bibfield  {title} {\bibinfo {title}
  {{Covariant Compton Amplitudes in Gravity with Classical Spin}},\ }\href
  {https://doi.org/10.1103/PhysRevLett.132.191603} {\bibfield  {journal}
  {\bibinfo  {journal} {Phys. Rev. Lett.}\ }\textbf {\bibinfo {volume} {132}},\
  \bibinfo {pages} {191603} (\bibinfo {year} {2024})},\ \Eprint
  {https://arxiv.org/abs/2309.11249} {arXiv:2309.11249 [hep-th]} \BibitemShut
  {NoStop}%
\bibitem [{\citenamefont {Bautista}\ \emph {et~al.}(2024)\citenamefont
  {Bautista}, \citenamefont {Bonelli}, \citenamefont {Iossa}, \citenamefont
  {Tanzini},\ and\ \citenamefont {Zhou}}]{Bautista:2023sdf}%
  \BibitemOpen
  \bibfield  {author} {\bibinfo {author} {\bibfnamefont {Y.~F.}\ \bibnamefont
  {Bautista}}, \bibinfo {author} {\bibfnamefont {G.}~\bibnamefont {Bonelli}},
  \bibinfo {author} {\bibfnamefont {C.}~\bibnamefont {Iossa}}, \bibinfo
  {author} {\bibfnamefont {A.}~\bibnamefont {Tanzini}},\ and\ \bibinfo {author}
  {\bibfnamefont {Z.}~\bibnamefont {Zhou}},\ }\bibfield  {title} {\bibinfo
  {title} {{Black hole perturbation theory meets CFT2: Kerr-Compton amplitudes
  from Nekrasov-Shatashvili functions}},\ }\href
  {https://doi.org/10.1103/PhysRevD.109.084071} {\bibfield  {journal} {\bibinfo
   {journal} {Phys. Rev. D}\ }\textbf {\bibinfo {volume} {109}},\ \bibinfo
  {pages} {084071} (\bibinfo {year} {2024})},\ \Eprint
  {https://arxiv.org/abs/2312.05965} {arXiv:2312.05965 [hep-th]} \BibitemShut
  {NoStop}%
\bibitem [{\citenamefont {Azevedo}\ \emph {et~al.}(2025)\citenamefont
  {Azevedo}, \citenamefont {Matamoros},\ and\ \citenamefont
  {Menezes}}]{Azevedo:2024rrf}%
  \BibitemOpen
  \bibfield  {author} {\bibinfo {author} {\bibfnamefont {T.}~\bibnamefont
  {Azevedo}}, \bibinfo {author} {\bibfnamefont {D.~E.~A.}\ \bibnamefont
  {Matamoros}},\ and\ \bibinfo {author} {\bibfnamefont {G.}~\bibnamefont
  {Menezes}},\ }\bibfield  {title} {\bibinfo {title} {{Compton scattering from
  superstrings}},\ }\href {https://doi.org/10.1007/JHEP01(2025)140} {\bibfield
  {journal} {\bibinfo  {journal} {JHEP}\ }\textbf {\bibinfo {volume} {01}},\
  \bibinfo {pages} {140}},\ \Eprint {https://arxiv.org/abs/2403.08899}
  {arXiv:2403.08899 [hep-th]} \BibitemShut {NoStop}%
\bibitem [{\citenamefont {Vazquez-Holm}\ and\ \citenamefont
  {Luna}(2025)}]{Vazquez-Holm:2025ztz}%
  \BibitemOpen
  \bibfield  {author} {\bibinfo {author} {\bibfnamefont {I.}~\bibnamefont
  {Vazquez-Holm}}\ and\ \bibinfo {author} {\bibfnamefont {A.}~\bibnamefont
  {Luna}},\ }\bibfield  {title} {\bibinfo {title} {{Bootstrapping Classical
  Spinning Compton Amplitudes with Colour-Kinematics}},\ }\href@noop {} {\
  (\bibinfo {year} {2025})},\ \Eprint {https://arxiv.org/abs/2503.22597}
  {arXiv:2503.22597 [hep-th]} \BibitemShut {NoStop}%
\bibitem [{\citenamefont {Galley}\ and\ \citenamefont
  {Hu}(2009)}]{Galley:2008ih}%
  \BibitemOpen
  \bibfield  {author} {\bibinfo {author} {\bibfnamefont {C.~R.}\ \bibnamefont
  {Galley}}\ and\ \bibinfo {author} {\bibfnamefont {B.~L.}\ \bibnamefont
  {Hu}},\ }\bibfield  {title} {\bibinfo {title} {{Self-force on extreme mass
  ratio inspirals via curved spacetime effective field theory}},\ }\href
  {https://doi.org/10.1103/PhysRevD.79.064002} {\bibfield  {journal} {\bibinfo
  {journal} {Phys. Rev. D}\ }\textbf {\bibinfo {volume} {79}},\ \bibinfo
  {pages} {064002} (\bibinfo {year} {2009})},\ \Eprint
  {https://arxiv.org/abs/0801.0900} {arXiv:0801.0900 [gr-qc]} \BibitemShut
  {NoStop}%
\bibitem [{\citenamefont {Detweiler}\ and\ \citenamefont
  {Whiting}(2003)}]{Detweiler:2002mi}%
  \BibitemOpen
  \bibfield  {author} {\bibinfo {author} {\bibfnamefont {S.~L.}\ \bibnamefont
  {Detweiler}}\ and\ \bibinfo {author} {\bibfnamefont {B.~F.}\ \bibnamefont
  {Whiting}},\ }\bibfield  {title} {\bibinfo {title} {{Selfforce via a Green's
  function decomposition}},\ }\href
  {https://doi.org/10.1103/PhysRevD.67.024025} {\bibfield  {journal} {\bibinfo
  {journal} {Phys. Rev. D}\ }\textbf {\bibinfo {volume} {67}},\ \bibinfo
  {pages} {024025} (\bibinfo {year} {2003})},\ \Eprint
  {https://arxiv.org/abs/gr-qc/0202086} {arXiv:gr-qc/0202086} \BibitemShut
  {NoStop}%
\bibitem [{\citenamefont {Gibbons}\ \emph {et~al.}(1993)\citenamefont
  {Gibbons}, \citenamefont {Rietdijk},\ and\ \citenamefont {van
  Holten}}]{Gibbons:1993ap}%
  \BibitemOpen
  \bibfield  {author} {\bibinfo {author} {\bibfnamefont {G.~W.}\ \bibnamefont
  {Gibbons}}, \bibinfo {author} {\bibfnamefont {R.~H.}\ \bibnamefont
  {Rietdijk}},\ and\ \bibinfo {author} {\bibfnamefont {J.~W.}\ \bibnamefont
  {van Holten}},\ }\bibfield  {title} {\bibinfo {title} {{SUSY in the sky}},\
  }\href {https://doi.org/10.1016/0550-3213(93)90472-2} {\bibfield  {journal}
  {\bibinfo  {journal} {Nucl. Phys. B}\ }\textbf {\bibinfo {volume} {404}},\
  \bibinfo {pages} {42} (\bibinfo {year} {1993})},\ \Eprint
  {https://arxiv.org/abs/hep-th/9303112} {arXiv:hep-th/9303112} \BibitemShut
  {NoStop}%
\bibitem [{\citenamefont {Bastianelli}\ \emph {et~al.}(2005)\citenamefont
  {Bastianelli}, \citenamefont {Benincasa},\ and\ \citenamefont
  {Giombi}}]{Bastianelli:2005vk}%
  \BibitemOpen
  \bibfield  {author} {\bibinfo {author} {\bibfnamefont {F.}~\bibnamefont
  {Bastianelli}}, \bibinfo {author} {\bibfnamefont {P.}~\bibnamefont
  {Benincasa}},\ and\ \bibinfo {author} {\bibfnamefont {S.}~\bibnamefont
  {Giombi}},\ }\bibfield  {title} {\bibinfo {title} {{Worldline approach to
  vector and antisymmetric tensor fields}},\ }\href
  {https://doi.org/10.1088/1126-6708/2005/04/010} {\bibfield  {journal}
  {\bibinfo  {journal} {JHEP}\ }\textbf {\bibinfo {volume} {04}},\ \bibinfo
  {pages} {010}},\ \Eprint {https://arxiv.org/abs/hep-th/0503155}
  {arXiv:hep-th/0503155} \BibitemShut {NoStop}%
\bibitem [{\citenamefont {Mogull}\ \emph {et~al.}(2021)\citenamefont {Mogull},
  \citenamefont {Plefka},\ and\ \citenamefont {Steinhoff}}]{Mogull:2020sak}%
  \BibitemOpen
  \bibfield  {author} {\bibinfo {author} {\bibfnamefont {G.}~\bibnamefont
  {Mogull}}, \bibinfo {author} {\bibfnamefont {J.}~\bibnamefont {Plefka}},\
  and\ \bibinfo {author} {\bibfnamefont {J.}~\bibnamefont {Steinhoff}},\
  }\bibfield  {title} {\bibinfo {title} {{Classical black hole scattering from
  a worldline quantum field theory}},\ }\href
  {https://doi.org/10.1007/JHEP02(2021)048} {\bibfield  {journal} {\bibinfo
  {journal} {JHEP}\ }\textbf {\bibinfo {volume} {02}},\ \bibinfo {pages}
  {048}},\ \Eprint {https://arxiv.org/abs/2010.02865} {arXiv:2010.02865
  [hep-th]} \BibitemShut {NoStop}%
\bibitem [{\citenamefont {Bonocore}\ \emph {et~al.}(2024)\citenamefont
  {Bonocore}, \citenamefont {Kulesza},\ and\ \citenamefont
  {Pirsch}}]{Bonocore:2024uxk}%
  \BibitemOpen
  \bibfield  {author} {\bibinfo {author} {\bibfnamefont {D.}~\bibnamefont
  {Bonocore}}, \bibinfo {author} {\bibfnamefont {A.}~\bibnamefont {Kulesza}},\
  and\ \bibinfo {author} {\bibfnamefont {J.}~\bibnamefont {Pirsch}},\
  }\bibfield  {title} {\bibinfo {title} {{Generalized Wilson lines and the
  gravitational scattering of spinning bodies}},\ }\href@noop {} {\  (\bibinfo
  {year} {2024})},\ \Eprint {https://arxiv.org/abs/2412.16049}
  {arXiv:2412.16049 [hep-th]} \BibitemShut {NoStop}%
\bibitem [{\citenamefont {Haddad}\ \emph {et~al.}(2024)\citenamefont {Haddad},
  \citenamefont {Jakobsen}, \citenamefont {Mogull},\ and\ \citenamefont
  {Plefka}}]{Haddad:2024ebn}%
  \BibitemOpen
  \bibfield  {author} {\bibinfo {author} {\bibfnamefont {K.}~\bibnamefont
  {Haddad}}, \bibinfo {author} {\bibfnamefont {G.~U.}\ \bibnamefont
  {Jakobsen}}, \bibinfo {author} {\bibfnamefont {G.}~\bibnamefont {Mogull}},\
  and\ \bibinfo {author} {\bibfnamefont {J.}~\bibnamefont {Plefka}},\
  }\bibfield  {title} {\bibinfo {title} {{Spinning bodies in general relativity
  from bosonic worldline oscillators}},\ }\href@noop {} {\  (\bibinfo {year}
  {2024})},\ \Eprint {https://arxiv.org/abs/2411.08176} {arXiv:2411.08176
  [hep-th]} \BibitemShut {NoStop}%
\bibitem [{\citenamefont {Gambino}\ \emph {et~al.}(2024)\citenamefont
  {Gambino}, \citenamefont {Pani},\ and\ \citenamefont
  {Riccioni}}]{Gambino:2024uge}%
  \BibitemOpen
  \bibfield  {author} {\bibinfo {author} {\bibfnamefont {C.}~\bibnamefont
  {Gambino}}, \bibinfo {author} {\bibfnamefont {P.}~\bibnamefont {Pani}},\ and\
  \bibinfo {author} {\bibfnamefont {F.}~\bibnamefont {Riccioni}},\ }\bibfield
  {title} {\bibinfo {title} {{Rotating metrics and new multipole moments from
  scattering amplitudes in arbitrary dimensions}},\ }\href
  {https://doi.org/10.1103/PhysRevD.109.124018} {\bibfield  {journal} {\bibinfo
   {journal} {Phys. Rev. D}\ }\textbf {\bibinfo {volume} {109}},\ \bibinfo
  {pages} {124018} (\bibinfo {year} {2024})},\ \Eprint
  {https://arxiv.org/abs/2403.16574} {arXiv:2403.16574 [hep-th]} \BibitemShut
  {NoStop}%
\bibitem [{\citenamefont {Myers}(2012)}]{Myers:2011yc}%
  \BibitemOpen
  \bibfield  {author} {\bibinfo {author} {\bibfnamefont {R.~C.}\ \bibnamefont
  {Myers}},\ }\bibinfo {title} {{Myers\textendash{}Perry black holes}},\ in\
  \href@noop {} {\emph {\bibinfo {booktitle} {{Black holes in higher
  dimensions}}}},\ \bibinfo {editor} {edited by\ \bibinfo {editor}
  {\bibfnamefont {G.~T.}\ \bibnamefont {Horowitz}}}\ (\bibinfo {year} {2012})\
  pp.\ \bibinfo {pages} {101--133},\ \Eprint {https://arxiv.org/abs/1111.1903}
  {arXiv:1111.1903 [gr-qc]} \BibitemShut {NoStop}%
\bibitem [{\citenamefont {Myers}\ and\ \citenamefont
  {Perry}(1986)}]{Myers:1986un}%
  \BibitemOpen
  \bibfield  {author} {\bibinfo {author} {\bibfnamefont {R.~C.}\ \bibnamefont
  {Myers}}\ and\ \bibinfo {author} {\bibfnamefont {M.~J.}\ \bibnamefont
  {Perry}},\ }\bibfield  {title} {\bibinfo {title} {{Black Holes in Higher
  Dimensional Space-Times}},\ }\href
  {https://doi.org/10.1016/0003-4916(86)90186-7} {\bibfield  {journal}
  {\bibinfo  {journal} {Annals Phys.}\ }\textbf {\bibinfo {volume} {172}},\
  \bibinfo {pages} {304} (\bibinfo {year} {1986})}\BibitemShut {NoStop}%
\bibitem [{\citenamefont {Bianchi}\ \emph {et~al.}(2025)\citenamefont
  {Bianchi}, \citenamefont {Gambino}, \citenamefont {Pani},\ and\ \citenamefont
  {Riccioni}}]{Bianchi:2024shc}%
  \BibitemOpen
  \bibfield  {author} {\bibinfo {author} {\bibfnamefont {M.}~\bibnamefont
  {Bianchi}}, \bibinfo {author} {\bibfnamefont {C.}~\bibnamefont {Gambino}},
  \bibinfo {author} {\bibfnamefont {P.}~\bibnamefont {Pani}},\ and\ \bibinfo
  {author} {\bibfnamefont {F.}~\bibnamefont {Riccioni}},\ }\bibfield  {title}
  {\bibinfo {title} {{Source multipoles and energy-momentum tensors for
  spinning black holes and other compact objects in arbitrary dimensions}},\
  }\href {https://doi.org/10.1103/PhysRevD.111.084013} {\bibfield  {journal}
  {\bibinfo  {journal} {Phys. Rev. D}\ }\textbf {\bibinfo {volume} {111}},\
  \bibinfo {pages} {084013} (\bibinfo {year} {2025})},\ \Eprint
  {https://arxiv.org/abs/2412.01771} {arXiv:2412.01771 [gr-qc]} \BibitemShut
  {NoStop}%
\bibitem [{\citenamefont {DeWitt}(1967)}]{DeWitt:1967ub}%
  \BibitemOpen
  \bibfield  {author} {\bibinfo {author} {\bibfnamefont {B.~S.}\ \bibnamefont
  {DeWitt}},\ }\bibfield  {title} {\bibinfo {title} {{Quantum Theory of
  Gravity. 2. The Manifestly Covariant Theory}},\ }\href
  {https://doi.org/10.1103/PhysRev.162.1195} {\bibfield  {journal} {\bibinfo
  {journal} {Phys. Rev.}\ }\textbf {\bibinfo {volume} {162}},\ \bibinfo {pages}
  {1195} (\bibinfo {year} {1967})}\BibitemShut {NoStop}%
\bibitem [{\citenamefont {'t~Hooft}\ and\ \citenamefont
  {Veltman}(1974)}]{tHooft:1974toh}%
  \BibitemOpen
  \bibfield  {author} {\bibinfo {author} {\bibfnamefont {G.}~\bibnamefont
  {'t~Hooft}}\ and\ \bibinfo {author} {\bibfnamefont {M.~J.~G.}\ \bibnamefont
  {Veltman}},\ }\bibfield  {title} {\bibinfo {title} {{One loop divergencies in
  the theory of gravitation}},\ }\href@noop {} {\bibfield  {journal} {\bibinfo
  {journal} {Ann. Inst. H. Poincare A Phys. Theor.}\ }\textbf {\bibinfo
  {volume} {20}},\ \bibinfo {pages} {69} (\bibinfo {year} {1974})}\BibitemShut
  {NoStop}%
\bibitem [{\citenamefont {Abbott}(1981)}]{Abbott:1980hw}%
  \BibitemOpen
  \bibfield  {author} {\bibinfo {author} {\bibfnamefont {L.~F.}\ \bibnamefont
  {Abbott}},\ }\bibfield  {title} {\bibinfo {title} {{The Background Field
  Method Beyond One Loop}},\ }\href
  {https://doi.org/10.1016/0550-3213(81)90371-0} {\bibfield  {journal}
  {\bibinfo  {journal} {Nucl. Phys. B}\ }\textbf {\bibinfo {volume} {185}},\
  \bibinfo {pages} {189} (\bibinfo {year} {1981})}\BibitemShut {NoStop}%
\bibitem [{\citenamefont {Abbott}(1982)}]{Abbott:1981ke}%
  \BibitemOpen
  \bibfield  {author} {\bibinfo {author} {\bibfnamefont {L.~F.}\ \bibnamefont
  {Abbott}},\ }\bibfield  {title} {\bibinfo {title} {{Introduction to the
  Background Field Method}},\ }\href@noop {} {\bibfield  {journal} {\bibinfo
  {journal} {Acta Phys. Polon. B}\ }\textbf {\bibinfo {volume} {13}},\ \bibinfo
  {pages} {33} (\bibinfo {year} {1982})}\BibitemShut {NoStop}%
\bibitem [{\citenamefont {Boulware}\ and\ \citenamefont
  {Brown}(1968)}]{Boulware:1968zz}%
  \BibitemOpen
  \bibfield  {author} {\bibinfo {author} {\bibfnamefont {D.~G.}\ \bibnamefont
  {Boulware}}\ and\ \bibinfo {author} {\bibfnamefont {L.~S.}\ \bibnamefont
  {Brown}},\ }\bibfield  {title} {\bibinfo {title} {{Tree Graphs and Classical
  Fields}},\ }\href {https://doi.org/10.1103/PhysRev.172.1628} {\bibfield
  {journal} {\bibinfo  {journal} {Phys. Rev.}\ }\textbf {\bibinfo {volume}
  {172}},\ \bibinfo {pages} {1628} (\bibinfo {year} {1968})}\BibitemShut
  {NoStop}%
\bibitem [{\citenamefont {Goldberger}\ and\ \citenamefont
  {Rothstein}(2006)}]{Goldberger:2004jt}%
  \BibitemOpen
  \bibfield  {author} {\bibinfo {author} {\bibfnamefont {W.~D.}\ \bibnamefont
  {Goldberger}}\ and\ \bibinfo {author} {\bibfnamefont {I.~Z.}\ \bibnamefont
  {Rothstein}},\ }\bibfield  {title} {\bibinfo {title} {{An Effective field
  theory of gravity for extended objects}},\ }\href
  {https://doi.org/10.1103/PhysRevD.73.104029} {\bibfield  {journal} {\bibinfo
  {journal} {Phys. Rev. D}\ }\textbf {\bibinfo {volume} {73}},\ \bibinfo
  {pages} {104029} (\bibinfo {year} {2006})},\ \Eprint
  {https://arxiv.org/abs/hep-th/0409156} {arXiv:hep-th/0409156} \BibitemShut
  {NoStop}%
\bibitem [{\citenamefont {Porto}(2016)}]{Porto:2016pyg}%
  \BibitemOpen
  \bibfield  {author} {\bibinfo {author} {\bibfnamefont {R.~A.}\ \bibnamefont
  {Porto}},\ }\bibfield  {title} {\bibinfo {title} {{The effective field
  theorist\textquoteright{}s approach to gravitational dynamics}},\ }\href
  {https://doi.org/10.1016/j.physrep.2016.04.003} {\bibfield  {journal}
  {\bibinfo  {journal} {Phys. Rept.}\ }\textbf {\bibinfo {volume} {633}},\
  \bibinfo {pages} {1} (\bibinfo {year} {2016})},\ \Eprint
  {https://arxiv.org/abs/1601.04914} {arXiv:1601.04914 [hep-th]} \BibitemShut
  {NoStop}%
\bibitem [{\citenamefont {Donoghue}\ \emph {et~al.}(2017)\citenamefont
  {Donoghue}, \citenamefont {Ivanov},\ and\ \citenamefont
  {Shkerin}}]{Donoghue:2017pgk}%
  \BibitemOpen
  \bibfield  {author} {\bibinfo {author} {\bibfnamefont {J.~F.}\ \bibnamefont
  {Donoghue}}, \bibinfo {author} {\bibfnamefont {M.~M.}\ \bibnamefont
  {Ivanov}},\ and\ \bibinfo {author} {\bibfnamefont {A.}~\bibnamefont
  {Shkerin}},\ }\bibfield  {title} {\bibinfo {title} {{EPFL Lectures on General
  Relativity as a Quantum Field Theory}},\ }\href@noop {} {\  (\bibinfo {year}
  {2017})},\ \Eprint {https://arxiv.org/abs/1702.00319} {arXiv:1702.00319
  [hep-th]} \BibitemShut {NoStop}%
\bibitem [{\citenamefont {Goldberger}(2022)}]{Goldberger:2022rqf}%
  \BibitemOpen
  \bibfield  {author} {\bibinfo {author} {\bibfnamefont {W.~D.}\ \bibnamefont
  {Goldberger}},\ }\bibfield  {title} {\bibinfo {title} {{Effective Field
  Theory for Compact Binary Dynamics}},\ }\href@noop {} {\  (\bibinfo {year}
  {2022})},\ \Eprint {https://arxiv.org/abs/2212.06677} {arXiv:2212.06677
  [hep-th]} \BibitemShut {NoStop}%
\bibitem [{\citenamefont {Emparan}\ and\ \citenamefont
  {Reall}(2008)}]{Emparan:2008eg}%
  \BibitemOpen
  \bibfield  {author} {\bibinfo {author} {\bibfnamefont {R.}~\bibnamefont
  {Emparan}}\ and\ \bibinfo {author} {\bibfnamefont {H.~S.}\ \bibnamefont
  {Reall}},\ }\bibfield  {title} {\bibinfo {title} {{Black Holes in Higher
  Dimensions}},\ }\href {https://doi.org/10.12942/lrr-2008-6} {\bibfield
  {journal} {\bibinfo  {journal} {Living Rev. Rel.}\ }\textbf {\bibinfo
  {volume} {11}},\ \bibinfo {pages} {6} (\bibinfo {year} {2008})},\ \Eprint
  {https://arxiv.org/abs/0801.3471} {arXiv:0801.3471 [hep-th]} \BibitemShut
  {NoStop}%
\bibitem [{\citenamefont {Frolov}\ \emph {et~al.}(2017)\citenamefont {Frolov},
  \citenamefont {Krtous},\ and\ \citenamefont {Kubiznak}}]{Frolov:2017kze}%
  \BibitemOpen
  \bibfield  {author} {\bibinfo {author} {\bibfnamefont {V.~P.}\ \bibnamefont
  {Frolov}}, \bibinfo {author} {\bibfnamefont {P.}~\bibnamefont {Krtous}},\
  and\ \bibinfo {author} {\bibfnamefont {D.}~\bibnamefont {Kubiznak}},\
  }\bibfield  {title} {\bibinfo {title} {{Black holes, hidden symmetries, and
  complete integrability}},\ }\href {https://doi.org/10.1007/s41114-017-0009-9}
  {\bibfield  {journal} {\bibinfo  {journal} {Living Rev. Rel.}\ }\textbf
  {\bibinfo {volume} {20}},\ \bibinfo {pages} {6} (\bibinfo {year} {2017})},\
  \Eprint {https://arxiv.org/abs/1705.05482} {arXiv:1705.05482 [gr-qc]}
  \BibitemShut {NoStop}%
\bibitem [{\citenamefont {Geroch}\ and\ \citenamefont
  {Traschen}(1986)}]{Geroch:1986jjl}%
  \BibitemOpen
  \bibfield  {author} {\bibinfo {author} {\bibfnamefont {R.~P.}\ \bibnamefont
  {Geroch}}\ and\ \bibinfo {author} {\bibfnamefont {J.~H.}\ \bibnamefont
  {Traschen}},\ }\bibfield  {title} {\bibinfo {title} {{Strings and Other
  Distributional Sources in General Relativity}},\ }\href
  {https://doi.org/10.1103/PhysRevD.36.1017} {\bibfield  {journal} {\bibinfo
  {journal} {Conf. Proc. C}\ }\textbf {\bibinfo {volume} {861214}},\ \bibinfo
  {pages} {138} (\bibinfo {year} {1986})}\BibitemShut {NoStop}%
\bibitem [{\citenamefont {Balasin}\ and\ \citenamefont
  {Nachbagauer}(1993)}]{Balasin:1993fn}%
  \BibitemOpen
  \bibfield  {author} {\bibinfo {author} {\bibfnamefont {H.}~\bibnamefont
  {Balasin}}\ and\ \bibinfo {author} {\bibfnamefont {H.}~\bibnamefont
  {Nachbagauer}},\ }\bibfield  {title} {\bibinfo {title} {{The energy-momentum
  tensor of a black hole, or what curves the Schwarzschild geometry?}},\ }\href
  {https://doi.org/10.1088/0264-9381/10/11/010} {\bibfield  {journal} {\bibinfo
   {journal} {Class. Quant. Grav.}\ }\textbf {\bibinfo {volume} {10}},\
  \bibinfo {pages} {2271} (\bibinfo {year} {1993})},\ \Eprint
  {https://arxiv.org/abs/gr-qc/9305009} {arXiv:gr-qc/9305009} \BibitemShut
  {NoStop}%
\bibitem [{\citenamefont {Jakobsen}\ \emph
  {et~al.}(2022{\natexlab{a}})\citenamefont {Jakobsen}, \citenamefont {Mogull},
  \citenamefont {Plefka},\ and\ \citenamefont {Steinhoff}}]{Jakobsen:2021zvh}%
  \BibitemOpen
  \bibfield  {author} {\bibinfo {author} {\bibfnamefont {G.~U.}\ \bibnamefont
  {Jakobsen}}, \bibinfo {author} {\bibfnamefont {G.}~\bibnamefont {Mogull}},
  \bibinfo {author} {\bibfnamefont {J.}~\bibnamefont {Plefka}},\ and\ \bibinfo
  {author} {\bibfnamefont {J.}~\bibnamefont {Steinhoff}},\ }\bibfield  {title}
  {\bibinfo {title} {{SUSY in the sky with gravitons}},\ }\href
  {https://doi.org/10.1007/JHEP01(2022)027} {\bibfield  {journal} {\bibinfo
  {journal} {JHEP}\ }\textbf {\bibinfo {volume} {01}},\ \bibinfo {pages}
  {027}},\ \Eprint {https://arxiv.org/abs/2109.04465} {arXiv:2109.04465
  [hep-th]} \BibitemShut {NoStop}%
\bibitem [{\citenamefont {Levi}\ and\ \citenamefont
  {Steinhoff}(2015)}]{Levi:2015msa}%
  \BibitemOpen
  \bibfield  {author} {\bibinfo {author} {\bibfnamefont {M.}~\bibnamefont
  {Levi}}\ and\ \bibinfo {author} {\bibfnamefont {J.}~\bibnamefont
  {Steinhoff}},\ }\bibfield  {title} {\bibinfo {title} {{Spinning gravitating
  objects in the effective field theory in the post-Newtonian scheme}},\ }\href
  {https://doi.org/10.1007/JHEP09(2015)219} {\bibfield  {journal} {\bibinfo
  {journal} {JHEP}\ }\textbf {\bibinfo {volume} {09}},\ \bibinfo {pages}
  {219}},\ \Eprint {https://arxiv.org/abs/1501.04956} {arXiv:1501.04956
  [gr-qc]} \BibitemShut {NoStop}%
\bibitem [{\citenamefont {Saketh}\ and\ \citenamefont
  {Vines}(2022)}]{Saketh:2022wap}%
  \BibitemOpen
  \bibfield  {author} {\bibinfo {author} {\bibfnamefont {M.~V.~S.}\
  \bibnamefont {Saketh}}\ and\ \bibinfo {author} {\bibfnamefont
  {J.}~\bibnamefont {Vines}},\ }\bibfield  {title} {\bibinfo {title}
  {{Scattering of gravitational waves off spinning compact objects with an
  effective worldline theory}},\ }\href
  {https://doi.org/10.1103/PhysRevD.106.124026} {\bibfield  {journal} {\bibinfo
   {journal} {Phys. Rev. D}\ }\textbf {\bibinfo {volume} {106}},\ \bibinfo
  {pages} {124026} (\bibinfo {year} {2022})},\ \Eprint
  {https://arxiv.org/abs/2208.03170} {arXiv:2208.03170 [gr-qc]} \BibitemShut
  {NoStop}%
\bibitem [{\citenamefont {Ben-Shahar}(2024)}]{Ben-Shahar:2023djm}%
  \BibitemOpen
  \bibfield  {author} {\bibinfo {author} {\bibfnamefont {M.}~\bibnamefont
  {Ben-Shahar}},\ }\bibfield  {title} {\bibinfo {title} {{Scattering of
  spinning compact objects from a worldline EFT}},\ }\href
  {https://doi.org/10.1007/JHEP03(2024)108} {\bibfield  {journal} {\bibinfo
  {journal} {JHEP}\ }\textbf {\bibinfo {volume} {03}},\ \bibinfo {pages}
  {108}},\ \Eprint {https://arxiv.org/abs/2311.01430} {arXiv:2311.01430
  [hep-th]} \BibitemShut {NoStop}%
\bibitem [{\citenamefont {Thorne}(1980)}]{Thorne:1980ru}%
  \BibitemOpen
  \bibfield  {author} {\bibinfo {author} {\bibfnamefont {K.~S.}\ \bibnamefont
  {Thorne}},\ }\bibfield  {title} {\bibinfo {title} {{Multipole Expansions of
  Gravitational Radiation}},\ }\href
  {https://doi.org/10.1103/RevModPhys.52.299} {\bibfield  {journal} {\bibinfo
  {journal} {Rev. Mod. Phys.}\ }\textbf {\bibinfo {volume} {52}},\ \bibinfo
  {pages} {299} (\bibinfo {year} {1980})}\BibitemShut {NoStop}%
\bibitem [{\citenamefont {Porto}\ and\ \citenamefont
  {Rothstein}(2008)}]{Porto:2008jj}%
  \BibitemOpen
  \bibfield  {author} {\bibinfo {author} {\bibfnamefont {R.~A.}\ \bibnamefont
  {Porto}}\ and\ \bibinfo {author} {\bibfnamefont {I.~Z.}\ \bibnamefont
  {Rothstein}},\ }\bibfield  {title} {\bibinfo {title} {{Next to Leading Order
  Spin(1)Spin(1) Effects in the Motion of Inspiralling Compact Binaries}},\
  }\href {https://doi.org/10.1103/PhysRevD.78.044013} {\bibfield  {journal}
  {\bibinfo  {journal} {Phys. Rev. D}\ }\textbf {\bibinfo {volume} {78}},\
  \bibinfo {pages} {044013} (\bibinfo {year} {2008})},\ \bibinfo {note}
  {[Erratum: Phys.Rev.D 81, 029905 (2010)]},\ \Eprint
  {https://arxiv.org/abs/0804.0260} {arXiv:0804.0260 [gr-qc]} \BibitemShut
  {NoStop}%
\bibitem [{\citenamefont {Vines}\ \emph {et~al.}(2016)\citenamefont {Vines},
  \citenamefont {Kunst}, \citenamefont {Steinhoff},\ and\ \citenamefont
  {Hinderer}}]{Vines:2016unv}%
  \BibitemOpen
  \bibfield  {author} {\bibinfo {author} {\bibfnamefont {J.}~\bibnamefont
  {Vines}}, \bibinfo {author} {\bibfnamefont {D.}~\bibnamefont {Kunst}},
  \bibinfo {author} {\bibfnamefont {J.}~\bibnamefont {Steinhoff}},\ and\
  \bibinfo {author} {\bibfnamefont {T.}~\bibnamefont {Hinderer}},\ }\bibfield
  {title} {\bibinfo {title} {{Canonical Hamiltonian for an extended test body
  in curved spacetime: To quadratic order in spin}},\ }\href
  {https://doi.org/10.1103/PhysRevD.104.029902} {\bibfield  {journal} {\bibinfo
   {journal} {Phys. Rev. D}\ }\textbf {\bibinfo {volume} {93}},\ \bibinfo
  {pages} {103008} (\bibinfo {year} {2016})},\ \bibinfo {note} {[Erratum:
  Phys.Rev.D 104, 029902 (2021)]},\ \Eprint {https://arxiv.org/abs/1601.07529}
  {arXiv:1601.07529 [gr-qc]} \BibitemShut {NoStop}%
\bibitem [{\citenamefont {Witzany}(2019)}]{Witzany:2019nml}%
  \BibitemOpen
  \bibfield  {author} {\bibinfo {author} {\bibfnamefont {V.}~\bibnamefont
  {Witzany}},\ }\bibfield  {title} {\bibinfo {title} {{Hamilton-Jacobi equation
  for spinning particles near black holes}},\ }\href
  {https://doi.org/10.1103/PhysRevD.100.104030} {\bibfield  {journal} {\bibinfo
   {journal} {Phys. Rev. D}\ }\textbf {\bibinfo {volume} {100}},\ \bibinfo
  {pages} {104030} (\bibinfo {year} {2019})},\ \Eprint
  {https://arxiv.org/abs/1903.03651} {arXiv:1903.03651 [gr-qc]} \BibitemShut
  {NoStop}%
\bibitem [{\citenamefont {Ramond}(2022)}]{Ramond:2022vhj}%
  \BibitemOpen
  \bibfield  {author} {\bibinfo {author} {\bibfnamefont {P.}~\bibnamefont
  {Ramond}},\ }\bibfield  {title} {\bibinfo {title} {{Symplectic mechanics of
  relativistic spinning compact bodies I.: Covariant foundations and
  integrability around black holes}},\ }\href@noop {} {\  (\bibinfo {year}
  {2022})},\ \Eprint {https://arxiv.org/abs/2210.03866} {arXiv:2210.03866
  [gr-qc]} \BibitemShut {NoStop}%
\bibitem [{\citenamefont {Ramond}(2024)}]{Ramond:2024ozy}%
  \BibitemOpen
  \bibfield  {author} {\bibinfo {author} {\bibfnamefont {P.}~\bibnamefont
  {Ramond}},\ }\bibfield  {title} {\bibinfo {title} {{On the integrability of
  extended test body dynamics around black holes}},\ }\href@noop {} {\
  (\bibinfo {year} {2024})},\ \Eprint {https://arxiv.org/abs/2402.02670}
  {arXiv:2402.02670 [gr-qc]} \BibitemShut {NoStop}%
\bibitem [{\citenamefont {Witzany}\ \emph {et~al.}(2024)\citenamefont
  {Witzany}, \citenamefont {Skoup\'y}, \citenamefont {Stein},\ and\
  \citenamefont {Tanay}}]{Witzany:2024ttz}%
  \BibitemOpen
  \bibfield  {author} {\bibinfo {author} {\bibfnamefont {V.}~\bibnamefont
  {Witzany}}, \bibinfo {author} {\bibfnamefont {V.}~\bibnamefont {Skoup\'y}},
  \bibinfo {author} {\bibfnamefont {L.~C.}\ \bibnamefont {Stein}},\ and\
  \bibinfo {author} {\bibfnamefont {S.}~\bibnamefont {Tanay}},\ }\bibfield
  {title} {\bibinfo {title} {{Actions of spinning compact binaries: Spinning
  particle in Kerr matched to dynamics at 1.5 post-Newtonian order}},\
  }\href@noop {} {\  (\bibinfo {year} {2024})},\ \Eprint
  {https://arxiv.org/abs/2411.09742} {arXiv:2411.09742 [gr-qc]} \BibitemShut
  {NoStop}%
\bibitem [{\citenamefont {Skoup\'y}\ and\ \citenamefont
  {Witzany}(2024)}]{Skoupy:2024uan}%
  \BibitemOpen
  \bibfield  {author} {\bibinfo {author} {\bibfnamefont {V.}~\bibnamefont
  {Skoup\'y}}\ and\ \bibinfo {author} {\bibfnamefont {V.}~\bibnamefont
  {Witzany}},\ }\bibfield  {title} {\bibinfo {title} {{Analytic solution for
  the motion of spinning particles in Kerr space-time}},\ }\href@noop {} {\
  (\bibinfo {year} {2024})},\ \Eprint {https://arxiv.org/abs/2411.16855}
  {arXiv:2411.16855 [gr-qc]} \BibitemShut {NoStop}%
\bibitem [{\citenamefont {Barack}(2009)}]{Barack:2009ux}%
  \BibitemOpen
  \bibfield  {author} {\bibinfo {author} {\bibfnamefont {L.}~\bibnamefont
  {Barack}},\ }\bibfield  {title} {\bibinfo {title} {{Gravitational self force
  in extreme mass-ratio inspirals}},\ }\href
  {https://doi.org/10.1088/0264-9381/26/21/213001} {\bibfield  {journal}
  {\bibinfo  {journal} {Class. Quant. Grav.}\ }\textbf {\bibinfo {volume}
  {26}},\ \bibinfo {pages} {213001} (\bibinfo {year} {2009})},\ \Eprint
  {https://arxiv.org/abs/0908.1664} {arXiv:0908.1664 [gr-qc]} \BibitemShut
  {NoStop}%
\bibitem [{\citenamefont {Pfenning}\ and\ \citenamefont
  {Poisson}(2002)}]{Pfenning:2000zf}%
  \BibitemOpen
  \bibfield  {author} {\bibinfo {author} {\bibfnamefont {M.~J.}\ \bibnamefont
  {Pfenning}}\ and\ \bibinfo {author} {\bibfnamefont {E.}~\bibnamefont
  {Poisson}},\ }\bibfield  {title} {\bibinfo {title} {{Scalar, electromagnetic,
  and gravitational selfforces in weakly curved space-times}},\ }\href
  {https://doi.org/10.1103/PhysRevD.65.084001} {\bibfield  {journal} {\bibinfo
  {journal} {Phys. Rev. D}\ }\textbf {\bibinfo {volume} {65}},\ \bibinfo
  {pages} {084001} (\bibinfo {year} {2002})},\ \Eprint
  {https://arxiv.org/abs/gr-qc/0012057} {arXiv:gr-qc/0012057} \BibitemShut
  {NoStop}%
\bibitem [{\citenamefont {Gralla}\ and\ \citenamefont
  {Lobo}(2022)}]{Gralla:2021qaf}%
  \BibitemOpen
  \bibfield  {author} {\bibinfo {author} {\bibfnamefont {S.~E.}\ \bibnamefont
  {Gralla}}\ and\ \bibinfo {author} {\bibfnamefont {K.}~\bibnamefont {Lobo}},\
  }\bibfield  {title} {\bibinfo {title} {{Self-force effects in
  post-Minkowskian scattering}},\ }\href
  {https://doi.org/10.1088/1361-6382/ac5d88} {\bibfield  {journal} {\bibinfo
  {journal} {Class. Quant. Grav.}\ }\textbf {\bibinfo {volume} {39}},\ \bibinfo
  {pages} {095001} (\bibinfo {year} {2022})},\ \bibinfo {note} {[Erratum:
  Class.Quant.Grav. 41, 179501 (2024)]},\ \Eprint
  {https://arxiv.org/abs/2110.08681} {arXiv:2110.08681 [gr-qc]} \BibitemShut
  {NoStop}%
\bibitem [{\citenamefont {Detweiler}(2001)}]{Detweiler:2000gt}%
  \BibitemOpen
  \bibfield  {author} {\bibinfo {author} {\bibfnamefont {S.~L.}\ \bibnamefont
  {Detweiler}},\ }\bibfield  {title} {\bibinfo {title} {{Radiation reaction and
  the selfforce for a point mass in general relativity}},\ }\href
  {https://doi.org/10.1103/PhysRevLett.86.1931} {\bibfield  {journal} {\bibinfo
   {journal} {Phys. Rev. Lett.}\ }\textbf {\bibinfo {volume} {86}},\ \bibinfo
  {pages} {1931} (\bibinfo {year} {2001})},\ \Eprint
  {https://arxiv.org/abs/gr-qc/0011039} {arXiv:gr-qc/0011039} \BibitemShut
  {NoStop}%
\bibitem [{\citenamefont {Iteanu}\ \emph {et~al.}(2025)\citenamefont {Iteanu},
  \citenamefont {Riva}, \citenamefont {Santoni}, \citenamefont {Savi\'c},\ and\
  \citenamefont {Vernizzi}}]{Iteanu:2024dvx}%
  \BibitemOpen
  \bibfield  {author} {\bibinfo {author} {\bibfnamefont {S.}~\bibnamefont
  {Iteanu}}, \bibinfo {author} {\bibfnamefont {M.~M.}\ \bibnamefont {Riva}},
  \bibinfo {author} {\bibfnamefont {L.}~\bibnamefont {Santoni}}, \bibinfo
  {author} {\bibfnamefont {N.}~\bibnamefont {Savi\'c}},\ and\ \bibinfo {author}
  {\bibfnamefont {F.}~\bibnamefont {Vernizzi}},\ }\bibfield  {title} {\bibinfo
  {title} {{Vanishing of quadratic Love numbers of Schwarzschild black
  holes}},\ }\href {https://doi.org/10.1007/JHEP02(2025)174} {\bibfield
  {journal} {\bibinfo  {journal} {JHEP}\ }\textbf {\bibinfo {volume} {02}},\
  \bibinfo {pages} {174}},\ \Eprint {https://arxiv.org/abs/2410.03542}
  {arXiv:2410.03542 [gr-qc]} \BibitemShut {NoStop}%
\bibitem [{\citenamefont {Cheung}\ \emph {et~al.}(2022)\citenamefont {Cheung},
  \citenamefont {Parra-Martinez},\ and\ \citenamefont
  {Sivaramakrishnan}}]{Cheung:2022pdk}%
  \BibitemOpen
  \bibfield  {author} {\bibinfo {author} {\bibfnamefont {C.}~\bibnamefont
  {Cheung}}, \bibinfo {author} {\bibfnamefont {J.}~\bibnamefont
  {Parra-Martinez}},\ and\ \bibinfo {author} {\bibfnamefont {A.}~\bibnamefont
  {Sivaramakrishnan}},\ }\bibfield  {title} {\bibinfo {title} {{On-shell
  correlators and color-kinematics duality in curved symmetric spacetimes}},\
  }\href {https://doi.org/10.1007/JHEP05(2022)027} {\bibfield  {journal}
  {\bibinfo  {journal} {JHEP}\ }\textbf {\bibinfo {volume} {05}},\ \bibinfo
  {pages} {027}},\ \Eprint {https://arxiv.org/abs/2201.05147} {arXiv:2201.05147
  [hep-th]} \BibitemShut {NoStop}%
\bibitem [{\citenamefont {Adamo}\ \emph {et~al.}(2018)\citenamefont {Adamo},
  \citenamefont {Casali}, \citenamefont {Mason},\ and\ \citenamefont
  {Nekovar}}]{Adamo:2017nia}%
  \BibitemOpen
  \bibfield  {author} {\bibinfo {author} {\bibfnamefont {T.}~\bibnamefont
  {Adamo}}, \bibinfo {author} {\bibfnamefont {E.}~\bibnamefont {Casali}},
  \bibinfo {author} {\bibfnamefont {L.}~\bibnamefont {Mason}},\ and\ \bibinfo
  {author} {\bibfnamefont {S.}~\bibnamefont {Nekovar}},\ }\bibfield  {title}
  {\bibinfo {title} {{Scattering on plane waves and the double copy}},\ }\href
  {https://doi.org/10.1088/1361-6382/aa9961} {\bibfield  {journal} {\bibinfo
  {journal} {Class. Quant. Grav.}\ }\textbf {\bibinfo {volume} {35}},\ \bibinfo
  {pages} {015004} (\bibinfo {year} {2018})},\ \Eprint
  {https://arxiv.org/abs/1706.08925} {arXiv:1706.08925 [hep-th]} \BibitemShut
  {NoStop}%
\bibitem [{\citenamefont {Adamo}\ \emph
  {et~al.}(2024{\natexlab{a}})\citenamefont {Adamo}, \citenamefont
  {Cristofoli}, \citenamefont {Ilderton},\ and\ \citenamefont
  {Klisch}}]{Adamo:2023cfp}%
  \BibitemOpen
  \bibfield  {author} {\bibinfo {author} {\bibfnamefont {T.}~\bibnamefont
  {Adamo}}, \bibinfo {author} {\bibfnamefont {A.}~\bibnamefont {Cristofoli}},
  \bibinfo {author} {\bibfnamefont {A.}~\bibnamefont {Ilderton}},\ and\
  \bibinfo {author} {\bibfnamefont {S.}~\bibnamefont {Klisch}},\ }\bibfield
  {title} {\bibinfo {title} {{Scattering amplitudes for self-force}},\ }\href
  {https://doi.org/10.1088/1361-6382/ad210f} {\bibfield  {journal} {\bibinfo
  {journal} {Class. Quant. Grav.}\ }\textbf {\bibinfo {volume} {41}},\ \bibinfo
  {pages} {065006} (\bibinfo {year} {2024}{\natexlab{a}})},\ \Eprint
  {https://arxiv.org/abs/2307.00431} {arXiv:2307.00431 [hep-th]} \BibitemShut
  {NoStop}%
\bibitem [{\citenamefont {Maggiore}(2007)}]{Maggiore:2007ulw}%
  \BibitemOpen
  \bibfield  {author} {\bibinfo {author} {\bibfnamefont {M.}~\bibnamefont
  {Maggiore}},\ }\href
  {https://doi.org/10.1093/acprof:oso/9780198570745.001.0001} {\emph {\bibinfo
  {title} {{Gravitational Waves. Vol. 1: Theory and Experiments}}}}\ (\bibinfo
  {publisher} {Oxford University Press},\ \bibinfo {year} {2007})\BibitemShut
  {NoStop}%
\bibitem [{\citenamefont {Jakobsen}\ \emph {et~al.}(2021)\citenamefont
  {Jakobsen}, \citenamefont {Mogull}, \citenamefont {Plefka},\ and\
  \citenamefont {Steinhoff}}]{Jakobsen:2021smu}%
  \BibitemOpen
  \bibfield  {author} {\bibinfo {author} {\bibfnamefont {G.~U.}\ \bibnamefont
  {Jakobsen}}, \bibinfo {author} {\bibfnamefont {G.}~\bibnamefont {Mogull}},
  \bibinfo {author} {\bibfnamefont {J.}~\bibnamefont {Plefka}},\ and\ \bibinfo
  {author} {\bibfnamefont {J.}~\bibnamefont {Steinhoff}},\ }\bibfield  {title}
  {\bibinfo {title} {{Classical Gravitational Bremsstrahlung from a Worldline
  Quantum Field Theory}},\ }\href
  {https://doi.org/10.1103/PhysRevLett.126.201103} {\bibfield  {journal}
  {\bibinfo  {journal} {Phys. Rev. Lett.}\ }\textbf {\bibinfo {volume} {126}},\
  \bibinfo {pages} {201103} (\bibinfo {year} {2021})},\ \Eprint
  {https://arxiv.org/abs/2101.12688} {arXiv:2101.12688 [gr-qc]} \BibitemShut
  {NoStop}%
\bibitem [{\citenamefont {Mougiakakos}\ \emph {et~al.}(2021)\citenamefont
  {Mougiakakos}, \citenamefont {Riva},\ and\ \citenamefont
  {Vernizzi}}]{Mougiakakos:2021ckm}%
  \BibitemOpen
  \bibfield  {author} {\bibinfo {author} {\bibfnamefont {S.}~\bibnamefont
  {Mougiakakos}}, \bibinfo {author} {\bibfnamefont {M.~M.}\ \bibnamefont
  {Riva}},\ and\ \bibinfo {author} {\bibfnamefont {F.}~\bibnamefont
  {Vernizzi}},\ }\bibfield  {title} {\bibinfo {title} {{Gravitational
  Bremsstrahlung in the post-Minkowskian effective field theory}},\ }\href
  {https://doi.org/10.1103/PhysRevD.104.024041} {\bibfield  {journal} {\bibinfo
   {journal} {Phys. Rev. D}\ }\textbf {\bibinfo {volume} {104}},\ \bibinfo
  {pages} {024041} (\bibinfo {year} {2021})},\ \Eprint
  {https://arxiv.org/abs/2102.08339} {arXiv:2102.08339 [gr-qc]} \BibitemShut
  {NoStop}%
\bibitem [{\citenamefont {Strominger}\ and\ \citenamefont
  {Zhiboedov}(2016)}]{Strominger:2014pwa}%
  \BibitemOpen
  \bibfield  {author} {\bibinfo {author} {\bibfnamefont {A.}~\bibnamefont
  {Strominger}}\ and\ \bibinfo {author} {\bibfnamefont {A.}~\bibnamefont
  {Zhiboedov}},\ }\bibfield  {title} {\bibinfo {title} {{Gravitational Memory,
  BMS Supertranslations and Soft Theorems}},\ }\href
  {https://doi.org/10.1007/JHEP01(2016)086} {\bibfield  {journal} {\bibinfo
  {journal} {JHEP}\ }\textbf {\bibinfo {volume} {01}},\ \bibinfo {pages}
  {086}},\ \Eprint {https://arxiv.org/abs/1411.5745} {arXiv:1411.5745 [hep-th]}
  \BibitemShut {NoStop}%
\bibitem [{\citenamefont {Jakobsen}\ \emph
  {et~al.}(2022{\natexlab{b}})\citenamefont {Jakobsen}, \citenamefont {Mogull},
  \citenamefont {Plefka},\ and\ \citenamefont {Steinhoff}}]{Jakobsen:2021lvp}%
  \BibitemOpen
  \bibfield  {author} {\bibinfo {author} {\bibfnamefont {G.~U.}\ \bibnamefont
  {Jakobsen}}, \bibinfo {author} {\bibfnamefont {G.}~\bibnamefont {Mogull}},
  \bibinfo {author} {\bibfnamefont {J.}~\bibnamefont {Plefka}},\ and\ \bibinfo
  {author} {\bibfnamefont {J.}~\bibnamefont {Steinhoff}},\ }\bibfield  {title}
  {\bibinfo {title} {{Gravitational Bremsstrahlung and Hidden Supersymmetry of
  Spinning Bodies}},\ }\href {https://doi.org/10.1103/PhysRevLett.128.011101}
  {\bibfield  {journal} {\bibinfo  {journal} {Phys. Rev. Lett.}\ }\textbf
  {\bibinfo {volume} {128}},\ \bibinfo {pages} {011101} (\bibinfo {year}
  {2022}{\natexlab{b}})},\ \Eprint {https://arxiv.org/abs/2106.10256}
  {arXiv:2106.10256 [hep-th]} \BibitemShut {NoStop}%
\bibitem [{\citenamefont {Di~Vecchia}\ \emph {et~al.}(2022)\citenamefont
  {Di~Vecchia}, \citenamefont {Heissenberg},\ and\ \citenamefont
  {Russo}}]{DiVecchia:2022owy}%
  \BibitemOpen
  \bibfield  {author} {\bibinfo {author} {\bibfnamefont {P.}~\bibnamefont
  {Di~Vecchia}}, \bibinfo {author} {\bibfnamefont {C.}~\bibnamefont
  {Heissenberg}},\ and\ \bibinfo {author} {\bibfnamefont {R.}~\bibnamefont
  {Russo}},\ }\bibfield  {title} {\bibinfo {title} {{Angular momentum of
  zero-frequency gravitons}},\ }\href {https://doi.org/10.1007/JHEP08(2022)172}
  {\bibfield  {journal} {\bibinfo  {journal} {JHEP}\ }\textbf {\bibinfo
  {volume} {08}},\ \bibinfo {pages} {172}},\ \Eprint
  {https://arxiv.org/abs/2203.11915} {arXiv:2203.11915 [hep-th]} \BibitemShut
  {NoStop}%
\bibitem [{\citenamefont {Veneziano}\ and\ \citenamefont
  {Vilkovisky}(2022)}]{Veneziano:2022zwh}%
  \BibitemOpen
  \bibfield  {author} {\bibinfo {author} {\bibfnamefont {G.}~\bibnamefont
  {Veneziano}}\ and\ \bibinfo {author} {\bibfnamefont {G.~A.}\ \bibnamefont
  {Vilkovisky}},\ }\bibfield  {title} {\bibinfo {title} {{Angular momentum loss
  in gravitational scattering, radiation reaction, and the Bondi gauge
  ambiguity}},\ }\href {https://doi.org/10.1016/j.physletb.2022.137419}
  {\bibfield  {journal} {\bibinfo  {journal} {Phys. Lett. B}\ }\textbf
  {\bibinfo {volume} {834}},\ \bibinfo {pages} {137419} (\bibinfo {year}
  {2022})},\ \Eprint {https://arxiv.org/abs/2201.11607} {arXiv:2201.11607
  [gr-qc]} \BibitemShut {NoStop}%
\bibitem [{\citenamefont {Herderschee}\ \emph {et~al.}(2023)\citenamefont
  {Herderschee}, \citenamefont {Roiban},\ and\ \citenamefont
  {Teng}}]{Herderschee:2023fxh}%
  \BibitemOpen
  \bibfield  {author} {\bibinfo {author} {\bibfnamefont {A.}~\bibnamefont
  {Herderschee}}, \bibinfo {author} {\bibfnamefont {R.}~\bibnamefont
  {Roiban}},\ and\ \bibinfo {author} {\bibfnamefont {F.}~\bibnamefont {Teng}},\
  }\bibfield  {title} {\bibinfo {title} {{The sub-leading scattering waveform
  from amplitudes}},\ }\href {https://doi.org/10.1007/JHEP06(2023)004}
  {\bibfield  {journal} {\bibinfo  {journal} {JHEP}\ }\textbf {\bibinfo
  {volume} {06}},\ \bibinfo {pages} {004}},\ \Eprint
  {https://arxiv.org/abs/2303.06112} {arXiv:2303.06112 [hep-th]} \BibitemShut
  {NoStop}%
\bibitem [{\citenamefont {Georgoudis}\ \emph
  {et~al.}(2024{\natexlab{a}})\citenamefont {Georgoudis}, \citenamefont
  {Heissenberg},\ and\ \citenamefont {Russo}}]{Georgoudis:2023eke}%
  \BibitemOpen
  \bibfield  {author} {\bibinfo {author} {\bibfnamefont {A.}~\bibnamefont
  {Georgoudis}}, \bibinfo {author} {\bibfnamefont {C.}~\bibnamefont
  {Heissenberg}},\ and\ \bibinfo {author} {\bibfnamefont {R.}~\bibnamefont
  {Russo}},\ }\bibfield  {title} {\bibinfo {title} {{An eikonal-inspired
  approach to the gravitational scattering waveform}},\ }\href
  {https://doi.org/10.1007/JHEP03(2024)089} {\bibfield  {journal} {\bibinfo
  {journal} {JHEP}\ }\textbf {\bibinfo {volume} {03}},\ \bibinfo {pages}
  {089}},\ \Eprint {https://arxiv.org/abs/2312.07452} {arXiv:2312.07452
  [hep-th]} \BibitemShut {NoStop}%
\bibitem [{\citenamefont {Bohnenblust}\ \emph
  {et~al.}(2024{\natexlab{b}})\citenamefont {Bohnenblust}, \citenamefont {Ita},
  \citenamefont {Kraus},\ and\ \citenamefont {Schlenk}}]{Bohnenblust:2023qmy}%
  \BibitemOpen
  \bibfield  {author} {\bibinfo {author} {\bibfnamefont {L.}~\bibnamefont
  {Bohnenblust}}, \bibinfo {author} {\bibfnamefont {H.}~\bibnamefont {Ita}},
  \bibinfo {author} {\bibfnamefont {M.}~\bibnamefont {Kraus}},\ and\ \bibinfo
  {author} {\bibfnamefont {J.}~\bibnamefont {Schlenk}},\ }\bibfield  {title}
  {\bibinfo {title} {{Gravitational Bremsstrahlung in black-hole scattering at
  $ \mathcal{O}\left({G}^3\right) $: linear-in-spin effects}},\ }\href
  {https://doi.org/10.1007/JHEP11(2024)109} {\bibfield  {journal} {\bibinfo
  {journal} {JHEP}\ }\textbf {\bibinfo {volume} {11}},\ \bibinfo {pages}
  {109}},\ \Eprint {https://arxiv.org/abs/2312.14859} {arXiv:2312.14859
  [hep-th]} \BibitemShut {NoStop}%
\bibitem [{\citenamefont {Adamo}\ \emph
  {et~al.}(2024{\natexlab{b}})\citenamefont {Adamo}, \citenamefont {Gonzo},\
  and\ \citenamefont {Ilderton}}]{Adamo:2024oxy}%
  \BibitemOpen
  \bibfield  {author} {\bibinfo {author} {\bibfnamefont {T.}~\bibnamefont
  {Adamo}}, \bibinfo {author} {\bibfnamefont {R.}~\bibnamefont {Gonzo}},\ and\
  \bibinfo {author} {\bibfnamefont {A.}~\bibnamefont {Ilderton}},\ }\bibfield
  {title} {\bibinfo {title} {{Gravitational bound waveforms from amplitudes}},\
  }\href {https://doi.org/10.1007/JHEP05(2024)034} {\bibfield  {journal}
  {\bibinfo  {journal} {JHEP}\ }\textbf {\bibinfo {volume} {05}},\ \bibinfo
  {pages} {034}},\ \Eprint {https://arxiv.org/abs/2402.00124} {arXiv:2402.00124
  [hep-th]} \BibitemShut {NoStop}%
\bibitem [{\citenamefont {Elkhidir}\ \emph {et~al.}(2024)\citenamefont
  {Elkhidir}, \citenamefont {O'Connell},\ and\ \citenamefont
  {Roiban}}]{Elkhidir:2024izo}%
  \BibitemOpen
  \bibfield  {author} {\bibinfo {author} {\bibfnamefont {A.}~\bibnamefont
  {Elkhidir}}, \bibinfo {author} {\bibfnamefont {D.}~\bibnamefont
  {O'Connell}},\ and\ \bibinfo {author} {\bibfnamefont {R.}~\bibnamefont
  {Roiban}},\ }\bibfield  {title} {\bibinfo {title} {{Supertranslations from
  Scattering Amplitudes}},\ }\href@noop {} {\  (\bibinfo {year} {2024})},\
  \Eprint {https://arxiv.org/abs/2408.15961} {arXiv:2408.15961 [hep-th]}
  \BibitemShut {NoStop}%
\bibitem [{\citenamefont {Georgoudis}\ \emph
  {et~al.}(2024{\natexlab{b}})\citenamefont {Georgoudis}, \citenamefont
  {Heissenberg},\ and\ \citenamefont {Russo}}]{Georgoudis:2024pdz}%
  \BibitemOpen
  \bibfield  {author} {\bibinfo {author} {\bibfnamefont {A.}~\bibnamefont
  {Georgoudis}}, \bibinfo {author} {\bibfnamefont {C.}~\bibnamefont
  {Heissenberg}},\ and\ \bibinfo {author} {\bibfnamefont {R.}~\bibnamefont
  {Russo}},\ }\bibfield  {title} {\bibinfo {title} {{Post-Newtonian multipoles
  from the next-to-leading post-Minkowskian gravitational waveform}},\ }\href
  {https://doi.org/10.1103/PhysRevD.109.106020} {\bibfield  {journal} {\bibinfo
   {journal} {Phys. Rev. D}\ }\textbf {\bibinfo {volume} {109}},\ \bibinfo
  {pages} {106020} (\bibinfo {year} {2024}{\natexlab{b}})},\ \Eprint
  {https://arxiv.org/abs/2402.06361} {arXiv:2402.06361 [hep-th]} \BibitemShut
  {NoStop}%
\bibitem [{\citenamefont {Bini}\ \emph
  {et~al.}(2024{\natexlab{b}})\citenamefont {Bini}, \citenamefont {Damour},
  \citenamefont {De~Angelis}, \citenamefont {Geralico}, \citenamefont
  {Herderschee}, \citenamefont {Roiban},\ and\ \citenamefont
  {Teng}}]{Bini:2024rsy}%
  \BibitemOpen
  \bibfield  {author} {\bibinfo {author} {\bibfnamefont {D.}~\bibnamefont
  {Bini}}, \bibinfo {author} {\bibfnamefont {T.}~\bibnamefont {Damour}},
  \bibinfo {author} {\bibfnamefont {S.}~\bibnamefont {De~Angelis}}, \bibinfo
  {author} {\bibfnamefont {A.}~\bibnamefont {Geralico}}, \bibinfo {author}
  {\bibfnamefont {A.}~\bibnamefont {Herderschee}}, \bibinfo {author}
  {\bibfnamefont {R.}~\bibnamefont {Roiban}},\ and\ \bibinfo {author}
  {\bibfnamefont {F.}~\bibnamefont {Teng}},\ }\bibfield  {title} {\bibinfo
  {title} {{Gravitational waveforms: A tale of two formalisms}},\ }\href
  {https://doi.org/10.1103/PhysRevD.109.125008} {\bibfield  {journal} {\bibinfo
   {journal} {Phys. Rev. D}\ }\textbf {\bibinfo {volume} {109}},\ \bibinfo
  {pages} {125008} (\bibinfo {year} {2024}{\natexlab{b}})},\ \Eprint
  {https://arxiv.org/abs/2402.06604} {arXiv:2402.06604 [hep-th]} \BibitemShut
  {NoStop}%
\bibitem [{\citenamefont {Warburton}\ \emph {et~al.}(2024)\citenamefont
  {Warburton}, \citenamefont {Wardell}, \citenamefont {Trestini}, \citenamefont
  {Henry}, \citenamefont {Pound}, \citenamefont {Blanchet}, \citenamefont
  {Durkan}, \citenamefont {Faye},\ and\ \citenamefont
  {Miller}}]{Warburton:2024xnr}%
  \BibitemOpen
  \bibfield  {author} {\bibinfo {author} {\bibfnamefont {N.}~\bibnamefont
  {Warburton}}, \bibinfo {author} {\bibfnamefont {B.}~\bibnamefont {Wardell}},
  \bibinfo {author} {\bibfnamefont {D.}~\bibnamefont {Trestini}}, \bibinfo
  {author} {\bibfnamefont {Q.}~\bibnamefont {Henry}}, \bibinfo {author}
  {\bibfnamefont {A.}~\bibnamefont {Pound}}, \bibinfo {author} {\bibfnamefont
  {L.}~\bibnamefont {Blanchet}}, \bibinfo {author} {\bibfnamefont
  {L.}~\bibnamefont {Durkan}}, \bibinfo {author} {\bibfnamefont
  {G.}~\bibnamefont {Faye}},\ and\ \bibinfo {author} {\bibfnamefont
  {J.}~\bibnamefont {Miller}},\ }\bibfield  {title} {\bibinfo {title}
  {{Comparison of 4.5PN and 2SF gravitational energy fluxes from quasicircular
  compact binaries}},\ }\href@noop {} {\  (\bibinfo {year} {2024})},\ \Eprint
  {https://arxiv.org/abs/2407.00366} {arXiv:2407.00366 [gr-qc]} \BibitemShut
  {NoStop}%
\bibitem [{\citenamefont {Mano}\ \emph
  {et~al.}(1996{\natexlab{a}})\citenamefont {Mano}, \citenamefont {Suzuki},\
  and\ \citenamefont {Takasugi}}]{Mano:1996mf}%
  \BibitemOpen
  \bibfield  {author} {\bibinfo {author} {\bibfnamefont {S.}~\bibnamefont
  {Mano}}, \bibinfo {author} {\bibfnamefont {H.}~\bibnamefont {Suzuki}},\ and\
  \bibinfo {author} {\bibfnamefont {E.}~\bibnamefont {Takasugi}},\ }\bibfield
  {title} {\bibinfo {title} {{Analytic solutions of the Regge-Wheeler equation
  and the postMinkowskian expansion}},\ }\href
  {https://doi.org/10.1143/PTP.96.549} {\bibfield  {journal} {\bibinfo
  {journal} {Prog. Theor. Phys.}\ }\textbf {\bibinfo {volume} {96}},\ \bibinfo
  {pages} {549} (\bibinfo {year} {1996}{\natexlab{a}})},\ \Eprint
  {https://arxiv.org/abs/gr-qc/9605057} {arXiv:gr-qc/9605057} \BibitemShut
  {NoStop}%
\bibitem [{\citenamefont {Mano}\ and\ \citenamefont
  {Takasugi}(1997)}]{Mano:1996gn}%
  \BibitemOpen
  \bibfield  {author} {\bibinfo {author} {\bibfnamefont {S.}~\bibnamefont
  {Mano}}\ and\ \bibinfo {author} {\bibfnamefont {E.}~\bibnamefont
  {Takasugi}},\ }\bibfield  {title} {\bibinfo {title} {{Analytic solutions of
  the Teukolsky equation and their properties}},\ }\href
  {https://doi.org/10.1143/PTP.97.213} {\bibfield  {journal} {\bibinfo
  {journal} {Prog. Theor. Phys.}\ }\textbf {\bibinfo {volume} {97}},\ \bibinfo
  {pages} {213} (\bibinfo {year} {1997})},\ \Eprint
  {https://arxiv.org/abs/gr-qc/9611014} {arXiv:gr-qc/9611014} \BibitemShut
  {NoStop}%
\bibitem [{\citenamefont {Mano}\ \emph
  {et~al.}(1996{\natexlab{b}})\citenamefont {Mano}, \citenamefont {Suzuki},\
  and\ \citenamefont {Takasugi}}]{Mano:1996vt}%
  \BibitemOpen
  \bibfield  {author} {\bibinfo {author} {\bibfnamefont {S.}~\bibnamefont
  {Mano}}, \bibinfo {author} {\bibfnamefont {H.}~\bibnamefont {Suzuki}},\ and\
  \bibinfo {author} {\bibfnamefont {E.}~\bibnamefont {Takasugi}},\ }\bibfield
  {title} {\bibinfo {title} {{Analytic solutions of the Teukolsky equation and
  their low frequency expansions}},\ }\href
  {https://doi.org/10.1143/PTP.95.1079} {\bibfield  {journal} {\bibinfo
  {journal} {Prog. Theor. Phys.}\ }\textbf {\bibinfo {volume} {95}},\ \bibinfo
  {pages} {1079} (\bibinfo {year} {1996}{\natexlab{b}})},\ \Eprint
  {https://arxiv.org/abs/gr-qc/9603020} {arXiv:gr-qc/9603020} \BibitemShut
  {NoStop}%
\bibitem [{\citenamefont {Mino}\ \emph {et~al.}(1997)\citenamefont {Mino},
  \citenamefont {Sasaki}, \citenamefont {Shibata}, \citenamefont {Tagoshi},\
  and\ \citenamefont {Tanaka}}]{Mino:1997bx}%
  \BibitemOpen
  \bibfield  {author} {\bibinfo {author} {\bibfnamefont {Y.}~\bibnamefont
  {Mino}}, \bibinfo {author} {\bibfnamefont {M.}~\bibnamefont {Sasaki}},
  \bibinfo {author} {\bibfnamefont {M.}~\bibnamefont {Shibata}}, \bibinfo
  {author} {\bibfnamefont {H.}~\bibnamefont {Tagoshi}},\ and\ \bibinfo {author}
  {\bibfnamefont {T.}~\bibnamefont {Tanaka}},\ }\bibfield  {title} {\bibinfo
  {title} {{Black hole perturbation: Chapter 1}},\ }\href
  {https://doi.org/10.1143/PTPS.128.1} {\bibfield  {journal} {\bibinfo
  {journal} {Prog. Theor. Phys. Suppl.}\ }\textbf {\bibinfo {volume} {128}},\
  \bibinfo {pages} {1} (\bibinfo {year} {1997})},\ \Eprint
  {https://arxiv.org/abs/gr-qc/9712057} {arXiv:gr-qc/9712057} \BibitemShut
  {NoStop}%
\bibitem [{\citenamefont {Ivanov}\ and\ \citenamefont
  {Zhou}(2023)}]{Ivanov:2022qqt}%
  \BibitemOpen
  \bibfield  {author} {\bibinfo {author} {\bibfnamefont {M.~M.}\ \bibnamefont
  {Ivanov}}\ and\ \bibinfo {author} {\bibfnamefont {Z.}~\bibnamefont {Zhou}},\
  }\bibfield  {title} {\bibinfo {title} {{Vanishing of Black Hole Tidal Love
  Numbers from Scattering Amplitudes}},\ }\href
  {https://doi.org/10.1103/PhysRevLett.130.091403} {\bibfield  {journal}
  {\bibinfo  {journal} {Phys. Rev. Lett.}\ }\textbf {\bibinfo {volume} {130}},\
  \bibinfo {pages} {091403} (\bibinfo {year} {2023})},\ \Eprint
  {https://arxiv.org/abs/2209.14324} {arXiv:2209.14324 [hep-th]} \BibitemShut
  {NoStop}%
\bibitem [{\citenamefont {Aminov}\ \emph {et~al.}(2023)\citenamefont {Aminov},
  \citenamefont {Arnaudo}, \citenamefont {Bonelli}, \citenamefont {Grassi},\
  and\ \citenamefont {Tanzini}}]{Aminov:2023jve}%
  \BibitemOpen
  \bibfield  {author} {\bibinfo {author} {\bibfnamefont {G.}~\bibnamefont
  {Aminov}}, \bibinfo {author} {\bibfnamefont {P.}~\bibnamefont {Arnaudo}},
  \bibinfo {author} {\bibfnamefont {G.}~\bibnamefont {Bonelli}}, \bibinfo
  {author} {\bibfnamefont {A.}~\bibnamefont {Grassi}},\ and\ \bibinfo {author}
  {\bibfnamefont {A.}~\bibnamefont {Tanzini}},\ }\bibfield  {title} {\bibinfo
  {title} {{Black hole perturbation theory and multiple polylogarithms}},\
  }\href {https://doi.org/10.1007/JHEP11(2023)059} {\bibfield  {journal}
  {\bibinfo  {journal} {JHEP}\ }\textbf {\bibinfo {volume} {11}},\ \bibinfo
  {pages} {059}},\ \Eprint {https://arxiv.org/abs/2307.10141} {arXiv:2307.10141
  [hep-th]} \BibitemShut {NoStop}%
\bibitem [{\citenamefont {Fucito}\ and\ \citenamefont
  {Morales}(2024)}]{Fucito:2023afe}%
  \BibitemOpen
  \bibfield  {author} {\bibinfo {author} {\bibfnamefont {F.}~\bibnamefont
  {Fucito}}\ and\ \bibinfo {author} {\bibfnamefont {J.~F.}\ \bibnamefont
  {Morales}},\ }\bibfield  {title} {\bibinfo {title} {{Post Newtonian emission
  of gravitational waves from binary systems: a gauge theory perspective}},\
  }\href {https://doi.org/10.1007/JHEP03(2024)106} {\bibfield  {journal}
  {\bibinfo  {journal} {JHEP}\ }\textbf {\bibinfo {volume} {03}},\ \bibinfo
  {pages} {106}},\ \Eprint {https://arxiv.org/abs/2311.14637} {arXiv:2311.14637
  [gr-qc]} \BibitemShut {NoStop}%
\bibitem [{\citenamefont {Ivanov}\ \emph {et~al.}(2024)\citenamefont {Ivanov},
  \citenamefont {Li}, \citenamefont {Parra-Martinez},\ and\ \citenamefont
  {Zhou}}]{Ivanov:2024sds}%
  \BibitemOpen
  \bibfield  {author} {\bibinfo {author} {\bibfnamefont {M.~M.}\ \bibnamefont
  {Ivanov}}, \bibinfo {author} {\bibfnamefont {Y.-Z.}\ \bibnamefont {Li}},
  \bibinfo {author} {\bibfnamefont {J.}~\bibnamefont {Parra-Martinez}},\ and\
  \bibinfo {author} {\bibfnamefont {Z.}~\bibnamefont {Zhou}},\ }\bibfield
  {title} {\bibinfo {title} {{Gravitational Raman Scattering in Effective Field
  Theory: A Scalar Tidal Matching at O(G3)}},\ }\href
  {https://doi.org/10.1103/PhysRevLett.132.131401} {\bibfield  {journal}
  {\bibinfo  {journal} {Phys. Rev. Lett.}\ }\textbf {\bibinfo {volume} {132}},\
  \bibinfo {pages} {131401} (\bibinfo {year} {2024})},\ \Eprint
  {https://arxiv.org/abs/2401.08752} {arXiv:2401.08752 [hep-th]} \BibitemShut
  {NoStop}%
\bibitem [{\citenamefont {Fucito}\ \emph {et~al.}(2025)\citenamefont {Fucito},
  \citenamefont {Morales},\ and\ \citenamefont {Russo}}]{Fucito:2024wlg}%
  \BibitemOpen
  \bibfield  {author} {\bibinfo {author} {\bibfnamefont {F.}~\bibnamefont
  {Fucito}}, \bibinfo {author} {\bibfnamefont {J.~F.}\ \bibnamefont
  {Morales}},\ and\ \bibinfo {author} {\bibfnamefont {R.}~\bibnamefont
  {Russo}},\ }\bibfield  {title} {\bibinfo {title} {{Gravitational wave forms
  for extreme mass ratio collisions from supersymmetric gauge theories}},\
  }\href {https://doi.org/10.1103/PhysRevD.111.044054} {\bibfield  {journal}
  {\bibinfo  {journal} {Phys. Rev. D}\ }\textbf {\bibinfo {volume} {111}},\
  \bibinfo {pages} {044054} (\bibinfo {year} {2025})},\ \Eprint
  {https://arxiv.org/abs/2408.07329} {arXiv:2408.07329 [hep-th]} \BibitemShut
  {NoStop}%
\bibitem [{\citenamefont {Cipriani}\ \emph {et~al.}(2025)\citenamefont
  {Cipriani}, \citenamefont {Di~Russo}, \citenamefont {Fucito}, \citenamefont
  {Morales}, \citenamefont {Poghosyan},\ and\ \citenamefont
  {Poghossian}}]{Cipriani:2025ikx}%
  \BibitemOpen
  \bibfield  {author} {\bibinfo {author} {\bibfnamefont {A.}~\bibnamefont
  {Cipriani}}, \bibinfo {author} {\bibfnamefont {G.}~\bibnamefont {Di~Russo}},
  \bibinfo {author} {\bibfnamefont {F.}~\bibnamefont {Fucito}}, \bibinfo
  {author} {\bibfnamefont {J.~F.}\ \bibnamefont {Morales}}, \bibinfo {author}
  {\bibfnamefont {H.}~\bibnamefont {Poghosyan}},\ and\ \bibinfo {author}
  {\bibfnamefont {R.}~\bibnamefont {Poghossian}},\ }\bibfield  {title}
  {\bibinfo {title} {{Resumming Post-Minkowskian and Post-Newtonian
  gravitational waveform expansions}},\ }\href@noop {} {\  (\bibinfo {year}
  {2025})},\ \Eprint {https://arxiv.org/abs/2501.19257} {arXiv:2501.19257
  [gr-qc]} \BibitemShut {NoStop}%
\bibitem [{\citenamefont {Correia}\ and\ \citenamefont
  {Isabella}(2024)}]{Correia:2024jgr}%
  \BibitemOpen
  \bibfield  {author} {\bibinfo {author} {\bibfnamefont {M.}~\bibnamefont
  {Correia}}\ and\ \bibinfo {author} {\bibfnamefont {G.}~\bibnamefont
  {Isabella}},\ }\bibfield  {title} {\bibinfo {title} {{The Born regime of
  gravitational amplitudes}},\ }\href@noop {} {\  (\bibinfo {year} {2024})},\
  \Eprint {https://arxiv.org/abs/2406.13737} {arXiv:2406.13737 [hep-th]}
  \BibitemShut {NoStop}%
\bibitem [{\citenamefont {De~Angelis}\ \emph {et~al.}(2024)\citenamefont
  {De~Angelis}, \citenamefont {Novichkov},\ and\ \citenamefont
  {Gonzo}}]{DeAngelis:2023lvf}%
  \BibitemOpen
  \bibfield  {author} {\bibinfo {author} {\bibfnamefont {S.}~\bibnamefont
  {De~Angelis}}, \bibinfo {author} {\bibfnamefont {P.~P.}\ \bibnamefont
  {Novichkov}},\ and\ \bibinfo {author} {\bibfnamefont {R.}~\bibnamefont
  {Gonzo}},\ }\bibfield  {title} {\bibinfo {title} {{Spinning waveforms from
  the Kosower-Maybee-O\textquoteright{}Connell formalism at leading order}},\
  }\href {https://doi.org/10.1103/PhysRevD.110.L041502} {\bibfield  {journal}
  {\bibinfo  {journal} {Phys. Rev. D}\ }\textbf {\bibinfo {volume} {110}},\
  \bibinfo {pages} {L041502} (\bibinfo {year} {2024})},\ \Eprint
  {https://arxiv.org/abs/2309.17429} {arXiv:2309.17429 [hep-th]} \BibitemShut
  {NoStop}%
\bibitem [{\citenamefont {Adamo}\ and\ \citenamefont
  {Gonzo}(2023)}]{Adamo:2022ooq}%
  \BibitemOpen
  \bibfield  {author} {\bibinfo {author} {\bibfnamefont {T.}~\bibnamefont
  {Adamo}}\ and\ \bibinfo {author} {\bibfnamefont {R.}~\bibnamefont {Gonzo}},\
  }\bibfield  {title} {\bibinfo {title} {{Bethe-Salpeter equation for classical
  gravitational bound states}},\ }\href
  {https://doi.org/10.1007/JHEP05(2023)088} {\bibfield  {journal} {\bibinfo
  {journal} {JHEP}\ }\textbf {\bibinfo {volume} {05}},\ \bibinfo {pages}
  {088}},\ \Eprint {https://arxiv.org/abs/2212.13269} {arXiv:2212.13269
  [hep-th]} \BibitemShut {NoStop}%
\bibitem [{\citenamefont {Monteiro}\ \emph {et~al.}(2021)\citenamefont
  {Monteiro}, \citenamefont {O'Connell}, \citenamefont {Peinador~Veiga},\ and\
  \citenamefont {Sergola}}]{Monteiro:2020plf}%
  \BibitemOpen
  \bibfield  {author} {\bibinfo {author} {\bibfnamefont {R.}~\bibnamefont
  {Monteiro}}, \bibinfo {author} {\bibfnamefont {D.}~\bibnamefont {O'Connell}},
  \bibinfo {author} {\bibfnamefont {D.}~\bibnamefont {Peinador~Veiga}},\ and\
  \bibinfo {author} {\bibfnamefont {M.}~\bibnamefont {Sergola}},\ }\bibfield
  {title} {\bibinfo {title} {{Classical solutions and their double copy in
  split signature}},\ }\href {https://doi.org/10.1007/JHEP05(2021)268}
  {\bibfield  {journal} {\bibinfo  {journal} {JHEP}\ }\textbf {\bibinfo
  {volume} {05}},\ \bibinfo {pages} {268}},\ \Eprint
  {https://arxiv.org/abs/2012.11190} {arXiv:2012.11190 [hep-th]} \BibitemShut
  {NoStop}%
\bibitem [{\citenamefont {Holstein}(2006)}]{Holstein:2006bh}%
  \BibitemOpen
  \bibfield  {author} {\bibinfo {author} {\bibfnamefont {B.~R.}\ \bibnamefont
  {Holstein}},\ }\bibfield  {title} {\bibinfo {title} {{Graviton Physics}},\
  }\href {https://doi.org/10.1119/1.2338547} {\bibfield  {journal} {\bibinfo
  {journal} {Am. J. Phys.}\ }\textbf {\bibinfo {volume} {74}},\ \bibinfo
  {pages} {1002} (\bibinfo {year} {2006})},\ \Eprint
  {https://arxiv.org/abs/gr-qc/0607045} {arXiv:gr-qc/0607045} \BibitemShut
  {NoStop}%
\bibitem [{\citenamefont {Bern}\ \emph {et~al.}(1994)\citenamefont {Bern},
  \citenamefont {Dixon}, \citenamefont {Dunbar},\ and\ \citenamefont
  {Kosower}}]{Bern:1994zx}%
  \BibitemOpen
  \bibfield  {author} {\bibinfo {author} {\bibfnamefont {Z.}~\bibnamefont
  {Bern}}, \bibinfo {author} {\bibfnamefont {L.~J.}\ \bibnamefont {Dixon}},
  \bibinfo {author} {\bibfnamefont {D.~C.}\ \bibnamefont {Dunbar}},\ and\
  \bibinfo {author} {\bibfnamefont {D.~A.}\ \bibnamefont {Kosower}},\
  }\bibfield  {title} {\bibinfo {title} {{One loop n point gauge theory
  amplitudes, unitarity and collinear limits}},\ }\href
  {https://doi.org/10.1016/0550-3213(94)90179-1} {\bibfield  {journal}
  {\bibinfo  {journal} {Nucl. Phys. B}\ }\textbf {\bibinfo {volume} {425}},\
  \bibinfo {pages} {217} (\bibinfo {year} {1994})},\ \Eprint
  {https://arxiv.org/abs/hep-ph/9403226} {arXiv:hep-ph/9403226} \BibitemShut
  {NoStop}%
\bibitem [{\citenamefont {Bern}\ \emph {et~al.}(1995)\citenamefont {Bern},
  \citenamefont {Dixon}, \citenamefont {Dunbar},\ and\ \citenamefont
  {Kosower}}]{Bern:1994cg}%
  \BibitemOpen
  \bibfield  {author} {\bibinfo {author} {\bibfnamefont {Z.}~\bibnamefont
  {Bern}}, \bibinfo {author} {\bibfnamefont {L.~J.}\ \bibnamefont {Dixon}},
  \bibinfo {author} {\bibfnamefont {D.~C.}\ \bibnamefont {Dunbar}},\ and\
  \bibinfo {author} {\bibfnamefont {D.~A.}\ \bibnamefont {Kosower}},\
  }\bibfield  {title} {\bibinfo {title} {{Fusing gauge theory tree amplitudes
  into loop amplitudes}},\ }\href
  {https://doi.org/10.1016/0550-3213(94)00488-Z} {\bibfield  {journal}
  {\bibinfo  {journal} {Nucl. Phys. B}\ }\textbf {\bibinfo {volume} {435}},\
  \bibinfo {pages} {59} (\bibinfo {year} {1995})},\ \Eprint
  {https://arxiv.org/abs/hep-ph/9409265} {arXiv:hep-ph/9409265} \BibitemShut
  {NoStop}%
\bibitem [{\citenamefont {Bern}\ \emph {et~al.}(1998)\citenamefont {Bern},
  \citenamefont {Dixon},\ and\ \citenamefont {Kosower}}]{Bern:1997sc}%
  \BibitemOpen
  \bibfield  {author} {\bibinfo {author} {\bibfnamefont {Z.}~\bibnamefont
  {Bern}}, \bibinfo {author} {\bibfnamefont {L.~J.}\ \bibnamefont {Dixon}},\
  and\ \bibinfo {author} {\bibfnamefont {D.~A.}\ \bibnamefont {Kosower}},\
  }\bibfield  {title} {\bibinfo {title} {{One loop amplitudes for e+ e- to four
  partons}},\ }\href {https://doi.org/10.1016/S0550-3213(97)00703-7} {\bibfield
   {journal} {\bibinfo  {journal} {Nucl. Phys. B}\ }\textbf {\bibinfo {volume}
  {513}},\ \bibinfo {pages} {3} (\bibinfo {year} {1998})},\ \Eprint
  {https://arxiv.org/abs/hep-ph/9708239} {arXiv:hep-ph/9708239} \BibitemShut
  {NoStop}%
\bibitem [{\citenamefont {Britto}\ \emph {et~al.}(2005)\citenamefont {Britto},
  \citenamefont {Cachazo},\ and\ \citenamefont {Feng}}]{Britto:2004nc}%
  \BibitemOpen
  \bibfield  {author} {\bibinfo {author} {\bibfnamefont {R.}~\bibnamefont
  {Britto}}, \bibinfo {author} {\bibfnamefont {F.}~\bibnamefont {Cachazo}},\
  and\ \bibinfo {author} {\bibfnamefont {B.}~\bibnamefont {Feng}},\ }\bibfield
  {title} {\bibinfo {title} {{Generalized unitarity and one-loop amplitudes in
  N=4 super-Yang-Mills}},\ }\href
  {https://doi.org/10.1016/j.nuclphysb.2005.07.014} {\bibfield  {journal}
  {\bibinfo  {journal} {Nucl. Phys. B}\ }\textbf {\bibinfo {volume} {725}},\
  \bibinfo {pages} {275} (\bibinfo {year} {2005})},\ \Eprint
  {https://arxiv.org/abs/hep-th/0412103} {arXiv:hep-th/0412103} \BibitemShut
  {NoStop}%
\bibitem [{\citenamefont {Brandhuber}\ \emph {et~al.}(2021)\citenamefont
  {Brandhuber}, \citenamefont {Chen}, \citenamefont {Travaglini},\ and\
  \citenamefont {Wen}}]{Brandhuber:2021eyq}%
  \BibitemOpen
  \bibfield  {author} {\bibinfo {author} {\bibfnamefont {A.}~\bibnamefont
  {Brandhuber}}, \bibinfo {author} {\bibfnamefont {G.}~\bibnamefont {Chen}},
  \bibinfo {author} {\bibfnamefont {G.}~\bibnamefont {Travaglini}},\ and\
  \bibinfo {author} {\bibfnamefont {C.}~\bibnamefont {Wen}},\ }\bibfield
  {title} {\bibinfo {title} {{Classical gravitational scattering from a
  gauge-invariant double copy}},\ }\href
  {https://doi.org/10.1007/JHEP10(2021)118} {\bibfield  {journal} {\bibinfo
  {journal} {JHEP}\ }\textbf {\bibinfo {volume} {10}},\ \bibinfo {pages}
  {118}},\ \Eprint {https://arxiv.org/abs/2108.04216} {arXiv:2108.04216
  [hep-th]} \BibitemShut {NoStop}%
\bibitem [{\citenamefont {Brandhuber}\ \emph {et~al.}(2023)\citenamefont
  {Brandhuber}, \citenamefont {Brown}, \citenamefont {Chen}, \citenamefont
  {De~Angelis}, \citenamefont {Gowdy},\ and\ \citenamefont
  {Travaglini}}]{Brandhuber:2023hhy}%
  \BibitemOpen
  \bibfield  {author} {\bibinfo {author} {\bibfnamefont {A.}~\bibnamefont
  {Brandhuber}}, \bibinfo {author} {\bibfnamefont {G.~R.}\ \bibnamefont
  {Brown}}, \bibinfo {author} {\bibfnamefont {G.}~\bibnamefont {Chen}},
  \bibinfo {author} {\bibfnamefont {S.}~\bibnamefont {De~Angelis}}, \bibinfo
  {author} {\bibfnamefont {J.}~\bibnamefont {Gowdy}},\ and\ \bibinfo {author}
  {\bibfnamefont {G.}~\bibnamefont {Travaglini}},\ }\bibfield  {title}
  {\bibinfo {title} {{One-loop gravitational bremsstrahlung and waveforms from
  a heavy-mass effective field theory}},\ }\href
  {https://doi.org/10.1007/JHEP06(2023)048} {\bibfield  {journal} {\bibinfo
  {journal} {JHEP}\ }\textbf {\bibinfo {volume} {06}},\ \bibinfo {pages}
  {048}},\ \Eprint {https://arxiv.org/abs/2303.06111} {arXiv:2303.06111
  [hep-th]} \BibitemShut {NoStop}%
\bibitem [{\citenamefont {Smirnov}\ and\ \citenamefont
  {Chuharev}(2020)}]{Smirnov:2019qkx}%
  \BibitemOpen
  \bibfield  {author} {\bibinfo {author} {\bibfnamefont {A.~V.}\ \bibnamefont
  {Smirnov}}\ and\ \bibinfo {author} {\bibfnamefont {F.~S.}\ \bibnamefont
  {Chuharev}},\ }\bibfield  {title} {\bibinfo {title} {{FIRE6: Feynman Integral
  REduction with Modular Arithmetic}},\ }\href
  {https://doi.org/10.1016/j.cpc.2019.106877} {\bibfield  {journal} {\bibinfo
  {journal} {Comput. Phys. Commun.}\ }\textbf {\bibinfo {volume} {247}},\
  \bibinfo {pages} {106877} (\bibinfo {year} {2020})},\ \Eprint
  {https://arxiv.org/abs/1901.07808} {arXiv:1901.07808 [hep-ph]} \BibitemShut
  {NoStop}%
\bibitem [{\citenamefont {Bjerrum-Bohr}\ \emph {et~al.}(2017)\citenamefont
  {Bjerrum-Bohr}, \citenamefont {Holstein}, \citenamefont {Donoghue},
  \citenamefont {Plant\'e},\ and\ \citenamefont
  {Vanhove}}]{Bjerrum-Bohr:2017dxw}%
  \BibitemOpen
  \bibfield  {author} {\bibinfo {author} {\bibfnamefont {N.~E.~J.}\
  \bibnamefont {Bjerrum-Bohr}}, \bibinfo {author} {\bibfnamefont {B.~R.}\
  \bibnamefont {Holstein}}, \bibinfo {author} {\bibfnamefont {J.~F.}\
  \bibnamefont {Donoghue}}, \bibinfo {author} {\bibfnamefont {L.}~\bibnamefont
  {Plant\'e}},\ and\ \bibinfo {author} {\bibfnamefont {P.}~\bibnamefont
  {Vanhove}},\ }\bibfield  {title} {\bibinfo {title} {{Illuminating Light
  Bending}},\ }\href {https://doi.org/10.22323/1.292.0077} {\bibfield
  {journal} {\bibinfo  {journal} {PoS}\ }\textbf {\bibinfo {volume}
  {CORFU2016}},\ \bibinfo {pages} {077} (\bibinfo {year} {2017})},\ \Eprint
  {https://arxiv.org/abs/1704.01624} {arXiv:1704.01624 [gr-qc]} \BibitemShut
  {NoStop}%
\bibitem [{\citenamefont {Gonzo}\ and\ \citenamefont
  {Ilderton}(2023)}]{Gonzo:2023cnv}%
  \BibitemOpen
  \bibfield  {author} {\bibinfo {author} {\bibfnamefont {R.}~\bibnamefont
  {Gonzo}}\ and\ \bibinfo {author} {\bibfnamefont {A.}~\bibnamefont
  {Ilderton}},\ }\bibfield  {title} {\bibinfo {title} {{Wave scattering event
  shapes at high energies}},\ }\href {https://doi.org/10.1007/JHEP10(2023)108}
  {\bibfield  {journal} {\bibinfo  {journal} {JHEP}\ }\textbf {\bibinfo
  {volume} {10}},\ \bibinfo {pages} {108}},\ \Eprint
  {https://arxiv.org/abs/2305.17166} {arXiv:2305.17166 [hep-th]} \BibitemShut
  {NoStop}%
\bibitem [{\citenamefont {Andersson}\ \emph {et~al.}(2021)\citenamefont
  {Andersson}, \citenamefont {Joudioux}, \citenamefont {Oancea},\ and\
  \citenamefont {Raj}}]{Andersson:2020gsj}%
  \BibitemOpen
  \bibfield  {author} {\bibinfo {author} {\bibfnamefont {L.}~\bibnamefont
  {Andersson}}, \bibinfo {author} {\bibfnamefont {J.}~\bibnamefont {Joudioux}},
  \bibinfo {author} {\bibfnamefont {M.~A.}\ \bibnamefont {Oancea}},\ and\
  \bibinfo {author} {\bibfnamefont {A.}~\bibnamefont {Raj}},\ }\bibfield
  {title} {\bibinfo {title} {{Propagation of polarized gravitational waves}},\
  }\href {https://doi.org/10.1103/PhysRevD.103.044053} {\bibfield  {journal}
  {\bibinfo  {journal} {Phys. Rev. D}\ }\textbf {\bibinfo {volume} {103}},\
  \bibinfo {pages} {044053} (\bibinfo {year} {2021})},\ \Eprint
  {https://arxiv.org/abs/2012.08363} {arXiv:2012.08363 [gr-qc]} \BibitemShut
  {NoStop}%
\bibitem [{\citenamefont {Bern}\ \emph {et~al.}(2022)\citenamefont {Bern},
  \citenamefont {Parra-Martinez}, \citenamefont {Roiban}, \citenamefont {Ruf},
  \citenamefont {Shen}, \citenamefont {Solon},\ and\ \citenamefont
  {Zeng}}]{Bern:2021yeh}%
  \BibitemOpen
  \bibfield  {author} {\bibinfo {author} {\bibfnamefont {Z.}~\bibnamefont
  {Bern}}, \bibinfo {author} {\bibfnamefont {J.}~\bibnamefont
  {Parra-Martinez}}, \bibinfo {author} {\bibfnamefont {R.}~\bibnamefont
  {Roiban}}, \bibinfo {author} {\bibfnamefont {M.~S.}\ \bibnamefont {Ruf}},
  \bibinfo {author} {\bibfnamefont {C.-H.}\ \bibnamefont {Shen}}, \bibinfo
  {author} {\bibfnamefont {M.~P.}\ \bibnamefont {Solon}},\ and\ \bibinfo
  {author} {\bibfnamefont {M.}~\bibnamefont {Zeng}},\ }\bibfield  {title}
  {\bibinfo {title} {{Scattering Amplitudes, the Tail Effect, and Conservative
  Binary Dynamics at O(G4)}},\ }\href
  {https://doi.org/10.1103/PhysRevLett.128.161103} {\bibfield  {journal}
  {\bibinfo  {journal} {Phys. Rev. Lett.}\ }\textbf {\bibinfo {volume} {128}},\
  \bibinfo {pages} {161103} (\bibinfo {year} {2022})},\ \Eprint
  {https://arxiv.org/abs/2112.10750} {arXiv:2112.10750 [hep-th]} \BibitemShut
  {NoStop}%
\bibitem [{\citenamefont {Kol}\ \emph {et~al.}(2022)\citenamefont {Kol},
  \citenamefont {O'connell},\ and\ \citenamefont {Telem}}]{Kol:2021jjc}%
  \BibitemOpen
  \bibfield  {author} {\bibinfo {author} {\bibfnamefont {U.}~\bibnamefont
  {Kol}}, \bibinfo {author} {\bibfnamefont {D.}~\bibnamefont {O'connell}},\
  and\ \bibinfo {author} {\bibfnamefont {O.}~\bibnamefont {Telem}},\ }\bibfield
   {title} {\bibinfo {title} {{The radial action from probe amplitudes to all
  orders}},\ }\href {https://doi.org/10.1007/JHEP03(2022)141} {\bibfield
  {journal} {\bibinfo  {journal} {JHEP}\ }\textbf {\bibinfo {volume} {03}},\
  \bibinfo {pages} {141}},\ \Eprint {https://arxiv.org/abs/2109.12092}
  {arXiv:2109.12092 [hep-th]} \BibitemShut {NoStop}%
\bibitem [{\citenamefont {Damgaard}\ \emph {et~al.}(2023)\citenamefont
  {Damgaard}, \citenamefont {Hansen}, \citenamefont {Plant\'e},\ and\
  \citenamefont {Vanhove}}]{Damgaard:2023ttc}%
  \BibitemOpen
  \bibfield  {author} {\bibinfo {author} {\bibfnamefont {P.~H.}\ \bibnamefont
  {Damgaard}}, \bibinfo {author} {\bibfnamefont {E.~R.}\ \bibnamefont
  {Hansen}}, \bibinfo {author} {\bibfnamefont {L.}~\bibnamefont {Plant\'e}},\
  and\ \bibinfo {author} {\bibfnamefont {P.}~\bibnamefont {Vanhove}},\
  }\bibfield  {title} {\bibinfo {title} {{Classical observables from the
  exponential representation of the gravitational S-matrix}},\ }\href
  {https://doi.org/10.1007/JHEP09(2023)183} {\bibfield  {journal} {\bibinfo
  {journal} {JHEP}\ }\textbf {\bibinfo {volume} {09}},\ \bibinfo {pages}
  {183}},\ \Eprint {https://arxiv.org/abs/2307.04746} {arXiv:2307.04746
  [hep-th]} \BibitemShut {NoStop}%
\bibitem [{\citenamefont {Parnachev}\ and\ \citenamefont
  {Sen}(2021)}]{Parnachev:2020zbr}%
  \BibitemOpen
  \bibfield  {author} {\bibinfo {author} {\bibfnamefont {A.}~\bibnamefont
  {Parnachev}}\ and\ \bibinfo {author} {\bibfnamefont {K.}~\bibnamefont
  {Sen}},\ }\bibfield  {title} {\bibinfo {title} {{Notes on AdS-Schwarzschild
  eikonal phase}},\ }\href {https://doi.org/10.1007/JHEP03(2021)289} {\bibfield
   {journal} {\bibinfo  {journal} {JHEP}\ }\textbf {\bibinfo {volume} {03}},\
  \bibinfo {pages} {289}},\ \Eprint {https://arxiv.org/abs/2011.06920}
  {arXiv:2011.06920 [hep-th]} \BibitemShut {NoStop}%
\bibitem [{\citenamefont {Cheung}\ \emph {et~al.}(2021)\citenamefont {Cheung},
  \citenamefont {Shah},\ and\ \citenamefont {Solon}}]{Cheung:2020gbf}%
  \BibitemOpen
  \bibfield  {author} {\bibinfo {author} {\bibfnamefont {C.}~\bibnamefont
  {Cheung}}, \bibinfo {author} {\bibfnamefont {N.}~\bibnamefont {Shah}},\ and\
  \bibinfo {author} {\bibfnamefont {M.~P.}\ \bibnamefont {Solon}},\ }\bibfield
  {title} {\bibinfo {title} {{Mining the Geodesic Equation for Scattering
  Data}},\ }\href {https://doi.org/10.1103/PhysRevD.103.024030} {\bibfield
  {journal} {\bibinfo  {journal} {Phys. Rev. D}\ }\textbf {\bibinfo {volume}
  {103}},\ \bibinfo {pages} {024030} (\bibinfo {year} {2021})},\ \Eprint
  {https://arxiv.org/abs/2010.08568} {arXiv:2010.08568 [hep-th]} \BibitemShut
  {NoStop}%
\bibitem [{\citenamefont {Fabbrichesi}\ \emph {et~al.}(1994)\citenamefont
  {Fabbrichesi}, \citenamefont {Pettorino}, \citenamefont {Veneziano},\ and\
  \citenamefont {Vilkovisky}}]{Fabbrichesi:1993kz}%
  \BibitemOpen
  \bibfield  {author} {\bibinfo {author} {\bibfnamefont {M.}~\bibnamefont
  {Fabbrichesi}}, \bibinfo {author} {\bibfnamefont {R.}~\bibnamefont
  {Pettorino}}, \bibinfo {author} {\bibfnamefont {G.}~\bibnamefont
  {Veneziano}},\ and\ \bibinfo {author} {\bibfnamefont {G.~A.}\ \bibnamefont
  {Vilkovisky}},\ }\bibfield  {title} {\bibinfo {title} {{Planckian energy
  scattering and surface terms in the gravitational action}},\ }\href
  {https://doi.org/10.1016/0550-3213(94)90361-1} {\bibfield  {journal}
  {\bibinfo  {journal} {Nucl. Phys. B}\ }\textbf {\bibinfo {volume} {419}},\
  \bibinfo {pages} {147} (\bibinfo {year} {1994})},\ \Eprint
  {https://arxiv.org/abs/hep-th/9309037} {arXiv:hep-th/9309037} \BibitemShut
  {NoStop}%
\bibitem [{\citenamefont {K\"alin}\ and\ \citenamefont
  {Porto}(2020{\natexlab{a}})}]{Kalin:2019rwq}%
  \BibitemOpen
  \bibfield  {author} {\bibinfo {author} {\bibfnamefont {G.}~\bibnamefont
  {K\"alin}}\ and\ \bibinfo {author} {\bibfnamefont {R.~A.}\ \bibnamefont
  {Porto}},\ }\bibfield  {title} {\bibinfo {title} {{From Boundary Data to
  Bound States}},\ }\href {https://doi.org/10.1007/JHEP01(2020)072} {\bibfield
  {journal} {\bibinfo  {journal} {JHEP}\ }\textbf {\bibinfo {volume} {01}},\
  \bibinfo {pages} {072}},\ \Eprint {https://arxiv.org/abs/1910.03008}
  {arXiv:1910.03008 [hep-th]} \BibitemShut {NoStop}%
\bibitem [{\citenamefont {K\"alin}\ and\ \citenamefont
  {Porto}(2020{\natexlab{b}})}]{Kalin:2019inp}%
  \BibitemOpen
  \bibfield  {author} {\bibinfo {author} {\bibfnamefont {G.}~\bibnamefont
  {K\"alin}}\ and\ \bibinfo {author} {\bibfnamefont {R.~A.}\ \bibnamefont
  {Porto}},\ }\bibfield  {title} {\bibinfo {title} {{From boundary data to
  bound states. Part II. Scattering angle to dynamical invariants (with
  twist)}},\ }\href {https://doi.org/10.1007/JHEP02(2020)120} {\bibfield
  {journal} {\bibinfo  {journal} {JHEP}\ }\textbf {\bibinfo {volume} {02}},\
  \bibinfo {pages} {120}},\ \Eprint {https://arxiv.org/abs/1911.09130}
  {arXiv:1911.09130 [hep-th]} \BibitemShut {NoStop}%
\bibitem [{\citenamefont {Gonzo}\ and\ \citenamefont
  {Shi}(2023)}]{Gonzo:2023goe}%
  \BibitemOpen
  \bibfield  {author} {\bibinfo {author} {\bibfnamefont {R.}~\bibnamefont
  {Gonzo}}\ and\ \bibinfo {author} {\bibfnamefont {C.}~\bibnamefont {Shi}},\
  }\bibfield  {title} {\bibinfo {title} {{Boundary to bound dictionary for
  generic Kerr orbits}},\ }\href {https://doi.org/10.1103/PhysRevD.108.084065}
  {\bibfield  {journal} {\bibinfo  {journal} {Phys. Rev. D}\ }\textbf {\bibinfo
  {volume} {108}},\ \bibinfo {pages} {084065} (\bibinfo {year} {2023})},\
  \Eprint {https://arxiv.org/abs/2304.06066} {arXiv:2304.06066 [hep-th]}
  \BibitemShut {NoStop}%
\bibitem [{\citenamefont {Stein}\ and\ \citenamefont
  {Warburton}(2020)}]{Stein:2019buj}%
  \BibitemOpen
  \bibfield  {author} {\bibinfo {author} {\bibfnamefont {L.~C.}\ \bibnamefont
  {Stein}}\ and\ \bibinfo {author} {\bibfnamefont {N.}~\bibnamefont
  {Warburton}},\ }\bibfield  {title} {\bibinfo {title} {{Location of the last
  stable orbit in Kerr spacetime}},\ }\href
  {https://doi.org/10.1103/PhysRevD.101.064007} {\bibfield  {journal} {\bibinfo
   {journal} {Phys. Rev. D}\ }\textbf {\bibinfo {volume} {101}},\ \bibinfo
  {pages} {064007} (\bibinfo {year} {2020})},\ \Eprint
  {https://arxiv.org/abs/1912.07609} {arXiv:1912.07609 [gr-qc]} \BibitemShut
  {NoStop}%
\bibitem [{\citenamefont {Caron-Huot}\ \emph {et~al.}(2025)\citenamefont
  {Caron-Huot}, \citenamefont {Correia}, \citenamefont {Isabella},\ and\
  \citenamefont {Solon}}]{Caron-Huot:2025tlq}%
  \BibitemOpen
  \bibfield  {author} {\bibinfo {author} {\bibfnamefont {S.}~\bibnamefont
  {Caron-Huot}}, \bibinfo {author} {\bibfnamefont {M.}~\bibnamefont {Correia}},
  \bibinfo {author} {\bibfnamefont {G.}~\bibnamefont {Isabella}},\ and\
  \bibinfo {author} {\bibfnamefont {M.}~\bibnamefont {Solon}},\ }\bibfield
  {title} {\bibinfo {title} {{Gravitational Wave Scattering via the Born
  Series: Scalar Tidal Matching to $\mathcal{O}(G^7)$ and Beyond}},\
  }\href@noop {} {\  (\bibinfo {year} {2025})},\ \Eprint
  {https://arxiv.org/abs/2503.13593} {arXiv:2503.13593 [hep-th]} \BibitemShut
  {NoStop}%
\bibitem [{\citenamefont {Alessio}\ \emph {et~al.}(2024)\citenamefont
  {Alessio}, \citenamefont {Di~Vecchia},\ and\ \citenamefont
  {Heissenberg}}]{Alessio:2024onn}%
  \BibitemOpen
  \bibfield  {author} {\bibinfo {author} {\bibfnamefont {F.}~\bibnamefont
  {Alessio}}, \bibinfo {author} {\bibfnamefont {P.}~\bibnamefont
  {Di~Vecchia}},\ and\ \bibinfo {author} {\bibfnamefont {C.}~\bibnamefont
  {Heissenberg}},\ }\bibfield  {title} {\bibinfo {title} {{Logarithmic soft
  theorems and soft spectra}},\ }\href
  {https://doi.org/10.1007/JHEP11(2024)124} {\bibfield  {journal} {\bibinfo
  {journal} {JHEP}\ }\textbf {\bibinfo {volume} {11}},\ \bibinfo {pages}
  {124}},\ \Eprint {https://arxiv.org/abs/2407.04128} {arXiv:2407.04128
  [hep-th]} \BibitemShut {NoStop}%
\bibitem [{\citenamefont {Bjerrum-Bohr}\ \emph {et~al.}(2025)\citenamefont
  {Bjerrum-Bohr}, \citenamefont {Chen}, \citenamefont {Eriksen},\ and\
  \citenamefont {Shah}}]{Bjerrum-Bohr:2025bqg}%
  \BibitemOpen
  \bibfield  {author} {\bibinfo {author} {\bibfnamefont {N.~E.~J.}\
  \bibnamefont {Bjerrum-Bohr}}, \bibinfo {author} {\bibfnamefont
  {G.}~\bibnamefont {Chen}}, \bibinfo {author} {\bibfnamefont {C.~J.}\
  \bibnamefont {Eriksen}},\ and\ \bibinfo {author} {\bibfnamefont
  {N.}~\bibnamefont {Shah}},\ }\bibfield  {title} {\bibinfo {title} {{The
  gravitational Compton amplitude from flat and curved spacetimes at second
  post-Minkowskian order}},\ }\href@noop {} {\  (\bibinfo {year} {2025})},\
  \Eprint {https://arxiv.org/abs/2506.19705} {arXiv:2506.19705 [hep-th]}
  \BibitemShut {NoStop}%
\end{thebibliography}%

\end{document}